\documentclass[12pt]{article}
    \usepackage{amssymb,amsmath,amsfonts,eurosym,geometry,ulem,graphicx,caption,color,setspace,sectsty,comment,footmisc,caption,natbib,pdflscape,subfigure,array,url}
    \usepackage{hyperref}
    \hypersetup{
        colorlinks=true,        
        linkcolor=blue,         
        citecolor=blue,         
        urlcolor=blue,          
        filecolor=blue          
    }
    \usepackage{chngcntr}
    \usepackage{booktabs}
    \usepackage[figuresright]{rotating}
    \usepackage{adjustbox}
    \usepackage{multirow}
    \usepackage{float}
    \usepackage{arydshln}
    \usepackage{pifont}
    \usepackage{pdflscape}
    \usepackage[auth-lg,affil-sl]{authblk}
    \usepackage{enumitem}
    \usepackage{mathabx}

    \usepackage{thmtools}
    \usepackage{thm-restate}
    \usepackage{cleveref}
    
    \usepackage{algorithm}
    \usepackage[indLines=true,noEnd=false]{algpseudocodex}

    \newenvironment{proofthm}[1]
    {\noindent\textbf{Proof of \textbf{Theorem \ref{#1}}.} }
    {\ \rule{0.5em}{0.5em} \vspace{\baselineskip}}
    \newenvironment{prooflmm}[1]
    {\noindent\textbf{Proof of \textbf{Lemma \ref{#1}}.} }
    {\ \rule{0.5em}{0.5em} \vspace{\baselineskip}}
    
    \newenvironment{proofprop}[1]
    {\noindent\textbf{Proof of \textbf{Proposition \ref{#1}}.} }
    {\ \rule{0.5em}{0.5em} \vspace{\baselineskip} }

    \newtheorem{proposition}{Proposition}
    \newtheorem{lemma}{Lemma}
    \newtheorem{theorem}{Theorem}
    \newtheorem{corollary}{Corollary}
    \newtheorem{definition}{Definition}
    \newtheorem{assumption}{Assumption}[section]
    \newtheorem{example}{Example}

    \newenvironment{remark*}{\vspace{0.5em}\noindent \textbf{{Remark.}} \itshape}{\vspace{0.5em}}

    \DeclareMathOperator*{\argmin}{argmin}

    \newcommand{\bs}{\boldsymbol}
    \newcommand{\mc}{\mathcal}
    
    \newcommand{\mb}{\mathbb}
    \newcommand{\mr}{\mathrm}

    \newcommand{\cp}{\stackrel{p}{\longrightarrow}}

    \newcommand{\cd}{\stackrel{d}{\longrightarrow}}

    \newcommand{\leqtext}[1]{\stackrel{\text{#1}}{\leq}}    
      
    \newcommand{\geqtext}[1]{\stackrel{\text{#1}}{\geq}}    
    \newcommand{\eqtext}[1]{\stackrel{\text{#1}}{=}}

    \makeatletter
    \newsavebox{\@brx}
    \newcommand{\llangle}[1][]{\savebox{\@brx}{\(\m@th{#1\langle}\)}%
    \mathopen{\copy\@brx\kern-0.5\wd\@brx\usebox{\@brx}}}
    \newcommand{\rrangle}[1][]{\savebox{\@brx}{\(\m@th{#1\rangle}\)}%
    \mathclose{\copy\@brx\kern-0.5\wd\@brx\usebox{\@brx}}}
    \makeatother

    \newcommand{\vt}[1]{{\vert\kern-0.25ex\vert #1 
    \vert\kern-0.25ex\vert}}

    \numberwithin{equation}{section}

\begin{document}

\title{Causal Inference under Dynamic Selection and Latent Heterogeneity \thanks{
I thank Ben Deaner, Aureo de Paula, Chen-Wei Hsiang, Dennis Kristensen, Ge Sun, Martin Weidner, 
Andrei Zeleneev, and the participants in the UCL Econometrics Brownbag Seminar, the Cemmap Seminar, and the 2026 PhD Research Day for their valuable comments.  %
    }}
\author{ Weisheng \textsc{Zhang}%
\thanks{%
 University College London: \textsf{weisheng.zhang.21@ucl.ac.uk}.}
}

    \maketitle

\begin{abstract}
    I study dynamic treatment effects in panel data under staggered adoption when treatment timing depends jointly on unobserved time-invariant heterogeneity and time-varying pretreatment covariates, including lagged outcomes. Untreated potential outcomes follow a nonparametric dynamic panel model that allows flexible interactions between time-varying covariates and latent heterogeneity. I use pretreatment outcome histories to find individuals with similar time-invariant latent factors, and the key requirement is that these histories are sufficiently informative about those latent factors. I develop an identification strategy for the dynamic average treatment effect on the treated (ATT) and propose kernel-based doubly robust estimators for the dynamic ATT. I further combine double cross-fitting with undersmoothing and show that, under suitable regularity conditions, the proposed estimators are $\sqrt{n}$-consistent, asymptotically normal, and asymptotically unbiased. The simulation study demonstrates that the proposed method provides accurate inference across a wide range of data-generating processes. I illustrate the method with an application to the U.S. family planning program studied by \citet{bailey2012reexamining} and reestimate its effect on fertility rates.

\end{abstract}


\thispagestyle{empty}
\clearpage

\setcounter{page}{1} 

\clearpage

    \section{Introduction}\label{sec:introduction}

    Self-selection is a fundamental challenge in causal inference  because treated and untreated units may  be systematically different. These differences include time-invariant characteristics, such as individual ability, as well as \emph{time-varying} characteristics, such as lagged outcomes. Causal inference methods using panel data can help address self-selection by accounting for selection on unobserved time-invariant heterogeneity (fixed effects). These causal panel methods, including difference-in-differences (DiD), synthetic control, and matrix completion, have become increasingly popular in empirical research. 

    In many empirical applications, treated and untreated units may also differ in time-varying covariates, such as lagged outcomes. A well-known empirical example is that workers often experience a decline in earnings before participating in a job-training program \citep{ashenfelter1985using}. In these applications, controlling for pretreatment time-varying covariates is necessary to disentangle the treatment effect from the dynamic effects arising from lagged outcomes. 

    However, the validity of most existing causal panel methods builds on the assumption that treatment selection depends \emph{only} on time-invariant characteristics. For example, \citet{ghanem2022selection} and \citet{marx2024parallel} show that, except in a few special cases, the parallel trends assumption in DiD fails when treatment depends on both time-invariant characteristics and lagged outcomes. This concern is not limited to DiD but arises in other causal panel methods when treatment assignment responds to time-varying pretreatment covariates \citep{arkhangelsky2024causal}.

    In this paper, I propose a new causal inference method using panel data when treatment selection depends jointly on time-invariant latent factors and time-varying pretreatment covariates. I establish nonparametric identification of the dynamic average treatment effect on the treated (ATT), develop a doubly robust estimator, and provide a corresponding inference procedure. The method is designed to accommodate staggered adoption, which is common in empirical applications.

    The proposed method allows the outcome process to depend on  time-varying covariates, including lagged outcomes.  Such dynamic dependence is ubiquitous in economics (e.g., due to intertemporal decision-making, adjustment costs, and persistent shocks) but is often abstracted away in causal panel methods~\citep{arkhangelsky2024causal}. 
    Instead, I allow past outcomes to affect current outcomes, capturing dynamic decision-making in staggered adoption settings and connecting the proposed framework to the broader panel econometrics literature~\citep[e.g.,][]{arellano1991some} that emphasizes the importance of dynamic dependence. 
    
    In addition, I do not impose functional-form restrictions on untreated potential outcomes. Many popular causal panel methods rely on linearity assumptions that may be too restrictive in applications. For example, DiD relies on the parallel trends assumption, and most synthetic control and matrix completion methods require that untreated potential outcomes follow a linear factor model. In contrast, my method builds on a nonparametric model in which outcomes depend flexibly on their past outcomes and time-invariant latent factors, eliminating concerns about the functional-form specification of untreated potential outcomes in empirical applications.

    The key technical challenge is that time-invariant latent factors are unobserved. Since untreated outcomes follow a possibly nonlinear model, these factors can neither be differenced out as in DiD nor estimated using standard linear factor methods. \citet{feng2023optimal}, \citet{deaner2025inferring}, and \citet{athey2025identification} address this challenge in a pure nonlinear factor model without dynamic dependence. They construct pairwise pseudo-distances from long pretreatment outcome histories. Each pseudo-distance measures the similarity between two individuals' pretreatment histories and serves as a proxy for the distance between latent factors. However, their analysis does not directly extend to dynamic models, because in such settings, outcome histories reflect not only the effects of time-invariant heterogeneity but also the dynamic effects of time-varying covariates.

    To accommodate dynamic dependence, I extend the pseudo-distance approach to dynamic panel models and show that the pairwise pseudo-distance remains informative about latent heterogeneity under weak dependence. The intuition is that the dynamic effects decay over time, whereas individual latent factors continue to affect the entire outcome path. As a result, individuals with similar latent factors exhibit similar long-run pretreatment histories, and the pseudo-distance remains informative about latent similarity even in the presence of dynamic dependence.

    The identification of the dynamic ATT uses matching based on time-varying covariates and the pseudo-distance. The key assumption is the informativeness condition, which requires that if the pseudo-distance between two individuals is small, then the distance between their latent factors is also small. Under this assumption, although time-invariant latent factors are unobserved, individuals with similar latent factors are identified using the pseudo-distance. I match treated units to untreated units with similar covariates and a small pseudo-distance to impute untreated counterfactual outcomes, and then identify the dynamic ATT. Under staggered adoption, this matching strategy is combined with a backward recursive procedure to accommodate dynamic selection. The identification result is new under selection on both time-invariant latent heterogeneity and time-varying covariates and requires only pairwise latent similarity rather than identification of the latent factors or their distribution.

    Based on this identification strategy, I propose a doubly robust estimator for the dynamic ATT. I use Nadaraya-Watson estimators based on both covariates and the pseudo-distance to nonparametrically estimate counterfactual outcomes and propensity scores. Asymptotic properties are established for both the nuisance parameter estimators and the dynamic ATT estimator. When pretreatment histories are sufficiently long, the nuisance parameter estimators achieve the same convergence rates as if time-invariant latent factors were observed.

    I further apply a novel double cross-fitting procedure to obtain valid inference under weaker conditions. Unlike standard cross-fitting  \citep{chernozhukov2018double}, double cross-fitting constructs the counterfactual outcome and propensity score estimators using separate training samples. This reduces the dependence between their estimation errors and, when combined with undersmoothing, yields a faster convergence rate for the proposed estimator. I establish the root-$n$ consistency, asymptotic normality, and asymptotic unbiasedness of the proposed estimator, and show that these properties hold under weaker dimension restrictions than those required for standard cross-fitting.
     
    Extensive simulation studies evaluate the finite-sample performance of the proposed estimator. The estimator performs well in terms of bias, standard deviation, and coverage across different data-generating processes and sample sizes, even when the number of time periods is moderate. Based on these simulation results, practical guidance on bandwidth selection is provided  for applied researchers.
    
    I apply the proposed method to the U.S. family planning programs studied by \citet{bailey2012reexamining} and reestimate their effects on fertility rates. First, I find strong evidence that counties with higher lagged fertility rates are more likely to be treated, which motivates the use of the proposed method.  Second, my method estimates statistically significant and persistent reductions in fertility. This finding is consistent with the results in \citet{bailey2012reexamining} and those obtained using DiD. Lastly, the estimated reductions are larger than the DiD estimates, and the gap widens over longer post-treatment horizons. This difference reflects the importance of accounting for dynamic selection based on lagged fertility rates.

    \paragraph{Literature review} This paper contributes to the large and rapidly growing literature on causal inference with panel data by allowing treatment selection to depend on time-varying covariates. I refer readers to \citet{de2023two} and \citet{roth2023s} for recent surveys of the DiD literature and to \citet{arkhangelsky2024causal} for a broader survey of causal inference with panel data. The implications of treatment selection for the validity of causal panel methods have received increasing attention. Earlier discussions of this issue in DiD include \citet{ashenfelter1985using} and \citet{abadie2005semiparametric}, and more recently, \citet{ghanem2022selection} and \citet{marx2024parallel} show that, except in a few special cases, the parallel trends assumption fails when treatment decisions respond to previous outcomes. \citet{arkhangelsky2024causal} further emphasize that many causal panel results are established under strict exogeneity, which rules out treatment selection based on time-varying outcome shocks. Although \citet{arkhangelsky2023large} show that this restriction can be relaxed for synthetic control under simultaneous-treatment, their arguments do not directly extend to staggered adoption.

    A related strand of the causal panel literature addresses treatment selection on time-varying covariates. For DiD, these methods rely on conditional parallel trends \citep{abadie2005semiparametric,wooldridge2025two}, with extensions to doubly robust estimation \citep{sant2020doubly}, high-dimensional covariates \citep{chang2020double}, and treatment-affected covariates \citep{caetano2022difference}. However, conditional parallel trends may fail when treatment selection depends jointly on observed covariates and time-invariant latent factors. For the dynamic treatment effect literature, \citet{lewis2020double} and \citet{viviano2026dynamic} allow selection on high-dimensional covariates but assume that treatment assignment is independent of time-invariant latent factors conditional on these covariates. In addition, \citet{marx2024heterogeneous} and \citet{botosaru2025time} focus on dynamic treatment effects under linear panel models. In contrast, the proposed approach uses a long pretreatment panel to accommodate selection on both time-varying covariates and time-invariant latent factors without imposing functional-form restrictions on either the propensity score or the untreated outcome process.
    Lastly, \citet{gulek2025synthetic} combine synthetic control and instrumental variables to address unmeasured time-varying confounding, while my analysis focuses on treatment selection based on observed time-varying covariates and does not require external instruments.

    Beyond the literature on causal panel methods, this paper connects to the broader econometric literature on panel-data models. This literature has paid close attention to individual heterogeneity and outcome dynamics in both linear settings \citep[e.g.,][]{arellano1991some,blundell1998initial,alvarez2003time} and nonlinear settings \citep[e.g.,][]{blundell2002individual,honore2000panel,bonhomme2012functional, bonhomme2015grouped, bonhomme2022discretizing}, but has primarily focused on estimating parametric models. By contrast, the causal panel literature allows for rich heterogeneity in treatment effects but often abstracts away from dynamic dependence in outcomes \citep{arkhangelsky2024causal}. The proposed method bridges these two literatures by allowing for both flexible interactions between time-varying covariates and latent heterogeneity and dynamic dependence in untreated potential outcomes, without imposing a parametric outcome model.

    The nonparametric model for untreated potential outcomes connects this paper to the nonparametric panel literature. For example, \citet{chernozhukov2013average}, \citet{hoderlein2012nonparametric}, and \citet{chernozhukov2026linear} study average structural and causal effects under time-homogeneity conditions. The proposed method allows for rich time heterogeneity in untreated potential outcomes without imposing time-homogeneity conditions. Another strand of literature uses observations from different periods as measurements or proxies for (possibly time-varying) latent confounders \citep[e.g.,][]{hu2012nonparametric,sasaki2015heterogeneity, deaner2018proxy}. 
    These methods only need short panels but rely on completeness conditions. In comparison, although the proposed method requires a long panel, it imposes no completeness conditions and avoids solving an inverse problem.

    \bigskip 

    Methodologically, this paper builds on the recent literature on nonlinear factor models for causal inference with panel data, particularly \citet{deaner2025inferring}.
    The pseudo-distance approach to measuring similarity between latent factors was first developed by~\citet{zhang2017estimating} for graphon estimation. Since then, different forms of the pseudo-distance have been used in a range of applications, including nonparametric graphon estimation \citep{zeleneev2020identification,nowakowicz2024nonparametric}, network imputation \citep{sun2026flexible}, controlling for unobserved heterogeneity using network data \citep{auerbach2022identification}, nonlinear factor model estimation \citep{feng2023optimal}, and causal inference \citep{feng2020causal, deaner2025inferring,athey2025identification,hoshino2024estimating}. However, the validity of most pseudo-distance methods is established only under pure factor models, which rule out dynamic dependence in panel settings. I extend the  pseudo-distance approach to dynamic models in which outcomes depend on both latent factors and lagged outcomes. This extension is new and is of independent interest for the study of dynamic nonseparable panel models beyond causal inference.

    This paper contributes to the double/debiased machine learning literature \citep[e.g.,][]{chernozhukov2018double} in causal inference by introducing a novel double cross-fitting procedure. Doubly robust estimation combines Neyman-orthogonal moments with cross-fitting to obtain valid inference on target parameters in the presence of nonparametric nuisance parameters.\footnote{
        In the causal panel literature, doubly robust methods have been developed for DiD and TWFE models \citep{arkhangelsky2022doubly,arkhangelsky2024design,sant2020doubly}, dynamic treatment effects \citep{chernozhukov2022automatic,lewis2020double}, and linear factor models \citep{abadie2024doubly}. More recently, \citet{feng2020causal} and \citet{deaner2025inferring} extend doubly robust methods to nonlinear factor models. The proposed estimator builds directly on the doubly robust framework of \citet{deaner2025inferring} and likewise requires only weak smoothness conditions.
    } 
    I apply double cross-fitting \citep{newey2018cross,mcclean2026double} to dynamic ATT estimation with nonparametric regressions on \emph{latent factors} and show that, when combined with undersmoothing, double cross-fitting achieves a faster convergence rate and valid inference under weaker dimensionality restrictions than standard cross-fitting. This result may be of independent interest for broader causal inference methods involving nonparametric nuisance parameters.

    \bigskip 
    
    The rest of the paper is organized as follows. Section~\ref{sec:selection} presents the basic setup and the selection mechanism. Section~\ref{sec:overview_method} provides an overview of the proposed method. Section~\ref{sec:identification} establishes identification. Section~\ref{sec:estimation} derives the asymptotic distribution of the proposed estimators. Section~\ref{sec:extension} extends the analysis to staggered adoption. Section~\ref{sec:simulation} provides simulation evidence, and Section~\ref{sec:empirical} applies the method to the setting studied by \citet{bailey2012reexamining} to reexamine the effects of U.S. family planning programs on fertility rates.

    \section{Selection and Model}\label{sec:selection}

    This section introduces the basic setup of this paper. I discuss the treatment-selection problem in panel-data causal inference, focusing on settings where treatment assignment depends on both pretreatment information and latent heterogeneity. I then introduce a general outcome model that accommodates individual and time fixed effects and dynamic dependence. To fix ideas, I first focus on block assignment settings where all treated units receive treatment at the same time period. I extend the analysis to staggered adoption in Section~\ref{sec:extension}.

\subsection{Data}

    The panel data ${(Y_{it},X_{it},D_i)}$ are indexed by $i=1,\ldots,n$ and $t=0,1,\ldots,T$. Here, $Y_{it}\in\mb{R}$ denotes the outcome, and $X_{it}\in\mb{R}^{d_X}$ denotes a vector of time-varying covariates that may affect treatment selection. A leading example is that $X_{it}$ includes lagged outcomes, e.g., $X_{it}:=Y_{it-1}$. Treatment status is denoted by $D_i\in \{0,1\}$.  Multidimensional time-invariant latent heterogeneity is captured by $\alpha_i\in\mc{A}\subset\mathbb{R}^{d_\alpha}$. I consider a simultaneous-treatment setting in which treatment is assigned only once, at time $T_0$. Treatment is absorbing, so once a unit is treated at $T_0$, it remains treated thereafter. I assume that $X_{iT_0}$ is realized before treatment assignment at $T_0$ and therefore constitutes pretreatment information.

    Let $Y_{it}(0)$ and $Y_{it}(1)$ denote the untreated and treated potential outcomes, respectively. I impose no spillover effects across units, so that unit $i$'s potential outcomes depend only on its own treatment status. All treated units receive treatment at period $T_0$. Then, under the non-anticipation condition, 
    \begin{align*}
        Y_{it}= 
        \begin{cases}
            Y_{it}(0), & t = 0, \ldots, T_0 - 1,\\
            D_iY_{it}(1)+(1-D_i)Y_{it}(0), & t = T_0, \ldots, T.
        \end{cases}
    \end{align*}
    
    I also allow for an unobserved time factor $\gamma_t$, which captures aggregate shocks, such as macroeconomic conditions, that affect all units in the same period. Letting $\Gamma_t := (\ldots,\gamma_{t-1},\gamma_t)$ denote the realized time-factor path up to time $t$, I assume that $(Y_{it}, X_{it} , \alpha_i, D_i)$ are i.i.d. across units conditional on fixed effects $\Gamma_{T}$. 

    \paragraph{Estimand} For notational simplicity, define $\mb{E}_T(\cdot):=\mb{E}(\cdot\mid\Gamma_T)$ and $\mb{P}_T(\cdot):=\mb{P}(\cdot\mid\Gamma_T)$ as the expectation and probability conditional on the realized time-factor path $\Gamma_T$. 
    For each $t = T_0, \ldots, T$, the estimand of interest is the dynamic ATT
    \footnote{
        The definition differs from the standard ATT, $\mb{E}\left(Y_{it}(1)-Y_{it}(0)\mid D_i=1\right)$, only because I condition explicitly on the realized time fixed effects. Since these effects are fixed over the sample period, this is purely notational.
    }:
        \begin{align*}
        \mr{ATT}(t) := \mb{E}_T\left(Y_{it}(1) - Y_{it}(0) \mid D_i = 1\right).
    \end{align*}

\subsection{Selection}

    Self-selection is common in economic analyses using observational data, and failing to account for such selection can lead to biased estimates of causal effects. Panel data provide a way to account for selection on unobserved time-invariant characteristics because they contain repeated observations for the same units. The recent and rapidly growing literature on causal panel methods, including DiD, synthetic control, and matrix completion methods, builds on this feature to control for selection on fixed effects.

    However,  most existing causal panel methods do not accommodate selection on time-varying covariates. For DiD, \citet{ghanem2022selection} and \citet{marx2024parallel} show that the parallel trends assumption fails when treatment decisions respond to both time-invariant characteristics and past outcomes. This concern is not limited to DiD but arises for many other causal panel methods \citep{arkhangelsky2024causal}. Although \citet{arkhangelsky2023large} show that selection on lagged outcomes is less relevant for synthetic control under simultaneous treatment, their argument does not directly extend to staggered adoption.

    \paragraph{Ashenfelter's dip} To illustrate why selection on past outcomes is problematic for panel causal methods, consider Ashenfelter's dip in the context of DiD \citep{ashenfelter1978estimating, ashenfelter1985using}. Labor economists find that participants often experience a decline in earnings before joining a job training program, which violates parallel trends. To see this,  suppose that untreated earnings are i.i.d. over time. For participants, since the decline in earnings is temporary, their earnings would \emph{recover} on average in the next period even without treatment. Therefore, 
    \begin{align*}
        \mb{E}(Y_{iT_0}(0) - Y_{iT_0-1}(0) \mid D_i = 1) >0. 
    \end{align*} 
    By the same logic, we also have $\mb{E}(Y_{iT_0}(0) - Y_{iT_0-1}(0) \mid D_i = 0) < 0$ for nonparticipants. These differences violate the parallel trends assumption and lead to an upward-biased DiD estimator. Therefore, controlling for selection on $Y_{iT_0-1}$ is necessary to disentangle the treatment effect from the dynamic recovery effects.

    \paragraph{Conditional parallel trends} A common practice in the DiD literature to allow for selection on pretreatment covariates is to impose the conditional parallel trends assumption \citep{abadie2005semiparametric, sant2020doubly, chang2020double, caetano2022difference, wooldridge2025two}. Assume $X_{it}:=Y_{it-1}$ for simplicity, This assumption requires that treated and untreated units with the same lagged outcomes $Y_{iT_0-1}$ have parallel trends, i.e.,
    \begin{align}\label{eq:CPT}
        \mb{E}\left( Y_{iT_0 }(0) - Y_{iT_0 - 1}(0)\mid D_i = 1, Y_{iT_0-1} \right) = \mb{E}\left( Y_{iT_0}(0) - Y_{iT_0 - 1}(0)\mid D_i = 0, Y_{iT_0-1}\right). 
    \end{align}
    However, \eqref{eq:CPT} may still fail when treatment selection depends jointly on $\alpha_i$ and $Y_{iT_0-1}$, even when untreated potential outcomes admit a \emph{static} TWFE form. To see this, suppose $Y_{it}(0) = \alpha_i + \gamma_t + \epsilon_{it}$, where $\epsilon_{it}$ is exogenous and independent across $t$ and $D_i \sim \mr{Logit}(\alpha_i - Y_{iT_0-1})$. Cancelling $Y_{iT_0-1}(0)$ from both sides of \eqref{eq:CPT} shows that the conditional parallel trends assumption holds if and only if 
    \begin{align}\label{eq:CPT_2}
        \mb{E}\left( Y_{iT_0 }(0)\mid D_i = 1, Y_{iT_0-1} \right) = \mb{E}\left( Y_{iT_0}(0) \mid D_i = 0, Y_{iT_0-1}\right).
    \end{align}
    However, the selection mechanism implies that treated units have higher $\alpha_i$ than untreated units with the same $Y_{iT_0-1}$. Therefore, the LHS of \eqref{eq:CPT_2} is greater than the RHS, violating the conditional parallel trends assumption.

    \bigskip

\begin{figure}[H]
        \centering
        \caption{\centering Finite-sample distributions: proposed estimator, standard DiD, and DR DiD}
        \includegraphics[width=0.9\textwidth]
            {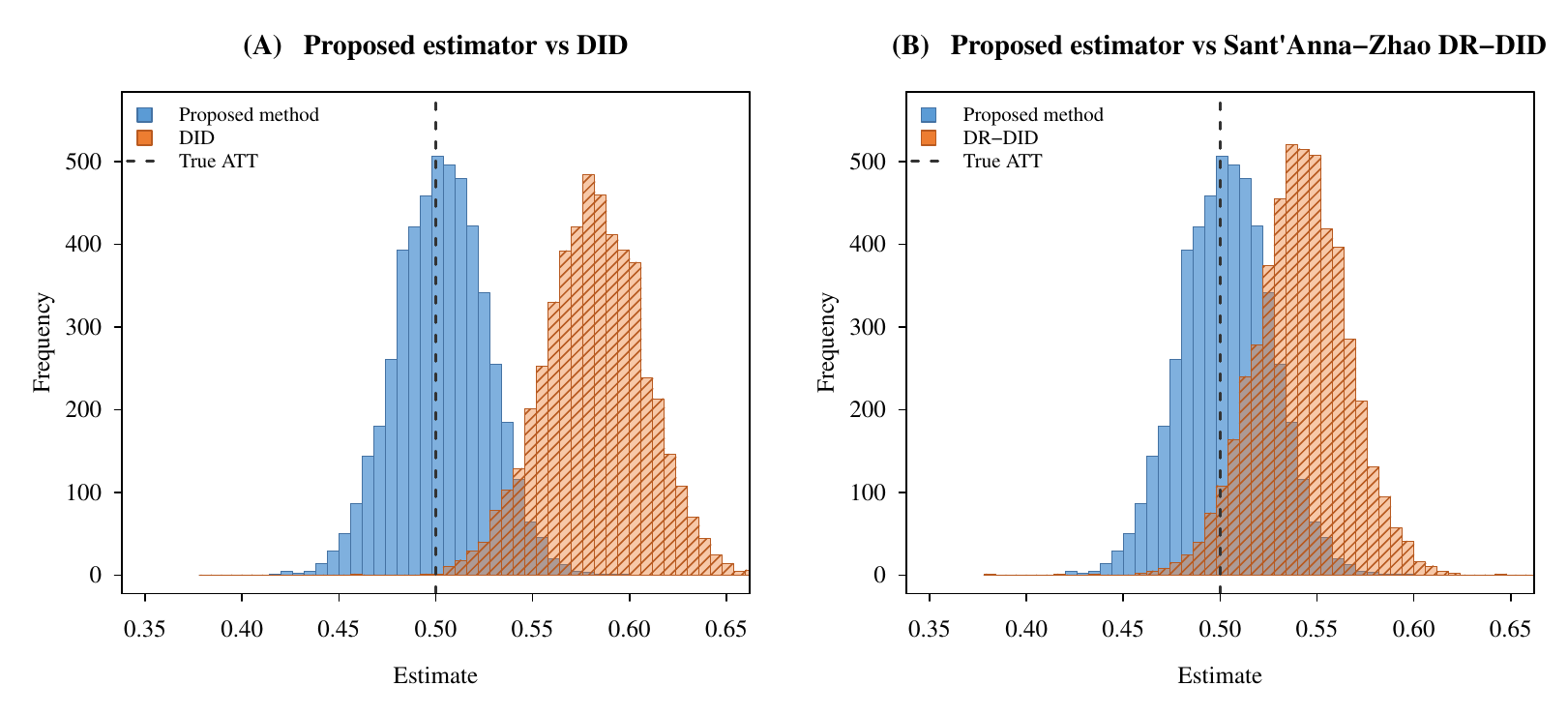}
        \vspace{0.4cm}
        \begin{minipage}{\textwidth}
            \footnotesize 
            \textbf{Note:} $(Y_{it}(0), Y_{it}(1))$ are generated by $Y_{it}(0) = \alpha_i + \gamma_t + \epsilon_{Y, it}$ and $Y_{it}(1) = Y_{it}(0) + 0.5$. $\alpha_i$ and $\gamma_t$ are independent random variables drawn from the uniform distribution on $(-\frac{1}{2}, \frac{1}{2})$. $\{\epsilon_{Y, it}\}$ are independent of $(\alpha_i, \gamma_t)$ and are i.i.d. across both dimensions, uniformly distributed on $(-\frac{1}{2}, \frac{1}{2})$. In addition, $D_i = \bs{1}\left( 2\alpha_i  - Y_{i T_0-1} +  \epsilon_{D, i} \geq 0\right)$, where $\{\epsilon_{D,i}\}$ are i.i.d. logistic random variables independent of $\{\alpha_i, \gamma_t, \epsilon_{Y,it}\}$. 
        \end{minipage}
        \label{fig:motivation}
    \end{figure}

    Figure~\ref{fig:motivation} presents a simple simulation to illustrate how treatment selection on fixed effects and lagged outcomes leads to biased estimates from both standard DiD and DiD based on conditional parallel trends, even when $Y_{it}(0)$ follows a static TWFE model and the treatment effect is homogeneous across units. The standard DiD estimator overestimates the treatment effect, as shown in Figure~\ref{fig:motivation}(A). In addition, the doubly robust DiD estimator proposed by \citet{sant2020doubly}, which relies on conditional parallel trends, is also heavily biased, as shown in Figure~\ref{fig:motivation}(B). By contrast, the distribution of the proposed estimator is centered on the true treatment effect.

    I formalize the selection mechanism as follows: 
    \begin{assumption}[Selection]\label{assumption:selection_simultaneous}
        For each $t  = T_0, \ldots, T$, 
        \begin{align}
            Y_{it}(0) \perp D_i
            \mid
            X_{iT_0}, \alpha_i, \Gamma_T.
        \end{align}
    \end{assumption}
    
    Assumption~\ref{assumption:selection_simultaneous} requires that, after conditioning on pretreatment covariates $X_{iT_0}$ and time-invariant heterogeneity $\alpha_i$, treatment assignment is independent of the untreated potential outcome $Y_{it}(0)$ for each $t=T_0,\ldots,T$. I also explicitly condition on $\Gamma_T$ to maintain the fixed-effect interpretation of the time factors. The conditional independence is imposed only on untreated potential outcomes, rather than on treated potential outcomes, because the estimand of interest is the ATT and only untreated counterfactual outcomes for treated individuals need to be imputed.

    Assumption~\ref{assumption:selection_simultaneous} extends the conventional selection-on-fixed-effects assumption
    by additionally allowing treatment selection to depend on time-varying pretreatment covariates. In addition, this assumption nests the sequential unconfoundedness assumption \citep{robins2000marginal,viviano2026dynamic,marx2024heterogeneous}, $Y_{it}(0) \perp D_i \mid X_{iT_0}$, as a special case in which treatment selection does not depend on unobserved fixed effects. Lastly, Assumption~\ref{assumption:selection_simultaneous} arises naturally in dynamic economic models \citep{heckman2007dynamic}, in which a forward-looking agent chooses treatment to maximize expected utility based on the state variables $(X_{iT_0},\alpha_i)$. 
    
    \begin{remark*}
        Since $X_{iT_0}$ is realized before treatment assignment, it cannot include potential outcomes at or after treatment assignment, such as $(Y_{iT_0}(0),Y_{iT_0}(1))$. This timing restriction therefore rules out anticipation effects (see the discussion in~\citet{roth2023s}) and Roy-type selection. In addition, Assumption~\ref{assumption:selection_simultaneous} rules out unobserved time-varying confounders that jointly affect treatment assignment and untreated potential outcomes. Addressing such confounding typically requires additional information, such as an external instrument. I refer readers to \citet{gulek2025synthetic}, who combine synthetic control and instrumental variables to address such confounding. 
    \end{remark*}

    The following assumption imposes a Markov restriction, under which earlier histories have no additional predictive power for future untreated outcomes or treatment assignment when conditioning on $(X_{iT_0},\alpha_i)$.
    \begin{assumption}[Markov]\label{assumption:surrogacy_simultaneous}
        Let $H_{it}  := \{(Y_{i\tau}, X_{i\tau })\}_{\tau \leq t}$ denote the history up to $t$. Assume 
        \begin{align*}
            (Y_{it}(0), D_i) \perp H_{iT_0 - 1} \mid X_{iT_0}, \alpha_i, \Gamma_T, \quad \forall \; t = T_0, \ldots, T. 
        \end{align*}
    \end{assumption} 
    This assumption is stronger than necessary and can be relaxed to conditional moment restrictions on $(Y_{it}(0), D_i)$\footnote{
        Specifically, let $Z_{it} = (Y_{it}(0),\alpha_i)'$, it suffices to require that, for each $t = T_0, \ldots, T$, 
        \begin{align*}
            \mb{E}_T (Z_{it} \mid H_{iT_0-1}, X_{iT_0}, \alpha_i ) =  \mb{E}_T (Z_{it}  \mid   X_{iT_0}, \alpha_i ), \text{ and }\mb{E}_T (Z_{it}Z_{it}' \mid H_{iT_0-1}, X_{iT_0}, \alpha_i ) =  \mb{E}_T (Z_{it}Z_{it}' \mid   X_{iT_0}, \alpha_i ). 
        \end{align*}
    }. 
    I retain the Markov restriction in this paper for ease of exposition.

\subsection{Dynamic nonparametric panel}
    
    I assume that the untreated outcome $Y_{it}(0)$ is generated  according to
    \begin{align}\label{eq:potential_outcome}
        Y_{it}(0) = f (X_{it}, \alpha_i, \gamma_t   ) + \epsilon_{it}, \quad \mb{E}_T(\epsilon_{it} \mid H_{it-1}, X_{it}, \alpha_i) = 0. 
    \end{align}
    Here, $f(\cdot)$ is a data-generating function,  $\alpha_i$ and $\gamma_t$ are unobserved individual and time fixed effects, respectively.   The exogeneity condition $\mb{E}_T(\epsilon_{it} \mid H_{it-1}, X_{it}, \alpha_i) = 0$ corresponds to the weak exogeneity condition in structural panel model literature that allows for feedbacks~\citep{bonhomme2025back}. The data-generating process in~\eqref{eq:potential_outcome} is flexible. I impose no parametric restrictions on the function $f$, allowing for rich interactions between individual-specific and time-specific heterogeneity. In addition, I do not specify the joint distribution of $X_{it}$ and the latent factors. 
    
    The model accommodates dynamic dependence in untreated potential outcomes when $X_{it}:=(Y_{it-1}(0),\ldots,Y_{it-p}(0))$. Dynamic dependence is ubiquitous in economics. Aggregate outcomes such as local employment and fertility rates often exhibit dynamic dependence because shocks are persistent over time, while individual outcomes may also exhibit state dependence, i.e., past employment may affect present employment \citep{heckman1981heterogeneity}. Such dynamic dependence is particularly relevant in staggered-adoption settings, as it allows researchers to disentangle treatment effects from dependence on lagged outcomes.

    \begin{remark*}
        The specification \eqref{eq:potential_outcome} can be motivated by a nonseparable panel model: 
        \begin{align*}
            Y_{it}(0) = h(X_{it}, \alpha_i, \gamma_t, \omega_{it}),  
        \end{align*}
        where $\omega_{it}$ is an unobserved shock. I allow its distribution to depend on $(X_{it}, \alpha_i, \gamma_t)$, which is more general than the time-stationarity condition  imposed in the nonseparable panel literature~\citep[e.g.,][]{chernozhukov2013average,chernozhukov2026linear}. If I further suppose that $\omega_{it}$ is independent of the previous history conditional on $(X_{it}, \alpha_i, \Gamma_T)$, i.e., $\omega_{it} \perp H_{it-1} \mid X_{it}, \alpha_i, \Gamma_T$, then $f(\cdot)$ in~\eqref{eq:potential_outcome} can be regarded as the projection of the untreated outcome $Y_{it}(0)$, and $\epsilon_{it}$ is the corresponding projection error. 
    \end{remark*}

    Model~\eqref{eq:potential_outcome} incorporates a number of important data-generating processes that are commonly used in the structural and causal panel literature. Here are some examples.

    \begin{example}[Linear panel with IFE]\label{example:interactive_fixed_effects}
        Consider the following linear panel model with covariates $X_{it}$ and interactive fixed effects: 
        \begin{equation}\label{eq:interactive_fixed_effects}
        \begin{aligned}
            Y_{it}(0) =  \beta' X_{it}  + \gamma_{t}'\alpha_i + \epsilon_{it}, \quad \mb{E}_{T} (\epsilon_{it} \mid H_{it-1}, X_{it} , \alpha_i) = 0. 
        \end{aligned}    
        \end{equation}
        Here, $\gamma_t$ is a vector of macroeconomic shocks, e.g., technology shocks and financial crises, and $\alpha_i$ captures heterogeneous responses to these shocks. For example, in labor economics, $Y_{it}$ represents the wage rate, $X_{it}$ represents experience, $\alpha_i$ is a vector of unobserved skills, and $\gamma_t$ contains the time-varying market returns to these skills. Changes in technology or labor-market conditions therefore affect workers differently according to their latent skill composition.
        
        Model~\eqref{eq:interactive_fixed_effects} is studied in~\citet{bai2009panel} and nests many commonly used untreated-outcome specifications in causal panel methods. For example, setting $\beta=0$ and imposing additive factor structure gives $Y_{it}(0)=\alpha_i+\gamma_t+\epsilon_{it}$, which is the untreated-outcome structure in standard DiD designs. Setting $\beta=0$ while allowing unrestricted interactive fixed effects gives $Y_{it}(0)=\alpha_i'\gamma_t+\epsilon_{it}$, which is typically assumed to support recently developed synthetic control and related factor-based causal panel methods \citep{arkhangelsky2021synthetic,athey2021matrix, bai2021matrix, chernozhukov2023inference}. 
    \end{example}

    \begin{example}[Dynamic linear panel with IFE]\label{example:dynamic_interactive_fixed_effects}
        Consider the following dynamic linear panel model with $X_{it}: = Y_{it-1}(0)$ and interactive fixed effects: 
        \begin{align}
            \label{eq:dynamic_interactive_fixed_effects}
            Y_{it}(0) = \rho Y_{it-1}(0)  + \alpha_i'\gamma_t + \epsilon_{it}, \quad \mb{E}_{T}(\epsilon_{it} \mid \{Y_{i\tau}(0)\}_{\tau \leq t-1}, \alpha_i ) = 0. 
        \end{align}
        Model~\eqref{eq:dynamic_interactive_fixed_effects}  captures outcome persistence through the lagged outcome $Y_{it-1}(0)$ and allows individuals to respond heterogeneously to time-varying aggregate shocks through the interactive term $\alpha_i'\gamma_t$. 
        This model is studied in \citet{moon2017dynamic}. A special case with additive fixed effects is $Y_{it}(0)=\rho Y_{it-1}(0)+ \alpha_i+\gamma_t+\epsilon_{it}$, which is a benchmark specification in the dynamic panel literature \citep{arellano1991some}. 
    \end{example}

    \begin{example}[Dynamic binary panel with IFE]\label{example:dynamic_nonlinear_interactive_fixed_effects}
        Consider a binary response panel model, 
        \begin{align}
            \label{eq:dynamic_nonlinear}
            Y_{it}(0) = \bs{1}(\beta Y_{it-1}(0)   + \alpha_i'\gamma_t- u_{it} \geq 0),
        \end{align}
        where the exogenous errors $\{u_{it}\}$ are independent (across $i$ and $t$) draws from a distribution with cumulative distribution function $F(\cdot)$.\footnote{E.g., $F(\cdot)$ can stand for the logistic or standard normal CDF in Logit and Probit models, respectively.}
        The binary panel \eqref{eq:dynamic_nonlinear} allows for heterogeneous responses of units to aggregate shocks and state dependence, e.g., women's labor-force participation decisions may depend on their employment histories, as those who have been out of the labor market for a long time may be less likely to re-enter. 
        When $F(\cdot)$ can stand for the logistic distribution, the data generating process $f(\cdot)$ as in~\eqref{eq:potential_outcome} takes the form 
        \begin{align*}
            f (Y_{it-1}(0), \alpha_i, \gamma_t )  = \frac{\exp\left(\beta Y_{it-1}(0)   + \alpha_i'\gamma_t\right)}{1 + \exp\left(\beta Y_{it-1}(0)   + \alpha_i'\gamma_t\right)}, 
        \end{align*}
        and $\epsilon_{it}$ is a centered Bernoulli error with $\mb{E}_T (\epsilon_{it} \mid \{Y_{i\tau}(0)\}_{\tau\leq t-1}, \alpha_i) = 0$.  
    \end{example}

    \begin{assumption}[Potential Outcome]\label{assumption:potential_outcome_simultaneous}
        $Y_{it}(0)$ evolves according to~\eqref{eq:potential_outcome} and satisfies:  
        \begin{enumerate}[label=(\roman*)]
            \item \label{item:potential_outcome_simultaneous_time_factor_exogeneity} \textbf{(Time factor exogeneity)} conditional on the individual latent factors $\alpha_i$ and the time-factor path $\Gamma_t$, $Y_{it}(0)$ is mean independent of future time factors, i.e., 
            \begin{align*}
                \mb{E}\left(Y_{it}(0) \mid X_{it},\alpha_i, \Gamma_T\right) = \mb{E}\left(Y_{it}(0) \mid  X_{it},\alpha_i, \Gamma_t\right), \quad \forall t = 1, \ldots, T.   
            \end{align*}
            \item  \label{item:potential_outcome_simultaneous_weak_dependence} \textbf{(Weak dependence)}  $(\ldots, Y_{iT_0-1}(0), Y_{iT_0}(0), \ldots )$ are conditionally weakly dependent,  such that $\sum_{s =-\infty}^{\infty} \left|\mr{Cov} \left(Y_{it}(0), Y_{is}(0)\mid  \alpha_{i} = \alpha,  \Gamma_T = \Gamma \right)\right| <\infty$ uniformly for any supported $\alpha$, any possible time path $\Gamma$, and any $t \in \mb{Z}$.
        \end{enumerate}
    \end{assumption}

    Assumption~\ref{assumption:potential_outcome_simultaneous}\ref{item:potential_outcome_simultaneous_time_factor_exogeneity} requires that future time factors provide no additional information for predicting $Y_{it}(0)$ when conditioning on $(\alpha_i, X_{it}, \Gamma_t)$. This restriction is mild and has natural interpretations. First, it reflects the principle that the future cannot affect the present. This assumption does not preclude forward-looking behavior based on expectations formed using the macroeconomic information set $\Gamma_t$. 
    Second, since $(\gamma_{t+1}, \gamma_{t + 2}, \ldots)$ capture aggregate conditions, they should not be affected by any single individual's behavior. 

    Assumption~\ref{assumption:potential_outcome_simultaneous}\ref{item:potential_outcome_simultaneous_weak_dependence} states that, conditional on the fixed effects $(\alpha_i, \Gamma_T)$, the serial dependence of the untreated outcome vanishes sufficiently fast such that the sum of the absolute conditional autocovariances is finite. This condition holds uniformly over any supported individual fixed effects $\alpha_i$ and any possible time path $\Gamma_T$. This conditional weak dependence requirement is also mild since persistence captured by the individual and time fixed effects, $\alpha_i$ and $\Gamma_T$, is unrestricted.

    The conditional weak dependence assumption is crucial to the proposed method. I discuss sufficient conditions under which it holds in Examples~\ref{example:interactive_fixed_effects}-\ref{example:dynamic_nonlinear_interactive_fixed_effects}.

    \setcounter{example}{0}

    \begin{example}[Linear panel with IFE - Continued]
        It is common in the literature to assume that $X_{it}$ also admits a factor structure \citep{bai2009panel}. For simplicity, set $d_X = 1$. Then,
        \begin{gather*}
            Y_{it}(0) = \beta X_{it} + \gamma_{Y, t}'\alpha_{Y, i} + \epsilon_{Y, it},  \quad  \mb{E}_{T} (\epsilon_{Y, it} \mid H_{it-1}, X_{it} , \alpha_i) = 0;  \\
            X_{it} = \gamma_{X, t}'\alpha_{X, i} + \epsilon_{X, it}, \quad \mb{E}_{T} (\epsilon_{X, it} \mid H_{it-1}, \alpha_i) = 0.
        \end{gather*}
        Here, $\alpha_{Y,i}$ and $\alpha_{X,i}$ are (possibly overlapping) subvectors of $\alpha_i$, and $\gamma_t := (\gamma_{Y,t}',\gamma_{X,t}')'$.
        Assumption~\ref{assumption:potential_outcome_simultaneous}\ref{item:potential_outcome_simultaneous_weak_dependence} holds if I additionally assume that the idiosyncratic terms $(\epsilon_{X,it}, \epsilon_{Y,it})$ are not persistent over time, i.e., for all $\alpha\in \mc{A}$, all possible time paths $\Gamma$, and any $t\in \mb{Z}$,
        \begin{align*}
            \sum_{s=-\infty}^{\infty} \left|\mr{Cov}\left(\epsilon_{Y,it}, \epsilon_{Y,is}\mid \alpha_i = \alpha, \Gamma_T = \Gamma\right)\right| <\infty,
            \quad
            \sum_{s=-\infty}^{\infty} \left|\mr{Cov}\left(\epsilon_{X,it}, \epsilon_{X,is}\mid \alpha_i = \alpha, \Gamma_T = \Gamma\right)\right| <\infty,
        \end{align*}
        which are standard in the factor-model literature. The result extends to multidimensional $X_{it}$ by imposing the same weak-dependence condition on each component of $\epsilon_{X,it}$.
    \end{example}

    \begin{example}[Dynamic linear panel with IFE - Continued]
        For the dynamic linear model with IFE, \eqref{eq:dynamic_interactive_fixed_effects} in Example~\ref{example:dynamic_interactive_fixed_effects}, it is straightforward to verify that Assumption~\ref{assumption:potential_outcome_simultaneous}\ref{item:potential_outcome_simultaneous_weak_dependence} holds if I additionally assume that (i) $|\rho| < 1$, and (ii) the idiosyncratic term $(\epsilon_{it})$ are not persistent over time, i.e., for all $\alpha\in \mc{A}$, all possible time paths $\Gamma$, and any $t\in \mb{Z}$,
        \begin{align*}
            \sum_{s=-\infty}^{\infty} \left|\mr{Cov}\left(\epsilon_{it}, \epsilon_{is}\mid \alpha_i = \alpha, \Gamma_T = \Gamma\right)\right| <\infty.
        \end{align*}
        The condition $|\rho|<1$ is standard in the time-series literature to ensure that the serial dependence decays geometrically over time. The result extends to higher-order lags as long as all roots of the autoregressive characteristic polynomial lie outside the unit circle.
    \end{example}

    \begin{example}[Dynamic nonlinear panel - Continued]
        For the dynamic model \eqref{eq:dynamic_nonlinear} in Example~\ref{example:dynamic_nonlinear_interactive_fixed_effects}, since $\{u_{it}\}_{t\in\mb{Z}}$ is independent across $t$, $\{Y_{it}(0)\}_{t\in\mb{Z}}$ is a time-inhomogeneous Markov process conditional on the fixed effects $(\alpha_i,\Gamma_T)$.
        
        If I further assume that (i) the cumulative distribution function $F(\cdot)$ is continuous and its derivative satisfies $F^{(1)}(\cdot)>0$, and (ii) the supports of $\alpha_i$ and $\gamma_t$ are uniformly bounded across $i$ and $t$, then each entry of the transition matrix is bounded away from $0$ and $1$ uniformly over all periods. This ensures that Assumption~\ref{assumption:potential_outcome_simultaneous}\ref{item:potential_outcome_simultaneous_weak_dependence} holds. The commonly used logit and probit specifications satisfy these restrictions on $F(\cdot)$. 
    \end{example}

    \section{Method}\label{sec:overview_method}
    
    This section provides an overview of the proposed method, which uses pretreatment histories to recover information about latent heterogeneity, and introduces a doubly robust estimator for the dynamic ATT. I focus on simultaneous treatment settings and formally extend the analysis to staggered adoption in Section~\ref{sec:extension}.

\subsection{Recovering latent heterogeneity} 

    The key technical challenge is that the individual fixed effects $\alpha_i$ are unobserved by econometricians. If $\alpha_i$ were observed, the problem would fall under selection on observables, and one could directly apply existing methods to estimate the dynamic ATT. Therefore, the problem of estimating dynamic $\mr{ATT}$ reduces to recovering information about the latent factors $\alpha_i$ from the observed information. 

    To see the basic idea, first consider a pure factor model without dynamics:
    \begin{align*}
        Y_{it}(0) = f(\alpha_i,\gamma_t) + \epsilon_{it}, \quad \mb{E}(\epsilon_{it}\mid \alpha_i, \Gamma_T) = 0.
    \end{align*}
    This is a nonlinear factor model in which the untreated outcome depends on an individual latent factor $\alpha_i$ and a time-varying aggregate factor $\gamma_t$, and $\{\epsilon_{it}\}$ are weakly dependent across time. If two individuals $i$ and $j$ have similar latent factors, then, for each $t$, the smoothness of $f(\cdot)$ implies that $f(\alpha_i,\gamma_t) \approx f(\alpha_j,\gamma_t)$. Thus, individuals with similar latent heterogeneity should have similar untreated histories up to idiosyncratic innovations. This observation provides the intuition for the reverse direction: similarity in pretreatment histories can be informative about similarity in latent factors. This is the logic behind the pseudo-distance approach studied in \citet{feng2023optimal} and \citet{deaner2025inferring}. 

    However, this idea does not directly extend to models with dynamic effects. This is because even if two individuals $i$ and $j$ have similar latent factors, the terms $f(X_{it}, \alpha_i,\gamma_t)$ and $f(X_{jt}, \alpha_j,\gamma_t)$ can still differ when their covariates are different. 
    Despite this complication, I show that, even in the presence of dynamic effects, the latent factors can still be recovered from pretreatment histories as long as the sequence is conditionally weakly dependent as in Assumption~\ref{assumption:potential_outcome_simultaneous}\ref{item:potential_outcome_simultaneous_weak_dependence}. Let us first consider the linear model in Example~\ref{example:dynamic_interactive_fixed_effects}: 
    \begin{align*}
        Y_{it}(0) = \rho Y_{it-1}(0)  + \alpha_i'\gamma_t + \epsilon_{it}, \quad \mb{E}(\epsilon_{it} \mid \{Y_{i\tau}(0)\}_{\tau\leq t-1}, \alpha_i, \Gamma_T) = 0. 
    \end{align*}
    I impose $|\rho|<1$ to ensure that the sequence is conditionally weakly dependent. Iterating the model backward gives, for any lag $L >0$, 
    \begin{align*}
        Y_{it}(0) =  \rho^{L} Y_{it-L}(0)  + \alpha_i' \sum_{\tau = 0}^{L-1}\rho^{\tau }\gamma_{t - \tau} + \sum_{\tau = 0}^{L-1}\rho^{\tau } \epsilon_{it-\tau}.  
    \end{align*}
    Letting $L\to\infty$, the first term vanishes under $|\rho|<1$. Hence,
    \begin{align*}
        Y_{it}(0) = \underbrace{\alpha_i' \sum_{\tau = 0}^{\infty}\rho^{\tau }\gamma_{t - \tau}}_{:= g(\alpha_i, \Gamma_t)} + \underbrace{\sum_{\tau = 0}^{\infty}\rho^{\tau} \epsilon_{it-\tau}}_{\tilde{\epsilon}_{it}} = g(\alpha_i, \Gamma_t) + \tilde{\epsilon}_{it}.  
    \end{align*}
    Thus, the dynamic linear model admits a transformed pure factor representation. The transformed innovation $\tilde{\epsilon}_{it}$ is weakly dependent over time conditional on $(\alpha_i,\Gamma_T)$, and satisfies $\mb{E}\left(\tilde{\epsilon}_{it} \mid \alpha_i,\Gamma_T\right) = 0$ by sequential exogeneity. 
    This implies that the pseudo-distance method based on pure factor models can also be applied to dynamic linear models. The intuition is that the dynamic effect of the initial condition vanishes, i.e., $\lim_{L\to\infty}\rho^L Y_{it-L}(0)=0$, and the behavior of the path is eventually driven by latent heterogeneity $\alpha_i$.

    I now extend the idea from the linear dynamic model to nonlinear cases. Motivated by the idea of backward iteration, I project $Y_{it}(0)$ onto the space of functions of $(\alpha_i, \Gamma_T)$:  
    \begin{align}\label{eq:potential_outcome_factor_representation}
        Y_{it}(0) = \underbrace{\mb{E}\left(Y_{it}(0) \mid \alpha_i, \Gamma_T \right)}_{:=g(\alpha_i, \Gamma_t)} + \underbrace{Y_{it}(0) -  \mb{E}\left(Y_{it}(0) \mid \alpha_i, \Gamma_T \right)}_{\tilde{\epsilon}_{it}} = g(\alpha_i, \Gamma_t) + \tilde{\epsilon}_{it}.  
    \end{align}
    Here, since future time factors provide no additional information about $Y_{it}(0)$ conditional on $(\alpha_i,\Gamma_t)$, the transformed pure factor representation $g(\alpha_i, \Gamma_t)$  depends only on time factors up to period $t$. In addition, the projection error $\tilde{\epsilon}_{it}$ is weakly dependent over time conditional on $(\alpha_i,\Gamma_T)$ and satisfies $\mb{E}\left(\tilde{\epsilon}_{it} \mid \alpha_i,\Gamma_T\right) = 0$. 
    Thus, I show that the dynamic nonlinear model also admits a transformed pure factor representation.  

    \begin{remark*}
        While the pure factor representation~\ref{eq:potential_outcome_factor_representation} lets you learn about similarities in $\alpha_i$, it does not get you the full way as $\tilde{\epsilon}$ is not mean independent of treatment conditional on $\alpha_i$ and $\Gamma_t$ alone.
    \end{remark*}
    
    Example~\ref{example:dynamic_nonlinear_interactive_fixed_effects} provides a concrete example of this transformed pure factor representation under nonlinear models. 

    \setcounter{example}{2}

    \begin{example}[Dynamic nonlinear panel - Continued]
        Suppose that $Y_{it}(0)$ evolves as in Example~\ref{example:dynamic_nonlinear_interactive_fixed_effects}  and let $F(\cdot)$ denote the cumulative distribution function of $u_{it}$. For notational simplicity, define $\rho_{it}: = F\left(\beta + \gamma_t' \alpha_i \right) - F \left( \gamma_t' \alpha_i \right)$ as the time-inhomogeneous discount factor, which satisfies $\sup_{i, t} |\rho_{it}| < 1$. It is straightforward to verify that one-step backward iteration gives 
        $g(\alpha_i, \Gamma_t) =  F \left(\gamma_t' \alpha_i \right) + \rho_{it}g(\alpha_i, \Gamma_{t-1})$. 
        Then, for any $L>0$, $L$-step backward iteration gives
        \begin{align*}
            g(\alpha_i, \Gamma_t) 
            = & \sum_{\tau = 0}^{L-1} \left(\prod_{s = 0}^{\tau-1}  \rho_{i, t-s}  \right) F \left(\gamma_{t-\tau}' \alpha_i \right)  + \left(\prod_{s = 0}^{L-1}  \rho_{i, t-s}\right) g(\alpha_i, \Gamma_{t-L}),  
        \end{align*}
        where the empty product is equal to one. Letting $L\to\infty$, the last term vanishes as $\sup_{i,t}|\rho_{it}|<1$. Hence,
        \begin{align*}
            g(\alpha_i, \Gamma_t) =\sum_{\tau = 0}^{\infty} \left(\prod_{s = 0}^{\tau-1}  \rho_{i, t-s}  \right) F \left(\gamma_{t-\tau}'\alpha_i \right). 
        \end{align*}
        A rigorous treatment is provided in the appendix.
    \end{example}

    The pure factor representation of the dynamic model allows us to compare individuals through their pretreatment histories. Intuitively, when pretreatment histories are informative about latent factors, closeness between the pretreatment histories of individuals $i$ and $j$ implies similarity between their latent factors, i.e., $\|\alpha_i - \alpha_j\|\approx 0$.  
    
    A natural measure of the difference between pretreatment histories is the squared $L_2$ distance. Specifically, for any $i, j$,  
    \begin{align*}
        \widehat{d}_{2, ij}^2 := \frac{1}{T_0} \sum_{t = 0}^{T_0-1}(Y_{it} - Y_{jt})^2. 
    \end{align*}
    However, $\widehat{d}_{2, ij}^2$ cannot serve as a good distance for comparing histories, because it captures not only differences in the factor component of the history but also the conditional variance of idiosyncratic innovations (see the discussion in Appendix~\ref{appendix_sub:informativeness_comparison}).
    As an alternative, I employ the pseudo-distance approach from the nonlinear factor literature \citep{feng2023optimal,deaner2025inferring}.  Specifically, for any $i, j$, define the pseudo-distance as: 
    \begin{align}\label{eq:definition_pseudo_distance}
        \widehat{d}_{ij} = \max_{\substack{k_1, k_2 =1, \ldots, n, \\ k_1, k_2 \neq  i, j}} \left|\frac{1}{T_0} \sum_{t = 0}^{T_0-1} (Y_{k_1t} - Y_{k_2t})(Y_{it} - Y_{jt})\right|. 
    \end{align}
    Unlike the $L_2$ distance, the pseudo-distance effectively filters  the idiosyncratic innovations by comparing the history difference between $i$ and $j$ against history differences between other individuals. It therefore provides a cleaner measure of similarity in the denoised histories.

    The key requirement for using the pseudo-distance $\widehat{d}_{ij}$ as a proxy for the latent distance $\|\alpha_i-\alpha_j\|$ is \emph{informativeness}: individuals who are close in terms of the pseudo-distance should also be close in their underlying latent factors. This condition is related to completeness conditions in nonparametric identification literature. Under the informativeness condition, matching on latent factors reduces to matching units with small pseudo-distances. Therefore, standard nonparametric methods, such as kernel regression or $k$-nearest-neighbor methods, can be used to impute missing potential outcomes and estimate propensity scores. I formally define the informativeness condition and discuss sufficient conditions under which it holds in Sections~\ref{sec:identification} and~\ref{sec:estimation}.

    \begin{remark*}\textbf{(Alternatives to long histories)} 
        This paper uses long pretreatment histories to recover time-invariant heterogeneity, consistent with synthetic control and matrix completion methods. However, this is not the only way to do so. For example, when the panel is short but rich cross-sectional variables are available, these variables can instead be used to recover latent heterogeneity. This idea is closely related to factor-augmented regression \citep{stock2002forecasting} and the method of \citet{feng2020causal}.
    \end{remark*}

\subsection{Estimator}

    For any $t = T_0, \ldots, T$, define the regression outcome and propensity score as 
    \begin{equation}\label{eq:expected_outcome_ps_simultaneous}
    \begin{aligned}
        m_{t}(x , \alpha ) := &\mb{E}_T(Y_{it}(0) \mid X_{iT_0} = x, \alpha_i = \alpha ), \\ 
        \pi(x, \alpha ) := &\mb{P}_T(D_i = 1 \mid X_{iT_0} = x, \alpha_i = \alpha ). 
    \end{aligned}  
    \end{equation}
    Here, the function $m_t:\mb{R}^{d_{X}}\times\mb{R}^{d_{\alpha}}\mapsto\mb{R}$ denotes the conditional expectation of the untreated outcome at time $t$ given pretreatment information\footnote{
        It is straightforward to verify that, for contemporaneous outcomes, $m_{T_0}(x, \alpha) = f(x, \alpha, \gamma_{T_0})$.  
    }, and the function $\pi:\mb{R}^{d_X}\times\mb{R}^{d_{\alpha}}\mapsto [0,1]$ denotes the propensity score given individuals' pretreatment information.

    For notational convenience, let $m_{it} := m_{t}(X_{iT_0}, \alpha_i)$ and $\pi_{i} := \pi (X_{iT_0}, \alpha_i)$. Under the selection mechanism in Assumption~\ref{assumption:selection_simultaneous}, an oracle doubly robust estimator for $\mr{ATT}(t)$ reads 
    \begin{align}\label{eq:oracle_DR_simultaneous}
        \widehat{\mr{ATT}(t)}^{\mr{oracle}} := \frac{1}{n_{1}} \sum_{i = 1}^{n} \left(D_{i}Y_{it} - \frac{(1 - D_i)\pi_iY_{it} + (D_i - \pi_i)m_{it}}{1 - \pi_i }\right), 
    \end{align}
    where $n_1:=\sum_{i=1}^{n} \bs{1}(D_i = 1)$. The estimator is oracle in the sense that it uses the true propensity score $\pi_i$ and outcome regression $m_{it}$, which are unknown in practice.

    \paragraph{Nadaraya-Watson estimator} If latent heterogeneity $\alpha_i$ were observable, researchers could apply the Nadaraya-Watson (NW) estimator to nonparametrically estimate the functions $m_t(\cdot,\cdot)$ and $\pi(\cdot,\cdot)$, and thus $m_{it}$ and $\pi_i$. %
    Since the pseudo-distance $\widehat{d}_{ij}$ serves as a proxy for the latent distance $\|\alpha_i-\alpha_j\|$, I replace the latent distance with the pseudo-distance $\widehat{d}_{ij}$ and obtain a feasible NW estimator. Specifically, for any $i,j=1,\ldots,n$, define the kernel weight
    \begin{align*}
        \widehat{K}_{h, ij} := K\left(\left(X_{jT_0} - X_{iT_0}\right)/ h \right) K\left(\widehat{d}_{ij}/ h\right). 
    \end{align*}
    Here, $K(\cdot)$ denotes the (product) kernel function and $h\rightarrow 0$ is the bandwidth. The hat notation emphasizes that the kernel weight is constructed using the estimated pseudo-distance $\widehat{d}_{ij}$. The Nadaraya-Watson estimators based on the pseudo-distance are given by
    \begin{align*}
        \widehat{m}_{it} = \sum_{j =1}^{n} (1 - D_j)Y_{jt}\widehat{K}_{h, ij} \big / \sum_{j =1}^{n} (1 - D_j) \widehat{K}_{h, ij}, \quad \widehat{\pi}_i =   \sum_{j =1}^{n} D_j \widehat{K}_{h, ij} \big / \sum_{j =1}^{n} \widehat{K}_{h, ij}. 
    \end{align*}
    Plugging these estimates into \eqref{eq:oracle_DR_simultaneous} yields a feasible doubly robust estimator: 
    \begin{equation}\label{eq:ATT_DR_simultaneous}
    \begin{aligned}
        \widehat{\mr{ATT}(t)} := \frac{1}{n_{1}} \sum_{i = 1}^{n} \left(D_{i}Y_{it} - \frac{(1 - D_i)\widehat{\pi}_iY_{it} + (D_i - \widehat{\pi}_i)\widehat{m}_{it}}{1 - \widehat{\pi}_i }\right).  
    \end{aligned}
    \end{equation}
    This estimator is doubly robust in the sense that it is consistent if either $\widehat{m}_{it}$ or $\widehat{\pi}_i$ is consistent. In the double machine learning literature, doubly robust estimators are combined with standard cross-fitting, which uses separate samples to obtain the nuisance estimates $(\widehat{m}_{it}, \widehat{\pi}_i)$ and evaluate the ATT score, to achieve faster convergence rate and make the bias arising from nuisance estimation negligible for inference \citep{chernozhukov2018double}.
    
    \paragraph{Double cross-fitting} Although standard cross-fitting eliminates overfitting bias, the estimation errors in $\widehat{m}_{it}$ and $\widehat{\pi}_i$ remain correlated because $(\widehat{m}_{it}, \widehat{\pi}_i)$ are estimated using the same subsample. I further employ a \emph{double} cross-fitting procedure \citep{newey2018cross, mcclean2026double} to estimate the outcome regressions and propensity scores on separate samples, thereby reducing the dependence between their estimation errors. 

    The double cross-fitting is implemented as follows. I randomly partition the sample into three separate folds, $\{\mc{I}_1, \mc{I}_2, \mc{I}_3\}$. For each unit $i$, let $\mc{I}(i)$ denote the fold containing unit $i$, and let $\mc{I}_m(i)$ and $\mc{I}_\pi(i)$ denote the folds used to estimate $\widehat{m}_{it}$ and $\widehat{\pi}_{i}$ for unit $i$, respectively\footnote{
        For example, when $i\in\mc{I}_1$, I use $\mc{I}_m(i)=\mc{I}_2$ to estimate the potential-outcome regression and obtain $\widehat{m}_{it}$, and use $\mc{I}_\pi(i)=\mc{I}_3$ to estimate the propensity score and obtain $\widehat{\pi}_i$ More broadly, if $i\in\mc{I}_s$, then $\mc{I}(i)=\mc{I}_s$, $\mc{I}_m(i)=\mc{I}_{1+(s\bmod 3)}$, and $\mc{I}_\pi(i)=\mc{I}_{1+((s+1)\bmod 3)}$.
    }. 
    In addition, different bandwidths $\{h_m,h_\pi\}$ are used to estimate $m_{it}$ and $\pi_i$, respectively. Theerefore, the NW estimators under double cross-fitting are given by 
    \begin{align*}
        \widehat{m}_{it} = \frac{\sum_{j\in \mc{I}_m(i)} (1 - D_j)Y_{jt}\widehat{K}_{h_m, ij}}{\sum_{j\in \mc{I}_m(i)} (1 - D_j) \widehat{K}_{h_m, ij}} , \quad \widehat{\pi}_i =  \frac{\sum_{j\in \mc{I}_\pi(i)} D_j \widehat{K}_{h_\pi, ij}}{\sum_{j\in \mc{I}_\pi (i)} \widehat{K}_{h_\pi, ij}}   .
    \end{align*}
    Plugging these estimates into \eqref{eq:ATT_DR_simultaneous} yields the proposed estimator for $\mr{ATT}(t)$. To highlight the main idea of the proposed algorithm, a simplified version that abstracts from cross-validation for selecting the bandwidths $\{h_m,h_\pi\}$ is summarized in Algorithm~\ref{alg:simultaneous_basic}. 

    I show that, compared with standard cross-fitting, combining double cross-fitting with undersmoothing yields faster convergence and relaxes the dimensionality restrictions associated with nonparametric nuisance estimation. This delivers a root-$n$ consistent, asymptotically normal, and asymptotically unbiased ATT estimator under more general conditions. These results are novel in the literature and contribute to the application of double machine learning to causal inference, particularly in settings with latent confounding.

    \begin{algorithm}[htbp]
        \caption{Doubly robust estimator for dynamic ATT (Basic Idea)}\label{alg:simultaneous_basic}
        \begin{algorithmic}
            \Require Kernel $K(\cdot)$, bandwidths $h_m, h_\pi$, and confidence level $1 - \alpha$.  
            \Ensure Dynamic ATT estimates $\widehat{\mr{ATT}}(t)$, standard errors $\widehat{\mr{se}}_t$, and level $1-\alpha$ confidence interval. 
                \State
                \State \textbf{Step 1: Sample-splitting} 
                \State Randomly partition the sample into $3$ disjoint  folds $\{\mc{I}_1, \mc{I}_2, \mc{I}_3\}$ of equal sizes. 
                \State\State \textbf{Step 2: Calculate similarity}  
                \State For each $i, j = 1, \ldots, n$, calculate the pseudo-distance using pretreatment outcomes: 
                \begin{align*}
                    \widehat{d}_{ij} = \max_{\substack{k_1, k_2 =1, \ldots, n, \\ k_1, k_2 \neq  i, j}} \left|\frac{1}{T_0} \sum_{t = 0}^{T_0-1} (Y_{k_1t} - Y_{k_2t})(Y_{it} - Y_{jt})\right|. 
                \end{align*} 
                \State \rule{0pt}{0.4em}
                \State\State \textbf{Step 3: Compute expected outcomes and propensity scores}   
                \State The Nadaraya-Watson estimators under double cross-fitting are given by
                \begin{align*}
                    \widehat{m}_{it} = \frac{\sum_{j\in \mc{I}_m(i)} (1 - D_j)Y_{jt}\widehat{K}_{h_m, ij}}{\sum_{j\in \mc{I}_m(i)} (1 - D_j) \widehat{K}_{h_m, ij}} , \quad \widehat{\pi}_i =    \frac{\sum_{j\in \mc{I}_\pi(i)} D_j \widehat{K}_{h_\pi, ij} }{\sum_{j\in \mc{I}_\pi (i)} \widehat{K}_{h_\pi, ij}}. 
                \end{align*}
                For any $i,j=1,\ldots,n$ and $h>0$, the kernel weight $\widehat{K}_{h, ij}$ is defined as:
                \begin{align*}
                    \widehat{K}_{h, ij} := K\left(\left(X_{jT_0} - X_{iT_0}\right)/ h \right) K\left(\widehat{d}_{ij}/ h\right). 
                \end{align*}
                \State\State \textbf{Step 4: Construct $\widehat{\mr{ATT}(t)}$, standard error $\widehat{\mr{se}}_t$, and confidence interval}   
                \State Let $n_1 :=\sum_{i=1}^{n} \bs{1}(D_i = 1)$, and compute 
                \begin{align*}
                    \widehat{\mr{ATT}(t)} = &\frac{1}{n_{1}} \sum_{i = 1}^{n} \left(D_{i}Y_{it} - \frac{(1 - D_i)\widehat{\pi}_iY_{it} + (D_i - \widehat{\pi}_i)\widehat{m}_{it}}{1 - \widehat{\pi}_i }\right), \\
                    \widehat{\mr{se}}_t = & \left(\frac{1}{n_1^2} \sum_{i=1}^{n} \left(D_{i}Y_{it} - \frac{(1 - D_i)\widehat{\pi}_iY_{it} + (D_i - \widehat{\pi}_i)\widehat{m}_{it}}{1 - \widehat{\pi}_i } -  D_i \widehat{\mr{ATT}(t)} \right)^2\right)^{1/2}. 
                \end{align*}
                The confidence interval is $\mr{CI}_t = [\widehat{\mr{ATT}}(t) \pm Z_{1 - \alpha/2} \cdot  \widehat{\mr{se}}_t ]$. 
        \end{algorithmic}
    \end{algorithm}

    \section{Theory for Identification}\label{sec:identification}

    This section establishes identification of the dynamic ATT. If $\alpha_i$ were observed or identified, the selection mechanism would reduce to selection on observables, and the dynamic ATT could be identified directly under standard conditions.  I show that, although $
    \alpha$ is unobserved, the pseudo-distance is informative about latent heterogeneity, thereby enabling a matching strategy that identifies the dynamic ATT.

    \paragraph{Pseudo-distance} I first discuss conditions under which the sample pseudo-distance $\widehat{d}_{ij}$ serves as a proxy for the latent distance. The argument proceeds in two steps: (i) I show that, as $n,T_0\to\infty$, $\widehat{d}_{ij}$ converges to a \emph{population} pseudo-distance $d(\alpha_i,\alpha_j)$ that depends only on the latent heterogeneity $\alpha_i$ and $\alpha_j$, and (ii) I introduce an informativeness condition under which the population pseudo-distance $d(\alpha_i,\alpha_j)$ reveals the latent distance $\|\alpha_i-\alpha_j\|$.
  
    Establishing the population limit of $\widehat{d}_{ij}$ requires the underlying time averages to converge as $T_0\to\infty$, which in turn requires restrictions on the long-run behavior of the time fixed effects.
    One way to formulate these restrictions is to condition on the path $\{\gamma_t\}_{t\in\mb Z}$ and impose restrictions directly on this path. This approach is consistent with the fixed-effect interpretation of $\gamma_t$ and is common in the factor model literature, but the resulting conditions are typically high-level and difficult to interpret, especially in nonlinear models.

    To focus on the main idea and provide more primitive conditions, I instead embed the realized time effects in a superpopulation model. Specifically, $\Gamma_{T} $ is viewed as one realization of an underlying stochastic process, and restrictions are imposed on this process to characterize the limiting behavior of the realized path as $T_0\to\infty$\footnote{
        This formulation does not change the fixed-effect interpretation in the analysis, because once the sample path is realized, the time effects are treated as fixed.
    }.  The superpopulation assumption serves only to provide transparent and easy-to-verify sufficient conditions for the required time-series limits. Importantly, all stochastic-process assumptions imposed below can be replaced by direct restrictions on the realized path that imply the same limiting results.

    \begin{assumption}[Pseudo distance]\label{assumption:identification_d_simultaneous}
            I assume that: 
            \begin{enumerate}[label=(\roman*)]
            \item \label{item:identification_d_simultaneous_panel_iid} \textbf{(Sampling)} Conditional on $\{\gamma_t\}_{t \in \mb{Z}}$, the panel $\{(Y_{it}, X_{it}, D_{i}, \alpha_i)\}_{i=1, \ldots, n, t = 1, \ldots, T}$ is i.i.d. across $i$, and $\{\gamma_t\}_{t \in \mb{Z}}$ is stationary and ergodic over time. The panel is large with $n\rightarrow\infty$ and the number of pretreatment periods $T_0\rightarrow\infty$.
            \item \label{item:identification_d_simultaneous_compact}  \textbf{(Compact)} The support of $\alpha$ is compact. 
            \item \label{item:identification_d_simultaneous_finite_moment} \textbf{(Moment)} $\mb{E}\left(  Y^2_{it}(0) \mid \alpha_i = \alpha, \Gamma_{t} = \Gamma , D_i = d \right) <\infty $ uniformly over all supported $\alpha$, all possible paths $\Gamma$,  $d\in\{0, 1\}$, and $t = 1, \ldots, T$.  
            \item \label{item:identification_d_simultaneous_ULLN}  \textbf{(Envelope)} The envelope of the factor structure $g(\alpha, \Gamma)$ (defined in~\eqref{eq:potential_outcome_factor_representation}) has a finite second moment, i.e.,  $\mb{E}\left(\sup_{\alpha \in \mc{A}} \left(g(\alpha, \Gamma) \right)^2\right) < \infty$.
            \item \label{item:identification_d_simultaneous_smoothness}  \textbf{(Smoothness)} The factor structure $g(\alpha, \Gamma)$ defined in~\eqref{eq:potential_outcome_factor_representation} is uniformly continuous in $\alpha$, i.e., for any $\epsilon >0$, there exists a $\delta >0$ such that 
            $$\sup_{\Gamma\in \mr{supp}(\Gamma_{T})} \sup_{\alpha, \alpha'\in \mc{A}, \| \alpha - \alpha' \| \leq \delta} | g(\alpha, \Gamma) - g(\alpha', \Gamma)| \leq \epsilon. $$ 
        \end{enumerate}
    \end{assumption}

    Assumption~\ref{assumption:identification_d_simultaneous}\ref{item:identification_d_simultaneous_panel_iid} requires panel data with $n\rightarrow\infty$ individuals and $T_0\rightarrow\infty$ pretreatment periods, and imposes conditional independence across individuals $i$ given the realized time fixed effects $\Gamma_{T}$. The conditional independence condition is weak and does not require time-homogeneity condition \citep{chernozhukov2013average,chernozhukov2026linear}. I also assume that $\{\gamma_t\}_{t\in\mb{Z}}$ is stationary and ergodic over time, so that $\{\Gamma_t\}_{t\in\mb Z}$ is also stationary and ergodic. Consequently, the distribution of $\Gamma_t$ is invariant over $t$. These restrictions, similar to the stochastic-process restrictions imposed in \citet{feng2023optimal} and \citet{deaner2025inferring}, ensure the convergence of the time averages in the sample pseudo-distance. These stationarity and ergodicity assumptions are stronger than necessary and serve as primitive conditions for the identification argument in the main text. Appendix~\ref{appendix:informativeness} discusses alternative conditions that allow for nonstationarity.

    Assumption~\ref{assumption:identification_d_simultaneous}\ref{item:identification_d_simultaneous_compact} is a standard compact-support condition on the latent heterogeneity.
    
    Assumption~\ref{assumption:identification_d_simultaneous}\ref{item:identification_d_simultaneous_finite_moment} imposes moment conditions on untreated potential outcomes. It requires that the conditional second moment $\mb{E}\left(Y^2_{it}(0) \mid \alpha_i=\alpha, \Gamma_t=\Gamma, D_i=d\right)$ be uniformly bounded over the support of $\alpha$, all possible realizations of $\Gamma$, $d\in\{0,1\}$, and all time periods $t=1,\ldots,T$. 
    It is worth noting that, although the presence of unbounded common time trends in $Y_{it}(0)$ violates Assumption~\ref{assumption:identification_d_simultaneous}\ref{item:identification_d_simultaneous_finite_moment}, the  analysis remains valid because the common trend cancels out in the cross-sectional differences used to construct the pseudo-distance, as discussed in Section~\ref{sec:overview_method}.

    Assumption~\ref{assumption:identification_d_simultaneous}\ref{item:identification_d_simultaneous_ULLN} requires that the envelope of $g^2(\cdot, \cdot)$ is integrable over the time-factor path $\Gamma$. The condition is mild in that it requires only a finite second moment of the envelope and does not require $g(\alpha,\Gamma)$ to be uniformly bounded. 

    Assumption~\ref{assumption:identification_d_simultaneous}\ref{item:identification_d_simultaneous_smoothness} imposes that   $g(\alpha,\Gamma)$ is uniformly continuous in $\alpha$ over all possible paths $\Gamma$. This assumption is necessary because it ensures that individuals with similar $\alpha$ behave similarly over a long time horizon.

    \begin{proposition}[Identification of $d(\alpha_i, \alpha_j)$]\label{prop:identification_d_simultaneous}     
        Under Assumptions~\ref{assumption:potential_outcome_simultaneous} and~\ref{assumption:identification_d_simultaneous}, for each  $i,j = 1, \ldots, n$, $\widehat d_{ij}\cp d(\alpha_i,\alpha_j)$, as $n,T_0\to\infty$, where $d(\alpha_i, \alpha_j)$ is given by
        \begin{equation}\label{eq:definition_d}
        \begin{aligned}
            d(\alpha_i,\alpha_j) : = \sup_{\alpha_1, \alpha_2 \in \mc{A}} \left| \int (g(\alpha_{1}, \Gamma) -g(\alpha_{2}, \Gamma) )(g(\alpha_{i}, \Gamma ) -g(\alpha_{j}, \Gamma) )\mr{d}\mb{P}(\Gamma) \right|. 
        \end{aligned}  
        \end{equation}
    \end{proposition}

    The identification of $d(\alpha_i,\alpha_j)$ follows directly from establishing that $\widehat d_{ij}\cp d(\alpha_i,\alpha_j)$. This constitutes an identification-by-construction approach (see the discussion in \citet{lewbel2019identification}). The population pseudo-distance $d(\alpha_i,\alpha_j)$ depends only on $(\alpha_i,\alpha_j)$ and does not depend on the realized outcome paths or the realization of the time fixed effects. In addition, it is finite, symmetric, and uniformly continuous in $(\alpha_i,\alpha_j)$, with $d(\alpha_i,\alpha_j)=0$ whenever $\alpha_i=\alpha_j$. 

    The integral in~\eqref{eq:definition_d} is taken with respect to the stationary distribution of the time-factor path $\Gamma$ because, under stationarity and ergodicity, the ergodic theorem implies that time averages along the realized path $\{\Gamma_t\}_{t\in\mb Z}$ converge to expectations under this stationary distribution.  The existence of the population pseudo-distance and the identification result, however, do not rely on stationarity or ergodicity. I provide the definition of $d(\alpha_i, \alpha_j)$ under more general conditions in Appendix~\ref{appendix_sub:definition_d_without_stationary_ergodicity}. 

    \begin{remark*}
        Proposition~\ref{prop:identification_d_simultaneous} identifies the $n(n-1)/2$ pairwise pseudo-distances $d(\alpha_i,\alpha_j)$   in the observed sample. The proposition does not identify the latent heterogeneity $\alpha_i$ itself or the functional form of $d(\cdot,\cdot)$ over the entire latent space. Nevertheless, I show that this pairwise information is sufficient for the identification of the dynamic ATT, as it allows us to identify and match units with similar latent heterogeneity.
    \end{remark*}

    \begin{remark*}\textbf{(Common time trend)} 
        The pseudo-distance remains well-defined even when $Y_{it}(0)$ contains a common time trend, in which case the factor representation of the data-generating process takes the form
        \begin{align*}
            Y_{it}(0)
            = \underbrace{\lambda_t + \tilde{g}(\alpha_i,\Gamma_t)}_{=g(\alpha_i,\Gamma_t)}
            + \tilde{\epsilon}_{it}
            = g(\alpha_i,\Gamma_t)+\tilde{\epsilon}_{it},
        \end{align*}
        where $\lambda_t$ is common across units and may be nonstationary and unbounded. The pseudo-distance proposed in this paper, unlike the pseudo-distances in~\citet{feng2023optimal}, \citet{feng2020causal},  and~\citet{deaner2025inferring}, accommodates such common trends without modification because these trends cancel in the cross-sectional differences entering construction~\eqref{eq:definition_pseudo_distance}.
    \end{remark*}
    
    \paragraph{Identification of $\mr{ATT}(t)$} 
    
    To use the identified pseudo-distance to recover latent similarity, I impose the following informativeness condition. 

    \begin{assumption}[Informativeness - identification]\label{assumption:informativeness_identification_simultaneous}
        Suppose the population pseudo-distance exists and for every $\varepsilon >0$, there exists a $\delta >0$, such that
        $$\sup_{\alpha_1, \alpha_2\in \mc{A}, d(\alpha_1, \alpha_2)\leq \delta}  \|\alpha_1 - \alpha_2 \| \leq \varepsilon. $$ 
    \end{assumption}

    This assumption is key to identifying dynamic ATT, which requires that, if two individuals are close in terms of $d(\alpha_i,\alpha_j)$, then their latent factors must also be close. Assumption~\ref{assumption:informativeness_identification_simultaneous}, analogous to the identification assumption in \citet[Assumption~3]{auerbach2022identification}, is mild and holds if distinct latent factors cannot generate exactly the same factor structure, i.e., for any $\alpha_1\neq\alpha_2$, $g(\alpha_1,\Gamma)$ and $g(\alpha_2,\Gamma)$ are not almost surely identical (see Lemma~\ref{lemma:sufficient_informativeness_identification_simultaneous} in Appendix~\ref{appendix_sub:informativeness_comparison} for a formal discussion). In addition, the dynamic panel models in Examples~\ref{example:dynamic_interactive_fixed_effects} and~\ref{example:dynamic_nonlinear_interactive_fixed_effects} satisfy the informativeness assumption under general conditions, and the discussion is postponed to Section~\ref{sec:estimation}. This assumption is called informativeness because, although $\alpha$ is unobserved, the pretreatment history contains sufficient information to recover similarity in latent heterogeneity through the population pseudo-distance $d(\alpha_i,\alpha_j)$.

    \begin{assumption}[Identification of ATT]\label{assumption:identification_ATT_simultaneous}
            I assume that: 
            \begin{enumerate}[label=(\roman*)]
            \item \label{item:identification_ATT_simultaneous_finite_moment} \textbf{(Moment)} $\mb{E}_{T}\left( | Y_{it}(0)| \mid D_i = 1\right) , \mb{E}_{T}\left( | Y_{it}(1)| \mid D_i = 1\right)<\infty $ for each $t = T_0, \ldots, T$. 
            \item \label{item:identification_ATT_simultaneous_smoothness}  \textbf{(Smoothness)} For each $t=T_0,\ldots,T$, the expected outcome $m_{t}(x,\alpha)$ defined in~\eqref{eq:expected_outcome_ps_simultaneous}, is uniformly continuous in $(x, \alpha)$, i.e., for any $\epsilon >0$, there exists $\delta >0$ such that for all $(x, \alpha) \in \mr{supp} (X_{iT_0}, \alpha_i )$,
            \begin{align*}
                \sup_{  \|(x', \alpha') - (x, \alpha) \| \leq \delta  } \left| m_{t}(x, \alpha ) - m_{t}(x', \alpha'  ) \right| \leq \epsilon. 
            \end{align*} 
            \item \label{item:identification_ATT_simultaneous_overlap}  \textbf{(Overlap)} $\mb{P}_T(D_i=1\mid X_{iT_0},\alpha_i)\in(0,1)$ a.s.
        \end{enumerate}
    \end{assumption}

    Assumption~\ref{assumption:identification_ATT_simultaneous}\ref{item:identification_ATT_simultaneous_finite_moment} imposes finite first-moment conditions on both untreated and treated potential outcomes. Assumption~\ref{assumption:identification_ATT_simultaneous}\ref{item:identification_ATT_simultaneous_smoothness} imposes uniform continuity on expected outcome $m_t(\cdot, \cdot)$. Uniform continuity of $m_t$ is a mild and standard condition commonly imposed for nonparametric identification of conditional expectations. 
    Assumption~\ref{assumption:identification_ATT_simultaneous}\ref{item:identification_ATT_simultaneous_overlap} imposes the standard overlap condition, which is commonly required in causal inference to identify counterfactual outcomes.

    The following theorem presents the identification result for the dynamic ATT. The key idea is that, although $\alpha$ is unobserved to econometricians, the identified population pseudo-distance reveals latent similarity. Therefore, the dynamic ATT is identified by matching units with similar latent heterogeneity and pretreatment history. 

    \begin{theorem}[Identification of ATT]\label{thm:identification_simultaneous}
        Suppose that the population pseudo-distance $d(\alpha_i,\alpha_j)$ is identified for each $i,j=1,\ldots,n$. Then, under Assumptions~\ref{assumption:selection_simultaneous}-\ref{assumption:potential_outcome_simultaneous}, \ref{assumption:informativeness_identification_simultaneous}, and \ref{assumption:identification_ATT_simultaneous}, $\mr{ATT}(t)$ is identified by
        \begin{align*}
            \mb{E}_T(Y_{it} \mid D_i = 1) -  \mb{E}_T\left( \lim_{\delta\downarrow 0}\mb{E}_T (Y_{jt} \mid X_{jT_0} = X_{iT_0}, d(\alpha_i, \alpha_j) \leq \delta, D_j = 0 ) \mid D_i = 1 \right). 
        \end{align*}
    \end{theorem}
    
    Theorem~\ref{thm:identification_simultaneous} shows that the dynamic ATT is identified using a matching strategy based on the pseudo-distance. 
    Identification of $\mr{ATT}(t)$ requires imputing the missing untreated counterfactual outcome for treated units, i.e., $\mb{E}_T\left(Y_{it}(0)\mid X_{iT_0},\alpha_i\right)$. I impute this counterfactual using matching. To elaborate, for each treated unit $i$, I match $i$ to untreated units $j$ with the same pretreatment covariates and a population pseudo-distance $d(\alpha_i,\alpha_j)$ no greater than $\delta$. For a fixed $\delta$, the imputed quantity is
    \begin{align*}
        \mb{E}_T(Y_{jt} \mid X_{jT_0} = X_{iT_0}, d(\alpha_i, \alpha_j) \leq \delta, D_j = 0 ). 
    \end{align*}
    By the informativeness condition, a small pseudo-distance implies similar latent heterogeneity. Together with the smoothness and selection conditions, this allows $\delta$ to approach zero, yielding 
    \begin{align*}
        \mb{E}_T\left(Y_{it}(0) \mid X_{iT_0}, \alpha_i\right) = \lim_{\delta\rightarrow 0} \mb{E}_T (Y_{jt} \mid X_{jT_0} = X_{iT_0}, d(\alpha_i, \alpha_j) \leq \delta, D_j = 0 ). 
    \end{align*}
    Finally, averaging the imputed counterfactual over the distribution of treated units yields the identification formula in Theorem~\ref{thm:identification_simultaneous}.
    
    My imputation strategy is closely related to the missing-outcome imputation approach in the panel-data causal inference literature, including DiD~\citep{borusyak2024revisiting}, synthetic control~\citep{abadie2003economic}, and factor-model imputation methods~\citep{bai2021matrix}. 
    I consider a more general nonlinear dynamic model, and as a result, the latent heterogeneity cannot be eliminated by differencing, as in standard DiD designs, and cannot be directly estimated using linear factor methods. The pseudo-distance identifies only pairwise latent similarity: it reveals whether two units have close latent types, but it does not identify the latent heterogeneity or its distribution. Nevertheless, Theorem~\ref{thm:identification_simultaneous} shows that even pairwise information is sufficient for identification.  

    The studies most closely related to this paper are \citet{feng2023optimal}, \citet{athey2025identification}, and \citet{deaner2025inferring}, which also consider imputation-based methods for nonlinear models. However, their frameworks allow selection to depend only on latent heterogeneity, whereas this paper extends the analysis to settings in which selection also depends on pretreatment covariates.

    \section{Theory for Estimation and Inference}\label{sec:estimation}

    This section establishes the asymptotic properties of the proposed Nadaraya-Watson estimators using the sample pseudo-distance to impute untreated outcomes and estimate propensity scores. I then provide sufficient conditions under which the doubly robust ATT estimator with double cross-fitting is $\sqrt{n}$-consistent, asymptotically unbiased, and asymptotically normal, thereby enabling valid inference for the dynamic ATT. 
    
    In this section, I focus on the case in which both the latent heterogeneity and the pretreatment covariates are continuously distributed. The analysis extends straightforwardly to settings where either $\alpha$ or $X$ has finite support.

\subsection{Asymptotic analysis of pseudo distance} 

    I impose two assumptions to ensure that a sufficiently small pseudo-distance $\widehat{d}_{ij}$ implies that $\alpha_i$ and $\alpha_j$ are close in the latent space. The first assumption characterizes the rate of convergence of $\widehat{d}_{ij}$ to its population counterpart $d(\alpha_i,\alpha_j)$.
    \begin{assumption}[Pseudo-distance approximation]\label{assumption:estimation_consistency_d}
        There exist constants $\lambda_1, \lambda_2 >0$ such that the following inequality holds wpa1:  
            \begin{align*}
                \max_{i,j \in \{1, \ldots, n\}} |\widehat{d}_{ij} - d(\alpha_i, \alpha_j) | \leq   \lambda_1 n^{-1/d_\alpha}\log (n)  + \lambda_2 T_0^{-1/2}\sqrt{\log(nT_0)},
            \end{align*}
    \end{assumption}
    Assumption~\ref{assumption:estimation_consistency_d} strengthens the pointwise convergence result used for identification by imposing a uniform convergence rate for $\widehat{d}_{ij}$. This assumption states that the estimation error between $\widehat{d}_{ij}$ and $d(\alpha_i,\alpha_j)$ is uniformly bounded by two terms in an asymptotic sense. The first term, $\lambda_1 n^{-1/d_\alpha}\sqrt{\log n}$, depends only on the dimension of the latent heterogeneity $d_\alpha$ and not on the length of the pretreatment history $p$. It captures the matching discrepancy in the latent characteristics $\alpha$, and its rate deteriorates as $d_\alpha$ increases. The second term, $\lambda_2 T_0^{-1/2}\sqrt{\log(nT_0)}$, captures the sampling error that arises because the pseudo-distance is constructed from noisy observed outcomes.

    The convergence rate in Assumption~\ref{assumption:estimation_consistency_d} is consistent with the indirect matching rates in \citet[Theorem~4.1]{feng2023optimal} and \citet[Lemma~A.1]{deaner2025inferring} for pure factor models, while the setting considered here is more general in allowing for dynamic dependence. I show that this condition is not restrictive even in the dynamic panel setting and provide sufficient conditions under which it holds (see Lemma~\ref{lemma:sufficient_estimation_consistency_d} in Appendix~\ref{appendix_sub:sufficient_consistency_d_without_proof}). Lastly, for notational simplicity, let $\delta_{n, T_0} := \lambda_1 n^{-1/d_\alpha}\log (n) + \lambda_2 T_0^{-1/2}\sqrt{\log(nT_0)}$ denote a uniform upper bound on these estimation errors. 

    \bigskip 

    The second assumption  is critical and requires that the population pseudo-distance $d (\alpha_i, \alpha_j)$ be
    informative to reveal the latent distance $\|\alpha_i - \alpha_j\|$. 
    \begin{assumption}[Informativeness-estimation]\label{assumption:informativeness_estimation}
       There exist constants $0 < \underline{\eta} < \overline{\eta} <\infty$ such that for any $\alpha_1, \alpha_2 \in \mc{A}$, $\underline{\eta} d (\alpha_1, \alpha_{2}) \leq \|\alpha_1 - \alpha_{2}\|\leq  \overline{\eta} d (\alpha_1, \alpha_{2})$. 
    \end{assumption}
    
    Assumption~\ref{assumption:informativeness_estimation} strengthens the informativeness condition used for identification (Assumption~\ref{assumption:informativeness_identification_simultaneous}) by requiring the latent distance to be linearly bounded by the population pseudo-distance. Assumption~\ref{assumption:informativeness_estimation} is analogous to the conditions in \citet[Assumption~4.1]{feng2023optimal} and \citet[Assumption~4(vi)]{deaner2025inferring}.
    The lower bound condition in Assumption~\ref{assumption:informativeness_estimation} is mild and can be directly verified when $g(\alpha,\Gamma)$ is Lipschitz continuous with respect to $\alpha$, uniformly over $\Gamma$. By contrast, verifying the upper bound condition can be challenging, especially in nonlinear models with dynamic effect. I provide sufficient conditions under which many commonly used dynamic panel models, including Examples~\ref{example:interactive_fixed_effects} and~\ref{example:dynamic_nonlinear_interactive_fixed_effects}, satisfy  the informativeness assumption.

    \setcounter{example}{1}

    \begingroup
    \renewcommand{\theexample}{1 and 2}
    \begin{example}[Linear model - Continued]
        Here I focus on Example~\ref{example:dynamic_interactive_fixed_effects}, since the verification for Example~\ref{example:interactive_fixed_effects} can be obtained by similar argument. 
        Suppose that $Y_{it}(0)$ follows $Y_{it}(0) = \rho Y_{it-1}(0) + \gamma_t'\alpha_i + \epsilon_{it}$ with sequentially mean-independent idiosyncratic error $\epsilon_{it}$. Then, under certain regularity conditions on $\mc{A}$, Assumption~\ref{assumption:informativeness_estimation} holds if (i) $|\rho|<1$, (ii) $\{\gamma_t\}_{t\in\mb{Z}}$ are stationary and ergodic over time, and (iii) the eigenvalues of $\mb{E}\left(\gamma_t\gamma_t'\right)$ are bounded above and bounded away from zero. Here, the condition $|\rho|<1$ ensures that the effects of past idiosyncratic shocks decay over time. In addition, the positive smallest eigenvalue of $\mb{E}\left(\gamma_t\gamma_t'\right)$,  which can be viewed as analogous to a strong-factor condition in the factor-model literature, ensures that each dimension of $\alpha$ makes an independent contribution to the factor component that cannot be perfectly explained by the remaining dimensions. A formal statement of this result is provided in Lemma~\ref{lemma:sufficient_informativeness_linear_IFE} in Appendix~\ref{appendix_sub:sufficient_informativeness_without_proof}.
    \end{example}
\endgroup

    \begin{example}[Dynamic nonlinear model - Continued]
        Suppose that $Y_{it}(0)$ evolves according to a dynamic Logit model as in Example~\ref{example:dynamic_nonlinear_interactive_fixed_effects}. Then, under certain regularity conditions on $\mc{A}$, Assumption~\ref{assumption:informativeness_estimation} holds if (i) $\{\gamma_t\}_{t\in\mb{Z}}$ are i.i.d. over time, (ii) the supports of $\alpha$ and $\gamma_t$ are bounded, and (iii) the largest eigenvalue of $\mb{E}\left(\gamma_t\gamma_t'\right)$ is finite and its smallest eigenvalue is strictly positive. %
        Similarly, the positive smallest eigenvalue of $\mb{E}\left(\gamma_t\gamma_t'\right)$ ensures that each dimension of $\alpha$ makes an independent contribution to the factor component that cannot be perfectly explained by the remaining dimensions. The same result also holds for dynamic Probit models. A formal statement is provided in Lemma~\ref{lemma:sufficient_informativeness_nonlinear_IFE}  in Appendix~\ref{appendix_sub:sufficient_informativeness_without_proof}. 
    \end{example}

\subsection{Asymptotic theory of $\widehat{\mr{ATT}(t)}$} 

    I impose the following regularity conditions to establish the asymptotic theory.  
    \begin{assumption}[Estimation]\label{assumption:estimation_simultaneous} I assume that 
        \begin{enumerate}[label=(\roman*)]
            \item \label{item:estimation_simultaneous_panel_weak_dependent} \textbf{(Sampling)} Conditional on $\Gamma_T$, the panel $\{(Y_{it},X_{it}, D_{i}, \alpha_i)\}_{i=1, \ldots, n, t = 1, \ldots, T}$ is i.i.d. across $i$. The panel is large with $n\rightarrow\infty$ and  $T_0\rightarrow\infty$.  
            \item \label{item:estimation_simultaneous_finite} \textbf{(Moment)} Suppose $X_{iT_0}$ is uniformly bounded across $i$. In addition, for some $\sigma>0$, $\mb{E}_T\left(\exp\left(\lambda(Y_{it}(0)-m_{it})\right)\mid X_{iT_0},\alpha_i\right)\leq \exp(\sigma^2\lambda^2/2)$ a.s. for all $\lambda\in\mb{R}$, all $i$, and all $t=T_0,\ldots,T$. Lastly, for some $\delta>0$, $\mb{E}_T\left(|Y_{it}(1)|^{2+\delta}\right)<\infty$ for all $t=T_0,\ldots,T$. 
            \item \label{item:estimation_simultaneous_compact}  \textbf{(Compactness)} The support of $\alpha$, $\mc{A}$, is compact.  There exist constants $\underline{c}_1, \overline{c}_1 >0$, such that, for any $\alpha \in \mc{A}$ and any $ r \in (0,  1)$, $\underline{c}_1 r^{d_{\alpha}} \leq \mb{P}\left(\|\alpha_i - \alpha\| \leq r \right) \leq \overline{c}_1 r^{d_{\alpha}}$. In addition, there exist constants $0<\underline{c}_2\leq\overline {c}_2<\infty$ such that, for any $(x,\alpha)\in\mr{supp}(X_{iT_0},\alpha_i) $ and any $ r \in (0,  1)$, $\underline{c}_2 r^{d_X + d_{\alpha}} \leq \mb{P}_T \left(\|X_{iT_0}-x \| \leq r,  \|\alpha_i-\alpha\|\leq r  \right)\leq  \overline c_2r^{d_X + d_{\alpha}}$. 
            \item \label{item:estimation_simultaneous_L_continuous} \textbf{(Lipschitz)} $m_{t}(x, \alpha)$ and $\pi (x, \alpha)$ are Lipschitz continuous on $\mr{supp} (X_{iT_0}, \alpha_i)$ for each $t = T_0, \ldots, T$.  
            \item \label{item:estimation_simultaneous_overlap} \textbf{(Common overlap)} There exist constants $\underline{p}, \overline{p}\in (0, 1)$ s.t.  $\underline{p} \leq  \pi( X_{iT_0}, \alpha_i)\leq \overline{p}$ a.s.  
            \item \label{item:estimation_simultaneous_kernel} \textbf{(Kernel)} The kernel $K: \mb{R}\mapsto \mb{R}_{+}$ is bounded by $\overline{K} >0$ and supported on $[-1, 1]$.   In addition, $K(0)>0$ and  $K(\cdot)$ is Lipschitz continuous with constant $L_K>0$. 
        \end{enumerate}
    \end{assumption}

    Assumption~\ref{assumption:estimation_simultaneous}\ref{item:estimation_simultaneous_panel_weak_dependent} requires that the panel data be i.i.d. given the time fixed effects $\Gamma_T$. %

    Assumption~\ref{assumption:estimation_simultaneous}\ref{item:estimation_simultaneous_finite} requires that the covariates $X_{iT_0}$ be uniformly bounded over units, which is a standard condition for regressors in nonparametric estimation. This assumption also requires $Y_{it}(0)$ to be uniformly sub-Gaussian conditional on covariates and latent factors. The sub-Gaussian assumption is analogous to \citet[Assumption~4(iii)]{deaner2025inferring} and facilitates the application of exponential concentration inequalities. The finite $(2+\delta)$-th moment condition on $Y_{it}(1)$ is used to apply the central limit theorem.  
    Note that the uniform boundedness requirement is imposed for technical convenience and is stronger than necessary. It suffices to require that $X_{iT_0}$ be uniformly bounded after removing an additive common time trend, since this trend cancels out under cross-sectional differencing even if it is unbounded.

    Assumption~\ref{assumption:estimation_simultaneous}\ref{item:estimation_simultaneous_compact} first imposes compactness on the support of $\alpha$ and requires that the distribution of $\alpha$ be neither locally too sparse nor too concentrated. That is, for any supported $\alpha$ and any small radius $r$, the probability that $\alpha_i$ falls within a ball of radius $r$ centered at $\alpha$ is of the same order as the volume of that ball. In addition, this assumption imposes the same restriction on the joint distribution of $(X_{iT_0}, \alpha)$ and requires that the joint distribution of $(X_{iT_0}, \alpha_i)$ be neither locally too sparse nor too concentrated. Assumption~\ref{assumption:estimation_simultaneous}\ref{item:estimation_simultaneous_compact} is common in establishing the uniform asymptotic properties of nonparametric kernel estimators and is automatically satisfied if (i) the support of $(X_{iT_0}, \alpha)$ is compact and convex, and (ii) the density of $\alpha$ and the joint density of $(X_{iT_0}, \alpha)$ are uniformly bounded above and bounded away from zero. 
    Lastly, although this assumption is primarily applicable to continuously distributed $\alpha$ and $X$, the density condition can be replaced by the corresponding condition on point-mass probabilities to accommodate the case in which either $\alpha$ or $X$ has finite support.

    Assumption~\ref{assumption:estimation_simultaneous}\ref{item:estimation_simultaneous_L_continuous} requires that both the expected outcomes and propensity scores be Lipschitz continuous functions of covariates and individual heterogeneity. This condition is mild and does not require differentiability. 

    Assumption~\ref{assumption:estimation_simultaneous}\ref{item:estimation_simultaneous_overlap} is a commonly adopted overlap condition in the causal inference literature.  Assumption~\ref{assumption:estimation_simultaneous}\ref{item:estimation_simultaneous_kernel} imposes standard regularity conditions on the kernel function. These conditions are satisfied by many commonly used kernels, including the Epanechnikov kernel. 

    \begin{proposition}[Convergence rates for counterfactual imputation]\label{prop:estimation_entry_simultaneous}
        Under Assumptions~\ref{assumption:selection_simultaneous}-\ref{assumption:potential_outcome_simultaneous} and  Assumptions~\ref{assumption:estimation_consistency_d}-\ref{assumption:estimation_simultaneous}, for any $h\in\{h_{\pi},h_m\}$ such that $h\rightarrow 0$, $nh^{d_\alpha+d_{X}}/\log n\rightarrow\infty$, and $\delta_{n,T_0}/h\rightarrow 0$, the following result holds for each $t = T_0, \ldots, T$, 
        \begin{align*}
            \max_{i = 1, \ldots, n}\left|\widehat{m}_{it} - m_{it} \right| = O_P\left(h_m +  \left(nh_m^{d_\alpha+d_{X}}\right)^{-1/2}\sqrt{\log (n)} \right).  
        \end{align*}
        In addition, 
        \begin{align*}
            \max_{i = 1, \ldots, n}\left|\widehat{\pi}_{i} - \pi_{i} \right| = O_P\left(h_{\pi} +  \left(nh_{\pi}^{d_\alpha+d_{X}}\right)^{-1/2}\sqrt{\log (n)} \right). 
        \end{align*}
    \end{proposition}

    Proposition~\ref{prop:estimation_entry_simultaneous} establishes the uniform convergence rates for counterfactual outcome estimators and propensity score estimators over all individuals. These convergence rates imply that, when the estimation error of the sample pseudo-distance is asymptotically negligible relative to the bandwidth $h$, the convergence rate of the estimator using $\widehat{d}_{ij}$ is identical (up to a logarithmic term) to that of the Nadaraya-Watson estimator if $\alpha$ were observed. 

    In standard practice, researchers set $h_m, h_\pi \asymp n^{-1/(d_\alpha+d_{X}+2)}$ to balance the bias and variance terms. The uniform convergence rates in this case are of order $n^{-1/(d_\alpha+d_{X}+2)}\log(n)$ and achieve Stone's optimal rate under Lipschitz continuity (up to a logarithmic term). Then, one typically employs standard single cross-fitting and plugs these nuisance-function estimators into the doubly robust estimator of the ATT. However, I show that the bandwidth choice that is rate-optimal for imputing counterfactual outcomes and estimating propensity scores is not necessarily optimal for estimating the ATT. By combining double cross-fitting with a different choice of bandwidths, I obtain a faster convergence rate for the ATT estimator and establish root-$n$ consistency, asymptotic normality, and asymptotic unbiasedness under a broader range of settings. The following Theorem formalizes this idea. 

    \begin{theorem}[Asymptotic theory for ATT estimator]\label{thm:estimation_ATT_simultaneous}
        Under conditions in Proposition~\ref{prop:estimation_entry_simultaneous}, for any $h\in\{h_{\pi},h_m\}$ such that $h\rightarrow 0$, $nh^{d_\alpha+d_{X}}/\log n\rightarrow\infty$, and $\delta_{n,T_0}/h\rightarrow 0$, the following result holds for each $t=T_0, \ldots, T$, 
        \begin{align*}
            \widehat{\mr{ATT}(t)} - \mr{ATT}( t) = \frac{1}{\sqrt{n}}  \mc{N}(0, V_t) + O_P\left(h_m \left(h_{\pi} + (nh_{\pi}^{d_{\alpha} + d_{X}})^{-1/2}\sqrt{\log(n)}\right)\right) + o_P(n^{-1/2}). 
        \end{align*}
        Here, $\mc{N}(\cdot,\cdot)$ denotes the normal distribution, and $V_t$ is the asymptotic variance of the oracle estimator.
    \end{theorem}
     
    Theorem~\ref{thm:estimation_ATT_simultaneous} is the main result of this paper. Ignoring logarithmic terms for simplicity, the theorem shows that the difference between the oracle estimator and the proposed doubly robust estimator using double cross-fitting is $O_P\left(h_m\left(h_{\pi}+(nh_{\pi}^{d_{\alpha}+d_{X}})^{-1/2}\right)\right)$.  
    Therefore, when this remainder term is $o_P(1/\sqrt n)$, the proposed estimator is asymptotically equivalent to the oracle estimator and achieves root-$n$ asymptotic normality and asymptotic unbiasedness. 

    Double cross-fitting achieves a faster convergence rate  than standard single cross-fitting. Under standard single cross-fitting, the remainder term is the product of nuisance estimation errors, and is of order $O_P\left(\left(h_m + (nh_{m}^{d_{\alpha}+d_{X}})^{-1/2}\right)\left(h_{\pi}+(nh_{\pi}^{d_{\alpha}+d_{X}})^{-1/2}\right)\right)$. Double cross-fitting eliminates the variance term $(nh_m^{d_{\alpha}+d_{X}})^{-1/2}$ from the remainder, which enables researchers to adopt an \emph{undersmoothing} choice of $h_m$ rather than the bandwidth that is optimal for estimating $m_{it}$ and achieves a faster convergence rate for the remainder term.

    The intuition behind this improvement is that double cross-fitting reduces the dependence between the outcome regression and propensity score estimators. Under single cross-fitting, $\widehat{m}_{it}$ and $\widehat{\pi}_{i}$ are constructed from the same sample and hence have correlated estimation errors. As a result, the product error is controlled by the Cauchy-Schwarz inequality, which depends on the convergence rates of both $\widehat{m}_{it}$ and $\widehat{\pi}_{i}$ in the $L_2$ norm. Double cross-fitting removes this dependence by estimating the two nuisance functions using separate samples. Therefore, the remainder term depends on the average bias of $\widehat{m}_{it}$ (which is of order $h_m$) rather than the $L_2$ norm of the estimation errors, allowing for undersmoothing the outcome regression. 

    \begin{corollary}[Inference]\label{corollary:estimation_ATT_simultaneous}
        Under conditions in Proposition~\ref{prop:estimation_entry_simultaneous}, set $h_\pi \asymp n^{-1/(d_{\alpha} + d_{X} + 2)}$ and $h_m \asymp n^{-1/(d_{\alpha} + d_{X})}(\log (n))^{2/(d_{\alpha} + d_{X})} $.  Then, for each $t=T_0, \ldots, T$ and $d_{\alpha}+ d_{X} \leq 3$, 
        \begin{align*}
            \sqrt n\left(\widehat{\mr{ATT}(t)}-\mr{ATT}(t)\right) \cd \mc{N}(0, V_t). 
        \end{align*}
        Here,  $V_t$ is the asymptotic variance of the oracle estimator.
    \end{corollary}
    Corollary~\ref{corollary:estimation_ATT_simultaneous} shows that, when $d_{\alpha}+d_{X}\leq 3$, combining double cross-fitting with undersmoothing yields a root-$n$ asymptotically normal and asymptotically unbiased ATT estimator, enabling valid inference on the dynamic ATT. I choose $h_\pi \asymp n^{-1/(d_{\alpha} + d_{X} + 2)}$, which is optimal for the MSE of the propensity score estimator, while aggressively undersmoothing $h_m$ because the outcome regression contributes to the remainder through its bias.  Corollary~\ref{corollary:estimation_ATT_simultaneous} expands the range of settings in which root-$n$ inference can be obtained relative to the  standard single cross-fitting, which establishes root-$n$ inference for the ATT estimator only when $d_{\alpha} + d_{X}=1$ (see \citet{deaner2025inferring}). 

    \begin{remark*}\textbf{(Other choices of bandwidths)} 
        Although Corollary~\ref{corollary:estimation_ATT_simultaneous} requires $h_m$ to be undersmoothed as aggressively as possible to achieve the fastest convergence rate of the remainder term, the bandwidth choice is not unique. It is straightforward to verify that there exists a range of bandwidth choices that make the remainder term $o_P(1/\sqrt{n})$ and therefore yield root-$n$ valid inference. I adopt this most aggressive undersmoothing choice because it leads to a concise theoretical characterization of the remainder term while also providing practical guidance for selecting bandwidths, which will be discussed later.
    \end{remark*}

    \begin{remark*}\textbf{(Discrete Covariates)} 
        Proposition~\ref{prop:estimation_entry_simultaneous}, Theorem~\ref{thm:estimation_ATT_simultaneous}, and Corollary~\ref{corollary:estimation_ATT_simultaneous} are primarily applicable to continuously distributed covariates. When $X_{it}$ takes discrete values, for example, in the binary response model as in Example~\ref{example:dynamic_nonlinear_interactive_fixed_effects}, the proposed estimator and the associated analysis remain valid, and the only difference is that $d_X$ does not enter the convergence rate asymptotically. Specifically, one only needs to replace all terms of the form $nh^{d_\alpha+d_X}$ with $nh^{d_\alpha}$ in the previous discussion to accommodate the discrete case. In addition, combining double cross-fitting with undersmoothing yields a root-$n$ asymptotically normal and asymptotically unbiased ATT estimator  when $d_{\alpha}\leq 3$. 
    \end{remark*}
    
    \begin{remark*}\textbf{(Asymptotics of $T_0$)} 
        Although Corollary~\ref{corollary:estimation_ATT_simultaneous} does not impose an explicit restriction on $T_0$, $T_0$ needs to grow sufficiently fast to achieve root-$n$ valid inference, as it affects the accuracy of $\widehat{d}_{ij}$ through $\delta_{n,T_0}$. A sufficient condition is that $T_0 \gg n^{2/(d_{\alpha}+d_{X})}$.
    \end{remark*}

    \begin{remark*}\textbf{(Parametric assumption)} 
        The proposed approach is nonparametric and does not impose functional form restrictions on the potential outcome or propensity score. If researchers are willing to impose parametric restrictions, for example, assuming that the potential outcome follows a linear dynamic model with interactive fixed effects or that the propensity score follows a logit model with a linear index, the dimensionality restrictions can be substantially relaxed. In addition, how to incorporate high-dimensional covariates into the outcome regression (see, e.g., \citet{viviano2026dynamic}), while controlling for individual fixed effects remains a promising direction for future research.
    \end{remark*}

\subsection{Bandwidth selection in practice}
    The theoretical bandwidth choice in Corollary~\ref{corollary:estimation_ATT_simultaneous} provides practical guidance for selecting $h_m$ and $h_\pi$ using data-driven methods. Since $h_\pi$ is set to minimize MSE of the propensity score estimator, I propose to use standard leave-one-out cross-validation. To elaborate, given $h >0$, for each $i =1, \ldots, n$, compute the within-sample leave-one-out estimator, 
    \begin{equation*}
        \widehat{\pi}_{i}^{(-i)}(h) = \sum_{i' \in \mc{I}(i)\setminus \{i\} }  D_{i'}  \widehat{K}_{h, ii'}  \big / \sum_{i' \in \mc{I}(i)\setminus \{i\}  }  \widehat{K}_{h, ii'}, 
    \end{equation*}
    and select the optimal bandwidth $h_{\pi}^*$ as $h_{\pi}^* =  \argmin_{h  > 0 } \frac{1}{n}\sum_{i=1}^{n}(D_i - \widehat{\pi}_{i}^{(-i)}(h)  )^2$. 

    Since Corollary~\ref{corollary:estimation_ATT_simultaneous} requires $h_m\asymp n^{-1/(d_{\alpha}+d_{X})}(\log(n))^{2/(d_{X} + d_{\alpha})} $ to be undersmoothed as aggressively as possible, this immediately means that the neighbors for each unit is of order $\sim (\log (n))^2$. Given any bandwidth $h>0$, for each $i$, I define the effective sample size as
    \begin{gather*}
        n_{i, \mr{eff}}(h) = \big(\sum_{i' \in \mc{I}(i)\setminus \{i\} }  (1 - D_{i'})  \widehat{K}_{h, ii'}\big)^2  \big /\sum_{i' \in \mc{I}(i)\setminus \{i\} }  (1 - D_{i'})  \widehat{K}^2_{h, ii'}, 
    \end{gather*}
    Following the theoretical guidance, a proper choice of $h_m$ should be as small as possible while maintaining a sufficient number of effective matches of order $\sim (\log (n))^2$. Therefore, in practice, I choose predetermined constants $\kappa>0$ and $\underline q\in(0,1]$ and select the smallest bandwidth that ensures at least a fraction $\underline q$ of observations have an effective sample size larger than $(\kappa \log(n))^2$: 
    \begin{align*}
        h_m^* = \min \left\{h >0: \frac{1}{n}\sum_{i=1}^{n}\bs{1}\left(n_{i,\mathrm{eff}}(h)>(\kappa \log(n))^2\right)\geq \underline q\right\}.
    \end{align*}
    Numerical simulations in Section~\ref{sec:simulation} show that $(\kappa=0.2, \underline{q} = 0.8 )$ performs well under various settings, and I recommend this value for empirical applications.

    \section{Extension: Staggered Adoption}\label{sec:extension}
    
    Section~\ref{sec:extension} extends the analysis to staggered adoptions, which are ubiquitous in empirical research. I adopt the idea of sequential unconfoundedness from \citet{robins2000marginal} and combine it with the pseudo-distance approach to identify and estimate dynamic ATT. The identification, estimation, and inference results developed for simultaneous treatment continue to hold under this extension with appropriate modifications.

\subsection{Setup and dynamic selection}
    ,
    Consider the panel data $\{(Y_{it}, X_{it}, D_{it}, \alpha_i)\}$ on $n$ units, indexed by $i=1, \ldots, n$, over $T$ periods, indexed by $t = 1, \ldots, T$. 
    The outcome variable is $Y_{it}\in\mb{R}$. The pretreatment covariates are $X_{it} \in \mb{R}^{d_X}$ and can include lagged outcomes, for example, $X_{it} = Y_{it-1}$. The treatment status is denoted by $D_i\in\{0,1\}$. I assume that treatment is \emph{absorbing}, i.e., $D_{it+1}\geq D_{it}$ for any $t$, in the sense that once a unit becomes treated, it remains treated in all subsequent periods. Let $G_i$ denote the first period in which unit $i$ receives the treatment, i.e., $D_{it} = \bs{1}\{t \ge G_i\}$.  For those individuals never treated in the observed time period, we set $G_i = \infty$. $\alpha_i$ is the unobservable individual fixed effect.

    To allow for dynamic treatment effects, potential outcomes are indexed by the entire treatment sequence $(d_1,...,d_T) \in \{0,1\}^T$, $Y_{it}(d_1,...,d_T)$. Since treatment is an absorbing state, the potential outcomes can be indexed by the first treatment period $G_i$ only. Here, we define $Y_{it}(g): = Y_{it}(\bs{0}_{g-1}, \bs{1}_{T-g+1})$, and $Y_{it}(\infty): = Y_{it}(\bs{0}_T)$. I follow the standard non-anticipation assumption such that 
    \begin{align*}
        Y_{it} = \left\{
        \begin{array}{lr}
            Y_{it}(\infty),  &  t < G_i\\
            Y_{it}(G_i),  & t\geq G_i
        \end{array}
        \right.
    \end{align*}
    For each $t \geq g \geq T_0$, the estimand of interest is the dynamic ATT: 
    \begin{align*}
        \mr{ATT}(g, t) := \mb{E}_T \left( Y_{it}(g) - Y_{it}(\infty) \mid G_i = g \right). 
    \end{align*} 
    It denotes the ATT at time $t$ for units that first receive the treatment in period $g$. As discussed in Section~\ref{sec:selection}, although the definition of the ATT here appears to differ from the standard definition, $\mb{E}(Y_{it}(g)-Y_{it}(\infty)\mid G_i=g)$, the two definitions are equivalent, and the difference is purely notational.

    \begin{assumption}[Selection]\label{assumption:selection_extension} Assume that for each $T_0\leq g\leq t\leq T$, 
    \begin{align*}
        Y_{it}(\infty) \perp \{G_{i} = g \} \mid X_{ig},  \alpha_i, \Gamma_{T}, G_i > g-1. 
    \end{align*}
    \end{assumption}

    The assumption extends Assumption~\ref{assumption:selection_extension} to staggered adoption, and is closely related to the sequential unconfoundedness assumption in the dynamic treatment effect literature. Assumption~\ref{assumption:selection_extension} requires that, for individuals who have not yet received the treatment before period $g$, treatment adoption at period $g$ is independent of future untreated potential outcomes, conditional on pretreatment covariates $X_{ig}$, individual fixed effects, and time factors. The conditional orthogonality is imposed only on untreated potential outcomes because the estimand of interest is the ATT, and only untreated outcomes need to be imputed.

    The following assumption imposes a Markov restriction. 
    \begin{assumption}[Markov]\label{assumption:surrogacy_extension}
        Let $H_{it}  := \{(Y_{i\tau}, X_{i\tau }, D_{i\tau})\}_{\tau \leq t}$ denote the history up to $t$,  
        \begin{align*}
            (Y_{it}(0), D_{it'} ) \perp H_{it' - 1} \mid X_{it'}, \alpha_i, \Gamma_T, G_i >t' -1, \quad \forall \; t = T_0, \ldots, T. 
        \end{align*}
    \end{assumption} 
    This assumption extends Assumption~\ref{assumption:surrogacy_simultaneous} to staggered adoption. It is stronger than necessary and can be relaxed to conditional moment restrictions on $(Y_{it}(0), D_{it'})$ (see the discussion in Section~\ref{sec:selection}). 
    Nevertheless, I retain the Markov restriction for ease of exposition.
    
    \begin{assumption}[Potential Outcome]\label{assumption:potential_outcome_extension}
        $Y_{it}(\infty)$ evolves according to~\eqref{eq:potential_outcome} and satisfies:  
        \begin{enumerate}[label=(\roman*)]
            \item \label{item:potential_outcome_extension_time_factor_exogeneity} \textbf{(Time factor exogeneity)} conditional on the individual latent factors $\alpha_i$ and the time-factor path $\Gamma_t$, $Y_{it}(\infty)$ is mean independent of future time factors, i.e., 
            \begin{align*}
                \mb{E}\left(Y_{it}(\infty) \mid X_{it}, \alpha_i, \Gamma_T\right) = \mb{E}\left(Y_{it}(\infty) \mid X_{it}, \alpha_i, \Gamma_t\right), \quad \forall t.  
            \end{align*}
            \item  \label{item:potential_outcome_extension_weak_dependence} \textbf{(Weak dependence)} $(\ldots, Y_{it-1}(\infty), Y_{it }(\infty), \ldots )$ are conditionally weakly dependent,  such that $\sum_{s =-\infty}^{\infty} \left|\mr{Cov} \left(Y_{it}(\infty), Y_{is}(\infty)\mid  \alpha_{i} = \alpha,  \Gamma_{T} = \Gamma \right)\right| <\infty$ uniformly for any supported $\alpha$, any possible time path $\Gamma$, and any $t \in \mb{Z}$. 
        \end{enumerate}
    \end{assumption}

    Assumption~\ref{assumption:potential_outcome_extension} is identical to Assumption~\ref{assumption:potential_outcome_simultaneous}, except that $Y_{it}(0)$ is replaced by $Y_{it}(\infty)$ to denote the untreated potential outcomes under staggered adoption.

    Similar to~\eqref{eq:expected_outcome_ps_simultaneous}, for each $T_0 \leq t'\leq t\leq T$ and $\alpha \in \mc{A}$, define 
    \begin{equation}\label{eq:expected_outcome_ps_extension}
    \begin{aligned}
        m_{t \mid t'}(x, \alpha ) := &\mb{E}_T(Y_{it}(\infty) \mid X_{it'} = x, \alpha_i = \alpha, G_i > t'-1), \\
        \pi_{t'}(x, \alpha ) := &\mb{P}(G_i=t' \mid X_{it'} = x, \alpha_i=\alpha, G_i>t'-1).
    \end{aligned}   
    \end{equation}
    The function $m_{t\mid t'}(x,\alpha)$ is the conditional expectation of the untreated outcome at period $t$ for units that remain untreated at time $t'-1$, given their pretreatment information summarized by $X_{it'} = x$ and latent heterogeneity $\alpha_i=\alpha$. 
    The function $\pi_{t'}(x,\alpha)$ is the propensity score of treatment adoption at time $t'$ for units that have not been treated by time $t'-1$, given observed covariates $X_{it'} = x$, and individual fixed effects $\alpha_i=\alpha$.

\subsection{Identification}

    The identification of the pseudo-distance and its informativeness under staggered adoption is identical to that under simultaneous treatment (see Assumptions~\ref{assumption:identification_d_simultaneous} and \ref{assumption:informativeness_identification_simultaneous} and Proposition~\ref{prop:identification_d_simultaneous}), except that $Y_{it}(0)$ is replaced by $Y_{it}(\infty)$ to accommodate staggered adoption. Therefore, I omit the discussion of pseudo-distance identification and informativeness and focus directly on the identification of the dynamic ATT.

    \begin{assumption}[Identification of ATT]\label{assumption:identification_ATT_extension}
            I assume that: 
            \begin{enumerate}[label=(\roman*)]
            \item \label{item:identification_ATT_extension_finite_moment} \textbf{(Moment)} For each $t\geq g \geq T_0$,  $\mb{E}_{T}\left( | Y_{it}(\infty)| \mid G_i = g \right), \mb{E}_{T}\left( | Y_{it}(g)| \mid G_i = g \right)<\infty $. 
            \item \label{item:identification_ATT_extension_smoothness}  \textbf{(Smoothness)} For each $T_0\leq t'\leq t\leq T$, $m_{t\mid t'}(x,\alpha)$ is uniformly continuous in $(x, \alpha)$,  i.e., for any $\epsilon >0$, there exists $\delta >0$ such that for all $(x, \alpha) \in \mr{supp} (X_{it'}, \alpha_i )$ 
            \begin{align*}
                \sup_{  \|(x', \alpha') - (x, \alpha) \| \leq \delta  } \left| m_{t\mid t'}(x, \alpha ) - m_{t\mid t'}(x', \alpha' ) \right| \leq \epsilon. 
            \end{align*}
            \item \label{item:identification_ATT_extension_overlap}  \textbf{(Overlap)} $\mb{P}_T(G_i = t\mid X_{it},\alpha_i, G_i >t-1)\in(0,1)$ a.s. 
        \end{enumerate}
    \end{assumption}

    Assumption~\ref{assumption:identification_ATT_extension}\ref{item:identification_ATT_extension_finite_moment} is standard and imposes finite first-moment conditions on both untreated and treated potential outcomes. Assumption~\ref{assumption:identification_ATT_extension}\ref{item:identification_ATT_extension_smoothness} extends Assumption~\ref{assumption:identification_ATT_simultaneous}\ref{item:identification_ATT_simultaneous_smoothness} by imposing uniform continuity on the  expected untreated outcomes $m_{t\mid t'}(\cdot, \cdot)$. Uniform continuity of $m_{t\mid t'}(\cdot, \cdot)$ is a mild and standard condition commonly imposed for nonparametric identification of conditional expectations.
    Assumption~\ref{assumption:identification_ATT_extension}\ref{item:identification_ATT_extension_overlap} strengthens Assumption~\ref{assumption:identification_ATT_simultaneous}\ref{item:identification_ATT_simultaneous_overlap} and is commonly required in causal inference to identify counterfactual outcomes.

    \begin{theorem}[Identification]\label{thm:identification_extension} 
        Suppose that the population pseudo-distance $d(\alpha_i,\alpha_j)$ is identified for each $i,j=1,\ldots,n$. Then, under Assumptions~\ref{assumption:selection_extension}-\ref{assumption:identification_ATT_extension} and Assumption~\ref{assumption:informativeness_identification_simultaneous}, the following identification results hold: (i) For each $t\geq T_0$ and each unit $i$ such that $G_i>t-1$, the contemporaneous counterfactual is identified by 
        \begin{align*}
            m_{t\mid t}(X_{it}, \alpha_i)= \lim_{\delta\downarrow 0}\mb{E}_{T}(Y_{jt} \mid X_{jt} = X_{it}, d(\alpha_j, \alpha_i)\leq \delta, G_j >t);
        \end{align*}
        (ii) for any $t' = t-1, \ldots, g$ and each unit $i$ such that $G_i>t' -1 $, the dynamic counterfactual expectation is identified recursively backward by
        \begin{align*}
            m_{t\mid t'}(X_{it'}, \alpha_i) = \lim_{\delta \downarrow 0} \mb{E}_{T}\left( m_{t\mid t'+1 } (X_{jt' + 1}, \alpha_j)\mid X_{jt'} = X_{it'}, d(\alpha_j , \alpha_i)\leq \delta, G_j >t' \right);
        \end{align*}
        and (iii) for any $t\geq g\geq T_0$, the dynamic ATT is identified by 
        \begin{align*}
            \mr{ATT}(g, t) = \mb{E}_{T}\left( Y_{it}\mid G_i = g\right) - \mb{E}_{T}\left( m_{t\mid g}(X_{ig}, \alpha_i) \mid G_i = g\right). 
        \end{align*}
    \end{theorem}

    Theorem~\ref{thm:identification_extension} extends the identification result to staggered-adoption settings and establishes the identification of $\mr{ATT}(g,t)$. It uses a pseudo-distance-based matching strategy with a backward recursive procedure to accommodate dynamic selection. 
    The theorem proceeds in three steps. First, it identifies the contemporaneous counterfactual expectation $m_{t\mid t}(\cdot,\cdot)$. Second, it identifies dynamic counterfactual expectations $m_{t\mid t'}(\cdot,\cdot)$ through a backward recursive procedure over $t'=t,t-1,\ldots,g$. Finally, averaging the imputed counterfactual $m_{t\mid g}(\cdot,\cdot)$ over the distribution of units being treated at $g$ identifies $\mr{ATT}(g,t)$. 
    
\subsection{Estimator}

    For notational simplicity, for each $t \geq t' \geq T_0$ and each $i$ who has not been treated at time $t'-1$, define its expected (counterfactual) potential outcome as $m_{i, t\mid t'} := m_{t \mid t'}(X_{it'}, \alpha_i )$. In addition, for each $t = T_0, \ldots, T$ and each $i$ such that $G_i>t-1$, define its propensity score as $\pi_{it} = \pi_t(X_{it}, \alpha_i  )$.  

    An oracle estimator for $\mr{ATT}(g,t)$ under staggered adoption takes the following doubly robust form:
    \begin{equation}\label{eq:ATT_DR_extension}
    \begin{aligned}
        \widehat{ATT(g, t)}^{\mr{oracle}} = \frac{1}{n_g} \sum_{i = 1}^{n} \Bigg[ & \bs{1}\left(G_i = g\right) \left(Y_{it} - {m}_{i, t\mid g}\right) \\
        & - \sum_{t' = g}^{t}\bs{1}\left(G_i > t'\right) \frac{{\pi}_{ig}}{1-  {\pi}_{ig}} \left(\prod_{r = g+1}^{t'} \frac{1}{1 - {\pi}_{ir}}\right)\left( {m}_{i, t\mid t'+1} - {m}_{i, t\mid t'} \right) \Bigg]. 
    \end{aligned}
    \end{equation}
    Here, (i) $n_g:=\sum_{i=1}^{n}\bs{1}(G_i=g)$ denotes the number of units first treated in period $g$; (ii) $m_{i,t\mid t+1}:=Y_{it}$ is defined for notational convenience; and (iii) an empty product is defined as one, so that $\prod_{r=g+1}^{t'}(1-\pi_{ir})^{-1}=1$ when $t'=g$. This is an oracle estimator in the sense that it uses the true propensity scores and outcome regressions, which are unknown in practice. As in the simultaneous treatment case, I use Nadaraya-Watson estimators based on the pseudo-distance to nonparametrically estimate $m_{i,t\mid t'}$ and $\pi_{it'}$, and then employ a \emph{double} cross-fitting procedure to achieve a faster convergence rate. 

    The full algorithm, similar to the identification strategy, involves a backward procedure and is tedious to present in the main text. Therefore, a simplified version that abstracts from cross-validation for selecting the bandwidths ${h_m,h_\pi}$ is summarized in Algorithm~\ref{alg:extension_basic} to highlight the main idea. The cross-validation procedure for bandwidth selection is discussed in the following text, and the full algorithm is presented in Algorithm~\ref{alg:extension_cv}.

\subsection{Estimation}

    The convergence rate of the pseudo-distance and its informativeness under staggered adoption are identical to those under simultaneous treatment (see Assumptions~\ref{assumption:estimation_consistency_d} and \ref{assumption:informativeness_estimation}), except that $Y_{it}(0)$ needs to be replaced by $Y_{it}(\infty)$ to accommodate dynamic selection. Therefore, I omit the discussion of the accuracy of the sampled pseudo-distance  and informativeness and focus directly on the inference on the dynamic ATT.

    \begin{assumption}[Estimation]\label{assumption:estimation_extension} I assume that 
        \begin{enumerate}[label=(\roman*)]
            \item \label{item:estimation_extension_panel_weak_dependent} \textbf{(Sampling)} Conditional on $\{\gamma_t\}_{t \in \mb{Z}}$, the panel $\{(Y_{it}, X_{it}, D_{it}, \alpha_i)\}_{i=1, \ldots, n, t = 1, \ldots, T}$ is i.i.d. across $i$. The panel is large with $n\rightarrow\infty$ and $T_0\rightarrow\infty$.  
            \item \label{item:estimation_extension_finite} \textbf{(Moment)} Suppose $X_{it}$ is uniformly bounded across $i$ and $t$. In addition, for some $\sigma>0$, $\mb{E}_T\left(\exp\left(\lambda(Y_{it}(\infty)-m_{i, t\mid t'})\right)\mid X_{it'},\alpha_i, G_i >t'-1\right)\leq \exp(\sigma^2\lambda^2/2)$ a.s. for all $\lambda\in\mb{R}$, all $i$, and all $t,t'$ satisfying $T_0\leq t'\leq t\leq T$. Lastly, for some $\delta>0$, $\mb{E}_T\left(|Y_{it}(g)|^{2+\delta}\right)<\infty$ for all $g,t$ satisfying $T_0\leq g\leq t\leq T$. 
            \item \label{item:estimation_extension_compact}  \textbf{(Compactness)}  The support of $\alpha$, $\mc{A}$, is compact. There exist constants $\underline{c}_1, \overline{c}_1 >0$, such that for any $\alpha \in \mc{A}$ and any $ r \in (0,  1)$, $\underline{c}_1 r^{d_{\alpha}} \leq \mb{P}\left(\|\alpha_i - \alpha\| \leq r \right) \leq \overline{c}_1 r^{d_{\alpha}}$. 
            In addition, for each $t = T_0, \ldots, T$, let $\mc S_{t-1}$ be the support of $(X_{it},\alpha_i)$  conditional on $\{G_i > t-1\}$. 
            There exist constants $0<\underline{c}_2\leq\overline {c}_2<\infty$ such that, for any $t = T_0, \ldots, T$, any $r\in (0, 1)$,  and any $(x,\alpha)\in\mc{S}_{t-1} $, 
            \begin{align*}
                \underline{c}_2 r^{d_X + d_{\alpha}} \leq \mb{P}\left(\|X_{it}-x \| \leq r, \|\alpha_i-\alpha\|\leq r, G_i > t-1  \right)\leq  \overline c_2 r^{d_X + d_{\alpha}}. 
            \end{align*}
            \item \label{item:estimation_extension_L_continuous} \textbf{(Lipschitz)} For each $T_0 \leq t'\leq t\leq T$, let $\mc S_{t'-1}$ be the support of $(X_{it'},\alpha_i)$ conditional on $\{G_i >t'-1\}$. Assume $m_{t \mid t'}(x, \alpha)$ and $\pi_{t'}(x, \alpha)$ are Lipschitz continuous on $\mc{S}_{t'-1}$. 
            \item \label{item:estimation_extension_overlap} \textbf{(Common overlap)} There exist constants  $\underline{p}, \overline{p}\in (0, 1)$ such that for any $t =  T_0, \ldots, T$, $\underline{p} \leq  \pi_{t}(X_{it}, \alpha_i)\leq \overline{p}$ a.s. 
            \item \label{item:estimation_extension_kernel} \textbf{(Kernel)} The kernel $K: \mb{R}\mapsto \mb{R}_{+}$ is bounded by $\overline{K}$ and supported on $[-1, 1]$.   In addition, $K(0)>0$ and  $K(\cdot)$ is Lipschitz continuous with constant $L_K>0$. 
        \end{enumerate}
    \end{assumption}

    Assumption~\ref{assumption:estimation_extension}\ref{item:estimation_extension_panel_weak_dependent} and Assumption~\ref{assumption:estimation_extension}\ref{item:estimation_extension_kernel} are identical to those conditions in the simultaneous treatment case. Assumption~\ref{assumption:estimation_extension}\ref{item:estimation_extension_finite} and \ref{item:estimation_extension_compact} extend  Assumption~\ref{assumption:estimation_simultaneous}\ref{item:estimation_simultaneous_finite} and \ref{item:estimation_simultaneous_compact} to staggered adoption, respectively.  
    Assumption~\ref{assumption:estimation_extension}\ref{item:estimation_extension_L_continuous} is similar to  Assumption~\ref{assumption:estimation_simultaneous}\ref{item:estimation_simultaneous_L_continuous}, but extends 
    continuity condition to the dynamic outcome regression functions $m_{t\mid t'}(\cdot,\cdot)$ and the propensity scores. Lastly, Assumption~\ref{assumption:estimation_extension}\ref{item:estimation_extension_overlap} is a common overlap condition imposed for treatment adoption at different periods.

    \begin{proposition}[Convergence rates for counterfactual imputation]\label{prop:estimation_entry_extension}
        Under Assumptions~\ref{assumption:estimation_consistency_d}-\ref{assumption:estimation_simultaneous}, \ref{assumption:selection_extension}-\ref{assumption:potential_outcome_extension}, and~\ref{assumption:estimation_extension}, for any $h\in\{h_{\pi},h_m\}$ such that $h\rightarrow 0$, $nh^{d_\alpha+d_{X}}/\log n\rightarrow\infty$, and $\delta_{n,T_0}/h\rightarrow 0$, we have that for each $T_0 \leq t'\leq t\leq T$, 
        \begin{align*}
            \max_{i: G_i >t'-1}\left|\widehat{m}_{i, t\mid t'} - m_{i, t\mid t'} \right| = O_p\left(h_m +  \left(nh_m^{d_\alpha+d_{X}}\right)^{-1/2}\sqrt{\log (n)} \right).  
        \end{align*}
        In addition, for each $t = T_0, \ldots, T$, 
        \begin{align*}
            \max_{i: G_i >t-1}\left|\widehat{\pi}_{it} - \pi_{it} \right| = O_p\left(h_{\pi} +  \left(nh_{\pi}^{d_\alpha+d_{X}}\right)^{-1/2}\sqrt{\log (n)} \right). 
        \end{align*}
    \end{proposition}
    Proposition~\ref{prop:estimation_entry_extension} establishes the convergence rates for counterfactual outcome estimators and propensity score estimators. The convergence rates are identical to those in the simultaneous treatment case.

    \begin{theorem}[Asymptotic normality]\label{thm:estimation_ATT_extension}
        Under conditions in Proposition~\ref{prop:estimation_entry_extension},  for each $T_0 \leq g \leq t\leq T$, 
        \begin{align*}
            \widehat{\mr{ATT}}(g, t) - \mr{ATT}(g, t) = \frac{1}{\sqrt{n}}  \mc{N}(0, V_{g, t}) + O_p\left(h_m \left(h_{\pi} + (nh_{\pi}^{d_{\alpha} + d_{X}})^{-1/2}\sqrt{\log(n)}\right)\right).  
        \end{align*}
        Here, $\mc{N}(\cdot,\cdot)$ denotes the normal distribution, and $V_{g, t}$ is the asymptotic variance of the oracle estimator.
    \end{theorem}
    Theorem~\ref{thm:estimation_ATT_extension} shows that, similar to the simultaneous treatment case, combining double cross-fitting with undersmoothing of $h_m$ yields a faster convergence rate than standard single cross-fitting.

    \begin{corollary}[Inference]\label{corollary:estimation_ATT_extension}
        Under conditions in Proposition~\ref{prop:estimation_entry_extension}, set $h_\pi \asymp n^{-1/(d_{\alpha} + d_{X} + 2)}$ and $h_m \asymp n^{-1/(d_{\alpha} + d_{X})}(\log (n))^{2/(d_X + d_{\alpha})} $. In addition, suppose $\delta_{n,T_0}/h_m \rightarrow 0$.  Then, for each $T_0\leq g \leq t\leq T$ and $d_{\alpha}+d_{X} \leq 3$, 
        \begin{align*}
            \sqrt n\left(\widehat{\mr{ATT}(g, t)}-\mr{ATT}(g, t)\right) \cd \mc{N}(0, V_{g, t}). 
        \end{align*}
        Here, $\mc{N}(\cdot,\cdot)$ denotes the normal distribution, and $V_{g, t}$ is the asymptotic variance of the oracle estimator.
    \end{corollary}
    Corollary~\ref{corollary:estimation_ATT_extension} shows that, when $d_{\alpha}+d_{X}\leq 3$, combining double cross-fitting with undersmoothing yields a root-$n$ asymptotically normal and asymptotically unbiased estimator of the dynamic ATT, enabling valid inference.

\subsection{Bandwidth selection in practice}
    The theoretical bandwidth choice in Corollary~\ref{corollary:estimation_ATT_extension} provides practical guidance for selecting $h_m$ and $h_\pi$ using data-driven methods. Similar to the simultaneous treatment case, the choice of $h_\pi$ optimizes the MSE of the propensity score estimator, while $h_m$ is aggressively undersmoothed. 
    
    I propose to use standard leave-one-out cross-validation for selecting $h_\pi$ and the effective number of matched neighbors for selecting $h_m$. The bandwidth selection procedure is almost identical to that in the simultaneous treatment case, except that it is implemented separately for each period. The full procedure is lengthy and is therefore omitted from the main text. The complete algorithm, including the bandwidth selection procedure, is provided in Algorithm~\ref{alg:extension_cv}.

    \section{Simulation}\label{sec:simulation}

    In this section, I illustrate the finite-sample properties of the proposed method under different DGPs, including dynamic models with additive fixed effects, dynamic models with interactive fixed effects, and nonlinear dynamic models.

    I report three estimators. (i) \textbf{DID}: the canonical difference-in-differences estimator. (ii) $\bs{\mr{DR2}}$: the doubly robust estimator based on a standard two-way sample split, with $m$ and $\pi$ estimated using the same training sample. The bandwidths for $m$ and $\pi$ are selected by leave-one-out cross-validation to minimize  mean squared errors. I also use $\mr{DR2}^*$ to denote the infeasible estimator that uses the true distance between individual latent factors rather than the pseudo-distance. (iii) $\bs{\mr{DR3}}$: the proposed estimator using a three-way sample split, with $m$ and $\pi$ estimated on separate training samples. I also use $\mr{DR3}^*$ to denote the infeasible estimator that uses the true distance between individual latent factors rather than the pseudo-distance. I vary $\kappa$ to examine how different choices affect the estimator's performance. 

\subsection{Dynamic panel with additive fixed effects}

    Consider the following data-generating process:
    \begin{equation}\label{eq:sim_TWFE_Y_simultaneous}
    \begin{aligned}
        Y_{it} = 
        \left\{
            \begin{array}{lr}
            \rho Y_{it-1} + \alpha_i + \gamma_t + \epsilon_{Y, it}, & t < T_0 \\
            \rho Y_{it-1} + \alpha_i + \gamma_t + 0.5 \cdot  D_i + \epsilon_{Y, it}, & t \geq T_0 \\
        \end{array}
        \right..  
    \end{aligned}   
    \end{equation} 
    The individual latent factors $\{\alpha_i\}_{i=1}^{n}$ and the time latent factors $\{\gamma_t\}_{t\leq T}$ are independent random variables drawn from the uniform distribution on $[-1/4,1/4]$. The error terms $\{\epsilon_{Y,it}\}_{i=1,\ldots,n,\;t=0,\ldots,T}$ are independent of the latent factors and are i.i.d. across both dimensions, following the uniform distribution on $[-1/2,1/2]$. I set $\rho=0.8$ to capture the dynamic effect. The contemporaneous ATT is $0.5$, and for each $t>T_0$, the dynamic treatment effect is $0.5\left(1+\rho+\ldots+\rho^{t-T_0}\right)$. Treatment assignment depends jointly on latent heterogeneity and the lagged outcome:
    \begin{align}\label{eq:sim_TWFE_selection_simultaneous}
        D_i = \bs{1}\left( \alpha_i/2  + Y_{i T_0 - 1}/2 +  \epsilon_{D, i} \geq 0\right). 
    \end{align}
    Here, the error terms $\{\epsilon_{D,i}\}_{i=1}^{n}$ are i.i.d. across individuals and are independent of the latent factors and $\{\epsilon_{Y,it}\}_{i=1,\ldots,n,\;t=0,\ldots,T}$. In the numerical designs, $\epsilon_{D,i}$ follows a standard logistic distribution. I vary $(N,T_0)\in\left\{(1000,20),(1000,40), (200,100),(1000,100),(2000,100)\right\}$ and perform $5{,}000$ replications for each design.

    Table~\ref{tab:TWFE_simultaneous} reports the simulation results for the contemporaneous ATT and shows that the proposed estimator performs well across all designs. Relative to DID and the standard doubly robust estimator, the proposed estimator substantially reduces bias and  has coverage close to the nominal 95\% level. In addition,  the proposed estimator using the pseudo-distance performs similarly to the infeasible DR3$^*$ using the true latent distance, indicating that the estimated pseudo-distance provides a good approximation to the true latent distance even when $T_0=20$. It is worth noting that, when $N=200$, DR2 has a smaller standard deviation and coverage closer to the nominal level, although the proposed estimator has a smaller absolute bias. This is because the double cross-fitting procedure leaves relatively few observations in each subsample. Finally, the performance of the proposed estimator is also stable across different values of $\kappa$, and I recommend $(\kappa,\underline{q})=(0.2,0.8)$ as the default choice.

\subsection{Dynamic panel with interactive fixed effects}

    Consider the following data-generating process:
    \begin{equation}\label{eq:sim_IFE_Y_simultaneous}
    \begin{aligned}
        Y_{it} = 
        \left\{
            \begin{array}{lr}
            \rho Y_{it-1} + \alpha_i  \gamma_t + \epsilon_{Y, it}, & t < T_0 \\
            \rho Y_{it-1} + \alpha_i  \gamma_t +  0.5 \cdot  D_i + \epsilon_{Y, it}, & t \geq T_0 \\
        \end{array}
        \right..  
    \end{aligned}   
    \end{equation}

    \begin{table}[H]
    \centering
    \caption{Simulation Results: Dynamic Panel with Additive Fixed Effects}\label{tab:TWFE_simultaneous}
        \begin{tabular}{cccccccccc}
        \toprule
            & DID   & $\mr{DR2}^*$   & DR2   & \multicolumn{3}{c}{$\mr{DR3}^*$ } & \multicolumn{3}{c}{DR3} \\
            \cmidrule(lr){5-7} \cmidrule(lr){8-10} &   &   &   &$\kappa=.15$ & {$\kappa=.2$} & {$\kappa=.25 $} &$\kappa=.15$ & {$\kappa=.2$} & {$\kappa=.25 $} \\
        \midrule
        $N= 1000, T_0= 20$ &       &       &       &       &       &       &       &       &  \\
        Bias  &   2.16    & 0.50  & 0.79  &  0.13  &  0.18  &  0.27  & 0.43  &  0.49  &  0.61  \\
        SD   &    1.94   & 2.05  & 2.10  &  2.47  &  2.40  & 2.30   & 2.53  &  2.43  & 2.33  \\
        Coverage  &     79.8   & 93.9 & 92.9 & 94.2   &  94.2  &  93.9  & 93.8  & 93.7   & 93.3 \\

        \addlinespace[5pt]
        $N= 1000, T_0= 40$ &       &     &       &       &       &       &       &       &  \\
        Bias  &  2.16     &  0.53       &  0.81     &   0.16    &    0.21   &   0.31    &    0.43   &   0.52    &  0.63 \\
        SD   &   1.94    &    2.05   &    2.11   &    2.47   &  2.39     &   2.29    &    2.40   &   2.40    &  2.33 \\
        Coverage &   79.8    &    93.6   &    92.8   &    94.2   &    93.9   &    94.0   &   94.4    &    94.2   &  93.7 \\
        
        \addlinespace[5pt]
        $N= 200, T_0= 100$ &       &       &       &       &       &       &       &       &  \\
        Bias  &  2.16     & 0.90  & 1.16   &  0.47  &  0.49  &  0.66  & 0.81  &   0.83  &  0.95  \\
        SD  &   1.94    & 5.37  & 5.44  &  6.75  & 6.92   &  6.77  &  7.06 &  7.01  & 6.86 \\
        Coverage &    79.8    & 93.9 & 93.5 &  93.1  &  92.8  & 93.0   &  93.6 &  93.5   & 93.3  \\

        \addlinespace[5pt]
        $N= 1000, T_0= 100$ &       &       &       &       &       &       &       &       &  \\
        Bias  &   2.16    & 0.50  & 0.71  &   0.11 &  0.17  &  0.27  & 0.35  &  0.44  &  0.55  \\
        SD   &    1.94   & 2.05  & 2.08  &  2.49  &  2.39  &  2.32   &  2.51  &  2.41  &  2.33  \\
        Coverage &   79.8     & 94.3 & 93.7 &  94.3  & 94.9   &  94.3  & 94.4  &  94.0  &  93.8 \\

        \addlinespace[5pt]
        $N= 2000, T_0= 100$  &       &       &       &       &       &       &       &       &  \\
        Bias  &  2.16     &   0.36    &   0.55    &  0.10  &  0.14   &  0.19  &  0.29 &   0.37 & 0.42 \\
        SD   &    1.94   &     1.44  &    1.46  &    1.69   &  1.62     &     1.57  &  1.72     &  1.63     &  1.59  \\
        Coverage  &  79.8  &    93.8    &    92.5   &  94.0  &  94.0  &  94.1 &  94.2 &  94.2  & 93.6 \\
        \bottomrule
        \end{tabular}%
        \vspace{0.3cm}
        \begin{minipage}{\textwidth}
            \footnotesize 
            \textbf{Note:} The table reports Monte Carlo results for the contemporaneous ATT based on $5000$ replications of the dynamic panel model with additive fixed effects specified in~\eqref{eq:sim_TWFE_Y_simultaneous} and the treatment selection mechanism specified in~\eqref{eq:sim_TWFE_selection_simultaneous}. I fix $\underline{q}=80\%$ and vary $\kappa \in \{.15,.2,.25\}$ when selecting the bandwidth $h_m^*$ according to Algorithm~\ref{alg:simultaneous_cv}. I report the bias, standard deviation, and coverage rate for the difference-in-differences estimator ($\mr{DID}$), the infeasible doubly robust estimator based on a standard two-way sample split and the true latent distance ($\mr{DR2}^*$), and its feasible counterpart based on the estimated pseudo-distance ($\mr{DR2}$). I also report the bias, standard deviation, and coverage rate for the infeasible estimator based on a three-way sample split and the true latent distance ($\mr{DR3}^*$), as well as the proposed estimator based on a three-way sample split and the estimated pseudo-distance ($\mr{DR3}$). 
            Biases and standard deviations are reported in units of $\bs{0.01}$, and coverage rates are reported in percentage points $\bs{1\%}$. 
        \end{minipage}
    \end{table}

    The individual latent factors $\{\alpha_i\}_{i=1}^{n}$ and the time latent factors $\{\gamma_t\}_{t\leq T}$ are independent random variables drawn from uniform distributions on $[-1,1]$ and $[-2,2]$, respectively. The error terms $\{\epsilon_{Y,it}\}_{i=1,\ldots,n,\;t=0,\ldots,T}$ are independent of the latent factors and are i.i.d. across both dimensions, following a uniform distribution on $[-1/2,1/2]$. I set $\rho=0.8$ to capture the dynamic effect. The contemporaneous ATT is $0.5$, and for each $t>T_0$, the dynamic treatment effect is $0.5\left(1+\rho+\ldots+\rho^{t-T_0}\right)$. Treatment assignment depends jointly on latent heterogeneity and the lagged outcome: 
    \begin{align}\label{eq:sim_IFE_selection_simultaneous}
        D_i = \bs{1}\left( \alpha_i/2  + Y_{i T_0 - 1}/2 +  \epsilon_{D, i} \geq 0\right). 
    \end{align}

    Here, the error terms $\{\epsilon_{D,i}\}_{i=1}^{n}$ are i.i.d. across individuals and are independent of the latent factors and $\{\epsilon_{Y,it}\}_{i=1,\ldots,n,\;t=0,\ldots,T}$. In the numerical designs, $\epsilon_{D,i}$ follows a standard logistic distribution. I vary $(N,T_0)\in\left\{(1000,20),(1000, 40), (200,100),(1000,100),(2000,100)\right\}$ and perform $5000$ replications for each design. 

    Table~\ref{tab:IFE_simultaneous} reports the simulation results for the contemporaneous ATT under interactive fixed effects (the DiD estimator is omitted because the DGP does not admit an additive two-way fixed-effects representation). The table shows that the proposed estimator reduces bias and achieves coverage close to the nominal 95\% level. Also, the proposed estimator performs similarly to the infeasible DR3$^*$ estimator, indicating that the estimated pseudo-distance provides a good approximation to the true latent distance even when $T_0=20$ under the interactive fixed effects model. When $N=200$, the performance of the proposed estimator is exceeded by that of the standard doubly robust estimator because the double cross-fitting procedure leaves relatively few observations in each subsample. Still, I recommend $(\kappa,\underline{q})=(0.2,0.8)$ as the default choice.

    \begin{table}[H]
    \centering
    \caption{Simulation Results: Dynamic Panel with Interactive Fixed Effects}\label{tab:IFE_simultaneous}
        \begin{tabular}{ccccccccc}
        \toprule
             & $\mr{DR2}^*$   & DR2   & \multicolumn{3}{c}{$\mr{DR3}^*$ } & \multicolumn{3}{c}{DR3} \\
            \cmidrule(lr){4-6} \cmidrule(lr){7-9} &     &   &$\kappa=.15$ & {$\kappa=.2$} & {$\kappa=.25$} & {$\kappa=.15$} & {$\kappa=.2$} & {$\kappa=.25$} \\
        \midrule
        $N= 1000, T_0= 20$ &             &       &       &       &       &       &       &  \\
        Bias  &         0.60  &  0.82 &   0.24    &   0.34    &    0.46  &   0.46    &  0.57     &  0.69 \\
        SD   &        2.27 & 2.56  &  2.70     &     2.61  &  2.53     &     3.00  &  2.90     & 2.82  \\
        Coverage          & 93.5 & 92.1 &    93.6   &    93.6   &  93.5     &   93.1    &    92.8   &  92.3 \\

        \addlinespace[5pt]
        $N= 1000, T_0= 40$ &             &       &       &       &       &       &       &  \\
        Bias  &         0.63  &  0.80 &   0.30    &   0.38    &    0.49  &   0.51    &  0.59     &  0.71 \\
        SD   &        2.29 & 2.49  &  2.70     &     2.65  &  2.56     &     2.87  &  2.81     & 2.76  \\
        Coverage          & 93.5 & 92.1 &    93.8   &    93.6   &  93.3     &   93.4    &    92.9   &  92.3 \\

        \addlinespace[5pt]
        $N= 200, T_0= 100$ &              &       &       &       &       &       &       &  \\
        Bias  &         1.06  &  1.16 &   0.86   &   0.90    &    1.07   &   0.91    &  0.94     & 1.18 \\
        SD   &         6.11 & 6.29  &   8.12    &   8.09    &   7.97    &  8.29     &   8.25    & 8.17 \\
        Coverage  &          93.9 & 93.8 &   92.9    &    92.6   &   92.4    &   92.8    &   92.8    &  92.7 \\

        \addlinespace[5pt]
        $N= 1000, T_0= 100$ &       &              &       &       &       &       &       &  \\
        Bias  &        0.62  & 0.75  &   0.30    &    0.36   &    0.50   &   0.41    &   0.50    &  0.62 \\
        SD   &       2.29  &  2.35  &    2.74   &    2.65   &     2.58  &    2.83   &   2.74    &  2.68 \\
        Coverage  &          93.3 & 92.7  &   93.7    &    93.5   &    93.1   &   93.8    &   93.4    & 92.6 \\

        \addlinespace[5pt]
        $N= 2000, T_0= 100$  &              &       &       &       &       &       &       &  \\
        Bias  &        0.40 &  0.52 &   0.13    &   0.20    &   0.27    &    0.25   &   0.32    &  0.40 \\
        SD   &        1.56 &  1.62 &    1.83   &   1.76    &  1.71     &   1.92    &   1.84    &  1.80 \\
        Coverage  &         93.7 & 92.6 &   94.2    &    93.9   &    93.7   &   93.8    &    93.3   &  93.2 \\
        \bottomrule
        \end{tabular}%
        \vspace{0.3cm}
        \begin{minipage}{\textwidth}
            \footnotesize 
            \textbf{Note:} The table reports Monte Carlo results for the contemporaneous ATT based on $5000$ replications of the dynamic panel model with interactive fixed effects specified in~\eqref{eq:sim_IFE_Y_simultaneous} and the treatment selection mechanism specified in~\eqref{eq:sim_IFE_selection_simultaneous}. I fix $\underline{q}=80\%$ and vary $\kappa \in \{.15,.2,.25\}$ when selecting the bandwidth $h_m^*$ according to Algorithm~\ref{alg:simultaneous_cv}. I report the bias, standard deviation, and coverage rate for  the infeasible doubly robust estimator based on a standard two-way sample split and the true latent distance ($\mr{DR2}^*$), and its feasible counterpart based on the estimated pseudo-distance ($\mr{DR2}$). I also report the bias, standard deviation, and coverage rate for the infeasible estimator based on a three-way sample split and the true latent distance ($\mr{DR3}^*$), as well as the proposed estimator based on a three-way sample split and the estimated pseudo-distance ($\mr{DR3}$). 
            Biases and standard deviations are reported in units of $\bs{0.01}$, and coverage rates are reported in percentage points $\bs{1\%}$. 
        \end{minipage}
    \end{table}

    \subsection{Nonlinear dynamic panel with fixed effects}

    Consider the following data-generating process:
    \begin{equation}\label{eq:sim_NL_Y_simultaneous}
    \begin{aligned}
        Y_{it} = 
        \left\{
            \begin{array}{lr}
            \bs{1}\left(-2 + \rho Y_{it-1} + 4 \alpha_i + \epsilon_{Y, it} \geq 0 \right), & t < T_0 \\
            \bs{1}\left(-2 + \rho Y_{it-1} + 4 \alpha_i + D_i + \epsilon_{Y, it} \geq 0 \right), & t \geq T_0 \\
        \end{array}
        \right..  
    \end{aligned}   
    \end{equation} 
    The individual latent factors $\{\alpha_i\}_{i=1}^{n}$ consist of independent random variables drawn from the uniform distribution on $[-1,1]$. The error terms $\{\epsilon_{Y,it}\}_{i=1,\ldots,n,\;t=0,\ldots,T}$ are independent of the latent factors and are i.i.d. across both dimensions, following the standard logistic distribution. I set $\rho=0.5$ to capture dynamic dependence. The contemporaneous ATT, computed using numerical simulation, is approximately $0.12$. Dynamic treatment effects for $t>T_0$ can also be calculated numerically. Treatment assignment depends jointly on latent heterogeneity and the lagged outcome:
    \begin{align}\label{eq:sim_NL_selection_simultaneous}
        D_i = \bs{1}\left( \alpha_i/2  + Y_{i T_0 - 1}/2 +  \epsilon_{D, i} \geq 0\right). 
    \end{align}
    Here, the error terms $\{\epsilon_{D,i}\}_{i=1}^{n}$ are i.i.d. across individuals and are independent of the latent factors and $\{\epsilon_{Y,it}\}_{i=1,\ldots,n,\;t=0,\ldots,T}$. In the numerical designs, $\epsilon_{D,i}$ follows the standard logistic distribution. I vary the sample sizes $(N,T_0)\in\left\{(1000,20),(200,100),(1000,100),(2000,100)\right\}$ and perform $5000$ replications for each design. 

    Table~\ref{tab:NL_simultaneous} reports the simulation results for the contemporaneous ATT under the nonlinear dynamic panel model (the DiD estimator is omitted because the DGP does not admit an additive two-way fixed-effects representation). The table shows that both the standard doubly robust estimator and the proposed estimator have small biases and achieve coverage close to the nominal 95\% level. The standard doubly robust estimator performs particularly well in this design because $Y_{it}$ is discrete, reducing the effective dimension of the nonparametric regression to one and allowing the estimator to be root-$N$ consistent and asymptotically unbiased \citep{deaner2025inferring}.  The proposed estimator  performs similarly to the infeasible DR3$^*$ estimator, indicating that the estimated pseudo-distance provides a good approximation to the true latent distance even when $T_0=20$. Finally, I recommend $(\kappa,\underline{q})=(0.2,0.8)$ as the default choice. 

    \bigskip

    Overall, the proposed estimator substantially reduces bias and achieves coverage rates close to the nominal level. It performs well across different DGPs and sample sizes. In particular, its performance remains strong when the number of pretreatment periods is moderate, such as $T_0=20$. This result is especially relevant because panels with a very large time dimension are uncommon in empirical applications. The simulations show that the performance of the proposed estimator is more sensitive to the cross-sectional sample size $N$ and deteriorates when $N$ is small.

    \begin{table}[H]
    \centering
    \caption{Simulation Results: Nonlinear Dynamic Panel with Fixed Effects}\label{tab:NL_simultaneous}
        \begin{tabular}{ccccccccc}
        \toprule
                & $\mr{DR2}^*$   & DR2   & \multicolumn{3}{c}{$\mr{DR3}^*$ } & \multicolumn{3}{c}{DR3} \\
            \cmidrule(lr){4-6} \cmidrule(lr){7-9} &  &      &$\kappa=.15$ & {$\kappa=.2$} & {$\kappa=.25$} & {$\kappa=.15$} & {$\kappa=.2$} & {$\kappa=.25$} \\
        \midrule
        $N= 1000, T_0= 20$ &             &       &       &       &       &       &       &  \\
        Bias         &   0.20  &  1.10  &  -0.31  &  -0.24  &  -0.02  &  -0.25   &   -0.05  &   0.38 \\
        SD           &  2.51  &  2.67  &  3.21   &  3.17   & 2.97   &  3.08   &  3.07   &  3.04  \\
        Coverage          &  95.1   &  93.1   &  94.2   &  94.3   &  94.9   &  94.0  &  94.1   &  94.1 \\

        \addlinespace[5pt]
        $N= 200, T_0= 100$ &              &       &       &       &       &       &       &  \\
        Bias   &         0.38  &  0.77   &   -0.48  &  -0.47   &  -0.38   &  -0.54   &  -0.50   &  -0.33  \\
        SD    &          6.72   &   6.96   &  9.90  &  9.89   &   9.83    &  10.01   &  10.01  & 10.01  \\
        Coverage   &         94.3   &   94.5  &   93.6   &   93.4  &   93.2  &  92.4  &  92.4  &  92.6 \\

        \addlinespace[5pt]
        $N= 1000, T_0= 100$ &              &       &       &       &       &       &       &  \\
        Bias   &          0.21  &  0.63  &   -0.31   &  -0.25  &  -0.06  & -0.09   &  -0.04  &  0.06 \\
        SD    &         2.56  &  2.58  &  3.19  &  3.14  &  2.99  &  3.08   &  3.06  &  3.03 \\
        Coverage          & 95.1 & 94.5  & 94.5 & 94.6 &  94.2 & 93.9 & 94.1 & 94.1  \\

        \addlinespace[5pt]
        $N= 2000, T_0= 100$         &       &       &       &       &       &       &       &  \\
        Bias  &           0.12   &    0.55   &  -0.06     &   -0.06    &   -0.06    &   -0.05    &  -0.04   & 0.00 \\
        SD   &    1.76    &  1.77     &   2.06    &   2.06    &  2.06     &   2.06    &  2.05     &  2.05  \\
        Coverage  &           95.1  &    94.5   &    94.8   &  94.8     &   94.6    &   94.2    &  94.3 & 94.2  \\ 
        \bottomrule
        \end{tabular}%
        \vspace{0.3cm}
        \begin{minipage}{\textwidth}
            \footnotesize 
            \textbf{Note:} The table reports Monte Carlo results for the contemporaneous ATT based on $5000$ replications of the nonlinear dynamic panel model with fixed effects specified in~\eqref{eq:sim_NL_Y_simultaneous} and the treatment selection mechanism specified in~\eqref{eq:sim_NL_selection_simultaneous}. I fix $\underline{q}=80\%$ and vary $\kappa \in \{.15,.2,.25\}$ when selecting the bandwidth $h_m^*$ according to Algorithm~\ref{alg:simultaneous_cv}. I report the bias, standard deviation, and coverage rate for the infeasible doubly robust estimator based on a standard two-way sample split and the true latent distance ($\mr{DR2}^*$), and its feasible counterpart based on the estimated pseudo-distance ($\mr{DR2}$). I also report the bias, standard deviation, and coverage rate for the infeasible estimator based on a three-way sample split and the true latent distance ($\mr{DR3}^*$), as well as the proposed estimator based on a three-way sample split and the estimated pseudo-distance ($\mr{DR3}$). 
            Biases and standard deviations are reported in units of $\bs{0.01}$, and coverage rates are reported in percentage points $\bs{1\%}$. 
        \end{minipage}
    \end{table}

    \section{Empirical Application}\label{sec:empirical}

    I apply the proposed method to the setting studied by \citet{bailey2012reexamining}, estimating the long-run effects of U.S. family planning programs on childbearing, allowing treatment assignment to depend on county fixed effects and pretreatment fertility rates.

    The original data is a county-year panel covers $3,037$ counties from $1959$ to $1988$. The outcome, $Y_{it}$, is the general fertility rate (GFR), defined as the number of live births per $1000$ women aged $15-44$ in county $i$ and year $t$. 
        Treatment timing is summarized in Table~\ref{tab:treatment_timing} and is measured by the fiscal year in which a county first received a recorded federal family planning grant. Counties were treated sequentially between 1965 and 1973, and those without a recorded grant during this period are coded as untreated. Following \citet{bailey2012reexamining}, I pool adjacent treatment years into three groups: $1965-1967$, $1968-1969$, and $1970-1973$\footnote{
        This is because estimating the ATT separately for each $g = 1965, \ldots, 1973$ is infeasible in practice: the number of counties first treated in some years (e.g., 1965 and 1973) is very small, making the corresponding nonparametric estimates unstable and imprecise. 
        In addition, since the analysis focuses on medium- and long-run effects, pooling adjacent cohorts sacrifices little timing variation while preserving the staggered rollout of the program.
    }.

    \begin{table}[H]
    \centering
    \caption{Treatment Timing}
    \label{tab:treatment_timing}
    \begin{tabular*}{0.85\textwidth}{@{\extracolsep{\fill}}lccc@{}}
        \toprule
        & Years & No. treated by year & Total \\
        \midrule
        Group 1 & $1965-1967$ & 6, \; 43, \; 74       & 123 \\
        Group 2 & $1968-1969$ & 52, \; 278          & 330 \\
        Group 3 & $1970-1973$ & 63, \;75, \;53, \;10   & 201 \\
        \bottomrule
    \end{tabular*}
    \end{table}

    These groups correspond to different stages of policy implementation. The first stage ($1965-67$) was the initial phase funded by the Community Action Program (CAP) with informal funding procedures. The second stage ($1968-69$) saw rapid expansion of the program, which the federal government had made a national emphasis program. The third stage ($1970-73$) saw further expansion and was more institutionalized under the 1970 Title X legislation, which allowed the Department of Health, Education, and Welfare to provide grants directly to local family planning organizations. These stages were heterogeneous in funding priorities and the agencies responsible for awarding grants, leading to differences in selection mechanisms and treatment effects, as discussed below.

    \paragraph{Pseudo distance}
    To obtain a longer pretreatment outcome history, I match the Bailey 
    replication sample to the \textit{U.S. County-Level Natality and Mortality
    Data, 1915--2007}\footnote{
        The data is publicly available at \href{https://www.icpsr.umich.edu/web/ICPSR/studies/36603}{https://www.icpsr.umich.edu/web/ICPSR/studies/36603}. 
    } assembled by \citet{BaileyEtAl2016Data}. This database provides
    county-year live births by the mother's county of residence and estimates of
    the female population aged $15-44$ since $1937$. This enables me to  construct the historical GFR from $1937$ to $1958$\footnote{
        For the overlapping period $1959$-$1988$, I just retain the GFR reported in the
    \citet{bailey2012reexamining} replication files rather than replacing it with the reconstructed
    series. 
    }. 
    The final panel contains $3,017$ county units observed from $1937$ to $1988$ with $28$ years of pretreatment outcomes for each county\footnote{
        Counties that could not be matched or lacked the required outcome history were excluded from the final sample. For example, Los Alamos was excluded because its outcome data are unavailable before $1950$.
    }. 

    I use the pretreatment outcomes from $1937-1964$ to calculate pseudo-distance between each pair of counties. Although someone may argue that there are structural changes in the series because the period span the Second World War and the baby boom, this does not affect the construction when the structural change can be captured by an  additive common time trend, as the common trend is differenced out by cross-sectional differencing. The results are similar when I use data from $1946-1964$ to calculate the pseudo-distance.

    \paragraph{Selection} The selection mechanism of the family planning program consists of two steps: first, local organizations decide whether to apply for the program, and  second, conditional on the applicaiton, the federal official devide whether and when to approve the applications. \citet{bailey2012reexamining} documents strong evidence that the selection depends on fixed effects, she documents that treated counties were larger and more urban and, on average, more educated and affluent than unfunded counties, which supports the selection on county fixed effects.  
    
    In addition, pretreatment fertility rates may affect treatment assignment through several channels. For example, counties with higher recent fertility rates may have greater demand for subsidized family planning services and, therefore, stronger incentives to apply for federal funding. This channel leads to a positive relationship between pretreatment fertility rates and selection into the program. On the other hand, federal officials may have been more likely to approve applications from counties with lower lagged fertility rates, as this would make it easier to claim the success of the program, generating a negative relationship between pretreatment fertility rates and selection into the program.

    I find strong reduced-form empirical evidence that lagged fertility rates affect selection into treatment. Specifically, for each group $g$ and each $t$ in the group, I estimate the following regression for units that are untreated at time \(t\):
    \begin{gather*}
                D_{it} \sim  \mr{Logit}\left(\beta'_g X_{it} + \alpha_i \right). 
            \end{gather*}
    
    In Table~\ref{tab:empirical_selection_lag_outcomes}, for each treatment group, I report logistic regressions of selection into treatment on either (1) the one-period lagged outcome or (2) the average of the three most recent lagged outcomes. The fixed effect is unobserved and is therefore proxied by urbanization, education, income, race, and population. Only the coefficient on the lagged outcome is reported.
    I find strong empirical evidence that lagged outcomes are associated with treatment for counties treated in $1965-67$, with negative and statistically significant coefficients across both specifications. The coefficients for $1968-69$ are not statistically significant, which may reflect the mixed selection channels (I also estimate year-specific regressions, reported in Table~\ref{tab:empirical_selection_lag_outcomes_year}, and find a negative and statistically significant coefficient for $1968$, indicating negative selection on lagged fertility rates in at least one year within the second group). In addition, I find a positive and statistically significant coefficient for $(1970-73)$ when using the one-period lagged outcome, but the coefficient is not statistically significant when using the three-period average. In summary, the reduced-form empirical evidence supports selection on lagged fertility rates, although the effect is heterogeneous over time.
    
    \begin{table}[h]
    \centering
    \caption{Selection on lagged outcomes}
        \begin{tabular}{lcccccc}
        \toprule
            & \multicolumn{2}{c}{ Treated $1965-67$}
    & \multicolumn{2}{c}{Treated $1968-69$}
    & \multicolumn{2}{c}{Treated  $1970-73$} \\
    \cmidrule(lr){2-3}\cmidrule(lr){4-5}\cmidrule(lr){6-7}
            & (1)   & (2) & (1)   & (2) & (1)   & (2) \\
        \bottomrule
        \addlinespace[5pt]
        $\widehat{\beta}_g$ &    $-2.08^{***}$  &  $-1.76^{***}$    & $0.46$   & $0.15$  &  $0.56^{**}$    & $0.46$ \\
            &   $(0.56)$    &   $(0.56)$    & $(0.34)$  & $(0.35)$  &  $(0.27)$     & $(0.37)$ \\
        $\mr{AME}$&   $-0.03$    &   $-0.02$    & $0.02$  & $0.01$  &  $0.01$     & $0.01$ \\
        \bottomrule
        \end{tabular}\label{tab:empirical_selection_lag_outcomes}
        \vspace{0.3cm}
        \begin{minipage}{\textwidth}
            \footnotesize 
            \textbf{Note:} This table reports the estimated logistic regression coefficients on lagged outcomes for counties first funded in 1965-67, 1968-69, and 1970-73. Columns (1) and (2) use the one-period lag and the average of the three preceding lags, respectively. Controls are added sequentially across rows. I control for urbanization (the percentage of the population living in urban areas in 1960), education (the percentage of the population with more than 12 years of education in 1960), income (the percentage of the population with an annual income below 3,000 US dollars), race (the percentage of the population that was nonwhite in 1960), and population (in 1960). Standard errors are reported in parentheses. $***$, $**$, and $*$   denote statistical significance at the 1\%, 5\%, and 10\% levels, respectively.
        \end{minipage}
    \end{table}%

    \paragraph{Dynamic dependence} I also find strong reduced-form empirical evidence supporting dynamic dependence in fertility rates. Specifically, I estimate the following linear panel model:
    \begin{gather*}
        Y_{it}(\infty)
        = \rho Y_{it-1}(\infty)
        + \alpha_i + \gamma_t + \epsilon_{it},
        \quad \forall t = 1965, \ldots, 1988.
    \end{gather*}
    I report the results using (i) the Arellano--Bond estimator \citep{arellano1991some} with different sets of instruments and (ii) a least-squares estimator with analytical \emph{bias correction}. Table~\ref{tab:dynamic_dependence} reports the results. Across specifications, I find positive and statistically significant dynamic dependence in fertility rates, even after controlling for unit and time fixed effects.

     \begin{table}[htbp]
    \centering
    \caption{Dynamic dependence}
    \setlength{\tabcolsep}{8pt}
    \renewcommand{\arraystretch}{1.0}

    \begin{tabular}{lcccc}
        \toprule
        & \multicolumn{3}{c}{Arellano--Bond} 
& Dynamic-panel  \\
\cmidrule(lr){2-4}
& IV lag 2
& IV lags $2-4$
& IV lags $2-6$
&  \\
        \midrule

        $\widehat{\rho}$
        & $0.21^{***}$
        & $0.24^{***}$
        & $0.34^{***}$
        & $0.50^{***}$ \\

        & $(0.03)$
        & $(0.04)$
        & $(0.04)$
        & $(0.02)$ \\

        \bottomrule
    \end{tabular}\label{tab:dynamic_dependence}
    \vspace{0.1em}

    \begin{minipage}{0.95\textwidth}
        \scriptsize
        \textit{Notes:}
        Standard errors are reported in parentheses.
        The first three columns report two-step
        Arellano--Bond estimator using the indicated
        lag windows. The final column reports the bias-corrected
        dynamic-panel within estimate.
        $^{***}p<0.01$.
    \end{minipage}
\end{table}

    \paragraph{Estimation}
    For estimation, I control for unobserved time-invariant heterogeneity using the pseudo-distance and for observed lagged fertility rates. I consider the following specifications\footnote{
        The average is calculated using fertility rates from the three years preceding the earliest treatment year in that group. For example, for the $1965-1967$ group, I use the average fertility rate over $1962-1964$.}. 
        I set $\kappa=0.2$, as suggested by the simulations. The estimates in \citet{bailey2012reexamining} are based on event-study regressions and are therefore potentially affected by negative weights, so I also report estimates obtained using the DiD estimator of~\citet{callaway2021difference}:

    \begin{center}
            \small
            \renewcommand{\arraystretch}{1.1}
            \setlength{\tabcolsep}{8pt}

            \begin{tabular}{lcc}
                \toprule
                & Pretreatment history & $X_{it}$ \\
                \midrule

                Model 1
                & $1937-1964$
                & $\displaystyle
                   (Y_{it-1}+Y_{it-2}+Y_{it-3})/3$ \\

                Model 2
                & $1937-1964$
                & $Y_{it-1}$ \\

                Model 3
                & $1946-1964$
                & $\displaystyle
                   (Y_{it-1}+Y_{it-2}+Y_{it-3})/3$ \\

                \bottomrule
            \end{tabular}
        \end{center}

        For each treatment group, I report ATT estimates beginning in the year after the latest treatment year in that group. For example, for the $1965-1967$ group, the reported ATT estimates begin in 1968.

\begin{figure}[H]
    \centering
    \includegraphics[width=1.0\textwidth]
        {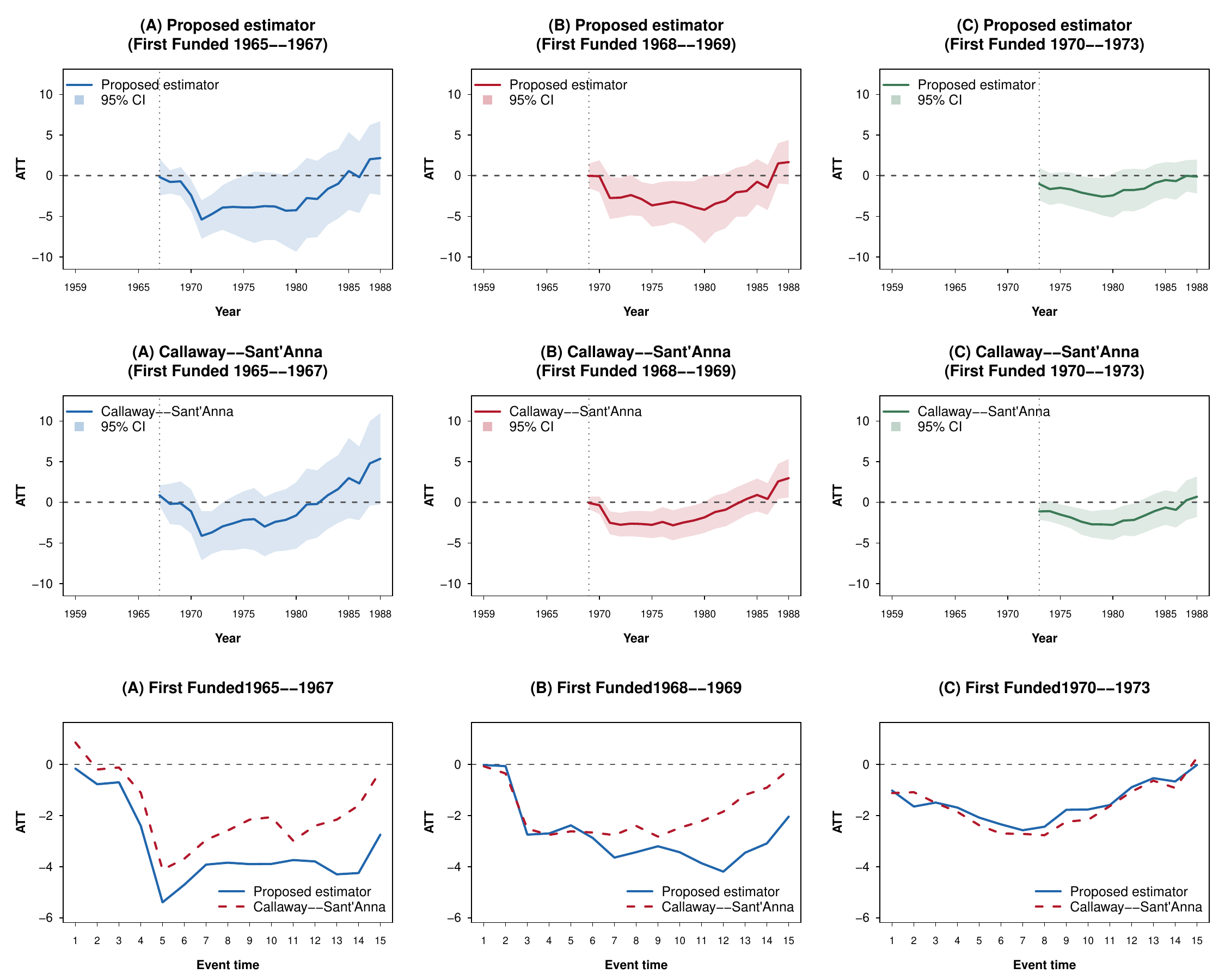}
    \caption{Weighted dynamic ATT (Model 1).}
    \label{fig:att_1}
    \vspace{0.3cm}
        \begin{minipage}{\textwidth}
            \footnotesize 
            \textbf{Note:} Following \citet{bailey2012reexamining}, the results are weighted by the number of women aged $15-44$ in $1970$. 
        \end{minipage}
    \end{figure}

        \paragraph{Results}
    Figure~\ref{fig:att_1} presents the dynamic ATT estimates and 95\% confidence intervals obtained using the proposed method and the DiD estimator of~\citet{callaway2021difference}. The first row displays the dynamic ATT estimates from the proposed method. The estimates show that the family planning program has a statistically significant negative effect on fertility. In addition, the long-run  effect is statistically significant for the group treated in 1968-69. The DiD estimates in the second row show similar results. 

    The last row in Figure~\ref{fig:att_1} compares estimates from the proposed method and the DiD estimator over the first $15$ post-treatment periods. The proposed method yields larger estimated reductions in fertility for the first two treatment groups, whereas the estimates are similar for the third group. For the first two groups, the gap becomes larger at longer post-treatment horizons. These results highlight the importance of accounting jointly for selection on lagged outcomes and their dynamic effects when evaluating policies.

    Table~\ref{tab:long_term_ATT} reports treatment effects averaged over three five-year windows and across the three treatment groups. The Callaway-Sant'Anna DiD estimates are similar to those reported by \citet{bailey2012reexamining}, but the proposed estimator produces more negative estimates than the Callaway-Sant'Anna DiD estimator in all three windows. The magnitude of the difference increases from 0.32 in years $1-5$ to 0.76 in years $6-10$ and then to 1.61 in years $11-15$. %

    \begin{table}[H]
    \centering
    \caption{ATT by Post-Treatment Horizon}
        \begin{tabular}{lccc}
        \toprule
            & \multicolumn{1}{l}{ATT (1 - 5)} & \multicolumn{1}{l}{ATT (6 - 10)} & \multicolumn{1}{l}{ATT (11 - 15)} \\
        \midrule
        Proposed estimator (Model 1)  & $-1.70^{**}$  & $-3.39^{***}$ & $-3.03^{**}$ \\
          & (0.67) & (1.01)  & (1.39)  \\
        Proposed estimator (Model 2)  & $-1.74^{***}$  & $-3.47^{***}$ & $-2.92^{**}$ \\
          & (0.59) & (0.90)  & (1.36)  \\
        Proposed estimator (Model 3)  & $-1.92^{***}$  & $-3.56^{***}$ & $-3.06^{**}$ \\
          & (0.69) & (1.06)  & (1.40)  \\
        DiD Callaway-Sant'anna  & $-1.37^{**}$ & $-2.63^{***}$ & $-1.42$ \\
         & (0.54) & (0.80) & (0.94) \\
        Difference (Model 1)  & -0.32 & -0.76 & -1.61 \\
        & (0.76) & (1.11) & (1.28) \\
        Difference (Model 2)  & -0.37 & -0.89 & -1.51 \\
        & (0.60) & (0.94) & (1.22) \\
        Difference (Model 2)  & -0.55 & -0.93 & -1.64 \\
        & (0.82) & (1.21) & (1.33) \\
        \bottomrule
        \end{tabular}%
    \label{tab:long_term_ATT}
    \begin{minipage}{\textwidth}
            \footnotesize 
            \textbf{Note:} Following \citet{bailey2012reexamining}, the results are weighted by the number of women aged $15-44$ in $1970$. 
        \end{minipage}
    \end{table}%

    \paragraph{Discussion} The differences between the two estimators can be explained by selection on lagged fertility rates and dynamic dependence. For counties first funded in $1965-67$, Table~\ref{tab:empirical_selection_lag_outcomes} indicates that counties with lower lagged fertility rates are more likely to be treated. Since these negative shocks to fertility rates are temporary, fertility in the treated counties would recover on average in the following period and generate increases in untreated fertility. DiD attributes this recovery effect to the program, generating upward bias and understating the reduction in fertility. In addition, since the recovery occurs gradually, it generates larger upward bias at longer horizons, consistent with the empirical results. 

    For the second group, although Table~\ref{tab:empirical_selection_lag_outcomes} does not provide evidence of selection on negative lagged fertility shocks, the regression for 1968 reported in Table~\ref{tab:empirical_selection_lag_outcomes_year} shows a negative coefficient. Thus, the insignificant coefficients in Table~\ref{tab:empirical_selection_lag_outcomes} may reflect offsetting selection effects across years. Hence, The same recovery mechanism explain the differences between the two estimators for the second group. For the third group, Table~\ref{tab:empirical_selection_lag_outcomes} shows evidence of selection on positive fertility shocks, while the two estimators produce similar treatment effect estimates, as shown in Figure~\ref{fig:att_1}. This similarity may be related to the slowdown in the decline of U.S. fertility rates after 1973.

    \bibliographystyle{aea}
    \bibliography{ref}

\clearpage 

\appendix

\numberwithin{proposition}{section}
\numberwithin{lemma}{section}
\numberwithin{theorem}{section}
\numberwithin{corollary}{section}
\numberwithin{example}{section}
\numberwithin{table}{section}
\numberwithin{algorithm}{section}
\numberwithin{definition}{section}
\numberwithin{assumption}{section}

    \section{Additional Empirical Results, Algorithm, and Simulations}\label{appendix:algorithm_simulation}

    For each  $t = 1965, \ldots, 1973$, I estimate the following regression for units that are untreated at time $t$:
    \begin{gather*}
                D_{it} \sim  \mr{Logit}\left(\beta'_t X_{it} + \alpha_i \right). 
            \end{gather*}
    
    In Table~\ref{tab:empirical_selection_lag_outcomes_year}, for each $t$, I report logistic regressions of selection into treatment on either (1) the one-period lagged outcome or (2) the average of the three most recent lagged outcomes. The fixed effect is unobserved and is therefore proxied by urbanization, education, income, race, and population. Only the coefficient on the lagged outcome is reported.
    \begin{table}[H]
    \centering
    \caption{Selection on lagged outcomes - suupplement} 
    \begin{tabular}{lccccc}
        \toprule
        Year & Treated
        & \multicolumn{2}{c}{Model (1)}
        & \multicolumn{2}{c}{Model (2)} \\
        \midrule
        \multicolumn{6}{l}{\textit{Initial CAP programs}} \\

        1965 & 6
        & $-5.01^{**}$ & $(1.99)$
        & $-4.19^{*}$  & $(2.14)$ \\

        1966 & 43
        & $-3.37$ & $(2.94)$
        & $-2.41$ & $(2.82)$ \\

        1967 & 74
        & $2.45$ & $(1.56)$
        & $2.34$ & $(1.61)$ \\

        \addlinespace
        \multicolumn{6}{l}{\textit{National-emphasis expansion}} \\

        1968 & 52
        & $-2.45$     & $(1.74)$
        & $-3.82^{*}$ & $(2.14)$ \\

        1969 & 278
        & $0.48$ & $(1.33)$
        & $1.27$ & $(1.21)$ \\

        \addlinespace
        \multicolumn{6}{l}{\textit{Title X expansion}} \\

        1970 & 63
        & $1.81$ & $(1.34)$
        & $1.54$ & $(1.42)$ \\

        1971 & 75
        & $1.99^{*}$  & $(1.20)$
        & $3.01^{**}$ & $(1.42)$ \\

        1972 & 53
        & $0.73$ & $(2.45)$
        & $0.76$ & $(2.81)$ \\

        1973 & 10
        & $-1.11$ & $(4.56)$
        & $-1.92$ & $(4.85)$ \\

        \bottomrule
    \end{tabular}\label{tab:empirical_selection_lag_outcomes_year}
    \vspace{0.1em}
    \begin{minipage}{0.90\textwidth}
        \scriptsize
        \textit{Notes:}
        Each row reports a separate weighted Logit regression.
        Coefficients and standard errors, reported in parentheses,
        are multiplied by 100.
        The regressions are weighted by the number of women aged 15-44 in 1970.
    \end{minipage}
\end{table}

    \begin{figure}[H]
    \centering
    \includegraphics[width=1.0\textwidth]
        {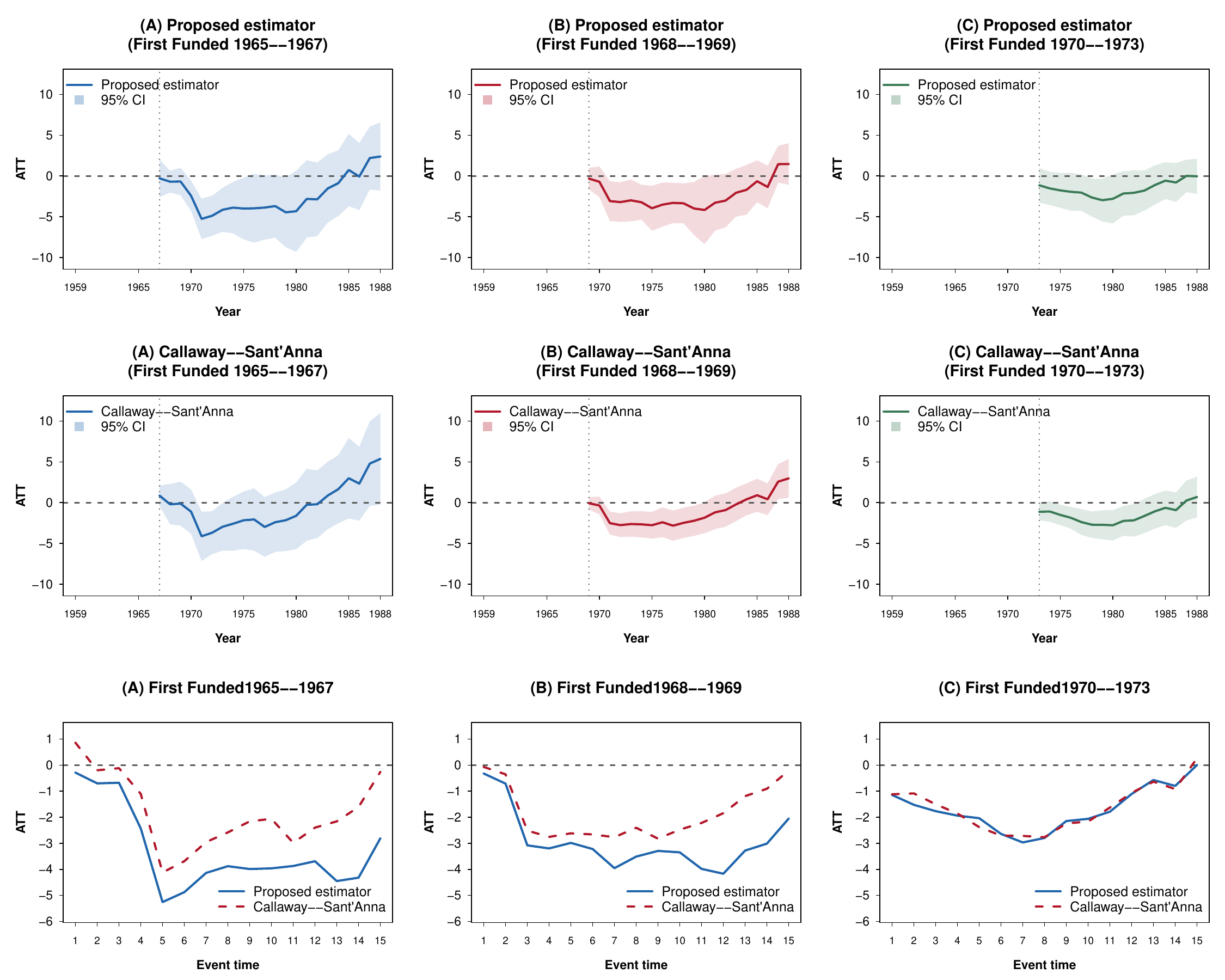}
    \caption{Weighted dynamic ATT (Model 2).}
    \label{fig:att_2}
    \vspace{0.3cm}
        \begin{minipage}{\textwidth}
            \footnotesize 
            \textbf{Note:} Following \citet{bailey2012reexamining}, the results are weighted by the number of women aged $15-44$ in $1970$. 
        \end{minipage}
    \end{figure}

    \begin{figure}[H]
    \centering
    \includegraphics[width=1.0\textwidth]
        {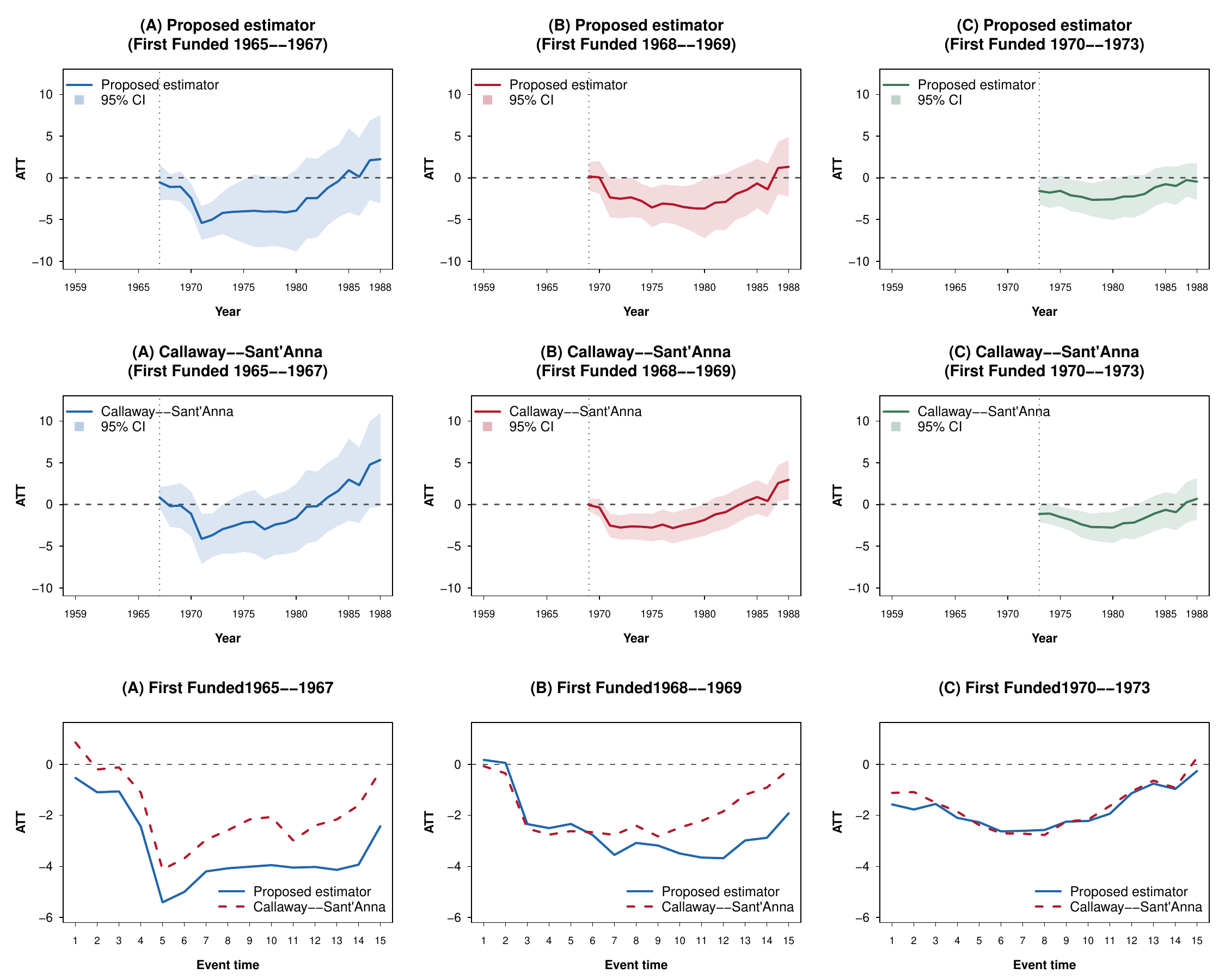}
    \caption{Weighted dynamic ATT (Model 3).}
    \label{fig:att_3}
    \vspace{0.3cm}
        \begin{minipage}{\textwidth}
            \footnotesize 
            \textbf{Note:} Following \citet{bailey2012reexamining}, the results are weighted by the number of women aged $15-44$ in $1970$. 
        \end{minipage}
    \end{figure}

    \begin{algorithm}[H]
        \caption{Doubly robust dynamic ATT estimator - simultaneous treatment}\label{alg:simultaneous_cv}
        \begin{algorithmic}
            \Require Kernel $K(\cdot)$, two lists of bandwidths, $\{h_{m1}, \ldots, h_{mb}\}$ and $\{h_{\pi 1}, \ldots, h_{\pi b}\}$, tuning parameters $\kappa >0 $ and $\underline{q} \in (0, 1]$,  and confidence level $1 -\alpha$.  

            \Ensure Dynamic ATT estimates $\widehat{\mr{ATT}}(t)$, stand errors $\widehat{\mr{se}}_t$,  $1-\alpha$ confidence interval.

                \State \textbf{Step 1: Sample-splitting} \text{ }Randomly partition samples into disjoint $\{\mc{I}_1, \mc{I}_2, \mc{I}_3\}$.   
                \State \textbf{Step 2: Calculate similarity}   For each $i, j = 1, \ldots, n$, calculate the pseudo-distance: 
                \Statex
                \makebox[\linewidth][c]{%
                $\displaystyle \widehat{d}_{ij} = \max_{\substack{k_1,k_2=1,\ldots,n,  k_1,k_2\neq i,j}}
                \left| \frac{1}{T_0} \sum_{t=0}^{T_0-1} (Y_{k_1t}-Y_{k_2t})(Y_{it}-Y_{jt})\right|.$
                }
                \State \textbf{Step 3: Cross validation}   
                \State For each $i,j=1,\ldots,n$, any $h>0$, $\widehat{K}_{h, ij} := K\left(\left(X_{jT_0} - X_{iT_0 }\right)/ h \right) K(\widehat{d}_{ij}/ h).$
                \State \textbf{(i) Leave-one-out cross validation for $\pi$. } For  each $i$, compute the leave-one-out predictors for each $h \in \{h_{\pi 1}, \ldots, h_{\pi b}\}$, 
                \Statex
                \makebox[\linewidth][c]{%
                $\displaystyle
                    \widetilde{\pi}_{i}^{(-i)}(h) = \sum_{i' \in \mc{I}(i)\setminus \{i\} }  D_{i'}  \widehat{K}_{h, ii'} \big / \sum_{i' \in \mc{I}(i)\setminus \{i\}  }  \widehat{K}_{h, ii'} ,$ 
                }
                and select the optimal bandwidth  $h_{\pi}^*$ as
                \Statex
                \makebox[\linewidth][c]{%
                $\displaystyle  h_{\pi}^*\in \argmin_{h\in \{h_{\pi 1}, \ldots, h_{\pi b}\} } \frac{1}{n}\sum_{i=1}^{n}(D_i - \widetilde{\pi}_{i}^{(-i)}(h)  )^2.$
                } 
                \State \textbf{(ii) Cross validation for $m$. } For $i$, compute effective sample size for each $h\in \{h_{m1}, \ldots, h_{mb}\}$, 
                \Statex
                \makebox[\linewidth][c]{%
                $\displaystyle
                    n_{i, \mr{eff}}(h) = \bigg(\sum_{i' \in \mc{I}(i)\setminus \{i\} }  (1 - D_{i'})  \widehat{K}_{h, ii'}\bigg)^2  \big/ \sum_{i' \in \mc{I}(i)\setminus \{i\} }  (1 - D_{i'})  \widehat{K}^2_{h, ii'}, $
                }
                and select $h_m^*$ as 
                 \Statex
                \makebox[\linewidth][c]{%
                $\displaystyle h_m^*\in \min \left\{h\in \{h_{m1}, \ldots, h_{mb}\}  :  \frac{1}{n} \sum_{i=1}^{n} \bs{1}\left(n_{i, \mr{eff}}(h) > (\kappa \log (n))^2 \right) \geq \underline{q} \right\}. $
                }

                \State\State \textbf{Step 4: Compute $\widehat{m}_{it}$ and $\widehat{\pi}_{i}$}   For each $i = 1, \ldots, n$, compute
                \begin{align*}
                    \widehat{m}_{it} = \frac{\sum_{j \in \mc{I}_{m}(i)}  (1-D_j) \widehat{K}_{h_m^*, ij}Y_{jt}  }{\sum_{j \in \mc{I}_{m}(i)  }  (1 - D_{j})\widehat{K}_{h_m^*, ij}}, \quad \widehat{\pi}_{i} = \frac{\sum_{j \in  \mc{I}_{\pi}(i) }  D_j \widehat{K}_{h_{\pi}^*, ij}  }{\sum_{j \in \mc{I}_{\pi}(i) }  \widehat{K}_{h_{\pi}^*, ij} }
                \end{align*} 

                \State\State \textbf{Step 5: Construct $\widehat{\mr{ATT}}(t)$ and its CI}   
                \State Let $n_1 :=\sum_{i=1}^{n} \bs{1}(D_i = 1)$, compute 
                \begin{gather*}
                    \widehat{\mr{ATT}}(t) = \frac{1}{n_1} \sum_{i}^{n} \left(D_{i}Y_{it} - \frac{(1 - D_i)\widehat{\pi}_iY_{it} + (D_i - \widehat{\pi}_i)\widehat{m}_{it}}{1 - \widehat{\pi}_i }\right), \\
                    \widehat{\mr{se}}_t = \left(\frac{1}{n_1^2} \sum_{i=1}^{n}\left(D_{i}Y_{it} - \frac{(1 - D_i)\widehat{\pi}_iY_{it} + (D_i - \widehat{\pi}_i)\widehat{m}_{it}}{1 - \widehat{\pi}_i } -  D_i \widehat{\mr{ATT}}(t) \right)^2\right)^{1/2}. 
                \end{gather*}
                The confidence interval is $\mr{CI}_t = [\widehat{\mr{ATT}}(t) \pm Z_{1 - \alpha/2} \cdot  \widehat{\mr{se}}_t ]$. 
        \end{algorithmic}
    \end{algorithm}

    \begin{algorithm}[H]
        \caption{Doubly robust dynamic ATT estimator - staggered adoption (Basic Idea)}\label{alg:extension_basic}
        \begin{algorithmic}
            \Require Kernel $K(\cdot)$, bandwidths $h_m, h_\pi$, and confidence level $1 - \alpha$.  
            \Ensure Dynamic ATT estimates $\widehat{\mr{ATT}(g, t)}$, stand errors $\widehat{\mr{se}}_t$, $1-\alpha$ confidence interval. 
                \State \textbf{Step 1: Sample-splitting} \text{ }Randomly partition samples into disjoint  $\{\mc{I}_1, \mc{I}_2, \mc{I}_3\}$. 
                \State \textbf{Step 2: Calculate similarity}  \text{ } For each $i,j=1,\ldots,n$, calculate the pseudo-distance:
                \Statex
                \makebox[\linewidth][c]{%
                $\displaystyle \widehat{d}_{ij} = \max_{\substack{k_1,k_2=1,\ldots,n,  k_1,k_2\neq i,j}}
                \left| \frac{1}{T_0} \sum_{t=0}^{T_0-1} (Y_{k_1t}-Y_{k_2t})(Y_{it}-Y_{jt})\right|.$
                }
                \State\State \textbf{Step 3: Compute expected outcomes and propensity scores}  
                \State For each $i,j=1,\ldots,n$, any $h>0$, and each $t'$, $\widehat{K}^{(t')}_{h, ij} := K\left(\left(X_{jt'} - X_{it' }\right)/ h \right) K(\widehat{d}_{ij}/ h).$ 
                \textbf{Initialization.} For each $j=1,\ldots,n$, set $\widetilde{m}_{j,t\mid t+1}:=Y_{jt}$. 
                \For{$t'=t,t-1,\ldots,g$}
                    \State For each $i=1,\ldots,n$, compute the Nadaraya-Watson estimators
                    \begin{align*}
                        \widehat{m}_{i, t\mid t'} = \frac{\sum_{ j \in \mc{I}_{m}(i) } \bs{1}\left(G_{j} > t' \right) \widetilde{m}_{j, t\mid t' + 1} \widehat{K}^{(t')}_{h_m, ij}}{\sum_{ j \in \mc{I}_{m} (i) } \bs{1}\left(G_{j} > t' \right) \widehat{K}^{(t')}_{h_m, ij}}, \quad \widehat{\pi}_{it'} = \frac{\sum_{j \in \mc{I}_{\pi}(i)} \bs{1}\left(G_j = t'\right) \widehat{K}^{(t')}_{h_{\pi}, ij} }{\sum_{j \in \mc{I}_{\pi}(i)} \bs{1}\left(G_j > t' -1 \right) \widehat{K}^{(t')}_{h_{\pi}, ij}},  
                    \end{align*}
                    wherefor each $j=1, \ldots, n$, the within-sample estimators are:
                    \Statex
                    \makebox[\linewidth][c]{%
                    $\displaystyle
                            \widetilde{m}_{j, t\mid t' } = \sum_{ j' \in \mc{I}(j) } \bs{1}\left(G_{j'} > t' \right) \widetilde{m}_{j', t\mid t' + 1} \widehat{K}^{(t')}_{h_m, jj'}  \bigg / \sum_{ j' \in \mc{I}(j) } \bs{1}\left(G_{j'} > t' \right) \widehat{K}^{(t')}_{h_m, jj'}. $
                    } 
                \EndFor 
                \State \textbf{Step 4: Construct $\widehat{\mr{ATT}(t)}$, standard error $\widehat{\mr{se}}_t$, and confidence interval}   
                \State Let $n_g :=\sum_{i=1}^{n} \bs{1}(G_i = g)$, compute 
                \begin{align*} 
                    \widehat{ATT(g, t)} = \frac{1}{n_g} \sum_{i = 1}^{n} \Bigg[ & \bs{1}\left(G_i = g\right) \left(Y_{it} - \widehat{m}_{i, t\mid g}\right) \\
                    & - \sum_{t' = g}^{t}\bs{1}\left(G_i > t'\right) \frac{\widehat{\pi}_{ig}}{1-  \widehat{\pi}_{ig}} \left(\prod_{r = g+1}^{t'} \frac{1}{1 - \widehat{\pi}_{ir}}\right)\left( \widehat{m}_{i, t\mid t'+1} - \widehat{m}_{i, t\mid t'} \right) \Bigg],  
                \end{align*}
                where an empty product is defined as one. Then, calculate  
                \begin{align*}
                    \widehat{\mr{se}}_t = \Bigg(\frac{1}{n_g^2} \sum_{i = 1}^{n} \Bigg[ & \bs{1}\left(G_i = g\right) \left(Y_{it} - \widehat{m}_{i, t\mid g} - \widehat{ATT(g, t)}\right) \\
                    & - \sum_{t' = g}^{t}\bs{1}\left(G_i > t'\right) \frac{\widehat{\pi}_{ig}}{1-  \widehat{\pi}_{ig}} \left(\prod_{r = g+1}^{t'} \frac{1}{1 - \widehat{\pi}_{ir}}\right)\left( \widehat{m}_{i, t\mid t'+1} - \widehat{m}_{i, t\mid t'} \right) \Bigg]^2 \Bigg)^{1/2}. 
                \end{align*}
                The confidence interval is $\mr{CI}_t = [\widehat{\mr{ATT}(t)} \pm Z_{1 - \alpha/2} \cdot  \widehat{\mr{se}}_t ]$. 
        \end{algorithmic}
    \end{algorithm}

    \begin{algorithm}[H]
        \caption{Doubly robust estimator for dynamic ATT under staggered adoption}\label{alg:extension_cv}
        \begin{algorithmic}
            \Require Kernel $K(\cdot)$, two lists of bandwidths, $\bs{h}_m :=\{h_{m1}, \ldots, h_{mb}\}$ and $\bs{h}_\pi:= \{h_{\pi 1}, \ldots, h_{\pi b}\}$, tuning parameters $\kappa >0 $ and $\underline{q} \in (0, 1]$,  and confidence level $1 - \alpha$.  

           \Ensure Dynamic ATT estimates $\widehat{\mr{ATT}(g, t)}$, stand errors $\widehat{\mr{se}}_t$, $1-\alpha$ confidence interval. 
                \State \textbf{Step 1: Sample-splitting} \text{ }Randomly partition samples into disjoint $\{\mc{I}_1, \mc{I}_2, \mc{I}_3\}$. 
                \State \textbf{Step 2: Calculate similarity}  \text{ } For each $i,j=1,\ldots,n$, calculate the pseudo-distance:
                \Statex
                \makebox[\linewidth][c]{%
                $\displaystyle \widehat{d}_{ij} = \max_{\substack{k_1,k_2=1,\ldots,n,  k_1,k_2\neq i,j}}
                \left| \frac{1}{T_0} \sum_{t=0}^{T_0-1} (Y_{k_1t}-Y_{k_2t})(Y_{it}-Y_{jt})\right|.$
                }
                \State\State \textbf{Step 3: Compute expected outcomes and propensity scores}  
                \State Given $h>0$, for each $i,j$ and each $t'$, define $\widehat{K}^{(t')}_{h, ij} := K\left(\left(X_{jt'} - X_{it'}\right)/ h \right) K\left(\widehat{d}_{ij}/ h\right)$. 
                
                \textbf{Initialization.} For each $j=1,\ldots,n$, set $\widetilde{m}_{j,t\mid t+1} (h):=Y_{jt}$ for any $h >0$, 
                \For{$t'=t,t-1,\ldots,g$}
                    \State \textbf{Cross-validation} For each $j=1,\ldots,n$, each $h_m\in \bs{h}_m$, and each $h_\pi\in \bs{h}_\pi$, compute the leave-one-out within-sample estimators and effective sample sizes 
                    \begin{align*}
                        \widetilde{\pi}_{it'}^{(-i)}(h_\pi ) = \frac{\sum_{i' \in \mc{I}(i)\setminus \{i\}} \bs{1}\left(G_{i'} = t'\right) \widehat{K}^{(t')}_{h_{\pi}, ii'} }{\sum_{i' \in \mc{I}(i)\setminus \{i\}} \bs{1}\left(G_{i'} > t' -1 \right) \widehat{K}^{(t')}_{h_{\pi}, ii'}}, \quad n_{it', \mr{eff}}(h_m) = \frac{\left(\sum_{i' \in \mc{I}(i)\setminus \{i\} }  \bs{1}(G_{i'} > t')  \widehat{K}_{h_m, ii'}^{(t')}\right)^2  }{ \sum_{i' \in \mc{I}(i)\setminus \{i\}  }   \bs{1}(G_{i'} > t')   \widehat{K}^{(t')2}_{h_m, ii'}}. 
                    \end{align*}
                    Then, $h_{\pi}^*(t')\in \argmin_{h_\pi \in \bs{h}_\pi}  \sum_{i=1}^{n}\bs{1}(G_i >t'-1) (\bs{1}(G_i =t') - \widetilde{\pi}_{it'}^{(-i)}(h_\pi ) )^2$ and  
                        $h_{m}^*(t') = \min \left\{h_m \in \bs{h}_\pi  :  \frac{1}{\sum_{i=1}^{n}\bs{1}(G_i >t'-1)} \sum_{i=1}^{n} \bs{1}\left(G_i >t'-1, n_{it', \mr{eff}}(h_m) > (\kappa \log(n))^2  \right) \geq \underline{q} \right\}.$
                        
                    \State 
                    \State \textbf{Calculate $\widehat{m}_{i, t\mid t'}$ and $\widehat{\pi}_{i t'}$} For each $i, j = 1, \ldots, n$, calculate $\widehat{m}_{i, t\mid t'}$,  $\widehat{\pi}_{i t'}$, and $\widetilde{m}_{j, t\mid t' }$: 
                    \begin{gather*}
                        \widehat{m}_{i, t\mid t'} = \frac{\sum_{ j \in \mc{I}_{m}(i) } \bs{1}\left(G_{j} > t' \right) \widetilde{m}_{j, t\mid t' + 1} \widehat{K}^{(t')}_{h_{m}^*(t'), ij}}{\sum_{ j \in \mc{I}_{m} (i) } \bs{1}\left(G_{j} > t' \right) \widehat{K}^{(t')}_{h_{m}^*(t'), ij}}, \quad \widehat{\pi}_{it'} = \frac{\sum_{j \in \mc{I}_{\pi}(i)} \bs{1}\left(G_j = t'\right) \widehat{K}^{(t')}_{h_{\pi}^*(t'), ij} }{\sum_{j \in \mc{I}_{\pi}(i)} \bs{1}\left(G_j > t' -1 \right) \widehat{K}^{(t')}_{h_{\pi}^*(t'), ij}},  \\
                        \widetilde{m}_{j, t\mid t' } = \sum_{ j' \in \mc{I}(j) } \bs{1}\left(G_{j'} > t' \right) \widetilde{m}_{j', t\mid t' + 1} \widehat{K}^{(t')}_{h_{m}^*(t'), jj'}  \bigg / \sum_{ j' \in \mc{I}(j) } \bs{1}\left(G_{j'} > t' \right) \widehat{K}^{(t')}_{h_{m}^*(t'), jj'}.  
                    \end{gather*}
                \EndFor 

                \State\State \textbf{Step 3: Construct $\widehat{\mr{ATT}(g, t)}$, $\widehat{\mr{se}}_t$, and CI. }   \text{ }$\widehat{ATT(g, t)}$ is obtained via
                \Statex
                    \makebox[\linewidth][c]{%
                    $\displaystyle 
                   \frac{1}{n_g} \sum_{i = 1}^{n} \Bigg[  \bs{1}\left(G_i = g\right) \left(Y_{it} - \widehat{m}_{i, t\mid g}\right)  - \sum_{t' = g}^{t}\bs{1}\left(G_i > t'\right) \frac{\widehat{\pi}_{ig}}{1-  \widehat{\pi}_{ig}} \left(\prod_{r = g+1}^{t'} \frac{1}{1 - \widehat{\pi}_{ir}}\right)\left( \widehat{m}_{i, t\mid t'+1} - \widehat{m}_{i, t\mid t'} \right) \Bigg],$}  
                where $n_g :=\sum_{i=1}^{n} \bs{1}(G_i = g)$, and an empty product is defined as one.   $\widehat{\mr{se}}_t $ is 
                \begin{align*}
                    \widehat{\mr{se}}_t = \Bigg(\frac{1}{n_g^2} \sum_{i = 1}^{n} \Bigg[ & \bs{1}\left(G_i = g\right) \left(Y_{it} - \widehat{m}_{i, t\mid g} - \widehat{ATT(g, t)}\right) \\
                    & - \sum_{t' = g}^{t}\bs{1}\left(G_i > t'\right) \frac{\widehat{\pi}_{ig}}{1-  \widehat{\pi}_{ig}} \left(\prod_{r = g+1}^{t'} \frac{1}{1 - \widehat{\pi}_{ir}}\right)\left( \widehat{m}_{i, t\mid t'+1} - \widehat{m}_{i, t\mid t'} \right) \Bigg]^2 \Bigg)^{1/2}. 
                \end{align*}
                The confidence interval is $\mr{CI}_t = [\widehat{\mr{ATT}(t)} \pm Z_{1 - \alpha/2} \cdot  \widehat{\mr{se}}_t ]$. 
        \end{algorithmic}
    \end{algorithm}

    \section{Discussion on Pseudo Distance and Informativeness Condition}\label{appendix:informativeness}

    In this section, I (i) discuss the relationships among the population pseudo-distance $d(\alpha_i, \alpha_j)$, the sample squared $L_2$ distance $\widehat{d}_{2, ij}^2$, and the population squared $L_2$ distance $d_2^2(\alpha_i, \alpha_j)$, as well as the sufficient conditions for the informativeness used in identification (Assumption~\ref{assumption:informativeness_identification_simultaneous}), (ii) give the definition of $d(\alpha_i, \alpha_j)$ when $\{\gamma_t\}_{t\in \mb{Z}}$ is neither stationary nor ergodic over time, (iii) provide a set of primitive sufficient conditions for Assumption~\ref{assumption:estimation_consistency_d}, and (iv) discuss the informativeness condition used in estimation (Assumption~\ref{assumption:informativeness_estimation}) and verify it under Example~\ref{example:dynamic_interactive_fixed_effects} and~\ref{example:dynamic_nonlinear_interactive_fixed_effects}.

\subsection{Pseudo distance and $L_2$ distance}\label{appendix_sub:informativeness_comparison} 
    Recall the definition of the sample squared $L_2$ distance $\widehat{d}_{2, ij}^2$:
    \begin{align*}
        \widehat{d}_{2, ij}^2 = \frac{1}{T_0}\sum_{t=0}^{T_0-1}(Y_{it} - Y_{jt})^2. 
    \end{align*}
    Under the regularity conditions in Assumption~\ref{assumption:identification_d_simultaneous}, define the population squared $L_2$ distance $d_2^2(\alpha_i,\alpha_j)$ as
    \begin{align*}
        d^2 (\alpha_i, \alpha_j ) =  & \int  \left(g(\alpha_i, \Gamma ) - g(\alpha_{j}, \Gamma_t )\right)^2 \mr{d}\mb{P}(\Gamma ). 
    \end{align*}
    Although the population squared $L_2$ distance $d_2^2(\alpha_i,\alpha_j)$ serves as a natural measure of the distance between $\alpha_i$ and $\alpha_j$ in the function space, it is infeasible to estimate directly from the sample squared $L_2$ distance $\widehat{d}_{2,ij}^2$.
    To see this, note that under the regularity conditions in Assumption~\ref{assumption:identification_d_simultaneous}, 
    \begin{align*}
        \widehat{d}_{2, ij}^2 = d_2^2(\alpha_i, \alpha_j) + \mb{E}_T\left(\widetilde{\epsilon}^2_{it} \mid \alpha_i \right) + \mb{E}_T\left(\widetilde{\epsilon}^2_{jt} \mid \alpha_j \right) + o_P(1). 
    \end{align*}
    The result follows from an argument similar to that used to establish the consistency of $\widehat{d}_{ij}$ in Appendix~\ref{appendix:identification}, and the proof is therefore omitted. The additional variance terms, $\mb{E}_T\left(\widetilde{\epsilon}^2_{it} \mid \alpha_i \right)$ and $\mb{E}_T\left(\widetilde{\epsilon}^2_{jt} \mid \alpha_j \right)$, do not vanish asymptotically and therefore contaminate the estimation. In general, these conditional variance terms depend on latent characteristics and cannot be removed, and only under homoskedasticity can one recover $d_2^2(\alpha_i, \alpha_j)$ up to an additive constant. 

    By definition of pseudo distance, it is straightforward for verify that the population pseudo distance provides an upper bound for $d_2^2(\alpha_i, \alpha_j)$: 
    \begin{align*}
        d(\alpha_i,\alpha_j) : = \sup_{\alpha_1, \alpha_2 \in \mc{A}} \left| \int (g(\alpha_{1}, \Gamma) -g(\alpha_{2}, \Gamma) )(g(\alpha_{i}, \Gamma ) -g(\alpha_{j}, \Gamma) )\mr{d}\mb{P}(\Gamma) \right| \geq d_2^2(\alpha_i, \alpha_j). 
    \end{align*}
    This relationship is useful for proving the following lemma: 
    \begin{lemma}[Sufficient conditions for Assumption~\ref{assumption:informativeness_identification_simultaneous}]\label{lemma:sufficient_informativeness_identification_simultaneous}
        Under Assumption~\ref{assumption:identification_d_simultaneous}, suppose further that for any $\alpha_1\neq\alpha_2$, $g(\alpha_1,\Gamma)$ and $g(\alpha_2,\Gamma)$ are not almost surely identical. Then Assumption~\ref{assumption:informativeness_identification_simultaneous} holds. 
    \end{lemma} 
    \begin{prooflmm}{lemma:sufficient_informativeness_identification_simultaneous}
        For each pair $\alpha_1, \alpha_2\in\mc{A}$ such that $\alpha_1 \neq \alpha_2$, since $g(\alpha_1,\Gamma)$ and $g(\alpha_2,\Gamma)$ are not almost surely identical, we have 
        \begin{align*}
            d(\alpha_1, \alpha_2) \geq d_2^2(\alpha_1, \alpha_2) >0. 
        \end{align*}
        By the continuity of $g(\cdot, \cdot)$ and envelope condition (see Assumption~\ref{assumption:identification_d_simultaneous}\ref{item:identification_d_simultaneous_ULLN} and \ref{item:identification_d_simultaneous_smoothness}), $d(\cdot, \cdot)$ is also continuous. In addition, for any $\epsilon>0$, since the set $\{(\alpha_1, \alpha_2) \in \mc{A}^2\mid \| \alpha_1 - \alpha_2\| \geq \epsilon \}$ is compact. It follows that for any $\epsilon >0$, there exists a constant $\delta_{\epsilon} > 0$ such that 
        \begin{align*}
            \inf_{(\alpha_1, \alpha_2) \in \mc{A}^2, \| \alpha_1 - \alpha_2\| \geq \epsilon  } d(\alpha_1, \alpha_2) \geq \delta_{\epsilon}. 
        \end{align*}
        It immediately implies that for any $\epsilon >0$, as long as $d(\alpha_1, \alpha_2) \leq \delta_{\epsilon}/2$, $\|\alpha_1 - \alpha_2 \| \leq \epsilon$. This completes the proof. 
    \end{prooflmm}

\subsection{Pseudo distance without stationarity and ergodicity}\label{appendix_sub:definition_d_without_stationary_ergodicity}
    When stationarity and ergodicity are not imposed, define $d(\alpha_i, \alpha_j)$ as  
    \begin{align}\label{eq:pseudo_distance_without_stationarity_ergodicity}
        d(\alpha_i, \alpha_j) : = \sup_{\alpha_1, \alpha_2 \in \mc{A}} \left| \lim_{T_0 \rightarrow \infty} \frac{1}{T_0}\sum_{t = 1}^{T_0}\mb{E}\left[ (g(\alpha_1, \Gamma_t) - g(\alpha_2, \Gamma_t))(g(\alpha_i, \Gamma_t) - g(\alpha_j, \Gamma_t))\right]  \right| 
    \end{align}
    It is straightforward to see that, if the limit in~\eqref{eq:pseudo_distance_without_stationarity_ergodicity} exists, then $d(\alpha_i, \alpha_j)$ does not depend on the realization of $\{\gamma_t\}_{t\in\mb{Z}}$. The existence of the limit in~\eqref{eq:pseudo_distance_without_stationarity_ergodicity} requires restrictions on the limiting distribution of $\gamma_t$, and a discussion of such restrictions is therefore omitted.

    In addition, it is straightforward to verify that, under Assumption~\ref{assumption:identification_d_simultaneous}\ref{item:identification_d_simultaneous_compact}-\ref{item:identification_d_simultaneous_smoothness} and the following conditions: (i) $\{\gamma_t\}$ is weakly dependent, and (ii) information from the distant past becomes asymptotically negligible, i.e.,
    \begin{align*}
            \sup_{i, t} \left| g(\alpha_i, \Gamma_t) - \mb{E} (g(\alpha_i, \Gamma_t) \mid \gamma_{t}, \ldots, \gamma_{t-m} ) \right| \rightarrow 0, \quad \text{ as } m\rightarrow\infty, 
    \end{align*}
    whenever the limit in~\eqref{eq:pseudo_distance_without_stationarity_ergodicity} exists, we have $\widehat{d}_{ij} \cp d(\alpha_i, \alpha_j)$.

\subsection{Sufficient conditions for Assumption~\ref{assumption:estimation_consistency_d}}\label{appendix_sub:sufficient_consistency_d_without_proof}
    I introduce the following assumptions: 
    \begin{assumption}\label{assumption:sufficient_estimation_consistency_d}
        I assume that 
        \begin{enumerate}[label=(\roman*)]
            \item \label{item:sufficient_estimation_consistency_d_weak_dependent} \textbf{(Mixing)} Conditional on $\{\gamma_t\}_{t \in \mb{Z}}$, the panel $\{(Y_{it}, X_{it}, D_{it}, \alpha_i)\}_{i=1, \ldots, n, t = 1, \ldots, T}$ is i.i.d. across $i$. 
            In addition, For any $\alpha\in \mc{A}$ and any $\Gamma \in \mr{supp}(\Gamma_{T})$, conditional on $(\alpha_i, \Gamma_{T}) = (\alpha, \Gamma)$, $\{Y_{it}(\infty)\}_{t \in \mb{Z}}$ are $\alpha$-mixing with mixing coefficient $\phi_i(\tau; \alpha, \Gamma)\rightarrow 0$, where 
            \begin{align*}
                \phi_i(\tau; \alpha, \Gamma) = \sup_{ t\in \mb{Z}} \sup_{A \in \mc{A}_{it},  B\in \mc{B}_{it + \tau}} \big | & \mb{P}(B \cap  A \mid \alpha_i = \alpha, \Gamma_{T} = \Gamma) \\ 
                & - \mb{P}(A \mid \alpha_i = \alpha, \Gamma_{T} = \Gamma )\mb{P}_T(B \mid \alpha_i = \alpha, \Gamma_{T} = \Gamma ) \big |. 
            \end{align*} 
            Here, $\mc{A}_{it}$ is the sigma-field generated by $\{Y_{i s}(\infty) \}_{s =- \infty}^{t-1}$, and $ \mc{B}_{it+\tau}$ is the sigma-field generated by $\{ Y_{i t + \tau }(\infty), \ldots, Y_{i T }(\infty) \}$.  $\phi_i(\tau; \alpha, \Gamma) $ exhibits a uniformly exponential decay rate:  there exists $\zeta_0 >0$  such that $\max_{i = 1, \ldots, n}\phi(\tau; \alpha, \Gamma) \leq e^{-\zeta_0 \tau }$ uniformly over $\alpha\in \mc{A}$ and $\Gamma \in \mr{supp}(\Gamma_{T})$. 
            \item \label{item:sufficient_estimation_consistency_d_gamma} \textbf{(Convergence)} $\{\gamma_t\}_{t \in \mb{Z}}$ is stationary and ergodic over time. 
            In addition, there exists a constant $C>0$ such that
            \begin{align*}
                \sup_{\alpha_1, \alpha_2, \alpha_i, \alpha_j \in \mc{A}}\bigg | & \frac{1}{T_0} \sum_{t=0}^{T_0-1} (g(\alpha_{1}, \Gamma_t) - g(\alpha_{2}, \Gamma_t) ) (g(\alpha_{i}, \Gamma_t) - g(\alpha_{j}, \Gamma_t)) \\
                & - \int (g(\alpha_{1}, \Gamma) - g(\alpha_{2}, \Gamma) ) (g(\alpha_{i}, \Gamma) - g(\alpha_{j}, \Gamma)) \mr{d}\mb{P}(\Gamma)\bigg | \leq C \sqrt{\frac{ \log (T_0) }{T_0}}
            \end{align*}
            with probability approaching to $1$. 
            \item \label{item:sufficient_estimation_consistency_d_finite} \textbf{(Boundedness)} $g(\cdot, \cdot)$ is uniformly bounded on the relavent supports. In addition, for some $\sigma>0$, $\mb{E}\left(\exp\left(\lambda(Y_{it}(\infty)-g(\alpha_i, \Gamma_t))\right)\mid \alpha_i,\Gamma_T\right)\leq \exp(\sigma^2\lambda^2/2)$ a.s. for all $\lambda\in\mb{R}$, all $i$ and $t$. 
            \item \label{item:sufficient_estimation_consistency_d_compact}  \textbf{(Compactness)} The support of $\alpha$, $\mc{A}$, is compact. There exist constants $\underline{c}_1, \overline{c}_1 >0$, such that for any $\alpha \in \mc{A}$ and any $r \in (0, 1)$, $\underline{c}_1 r^{d_{\alpha}} \leq \mb{P}\left(\|\alpha_i - \alpha \| \leq r \right) \leq \overline{c}_1 r^{d_{\alpha}}$. 
            \item \label{item:sufficient_estimation_consistency_d_L_continuous} \textbf{(Lipschitz)} There exists a constant $L_g$ such that for any $\alpha_1, \alpha_2 \in \mc{A}$, 
            \begin{align*}
                \sup_{\Gamma \in \mr{supp}(\Gamma_{T})}  |g(\alpha_1, \Gamma) - g(\alpha_2, \Gamma) | \leq L_g \|\alpha_1 - \alpha_2\|_{\infty}. 
            \end{align*}
        \end{enumerate} 
    \end{assumption}
    
    Assumption~\ref{assumption:sufficient_estimation_consistency_d}\ref{item:sufficient_estimation_consistency_d_weak_dependent} strengthens the weak-dependence condition in Assumption~\ref{assumption:potential_outcome_extension}\ref{item:potential_outcome_extension_weak_dependence} by imposing conditional $\alpha$-mixing with mixing coefficients that decay uniformly at an exponential rate. Assumption~\ref{assumption:sufficient_estimation_consistency_d}\ref{item:sufficient_estimation_consistency_d_gamma} strengthens the ergodic convergence theorem by imposing an explicit uniform convergence rate. It is satisfied when the influence of distant time factors on $g(\alpha,\Gamma_t)$ decays sufficiently fast. Assumption~\ref{assumption:sufficient_estimation_consistency_d}\ref{item:sufficient_estimation_consistency_d_L_continuous} imposes Lipschitz continuity, which is stronger than the uniform continuity condition in Assumption~\ref{assumption:identification_d_simultaneous}\ref{item:identification_d_simultaneous_smoothness}.  

    \begin{lemma}[Sufficient conditions for Assumption~\ref{assumption:estimation_consistency_d}]\label{lemma:sufficient_estimation_consistency_d}
        Assumption~\ref{assumption:sufficient_estimation_consistency_d} implies Assumption~\ref{assumption:estimation_consistency_d}.
    \end{lemma}

\subsection{Discussion on Assumption~\ref{assumption:informativeness_estimation}}\label{appendix_sub:sufficient_informativeness_without_proof}

    Verifying the lower bound $\underline{\eta} d (\alpha_1, \alpha_{2}) \leq \|\alpha_1- \alpha_2\|$  is straightforward. When $g$ is Lipschitz as in~Assumption~\ref{assumption:sufficient_estimation_consistency_d}\ref{item:sufficient_estimation_consistency_d_L_continuous}, 
    \begin{align*}
       d(\alpha_i,\alpha_j) : = &\sup_{\alpha_1, \alpha_2 \in \mc{A}} \left| \int (g(\alpha_{1}, \Gamma) -g(\alpha_{2}, \Gamma) )(g(\alpha_{i}, \Gamma ) -g(\alpha_{j}, \Gamma) )\mr{d}\mb{P}(\Gamma) \right| \\
       \leqtext{(i)} &  \sup_{\alpha_1, \alpha_2 \in \mc{A}} \left( \int (g(\alpha_{1}, \Gamma) -g(\alpha_{2}, \Gamma) )^2 \mr{d}\mb{P}(\Gamma)\right)^{1/2}  L_g\sqrt{d_{\alpha}} \|\alpha_i- \alpha_j\|, 
    \end{align*}
    where (i) follows from the Cauchy-Schwarz inequality and the Lipschitz continuity of $g$. Thus, the lower bound holds if the second moment of $g$ is uniformly bounded over $\mc{A}$, i.e., 
    \begin{align*}
        \sup_{\alpha \in \mc{A}}  \int g^2(\alpha, \Gamma) \mr{d}\mb{P}(\Gamma) <\infty. 
    \end{align*}

    To verify the upper bound, I introduce the following non-redundancy condition: 
    \begin{definition}[Nonredundancy]\label{definition:nonredundancy}
        We say that $\mathcal{A}\subseteq \mathbb{R}^d$ is \emph{effective} if there does not exist a nontrivial partition of coordinates $\mc{I}_1, \mc{I}_2 \subseteq \{1,\dots,d\}$ with $\mc{I}_1 \cap \mc{I}_2 = \emptyset$, $\mc{I}_2 \neq \emptyset$, and a measurable function $q : \mathbb{R}^{|\mc{I}_1|} \to \mathbb{R}^{|\mc{I}_2|}$ such that
        \begin{align*}
            \alpha_{\mc{I}_2} = q(\alpha_{\mc{I}_1}), \quad \forall \alpha \in \mathcal{A}.
        \end{align*}
    \end{definition}
    That is, no subset of coordinates of $\alpha$ can be written as a deterministic function of the remaining coordinates on $\mathcal{A}$, or equivalently,  $\mc{A}$ is not contained in a lower-dimensional manifold. By definition, nonredundancy immediately implies that $\mr{affine}(\mc{A}) = \mb{R}^{d_{\alpha}}$. 

    \begin{lemma}[Linear dynamic panel with interactive fixed effects]\label{lemma:sufficient_informativeness_linear_IFE}
        Suppose 
        \begin{gather*}
            Y_{it}(\infty) = \rho Y_{it-1}(\infty) + \gamma_t' \alpha_i + \epsilon_{it}, \\
            \mb{E}\left(\epsilon_{it} \mid Y_{it-1}, \alpha_i, \Gamma_{T}\right) = 0, \quad \mb{E}\left(\epsilon_{it}^2 \mid Y_{it-1}, \alpha_i, \Gamma_{T}\right) <\infty. 
        \end{gather*}
        Then, the informativeness condition (Assumption~\ref{assumption:informativeness_estimation}) holds if (i) $\mc{A}$ is nonredundant, compact, and convex, (ii) $|\rho| <1$, (iii) $\gamma_t$ is ergodic and stationary, and (iv) $0 < s_{\min}(\mb{E}(\gamma_t\gamma_t')) \leq s_{\max}(\mb{E}(\gamma_t\gamma_t'))<\infty $, where $s_{\min}(\cdot)$ and $s_{\max}(\cdot)$ denote the smallest and largest eigenvalue, respectively.   
    \end{lemma}

    The condition that $\gamma_t$ is ergodic and stationary seems restrictive, but it is imposed to simplify the technical discussion. It is straightforward to show that, Lemma~\ref{lemma:sufficient_informativeness_linear_IFE} still holds if I replace conditions (iii) and (iv) with 
    \begin{align*}
        \frac{1}{T_0} \sum_{t=0}^{T_0-1} \gamma_t\gamma_t' \text{ exists,  and } 0 < s_{\min}\left(\mb{E}\left(\frac{1}{T_0} \sum_{t=0}^{T_0-1} \gamma_t\gamma_t'\right)\right) \leq s_{\max}\left(\mb{E}\left(\frac{1}{T_0} \sum_{t=0}^{T_0-1} \gamma_t\gamma_t'\right)\right)<\infty.  
    \end{align*}
    
    \bigskip 

    For nonlinear dynamic panel models discussed in Example~\ref{example:dynamic_nonlinear_interactive_fixed_effects}, recall that the outcome is generated according to 
    \begin{align*}
            Y_{it}(\infty) = \bs{1}\left(\beta Y_{it-1}(\infty) + \gamma_t'\alpha_i -u_{it}\geq 0\right), 
    \end{align*}
    and let $F(\cdot)$ denote the cumulative distribution function of $u_{it}$. Still, we assume that $u_{it}$ is independent of past histories conditional on all fixed effects. Note that the outcome process is also written as 
    \begin{align*}
        Y_{it}(\infty) = F\left( \beta Y_{it-1}(\infty) + \gamma_t'\alpha_i \right) + \epsilon_{it}. 
    \end{align*}
    For notational simplicity, define $g_{t}(\alpha_i): = g(\alpha_i, \Gamma_t) = \mb{E}\left(Y_{it}(\infty) \mid \alpha_i, \Gamma_t\right)$. 
        In addition, let $\rho_t(\alpha_i) := F\left(\beta  + \gamma_t'\alpha_i \right) - F\left( \gamma_t'\alpha_i \right)$ denote the stochastic discount factor. Then,   
        \begin{align*}
            g_t(\alpha_i) = & (1 - g_{t-1}(\alpha_i))F\left(\gamma_t'\alpha_i \right)  + g_{t-1}(\alpha_i) F\left(\beta  + \gamma_t'\alpha_i \right) = F\left(\gamma_t'\alpha_i \right) + \rho_t( \alpha_i ) g_{t-1}(\alpha_i). 
        \end{align*}
    Let $f_t: = \mb{E}\left(Y_{it} \mid  \alpha_i, \Gamma \right)$. Let $\rho(\gamma_t'\alpha_i )$ denote $\Gamma\left(1 + \gamma_t'\alpha_i \right) - \Gamma\left( \gamma_t'\alpha_i \right)$.  Then,  
    \begin{equation}\label{eq:nonlinear_model_recursive_transformation}
    \begin{aligned}
        g_t(\alpha_i) = & F \left(\gamma_t'\alpha_i \right) + \rho_t( \alpha_i ) g_{t-1}(\alpha_i)  \\
        = & F\left(\gamma_t'\alpha_i \right) + \rho_{t}(\alpha_i ) \left(F\left(\gamma_{t-1}'\alpha_i \right) + \rho_{t-1}(\alpha_i )g_{t-2}(\alpha_i) \right)  \\
        = & F\left(\gamma_t'\alpha_i \right) + \rho_t( \alpha_i ) F \left(\gamma_{t-1}'\alpha_i \right) + \rho_{t}(\alpha_i )\rho_{t-1}(\alpha_i ) F_{t-2}(\alpha_i) \\
        = & \sum_{\tau = 0}^{\infty} \left(\prod_{s = 0}^{\tau-1} \rho_{t-s}( \alpha_i)  \right) F\left(\gamma_{t-\tau}'\alpha_i \right)
    \end{aligned}   
    \end{equation}

    \begin{lemma}[Nonlinear dynamic panel with interactive fixed effects]\label{lemma:sufficient_informativeness_nonlinear_IFE} Suppose 
        \begin{align*}
            Y_{it}(\infty) = \bs{1}\left(\beta Y_{it-1}(\infty) + \gamma_t'\alpha_i -u_{it}\geq 0\right). 
        \end{align*}
        The error terms $\{u_{it}\}_{i=1,\ldots,n,\;t\in\mb{Z}}$ are independent of $\{Y_{i\tau}(\infty)\}_{i=1,\ldots,n,\;\tau\leq t-1}$ and the fixed effects. In addition, they are i.i.d. across time and units. Let $F(\cdot)$ denote the cumulative distribution function of $u_{it}$. 

        Then, the informativeness condition (Assumption~\ref{assumption:informativeness_estimation}) holds if (i) $F(\cdot)$ is twice continuously differentiable and $F^{(1)}(\cdot)$ on the index space; (ii) $\mc{A}$ is nonredundant, compact, and convex; (iii) $\{\gamma_t\}_{t\in\mb{Z}}$ are i.i.d. over time, with bounded $\mr{supp}(\gamma_t)$; and (iv) $s_{\min}\left(\mb{E}(\gamma_t\gamma_t')\right)>0$,
        where $s_{\min}(\cdot)$ denotes the smallest eigenvalue. 
    \end{lemma}

    \clearpage

    \section{Proofs of Identification}\label{appendix:identification}

    We provide only the proof of Theorem~\ref{thm:identification_extension}, as it extends Theorem~\ref{thm:identification_simultaneous} by incorporating staggered adoptions. 

    \bigskip

    \begin{proofthm}{thm:identification_extension}
        We first prove the identification of the population pseudo-distance in large-$n$, large-$T_0$ panel data by showing that the sample pseudo-distance $\widehat d_{ij}$ converges in probability to $d(\alpha_i,\alpha_j)$. We then establish the identification of the contemporaneous ATT. Finally, we show how to identify the dynamic ATT using a recursive argument. 

        \paragraph{Identification of $d(\alpha_i, \alpha_j)$} For notationally simplicity, for any $\alpha_1, \alpha_2 \in \mc{A}$  and any $\Gamma \in \mr{supp}(\Gamma_{T})$,  define $\Delta(\alpha_1, \alpha_2, \Gamma) := (g(\alpha_1, \Gamma) - g(\alpha_2, \Gamma))$. 
        Note that for any $i, j\in 1, \ldots, n$, 
        \begin{align*}
            & \left| \widehat{d}_{ij} - d(\alpha_i, \alpha_j)\right| \\ \leq  &  \underbrace{\max_{k_1, k_2 \in \{1, \ldots, n\}\setminus \{i, j\} }\left| \frac{1}{T_0} \sum_{t=0}^{T_0-1} \Delta(\alpha_{k_1}, \alpha_{k_2}, \Gamma_t) \Delta(\alpha_i, \alpha_j, \Gamma_t) - \mb{E}\left(\Delta(\alpha_{k_1}, \alpha_{k_2}, \Gamma) \Delta(\alpha_i, \alpha_j, \Gamma) \mid \alpha_{k_1}, \alpha_{k_2}, \alpha_i, \alpha_j\right)\right|}_{A_1} \\
            & + \underbrace{\left| \max_{k_1, k_2 \in \{1, \ldots, n\}\setminus \{i, j\} }  \mb{E}\left(\Delta(\alpha_{k_1}, \alpha_{k_2}, \Gamma) \Delta(\alpha_i, \alpha_j, \Gamma) \mid \alpha_{k_1}, \alpha_{k_2}, \alpha_i, \alpha_j \right) - d(\alpha_i, \alpha_j) \right|}_{A_2} \\
            & +  \underbrace{ \max_{k_1, k_2 \in \{1, \ldots, n\}\setminus \{i, j\} } \left|\frac{1}{T_0} \sum_{t=0}^{T_0-1}\Delta(\alpha_i, \alpha_j, \Gamma_t) (\widetilde{\epsilon}_{k_1 t} - \widetilde{\epsilon}_{k_2 t}) \right|}_{A_3} + \underbrace{\max_{k_1, k_2 \in \{1, \ldots, n\}\setminus \{i, j\} } \left|\frac{1}{T_0} \sum_{t=0}^{T_0-1}  \Delta(\alpha_{k_1}, \alpha_{k_2}, \Gamma_t)(\widetilde{\epsilon}_{i t} - \widetilde{\epsilon}_{j t})\right|}_{A_4}  \\
            & +  \underbrace{\max_{k_1, k_2 \in \{1, \ldots, n\}\setminus \{i, j\} } \left|\frac{1}{T_0} \sum_{t=0}^{T_0-1} (\widetilde{\epsilon}_{it} - \widetilde{\epsilon}_{jt}) (\widetilde{\epsilon}_{k_1 t} - \widetilde{\epsilon}_{k_2 t})\right| }_{A_5}
        \end{align*}
        For the first term $A_1$, since (i) $\mc{A}$ is compact, (ii) fixed $\alpha_1, \alpha_2 \in \mc{A}$, $\Delta(\alpha_1, \alpha_2,  \Gamma_t)$ is stationary and ergodic over time by Assumption~\ref{assumption:identification_d_simultaneous}\ref{item:identification_d_simultaneous_panel_iid}, (iii) $\Delta(\alpha_1, \alpha_2,  \Gamma_t)$ is continuous in $(\alpha_1, \alpha_2)$  for all $\Gamma_t$ by Assumption~\ref{assumption:identification_d_simultaneous}\ref{item:identification_d_simultaneous_smoothness}, and (iv) the envelope condition in Assumption~\ref{assumption:identification_d_simultaneous}\ref{item:identification_d_simultaneous_ULLN}, 
        \begin{align*}
            \mb{E}\left(\sup_{ \alpha_1, \alpha_2 }\left| \Delta(\alpha_{k_1}, \alpha_{k_2}, \Gamma )\Delta(\alpha_i, \alpha_j, \Gamma)\right| \right)\leq 4 \mb{E}\left(\sup_{\alpha}\left| g(\alpha, \Gamma)\right|^2 \right) <\infty,  
        \end{align*}
        we employ the uniform law of large numbers of ergodic and stationary process (\citet[Lemma~7.2]{hayashi2011econometrics}) to show that $A_1 =o_P(1)$.  
        
        For the second term, since $\mb{E}\left(\Delta(\alpha_{k_1}, \alpha_{k_2}, \Gamma) \Delta(\alpha_i, \alpha_j, \Gamma) \mid \alpha_{k_1}, \alpha_{k_2}, \alpha_i, \alpha_j\right)$ is continuous in $(\alpha_{k_1}, \alpha_{k_2})$ when fixing $(\alpha_i, \alpha_j)$ (see \citet[Lemma~7.2]{hayashi2011econometrics}), then, by the compactness of $\mc{A}$ in Assumption~\ref{assumption:identification_d_simultaneous}\ref{item:identification_d_simultaneous_compact}, it immediately follows that
        \begin{align*}
             \max_{k_1, k_2 \in \{1, \ldots, n\}\setminus \{i, j\} }  \mb{E}\left(\Delta(\alpha_{k_1}, \alpha_{k_2}, \Gamma) \Delta(\alpha_i, \alpha_j, \Gamma) \mid \alpha_{k_1}, \alpha_{k_2}, \alpha_i, \alpha_j \right) \cp \sup_{\alpha_1, \alpha_2\in \mc{A}} \int \Delta(\alpha_{1}, \alpha_{2}, \Gamma)\Delta(\alpha_{i}, \alpha_{j}, \Gamma)\mr{d}\mb{P}(\Gamma). 
        \end{align*}
        In other words, the sample maximum over $(\alpha_{k_1},\alpha_{k_2})$ converges in probability to the population supremum over $(\alpha_1,\alpha_2)\in\mc{A}^2$. Therefore, $A_2 = o_P(1)$. 

        For $A_3$, Since $\mb{E}\left(\widetilde{\epsilon}_{k_1t} \mid \{\alpha_i\}_{i=1}^{n}, \Gamma_{T}\right) = \mb{E}\left(\widetilde{\epsilon}_{k_2 t} \mid \{\alpha_i\}_{i=1}^{n}, \Gamma_{T}\right) =  0$,  we have 
        \begin{align*}
            &  \mr{Var}\left(\frac{1}{T_0} \sum_{t=0}^{T_0-1}\Delta(\alpha_i, \alpha_j, \Gamma_t)(\widetilde{\epsilon}_{k_1 t} -\widetilde{\epsilon}_{k_2 t}) \mid \{\alpha_i\}_{i=1}^{n}, \Gamma_{T} \right)\\
            =  &   \frac{1}{T^2_0} \sum_{s, t = 1, \ldots, T_0} \Delta(\alpha_i, \alpha_j, \Gamma_s)\Delta(\alpha_i, \alpha_j, \Gamma_t)\mb{E} \left(\widetilde{\epsilon}_{k_1 t}\widetilde{\epsilon}_{k_1 s} + \widetilde{\epsilon}_{k_2 t}\widetilde{\epsilon}_{k_2 s}\mid  \{\alpha_i\}_{i=1}^{n}, \Gamma_{T} \right) \\
            \leq & \frac{1}{ T^2_0} \sum_{s, t = 1, \ldots, T_0 } \left|\Delta(\alpha_i, \alpha_j, \Gamma_s)\Delta(\alpha_i, \alpha_j, \Gamma_t)\right| \left|\mb{E} \left(\widetilde{\epsilon}_{k_1 t}\widetilde{\epsilon}_{k_1 s} + \widetilde{\epsilon}_{k_2 t}\widetilde{\epsilon}_{k_2 s}\mid  \{\alpha_i\}_{i=1}^{n}, \Gamma_{T} \right)\right| \\
            \leq & \left(2 \max_{t = 1, \ldots, T_0}\Delta^2(\alpha_i, \alpha_j, \Gamma_t) \right)\frac{1}{T^2_0} \sum_{t = 1}^{T_0} \sum_{s = 1}^{T_0} \left(  \left|\mb{E} \left(\widetilde{\epsilon}_{k_1 t}\widetilde{\epsilon}_{k_1s}\mid  \alpha_{k_1}, \Gamma_{T} \right)\right| + \left| \mb{E} \left(\widetilde{\epsilon}_{k_2 t}\widetilde{\epsilon}_{k_2s}\mid  \alpha_{k_2}, \Gamma_{T} \right)\right|  \right), 
        \end{align*}
        The weak dependence condition in  Assumption~\ref{assumption:potential_outcome_extension}\ref{item:potential_outcome_extension_weak_dependence} requires that  
        \begin{align*}
            \frac{1}{T_0}\sum_{t = 1}^{T_0} \sum_{s = 1}^{T_0}\left|\mb{E} \left(\widetilde{\epsilon}_{k_1 t}\widetilde{\epsilon}_{k_1s}\mid  \alpha_{k_1}, \Gamma_{T} \right)\right| + \left| \mb{E} \left(\widetilde{\epsilon}_{k_2 t}\widetilde{\epsilon}_{k_2s}\mid  \alpha_{k_2}, \Gamma_{T} \right)\right| <\infty
        \end{align*}
        In addition, it is straightforward to verify that, by moment restriction in Assumption~\ref{assumption:identification_d_simultaneous}\ref{item:identification_d_simultaneous_finite_moment},  
        \begin{align}
            \max_{t = 1, \ldots, T_0}\Delta^2(\alpha_i, \alpha_j, \Gamma_t) < \leq \max_{t = 1, \ldots, T_0}\mb{E}\left(Y^2_{it}(\infty) \mid \alpha_i, \Gamma_{T} \right) < \infty. 
        \end{align}
        Then, it follows that there exists a constant $C_1$ (irrelevant of the realization of $(\{\alpha_i\}_{i=1}^{n}, \Gamma_{T})$) such that 
        \begin{align*}
            \mr{Var}\left(\frac{1}{T_0} \sum_{t=0}^{T_0-1}\Delta(\alpha_i, \alpha_j, \Gamma_t)(\widetilde{\epsilon}_{k_1 t} -\widetilde{\epsilon}_{k_2 t}) \mid \{\alpha_i\}_{i=1}^{n}, \Gamma_{T}  \right) \leq \frac{C_1}{T_0}
        \end{align*}
        Then, by Markov inequality, for any $\delta >0$, 
        \begin{align*}
            \mb{P}\left(\left| A_3\right| \geq \delta \mid \{\alpha_i\}_{i=1}^{n}, \Gamma_{T}\right) = & \mb{P}\left(\bigcup_{k_1, k_2} \left\{ \left| \frac{1}{T_0} \sum_{t=0}^{T_0-1}\Delta(\alpha_i, \alpha_j, \Gamma_t)(\widetilde{\epsilon}_{k_1 t} -\widetilde{\epsilon}_{k_2 t})\right|  \geq \delta \right\} \mid \{\alpha_i\}_{i=1}^{n}, \Gamma_{T}\right) \\
            \leq & \sum_{k_1, k_2=1, \ldots, n}\mb{P}\left(  \left| \frac{1}{T_0} \sum_{t=0}^{T_0-1}\Delta(\alpha_i, \alpha_j, \Gamma_t)(\widetilde{\epsilon}_{k_1 t} -\widetilde{\epsilon}_{k_2 t}) \right| \geq \delta  \mid \{\alpha_i\}_{i=1}^{n}, \Gamma_{T}\right) \\
            \leq & n^2 \frac{C_1}{T_0 \delta^2}. 
        \end{align*}
        Since the inequalities above does not depend on the realization of $\Gamma_{T}$, we integrate them out to obtain 
        \begin{align}
            \mb{P}\left(\left| A_3\right| \geq \delta \right) \leq n^2 \frac{C_1 }{T_0 \delta^2}. 
        \end{align}
        The bound above implies that $A_3=o_P(1)$ whenever $n^2/T_0\rightarrow0$. This relative-rate restriction, however, is not essential for identification. Indeed, even when $n^2/T_0\nrightarrow0$, it suffices to construct the sample maximum using a subsample of $n_{T_0}$ individuals such that $n_{T_0}\rightarrow\infty$ and $n_{T_0}=o(\sqrt{T_0})$. The sample maximum over this subsample still converges to the population supremum, which implies that $A_3=o_P(1)$. 
        Applying the same argument on $A_4$ yields $A_4 = o_P(1)$. 

        For the last term $A_5$, note that $\mb{E}\left((\widetilde{\epsilon}_{it} - \widetilde{\epsilon}_{jt}) (\widetilde{\epsilon}_{k_1 t} - \widetilde{\epsilon}_{k_2 t}) \mid \{\alpha_i\}_{i=1}^{n}, \Gamma_{T}\right) = 0$, and 
        \begin{align*}
            & \mr{Var}\left(\frac{1}{T_0} \sum_{t=0}^{T_0-1} (\widetilde{\epsilon}_{it} - \widetilde{\epsilon}_{jt}) (\widetilde{\epsilon}_{k_1 t} - \widetilde{\epsilon}_{k_2 t}) \mid \{\alpha_i\}_{i=1}^{n}, \Gamma_{T} \right) \\
            = & \frac{1}{T^2_0} \sum_{s, t = 1, \ldots, T_0} \mb{E} \left(\widetilde{\epsilon}_{i t}\widetilde{\epsilon}_{is} + \widetilde{\epsilon}_{jt}\widetilde{\epsilon}_{j s}\mid  \{\alpha_i\}_{i=1}^{n}, \Gamma_{T} \right) \mb{E} \left(\widetilde{\epsilon}_{k_1 t}\widetilde{\epsilon}_{k_1 s} + \widetilde{\epsilon}_{k_2 t}\widetilde{\epsilon}_{k_2 s}\mid  \{\alpha_i\}_{i=1}^{n}, \Gamma_{T} \right)\\
            \leq & \max_{t = 1\ldots, T_0} \left(\mb{E} \left(\widetilde{\epsilon}_{i t}\widetilde{\epsilon}_{is} + \widetilde{\epsilon}_{jt}\widetilde{\epsilon}_{j s}\mid  \{\alpha_i\}_{i=1}^{n}, \Gamma_{T} \right)\right)  \frac{1}{T^2_0} \sum_{t = 1}^{T_0} \sum_{s = 1}^{T_0} \left(  \left|\mb{E} \left(\widetilde{\epsilon}_{k_1 t}\widetilde{\epsilon}_{k_1s}\mid  \alpha_{k_1}, \Gamma_{T} \right)\right| + \left| \mb{E} \left(\widetilde{\epsilon}_{k_2 t}\widetilde{\epsilon}_{k_2s}\mid  \alpha_{k_2}, \Gamma_{T} \right)\right|  \right)  
        \end{align*}
        Since  for any realization of $(\{\alpha_i\}_{i=1}^{n}, \Gamma_{T} )$, (i) $\max_{t = 1\ldots, T_0} \left(\mb{E} \left(\widetilde{\epsilon}_{i t}\widetilde{\epsilon}_{is} + \widetilde{\epsilon}_{jt}\widetilde{\epsilon}_{j s}\mid  \{\alpha_i\}_{i=1}^{n}, \Gamma_{T} \right)\right)  <\infty$ by Assumption~\ref{assumption:identification_d_simultaneous}\ref{item:identification_d_simultaneous_finite_moment}, and $\frac{1}{T_0} \sum_{t = 1}^{T_0} \sum_{s = 1}^{T_0} \left(  \left|\mb{E} \left(\widetilde{\epsilon}_{k_1 t}\widetilde{\epsilon}_{k_1s}\mid  \alpha_{k_1}, \Gamma_{T} \right)\right| + \left| \mb{E} \left(\widetilde{\epsilon}_{k_2 t}\widetilde{\epsilon}_{k_2s}\mid  \alpha_{k_2}, \Gamma_{T} \right)\right|  \right) <\infty$ by the weak dependence condition in Assumption~\ref{assumption:potential_outcome_extension}\ref{item:potential_outcome_extension_weak_dependence}. There exists a constant $C_2$ (irrelevant of the realization of $(\{\alpha_i\}_{i=1}^{n}, \Gamma_{T})$) such that 
        \begin{align*}
            \mr{Var}\left(\frac{1}{T_0} \sum_{t=0}^{T_0-1}(\widetilde{\epsilon}_{i t} -\widetilde{\epsilon}_{j t})(\widetilde{\epsilon}_{k_1 t} -\widetilde{\epsilon}_{k_2 t}) \mid \{\alpha_i\}_{i=1}^{n}, \Gamma_{T}  \right) \leq \frac{C_2}{T_0}
        \end{align*}
        Then, by Markov inequality, for any $\delta >0$, 
        \begin{align*}
            \mb{P}\left(\left| A_5\right| \geq \delta \mid \{\alpha_i\}_{i=1}^{n}, \Gamma_{T}\right) = & \mb{P}\left(\bigcup_{k_1, k_2} \left\{ \left| \frac{1}{T_0} \sum_{t=0}^{T_0-1}(\widetilde{\epsilon}_{i t} -\widetilde{\epsilon}_{j t})(\widetilde{\epsilon}_{k_1 t} -\widetilde{\epsilon}_{k_2 t})\right|  \geq \delta \right\} \mid \{\alpha_i\}_{i=1}^{n}, \Gamma_{T}\right) \\
            \leq & \sum_{k_1, k_2=1, \ldots, n}\mb{P}\left(  \left| \frac{1}{T_0} \sum_{t=0}^{T_0-1}(\widetilde{\epsilon}_{i t} -\widetilde{\epsilon}_{j t})(\widetilde{\epsilon}_{k_1 t} -\widetilde{\epsilon}_{k_2 t}) \right| \geq \delta   \mid \{\alpha_i\}_{i=1}^{n}, \Gamma_{T}\right) \\
            \leq & n^2 \frac{C_2}{T_0 \delta^2}. 
        \end{align*}
        Since the inequalities above does not depend on the realization of $(\{\alpha_i\}_{i=1}^{n}, \Gamma_{T})$, we integrate them out to obtain 
        \begin{align}
            \mb{P}\left(\left| A_5\right| \geq \delta \right) \leq n^2 \frac{C_1 }{T_0 \widetilde{\epsilon}^2}. 
        \end{align}
        The bound above implies that $A_5=o_P(1)$ whenever $n^2/T_0\rightarrow0$. This relative-rate restriction, however, as we discussed previously, is not essential for identification. Because even when $n^2/T_0\nrightarrow0$, it suffices to construct the sample maximum using a subsample of $n_{T_0}$ individuals such that $n_{T_0}\rightarrow\infty$ and $n_{T_0}=o(\sqrt{T_0})$. The sample maximum over this subsample still converges to the population supremum, which implies that $A_3=o_P(1)$. 

        Therefore, combining the probability bounds for $A_1$--$A_5$, we have that, for any $i,j = 1, \ldots, n$
        \begin{align*}
            \widehat{d}_{ij} - d(\alpha_i, \alpha_j) = o_P(1),  
        \end{align*}
        which immediately implies that $d(\alpha_i, \alpha_j)$ is identified by the construction of $\widehat{d}_{ij}$. 

        \paragraph{Identification of $\mr{ATT}(t, t)$} For any $t \geq T_0$, we have 
        \begin{align*}
            \mr{ATT}(t, t) = \mb{E}(Y_{it}(t) \mid G_i = t)  - \mb{E}(Y_{it}(\infty) \mid G_i = t)
        \end{align*}
        Since $\mb{E}(Y_{it}(t) \mid G_i = t)$ is identified from the sample, we only need to identify counterfactual $\mb{E}(Y_{it}(\infty) \mid G_i = t)$.  Note that by iterated expectation, 
        \begin{align*}
            \mb{E}\left(Y_{it}(\infty) \mid G_i = t \right) = & \mb{E}\left(\mb{E}\left(Y_{it}(\infty) \mid  X_{it}, \alpha_i,  G_i >t-1, G_i = t \right) \mid G_i = t\right) \\
            \eqtext{(i)} & \mb{E}\left(\mb{E}\left(Y_{it}(\infty) \mid  X_{it}, \alpha_i, G_i >t \right) \mid G_i = t\right) \\
            = & \mb{E}\left( m_{t\mid t}\left(X_{it}, \alpha_i \right) \mid G_i = t\right), 
        \end{align*}
        where (i) follows from Assumption~\ref{assumption:selection_extension}.  
        Let $\mc{S}_{t-1}$ be the support of $(X_{it}, \alpha_i)$ conditional on $\{G_i > t-1\}$. It is straightforward to verify that, by the overlap condition in Assumption~\ref{assumption:identification_ATT_extension}\ref{item:identification_ATT_extension_overlap}, $\mc{S}_{t-1}$ is also the support of $(X_{it}, \alpha_i)$ conditional on $\{G_i > t\}$ and conditional on $\{G_i = t\}$. Then, by the uniform continuity of $m_{t\mid t}(\cdot, \cdot)$ (Assumption~\ref{assumption:identification_ATT_extension}\ref{item:identification_ATT_extension_smoothness}), for any $\epsilon>0$, there exists a $\widetilde{\delta} >0$ such that
        \begin{align*}
            &\sup_{(x, \alpha) \in \mc{S}_{t-1} }\left| \mb{E}\left(Y_{jt}(\infty) \mid X_{jt} = x, \|\alpha_j, \alpha\| \leq \widetilde{\delta}, G_j>t \right) - m_{t\mid t}(x , \alpha)\right|   \leq \epsilon. 
        \end{align*} 
        Therefore, combining the inequality above and the informativeness condition in Assumption~\ref{assumption:informativeness_identification_simultaneous} implies that there exist a $\delta >0$ such that
        \begin{align*}
            &\sup_{(x, \alpha) \in \mc{S}_{t-1} } \left| \mb{E}\left(Y_{jt}(\infty) \mid X_{jt} = x, d(\alpha_j, \alpha) \leq \delta, G_j>t \right) - m_{t\mid t}(x, \alpha)\right|   \leq \epsilon. 
        \end{align*}
        It immediately follows that for any $i$ such that $G_i  >t-1$, the inequality below holds wpa1, 
        \begin{align*}
            \left| \mb{E}\left(Y_{jt}(\infty) \mid X_{jt} = X_{it}, d(\alpha_j, \alpha_i) \leq \delta , G_j >t \right) - m_{t\mid t}(X_{it}, \alpha_i )\right|  \leq \epsilon
        \end{align*}
        Since $d(\alpha_j, \alpha_i)$ is identified, and $\mb{E}\left(Y_{jt}(\infty) \mid X_{jt} = X_{it}, d(\alpha_j, \alpha_i) \leq \delta , G_j >t \right)$ is directly identified by $\mb{E}\left(Y_{jt} \mid X_{jt} = X_{it}, d(\alpha_j, \alpha_i) \leq \delta , G_j >t \right)$, for any $i$ such that $G_i >t-1$,  the contemporaneous counterfactual expectation is identified by 
        \begin{align}\label{eq:thm_identification_outcome_tt}
            m_{t\mid t}(X_{it}, \alpha_i ) = \lim_{\delta \downarrow 0 }\mb{E}\left(Y_{jt} \mid X_{jt} = X_{it}, d(\alpha_j, \alpha_i) \leq \delta , G_j >t \right). 
        \end{align} 
        This completes the proof of part (i) in Theorem~\ref{thm:identification_extension}.

        Averaging the imputed counterfactual over the distribution of treated units in \eqref{eq:thm_identification_outcome_tt} yields 
        \begin{align*}
            \left| \mb{E}\left(Y_{it}(\infty) \mid G_i = t \right) - \mb{E}\left(\mb{E}\left(Y_{jt}(\infty) \mid X_{jt} = X_{it}, d(\alpha_j, \alpha_i) \leq \delta , G_j >t\right) \mid G_i = t \right) \right| \leq \epsilon \\
            \Rightarrow \mb{E}\left(Y_{it}(\infty) \mid G_i = t \right) = \mb{E}\left(\lim_{\delta\downarrow 0}\mb{E}\left(Y_{jt}(\infty) \mid X_{jt} = X_{it}, d(\alpha_j, \alpha_i) \leq \delta , G_j >t\right) \mid G_i = t \right) \\
            \Rightarrow \mb{E}\left(Y_{it}(\infty) \mid G_i = t \right) = \mb{E}\left(\lim_{\delta\downarrow 0}\mb{E}\left(Y_{jt} \mid X_{jt} = X_{it}, d(\alpha_j, \alpha_i) \leq \delta , G_j >t\right) \mid G_i = t \right) 
        \end{align*}
        This give the contemporaneous ATT identification formula in Theorem~\ref{thm:identification_simultaneous}. %

        \paragraph{Identification of $\mr{ATT}(g, t)$} The proof follows a backward induction. For any $t > t' \geq g $, 
        we show that when $m_{t\mid t' + 1}(X_{it'+1}, \alpha_i)$ is identified for any $i$ such that $G_i > t'$, then $m_{t\mid t'}(X_{it'}, \alpha_i)$ is also identified for any $i$ such that $G_i > t'-1$. Now, assume that $m_{t\mid t' + 1}(X_{it' + 1}, \alpha_i)$ is identified for any $i$ such that $G_i > t'$. Note that for any $(x, \alpha) \in \mc{S}_{t'-1}$, 
        \begin{align*}
            m_{t\mid t' }(x, \alpha) = & \mb{E}\left(Y_{jt}(\infty)\mid X_{jt'} = x, \alpha_j = \alpha, G_j > t' \right) \\
            \eqtext{(i)} & \mb{E}\left(\mb{E}\left( Y_{jt}(\infty)\mid X_{jt'+ 1}, X_{jt'} = x, \alpha_j, G_j >t' \right)\mid X_{jt'} = x, \alpha_j = \alpha, G_j >t' \right) \\
            \eqtext{(ii)} & \mb{E}\left(\mb{E}\left( Y_{jt}(\infty)\mid X_{jt'+ 1}, \alpha_j, G_j >t' \right)\mid X_{jt'} = x, \alpha_j = \alpha, G_j >t' \right) \\
            \eqtext{(ii)} & \mb{E}\left(\mb{E}\left( Y_{jt}(\infty)\mid X_{jt' + 1}, \alpha_j, G_j >t'+ 1 \right)\mid X_{jt'} = x, \alpha_j = \alpha, G_j >t'  \right) \\
            = & \mb{E}\left( m_{t\mid t' + 1}(X_{jt' + 1}, \alpha_j) \mid X_{jt'} = x, \alpha_j = \alpha, G_j >t'  \right)
        \end{align*}
        Here, (i) comes from the law of iterated expectation, (ii) follows from mean-independence assumption in Assumption~\ref{assumption:surrogacy_extension}, and (iii) holds because of  selection mechanism in Assumption~\ref{assumption:selection_extension}.  And $m_{t\mid t' + 1}(X_{jt' + 1}, \alpha_j)$ is identified for each $j$ such that $G_i >t'$ by assumption.

        Then, by the informativeness condition in Assumption~\ref{assumption:informativeness_identification_simultaneous} and the uniform continuity of $m_{t\mid t'}$ (Assumption~\ref{assumption:identification_ATT_extension}\ref{item:identification_ATT_extension_smoothness}), for any $\epsilon >0$, there exists $\delta >0$ such that for all $(x, \alpha) \in \mc{S}_{t' - 1}$, 
        \begin{align*}
            \left| \mb{E}\left(m_{t\mid t' + 1}(X_{jt' + 1}, \alpha_j) \mid X_{jt'} = x, d(\alpha_j, \alpha)\leq \delta, G_j > t' \right) - m_{t\mid t' }(x, \alpha)\right| \leq \epsilon, 
        \end{align*}
        which implies that the following inequality holds with probability $1$ for any $i$ such that $G_i >t'-1$: 
        \begin{align*}
            \left|  \mb{E}\left( m_{t\mid t' + 1}(X_{jt' + 1}, \alpha_j) \mid X_{jt'} = X_{it' + 1}, d(\alpha_j, \alpha_i)\leq \delta, G_j >t'  \right) - m_{t\mid t' }( X_{it'}, \alpha_i) \right|  \leq \epsilon
        \end{align*}
        Then, for any $i$ such that $G_i >t'-1$, $m_{t\mid t' }( X_{it'}, \alpha_i) $ is identified by 
        \begin{align*}
            m_{t\mid t' }( X_{it'}, \alpha_i) = \lim_{\delta \downarrow 0}\mb{E}\left( m_{t\mid t' + 1}(X_{it'+1}, \alpha_j) \mid X_{jt'} = X_{it'}, d(\alpha_j, \alpha_i)\leq \delta, G_j >t'  \right). 
        \end{align*}

        Since $m_{t\mid t }( X_{it}, \alpha_i) $ is identified for any $i$ such that $G_i >t-1$, we iterate the argument backward from $t$ to $g$ and obtain that $m_{t\mid g }( X_{ig}, \alpha_i)$ is identified for any $i$ such that $G_i >g-1$. This completes the proof for the second part in Theorem~\ref{thm:identification_extension}. 
        Lastly, averaging the $m_{t\mid g }( X_{ig}, \alpha_i)$ over the distribution of units with $G_i = g$ yields 
        \begin{align*}
            \mb{E}\left(Y_{it}(\infty) \mid G_i = g\right) = \mb{E}\left( m_{t\mid g }( X_{ig}, \alpha_i) \mid G_i = g\right)
        \end{align*}
        $\mr{ATT}(g, t)$ is identified by 
        \begin{align*}
            \mr{ATT}(g, t) = \mb{E}\left(Y_{it} \mid G_i = g\right) -  \mb{E}\left( m_{t\mid g }( X_{ig}, \alpha_i) \mid G_i = g\right). 
        \end{align*}
        This completes the proof. 
    \end{proofthm}

\clearpage

    \section{Proofs of Estimation}\label{appendix:estimation}
    
    For notational simplicity, for each $i=1,\ldots,n$, let $\mc{I}(i)$ denote the subsample containing unit $i$, and let $\mc{I}_m(i)$ and $\mc{I}_\pi(i)$ denote the subsamples used to estimate the potential-outcome regression and the propensity score for unit $i$, respectively. More precisely, if $i\in\mc{I}_s$, then $\mc{I}(i)=\mc{I}_s$, $\mc{I}_m(i)=\mc{I}_{1 + (s\bmod 3 )}$, and $\mc{I}_\pi(i)=\mc{I}_{1+((s+1)\bmod 3)}$.

    For notational simplicity, for any $i, j = 1, \ldots, n$, any bandwidth $h>0$, and any $t\geq t'-1 \geq T_0-1$, define 
    \begin{align*}
        \widehat{W}^{(t, t')}_{h, ij} = \bs{1}(G_j > t) \widehat{K}^{(t')}_{h, ij}
    \end{align*}
    To simplify the notation, we also assume, without loss of generality, that the kernel function is bounded by $\overline{K}<1$, so that the product kernel satisfies $|\widehat{W}^{(t,t')}_{h,ij}|<\overline{K}$.
    Then, for each $i \in \mc{I}_s$, the estimated propensity score $\widehat{\pi}_{it}$ can be rewritten as
    \begin{align*}
        \widehat{\pi}_{it} =   \frac{\sum_{j \in \mc{I}_{s_p} } \bs{1}(G_j = t)  \widehat{W}^{(t-1, t)}_{h_\pi, ij} }{\sum_{j \in \mc{I}_{s_p} }    \widehat{W}^{(t-1, t)}_{h_\pi, ij}  }. 
    \end{align*}
    In addition, for any $i \in \mc{I}_s$, $j \in \mc{I}_{s_m}$, and for any $t > g \geq T_0$, we can rewrite $\widehat{m}_{j, t \mid t'}$ in a recursive way: 
    \begin{align*}
        \widetilde{m}_{j, t \mid t} =  & \sum_{j' \in \mc{I}_{s_m} } Y_{j't}\widehat{W}^{(t, t)}_{h_m , jj'} \big / \sum_{j' \in \mc{I}_{s_m} } \widehat{W}^{(t, t)}_{h_m , jj'}, \\
        \widetilde{m}_{j, t \mid t-1} =  &  \sum_{j' \in \mc{I}_{s_m} } \widehat{m}_{j, t \mid t} \widehat{W}^{(t-1, t-1)}_{h_m , jj'} \big / \sum_{j' \in \mc{I}_{s_m} } \widehat{W}^{(t-1, t-1)}_{h_m , jj'} , \\
        & \vdots \\
        \widetilde{m}_{j, t \mid g + 1} = & \sum_{j' \in \mc{I}_{s_m} } \widehat{m}_{j, t \mid g + 2 }\widehat{W}^{(g + 1, g + 1)}_{h_m , jj'}  \big /  \sum_{j' \in \mc{I}_{s_m} } \widehat{W}^{(g + 1, g + 1)}_{h_m , jj'}, 
    \end{align*}
    and 
    \begin{align*}
        \widehat{m}_{i, t \mid t} =  & \sum_{j\in \mc{I}_{s_m} } Y_{jt}\widehat{W}^{(t, t)}_{h_m , ij}  \big / \sum_{j \in \mc{I}_{s_m} } \widehat{W}^{(t, t)}_{h_m , ij} , \\
        \widehat{m}_{i, t \mid t-1} =  & \sum_{j\in \mc{I}_{s_m} } \widetilde{m}_{j, t \mid t} \widehat{W}^{(t-1, t-1)}_{h_m , ij}  \big / \sum_{j \in \mc{I}_{s_m} } \widehat{W}^{(t-1, t-1)}_{h_m , ij} , \\
        & \vdots \\
        \widehat{m}_{i, t \mid g} = & \ \sum_{j \in \mc{I}_{s_m} } \widetilde{m}_{j, t \mid g + 1}\widehat{W}^{(g, g)}_{h_m , ij }  \big / \sum_{j \in \mc{I}_{s_m} } \widehat{W}^{(g, g)}_{h_m , ij} 
    \end{align*}
    Also, for notational simplicity, we use $\mc{A}$ to denote the event
    \begin{align}\label{eq:definition_A}
        \mc{A}:=\left\{\max_{i, j=1, \ldots, n} | \widehat{d}_{ij} - d(\alpha_i, \alpha_j)| \leq \delta_{n, T_0}\right\}. 
    \end{align}
    By Assumption~\ref{assumption:estimation_consistency_d}, $\lim_{n, T_0 \rightarrow \infty} \mb{P}\left(\mc{A}\right) = 1$. Let $\phi := \{\{\alpha_i\}_{i=1, \ldots, N}, \Gamma_t\}$ denote the set of factors. Then, It is straightforward to verify that 
    $\mc{A}$ belongs to the $\sigma$-field generated by $\phi$ and pretreatment histories $\{(X_{it}, Y_{it})\}_{i=1, \ldots, N, t = 0, \ldots, T_0-1}$. Let $\mc{F}_t$ denote the collection of fixed effects and information available at period $t$, i.e., 
    \begin{align*}
        \mc{F}_t :=\{\phi, \{(Y_{i\tau}, X_{i\tau}, D_{i\tau})\}_{i=1, \ldots, n, \tau = 1, \ldots, t}\}.
    \end{align*}
        Lastly, for notational simplicity, I write  
    \begin{align*}
        Y_{it}(\infty) = & m_{t \mid t'}(X_{it'} , \alpha_i ) + u_{i, t \mid t'}, \quad &\forall\text{ } T_0 \leq t'\leq t\leq T, \text{ and } G_i >t'-1,  \\
        \quad D_{it'} = & \pi_{t'} (X_{it'} , \alpha_i  ) + e_{it'}, \quad &\forall\text{ } T_0 \leq t' \leq T, \text{ and } G_i >t' -1. 
    \end{align*} 
    Here,  $u_{i, t\mid t'}$ and $e_{it'}$ are projection errors satisfying $\mb{E}_T(u_{i, t\mid t'} \mid X_{it'}, \alpha_i, G_i >t' -1 ) = \mb{E}_T(e_{it'} \mid X_{it'}, \alpha_i, G_i >t'-1 ) = 0$. In addition, when $t = t'$, we use $u_{it}:=u_{i, t\mid t'}$.

    \vspace{0.5cm}
    
    \begin{lemma}\label{lemma:kernel_hat_d}
        Under conditions in Theorem~\ref{thm:estimation_ATT_extension},  let $\delta_{n, T_0} :=  \lambda_1 n^{-1/d_\alpha} \log (n) + \lambda_2 T_0^{-1/2}\sqrt{\log(nT_0)} $ be the estimation error of $\widehat{d}_{ij}$ stated as in Assumption~\ref{assumption:estimation_consistency_d}. Then, For any  $h>0$ such that $\delta_{n, T_0}/h\rightarrow 0$, and for all 
        $i, j =1, \ldots, n$, if $\widehat{d}_{ij} \leq h$, then 
        \begin{align*}
            \|\alpha_i -\alpha_{j} \| \leq 2 \overline{\eta} h 
        \end{align*}
        holds with probability approaching $1$, where $\eta$ is defined as in Assumption~\ref{assumption:informativeness_estimation}. 
    \end{lemma}
    \begin{prooflmm}{lemma:kernel_hat_d}
        Note that for all 
        $i, j =1, \ldots, n$, since $\delta_{n, T_0}/h\rightarrow 0$, the following inequality holds with probability approach to $1$, 
        \begin{align*}
            \left\{\widehat{d}_{ij} \leq h  \right\} \subseteq \left\{d(\alpha_i, \alpha_{j})  \leq h + \delta_{n, T_0} \right\} \subseteq \left\{d(\alpha_i, \alpha_{j})  \leq 2 h \right\}. 
        \end{align*}
        Also, by informativeness condition (see Assumption~\ref{assumption:informativeness_estimation}), we have 
        \begin{align*}
            \left\{d(\alpha_i, \alpha_{j})  \leq 2 h \right\} \subseteq \left\{\|\alpha_i - \alpha_{j}\|  \leq 2\overline{\eta} h \right\}. 
        \end{align*}
        This completes the proof. 
    \end{prooflmm}

\subsection{Proofs for Proposition~\ref{prop:estimation_entry_simultaneous} and Proposition~\ref{prop:estimation_entry_extension}} 
    
    It sufices to prove Proposition~\ref{prop:estimation_entry_extension} because Proposition~\ref{prop:estimation_entry_simultaneous} is a special case of Proposition~\ref{prop:estimation_entry_extension}. 

    \begin{proofprop}{prop:estimation_entry_extension} 
        We establish the proposition in three steps. We first derive the uniform convergence rate of the contemporaneous outcome-regression estimator, then establish the corresponding result for the propensity-score estimator, and finally extend the outcome-regression result to the recursive dynamic estimator by backward induction.
        \paragraph{Bounds for $\max_{i=1, \ldots, n}\left| \widehat{m}_{i,t\mid t}-m_{i,t\mid t}\right|$} For each $i=1, \ldots, n$, consider the following decomposition:
        \begin{align*}
            \widehat{m}_{i, t \mid t} - m_{i, t\mid t}  
            = & \underbrace{\sum_{j\in \mc{I}_{m}(i)}  (m_{j, t\mid t} - m_{i, t\mid t})\widehat{W}^{(t, t)}_{h_m , ij} \big / \sum_{j \in \mc{I}_{m}(i) } \widehat{W}^{(t, t)}_{h_m , ij}}_{:=B_{h_m, it}}  +  \underbrace{\sum_{j \in \mc{I}_{m}(i) } u_{jt}\widehat{W}^{(t, t)}_{h_m , ij}  \big / \sum_{j \in \mc{I}_{m}(i) } \widehat{W}^{(t, t)}_{h_m , ij} }_{:=V_{h_m, it}}, 
        \end{align*}
        and the estimation errors of counterfactual potential outcomes are bounded by:
        \begin{align*}
             \max_{i=1, \ldots, n } \left| \widehat{m}_{i, t \mid t} - m_{i, t\mid t}\right| \leq  \max_{i=1, \ldots, n } \left| B_{h_m, it} \right|  + \max_{i=1, \ldots, n } \left|V_{h_m, it}\right|. 
        \end{align*}
        To establish the bound for $\max_{i=1, \ldots, n } \left| B_{h_m, it} \right|$, note that the following bound holds wpa1: 
        \begin{equation}\label{eq:proposition_entry_simultaneous_B}
        \begin{aligned}
            & \max_{i=1, \ldots, n } \left| B_{h_m, it} \right| \\ \eqtext{(i)} & \max_{i=1, \ldots, n } \left| \sum_{j\in \mc{I}_{m}(i)} \bs{1}\left(\|X_{it} - X_{jt}\|_{\infty}, \widehat{d}_{ij} \leq h\right) (m_{j, t\mid t} - m_{i, t\mid t})\widehat{W}^{(t, t)}_{h_m , ij} \big / \sum_{j\in \mc{I}_{m}(i) } \widehat{W}^{(t, t)}_{h_m , ij}\right|
            \\
            \leqtext{(ii)} & \max_{i=1, \ldots, n } \left| \sum_{j\in \mc{I}_{m}(i)} \bs{1}\left(\|X_{it} - X_{jt}\|_{\infty}\leq h, \|\alpha_i - \alpha_j\| \leq \frac{2}{\overline{\eta}}h\right) (m_{j, t\mid t} - m_{i, t\mid t})\widehat{W}^{(t, t)}_{h_m , ij} \big / \sum_{j\in \mc{I}_{m}(i) } \widehat{W}^{(t, t)}_{h_m , ij}\right| \\
            \leqtext{(ii)} & L_m(d_{\alpha} + d_{X}) (1 +2\overline{\eta} ) h.   
        \end{aligned}
        \end{equation}
        Here, (i) follows from Assumption~\ref{assumption:estimation_extension}\ref{item:estimation_extension_kernel}, which requires that the kernel function be supported on $[-1,1]$. Inequality (ii) follows from Lemma~\ref{lemma:kernel_hat_d}, and (iii) follows from the Lipschitz continuity of $m_{t\mid t'}(\cdot)$ imposed in Assumption~\ref{assumption:estimation_extension}\ref{item:estimation_extension_L_continuous}, where $L_m$ denotes its Lipschitz constant. To establish the bound for $\max_{i=1, \ldots, n } \left| V_{h_m, it} \right|$, note that the following inequality holds wpa1 by Lemma~\ref{lemma:numerator}: 
        \begin{align*}
            \max_{i=1, \ldots, n } \left| V_{h_m, it} \right| \leq & \left(\min_{i=1, \ldots, n } \sum_{j \in \mc{I}_{m}(i) }\widehat{W}^{(t, t)}_{h_m , ij}\right)^{-1} \max_{i=1, \ldots, n } \left| \sum_{j\in \mc{I}_{m}(i) } u_{jt}\widehat{W}^{(t, t)}_{h_m , ij} \right| \\
            \leq & \frac{\max_{i=1, \ldots, n } \left| \sum_{j\in \mc{I}_{m}(i) } u_{jt}\widehat{W}^{(t, t)}_{h_m , ij} \right|}{\underline{C}(t, t)nh^{d_{\alpha} + d_{X}}}. 
        \end{align*}
        In addition, since the event $\mc{A}$ (defined as in~\eqref{eq:definition_A}) and $\{\widehat{d}_{ij}\}_{i,j}$ belong to the $\sigma$-field generated by the set of factors $\phi$ and pretreatment histories $\{(X_{it}, Y_{it})\}_{i=1, \ldots, n, t = 0, \ldots, T_0-1}$, it is straightforward to verify that (i) $\mb{E}\left(u_{jt}  \mid  \{X_{it}\}_{i=1}^{n}, \mc{F}_{t-1}, \{\widehat{W}^{(t, t)}_{h_m , ij}\}_{i, j\in \mc{I}_{m}(i)}  \right) = 0$ by Assumption~\ref{assumption:selection_extension} and \ref{assumption:surrogacy_extension}, and (ii) $u_{jt}$ is independent across $j$ conditional on $(\{X_{it}\}_{i=1}^{n}, \mc{F}_{t-1}, \{\widehat{W}^{(t, t)}_{h_m , ij}\}_{i, j\in \mc{I}_{m}(i)})$ because of  Assumption~\ref{assumption:selection_extension}\ref{item:estimation_extension_panel_weak_dependent}, (iii) given $(\{X_{it}\}_{i=1}^{n}, \mc{F}_{t-1}, \{\widehat{W}^{(t, t)}_{h_m , ij}\}_{i, j\in \mc{I}_{m}(i)})$, $u_{jt}$ is conditionally sub-Gaussian by Markov  and conditional sub-Gaussian assumptions (Assumption~\ref{assumption:surrogacy_extension} and \ref{assumption:estimation_extension}\ref{item:estimation_extension_finite}).  
        Therefore, we employ the Hoeffding's inequality (Lemma~\ref{lemma:hoeffding}) to show that for each $i$ and for any $\epsilon>0$, 
        \begin{align*}
            \mb{P}\left( \left| \sum_{j\in \mc{I}_{m}(i) } u_{jt}\widehat{W}^{(t, t)}_{h_m , ij} \right| \geq \epsilon \mid \{X_{it}\}_{i=1}^{n}, \mc{F}_{t-1}, \{\widehat{W}^{(t, t)}_{h_m , ij}\}_{i, j\in \mc{I}_{m}(i)} \right) \leq 2 \exp\left(\frac{ -\epsilon^2 }{2 \sigma^2 V_n }\right),  
        \end{align*}
        where $V_n:= \max_{i=1, \ldots, n} \sum_{j\in \mc{I}_m(i)}\left(\widehat{W}^{(t, t)}_{h_m , ij}\right)^2 $. 
        Therefore, we obtain 
        \begin{align*}
            & \mb{P}\left(\max_{i =1, \ldots, n} \left| \sum_{j\in \mc{I}_{m}(i) } u_{jt}\widehat{W}^{(t, t)}_{h_m , ij} \right| \geq \epsilon \mid \{X_{it}\}_{i=1}^{n}, \mc{F}_{t-1}, \{\widehat{W}^{(t, t)}_{h_m , ij}\}_{i, j\in \mc{I}_{m}(i)} \right) \\
            = & \mb{P}\left(\bigcup_{i =1, \ldots, n} \left\{\left| \sum_{j\in \mc{I}_{m}(i) } u_{jt}\widehat{W}^{(t, t)}_{h_m , ij} \right| \geq \epsilon \right\}  \mid \{X_{it}\}_{i=1}^{n}, \mc{F}_{t-1}, \{\widehat{W}^{(t, t)}_{h_m , ij}\}_{i, j\in \mc{I}_{m}(i)} \right) \\
            \leq & \sum_{i  = 1, \ldots, n} \mb{P}\left( \left| \sum_{j\in \mc{I}_{m}(i) } u_{jt}\widehat{W}^{(t, t)}_{h_m , ij} \right| \geq \epsilon \mid \{X_{it}\}_{i=1}^{n}, \mc{F}_{t-1}, \{\widehat{W}^{(t, t)}_{h_m , ij}\}_{i, j\in \mc{I}_{m}(i)} \right) \\
            \leqtext{(i)} & 2n \exp\left(\frac{ -\epsilon^2 }{2 \sigma^2 V_n  }\right),
        \end{align*}
        where (i) follows from Hoeffding's inequality discussed above. Also, note that $V_n$ is bounded following inequality wpa1:
        \begin{align*}
            V_n \leqtext{(i)} &  \overline{K} \max_{i=1, \ldots, n} \sum_{j\in \mc{I}_m(i)}\widehat{W}^{(t, t)}_{h_m , ij} \leqtext{(ii)} \overline{K} \overline{C}(t, t) nh^{d_{\alpha}+ d_{X}}.  
        \end{align*}
        Here, (i) follows from Assumption~\ref{assumption:estimation_extension}\ref{item:estimation_extension_kernel} and (ii) follows from Lemma~\ref{lemma:numerator}. 
                
        Then, there exists a sufficiently large constant $C_1 >0$ such that when $\epsilon = C_1 \sqrt{nh^{d_{\alpha} + d_{X}} \log (n)}$, \footnote{
            Here we integrate $(\{X_{it}\}_{i=1}^{n}, \mc{F}_{t-1}, \{\widehat{W}^{(t, t)}_{h_m , ij}\}_{i, j\in \mc{I}_{m}(i)})$ to obtain unconditional probability. 
        } 
        \begin{align*}
            & \lim_{n\rightarrow \infty}\mb{P}\left(\max_{i  = 1, \ldots, n} \left| \sum_{j\in \mc{I}_{m}(i) } u_{jt}\widehat{W}^{(t, t)}_{h_m , ij} \right| \geq C_1 \sqrt{nh^{d_{\alpha} + d_{X}} \log (n)}   \right) =0. 
        \end{align*}
        Therefore, the following inequality holds wpa1: 
        \begin{equation}\label{eq:proposition_entry_simultaneous_V}
        \begin{aligned}
            \max_{i  = 1, \ldots, n} \left| V_{h_m, it} \right| 
            \leq \frac{ C_1 }{\underline{C}(t, t)} \left(nh_m^{d_\alpha + d_{X}}\right)^{-1/2}\sqrt{\log (n)}. 
        \end{aligned}
        \end{equation}
        Combining \eqref{eq:proposition_entry_simultaneous_B} and \eqref{eq:proposition_entry_simultaneous_V} yields 
        \begin{equation}\label{eq:proposition_entry_simultaneous_m_i}
        \begin{aligned}
            \max_{i  = 1, \ldots, n}\left|\widehat{m}_{i, t \mid t} - m_{i, t\mid t} \right| = O_P\left(h_m +  \left(nh_m^{d_\alpha+ d_{X}}\right)^{-1/2}\sqrt{\log (n)} \right)
        \end{aligned}
        \end{equation}
        By similar argument, we can also show that for each $j =1, \ldots, n$, the within-sample estimator $\widetilde{m}_{j, t \mid t}$ admits a similar error bound, 
        \begin{equation}\label{eq:proposition_entry_simultaneous_m_j}
        \begin{aligned}
            \max_{j  = 1, \ldots, n}\left|\widetilde{m}_{j, t \mid t} - m_{j, t\mid t} \right| = O_P\left(h_m +  \left(nh_m^{d_\alpha+ d_{X}}\right)^{-1/2}\sqrt{\log (n)} \right), 
        \end{aligned}
        \end{equation}
        which is crucial in obtaining the error bound for recursive estimator in the following analysis. 

        \paragraph{Bounds for $\max_{i\in \mc{I}_{s}}\left| \widehat{\pi}_{it'}-\pi_{it'}\right|$} The uniform error bound for $\widehat{\pi}_{it'}$ can be established similarly, and the proof is omitted. 

        \paragraph{Bounds for $\max_{i\in \mc{I}_{s}}\left| \widehat{m}_{i, t\mid g}-m_{i, t\mid g}\right|$}   We establish the result by backward induction. The induction statement is that, fixing $t' < t$, the within-sample estimator satisfies
        \begin{align*}
            \max_{j  = 1, \ldots, n} \left|\widetilde{m}_{j, t \mid t'+1} - m_{j, t\mid t'+1} \right| = O_P\left(h_m +  \left(nh_m^{d_\alpha+ d_{X}}\right)^{-1/2}\sqrt{\log (n)} \right). 
        \end{align*}
        For the base case $t'=t-1$, this has been established in~\eqref{eq:proposition_entry_simultaneous_m_j}. 
        Then, for any $i = 1, \ldots, n$ and fixing $ t' < t$,  consider the following decomposition: 
        \begin{align*}
            \widehat{m}_{i, t\mid t'}-m_{i, t\mid t'} = & \frac{\sum_{j\in \mc{I}_{m}(i) } \left(\widetilde{m}_{j, t \mid t' + 1} - m_{j, t \mid t' + 1}\right) \widehat{W}^{(t', t')}_{h_m , ij}  }{ \sum_{j\in \mc{I}_{m}(i) } \widehat{W}^{(t', t')}_{h_m , ij} } + \frac{\sum_{j\in \mc{I}_{m}(i) } \left(m_{j, t \mid t'+1} - m_{j, t \mid t'}\right) \widehat{W}^{(t', t')}_{h_m , ij}  }{ \sum_{j\in \mc{I}_{m}(i)} \widehat{W}^{(t', t')}_{h_m , ij} } \\
            & + \frac{\sum_{j\in \mc{I}_{m}(i) } \left(m_{j, t \mid t'} - m_{i, t \mid t'}\right) \widehat{W}^{(t', t')}_{h_m , ij}  }{ \sum_{j\in \mc{I}_{m}(i) } \widehat{W}^{(t', t')}_{h_m , ij} }. 
        \end{align*}
        By the induction hypothesis,  $\max_{j =1, \ldots, n}\left|\widetilde{m}_{j, t \mid t' + 1} - m_{j, t\mid t' + 1} \right| = O_P\left(h_m +  \left(nh_m^{d_\alpha+ d_{X}}\right)^{-1/2}\sqrt{\log (n)} \right)$, it follows that the first term is bounded by
        \begin{equation}\label{eq:proposition_entry_dynamic_1}
        \begin{aligned}
            \max_{i = 1, \ldots, n}\left| \frac{\sum_{j\in \mc{I}_{m}(i)  } \left(\widetilde{m}_{j, t \mid t' + 1} - m_{j, t \mid t' + 1}\right) \widehat{W}^{(t', t')}_{h_m , ij}  }{ \sum_{j\in \mc{I}_{m}(i) } \widehat{W}^{(t', t')}_{h_m , ij} } \right|  = O_P\left(h_m +  \left(nh_m^{d_\alpha+ d_{X}}\right)^{-1/2}\sqrt{\log (n)} \right)
        \end{aligned}
        \end{equation}
        
        For the second term, note that since $\mb{E}\left(m_{j, t \mid t'+1} - m_{j, t \mid t'} \mid \{X_{it'}\}_{i=1}^{n}, \phi, \mc{F}_{t'-1} \right) = 0$ (by Assumption~\ref{assumption:selection_extension} and \ref{assumption:surrogacy_extension}), we apply the same argument used to establish~\eqref{eq:proposition_entry_simultaneous_V} to obtain that the second term is bounded by 
        \begin{equation}\label{eq:proposition_entry_dynamic_2}
        \begin{aligned}
            \max_{i=1, \ldots, n}\left| \frac{\sum_{j\in \mc{I}_{m}(i) } \left(m_{j, t \mid t'+1} - m_{j, t \mid t'}\right) \widehat{W}^{(t', t')}_{h_m , ij}  }{ \sum_{j\in \mc{I}_{m}(i) } \widehat{W}^{(t', t')}_{h_m , ij} } \right|  = O_P\left(\left(nh_m^{d_\alpha+ d_{X}}\right)^{-1/2}\sqrt{\log (n)}\right)
        \end{aligned}
        \end{equation}
        Last, applying the same argument used to establish~\eqref{eq:proposition_entry_simultaneous_B}, we obtain that the third term is bounded by 
        \begin{equation}\label{eq:proposition_entry_dynamic_3}
        \begin{aligned}
            \max_{i = 1, \ldots, n}\left| \frac{\sum_{j\in \mc{I}_{m}(i) } \left(m_{j, t \mid t'} - m_{i, t \mid t'}\right) \widehat{W}^{(t', t')}_{h_m , ij}  }{ \sum_{j\in \mc{I}_{m}(i) } \widehat{W}^{(t', t')}_{h_m , ij} } \right|  = O_P\left(h_m\right)
        \end{aligned}
        \end{equation}

        Combining \eqref{eq:proposition_entry_dynamic_1}, \eqref{eq:proposition_entry_dynamic_2}, and \eqref{eq:proposition_entry_dynamic_3} yields 
        \begin{align*}
            \max_{i=1, \ldots, n}\left| \widehat{m}_{i, t\mid t'}-m_{i, t\mid t'}\right| = O_P\left(h_m +  \left(nh_m^{d_\alpha+ d_{X}}\right)^{-1/2}\sqrt{\log (n)} \right). 
        \end{align*}
        By similar argument, we can also show that for each $j=1, \ldots, n$, the in-sample estimator $\widehat{m}_{j, t \mid t'}$ admits a similar error bound, 
        \begin{align*}
            \max_{j  = 1, \ldots, n}\left|\widetilde{m}_{j, t \mid t'} - m_{j, t\mid t'} \right| = O_P\left(h_m +  \left(nh_m^{d_\alpha+ d_{X}}\right)^{-1/2}\sqrt{\log (n)} \right). 
        \end{align*}
        This establishes the induction step. Since the result holds for the base case $t'=t - 1$, backward induction implies that it holds for every $t' < t$. This completes the proof.         
    \end{proofprop}

\subsection{Proofs for Theorem~\ref{thm:estimation_ATT_extension} and Theorem } 

    \begin{proofthm}{thm:estimation_ATT_extension} 
        We first establish the result for the contemporaneous ATT and then extend it to the dynamic ATT under staggered adoption.

        \paragraph{Contemporaneous ATT} Without loss of generality, we assume no unit is treated before $t$, i.e., $t = T_0$ and $\{G_i >t-1\} = \{1, \ldots, n\}$. Consider the following decomposition of contemporaneous ATT estimator $\widehat{\mr{ATT}(t, t)}$: 
        \begin{align*}
            \widehat{\mr{ATT}(t, t)} = & \underbrace{\frac{1}{n_t} \sum_{i =1 }^{n}  \left(\bs{1}(G_i = t)Y_{it} -   \frac{ \bs{1}(G_i > t) Y_{it} \pi_{it} + (\bs{1}(G_i = t) - \pi_{it})m_{i, t\mid t}}{ 1- \pi_{it} }   \right)}_{:=Q_1} \\
            & - \underbrace{\frac{1}{n_t} \sum_{i=1}^{n}  \frac{\bs{1}(G_i >t)(\widehat{\pi}_{it} - \pi_{it}) }{(1 - \pi_{it})(1 - \widehat{\pi}_{it})}  u_{it}}_{:=Q_2} - \underbrace{ \frac{1}{n_t} \sum_{i =1 }^{n}   \frac{\widehat{m}_{i, t\mid t} - m_{i, t\mid t} }{1 - \widehat{\pi}_{it}} e_{it}}_{:=Q_3}  \\
            & + \underbrace{\frac{1}{n_t} \sum_{i =1 }^{n}  \frac{\widehat{\pi}_{it} - \pi_{it}}{1 - \widehat{\pi}_{it}}(\widehat{m}_{i, t\mid t} - m_{i, t\mid t} ) }_{:=Q_4}.  
        \end{align*}
        Here, $n_t = \sum_{i=1}^{n}\bs{1}(G_i = t)$, $Q_1$ stands for the oracle estimator,  and $Q_2, Q_3, Q_4$ are estimation errors introduced by nonparametric estimation. 
        Each of $Q_2$, $Q_3$, and $Q_4$ is the sum of contributions from the three folds. Since the same argument applies to each fold, it suffices to establish the bounds for one fold. Therefore, we focus on $i\in\mc{I}_1$ and always assume that unit $1$ belongs to $\mc{I}_1$ in the following derivations whenever this simplifies the analysis.

        \subparagraph{Step 1. Bound for $Q_2$} It is straightforward to verify that, for each $i \in \mc{I}_1$,  
        \begin{align*}
            \mb{E}\left( u_{it} \mid  \{X_{it}\}_{i=1}^{n}, \mc{F}_{t-1}, \left\{\frac{\bs{1}(G_i >t)(\widehat{\pi}_{it} - \pi_{it}) }{(1 - \pi_{it})(1 - \widehat{\pi}_{it})}\right\}_{i\in \mc{I}_1} \right) = 0
        \end{align*}
        by sample-splitting, Assumption~\ref{assumption:selection_extension}, and \ref{assumption:surrogacy_extension}. In addition, 
        \begin{align*}
            & \mr{Var}\left(Q_2 \mid \{X_{it}\}_{i=1}^{n}, \mc{F}_{t-1}, \left\{\frac{\bs{1}(G_i >t)(\widehat{\pi}_{it} - \pi_{it}) }{(1 - \pi_{it})(1 - \widehat{\pi}_{it})}\right\}_{i\in \mc{I}_1}  \right) \\
            \leqtext{(i)} & \frac{9 \sigma^2}{n^2} \sum_{i\in \mc{I}_1 }\frac{\bs{1}(G_i >t)(\widehat{\pi}_{it} - \pi_{it})^2 }{(1 - \pi_{it})^2(1 - \widehat{\pi}_{it})^2} \eqtext{(ii)} o_P\left(1/n\right). 
        \end{align*}
        Here, (i) holds because $u_{it}$ is conditional sub-Gaussian (Assumption~\ref{assumption:estimation_extension}\ref{item:estimation_extension_finite}) and independent across $i$ conditional on $\left(\{X_{it}\}_{i=1}^{n}, \mc{F}_{t-1}, \left\{\frac{\bs{1}(G_i >t)(\widehat{\pi}_{it} - \pi_{it}) }{(1 - \pi_{it})(1 - \widehat{\pi}_{it})}\right\}_{i\in \mc{I}_1} \right)$ (see Assumption~\ref{assumption:selection_extension}\ref{item:estimation_extension_panel_weak_dependent}), (ii) follows from the common overlap condition (Assumption~\ref{assumption:estimation_extension}\ref{item:estimation_extension_overlap})  and the consistency of $\widehat{\pi}_{it}$ (Proposition~\ref{prop:estimation_entry_extension}). Therefore, we can conclude that 
        \begin{equation}\label{eq:thm_ATT_contemporaneous_Q_2}
        \begin{aligned}
            Q_2 = o_P\left(1/\sqrt{n}\right). 
        \end{aligned}
        \end{equation}
        
        \subparagraph{Step 3. Bound for $Q_3$} We can use a similar argument in obtaining \eqref{eq:thm_ATT_contemporaneous_Q_2} to establish the bound for $Q_3$. Specifically, 
        \begin{equation}\label{eq:thm_ATT_contemporaneous_Q_3}
        \begin{aligned}
            Q_3 = o_P\left(1/\sqrt{n}\right). 
        \end{aligned}
        \end{equation}
        The proof is omitted for simplicity. 
        
        \subparagraph{Step 3. Bound for $Q_4$} Before deriving the probability bound for $Q_4$, consider the following decomposition of $\widehat{m}_{i, t\mid t}$ and $\widehat{\pi}_{it}$:
        \begin{equation}\label{eq:decomposition_m_it}
        \begin{aligned}
            \widehat{m}_{i, t\mid t} -  m_{i, t\mid t}  
            = & \underbrace{ \frac{\sum_{j\in \mc{I}_{m}(i)}  (m_{j, t\mid t} - m_{i, t\mid t})\widehat{W}^{(t, t)}_{h_m , ij} }{\sum_{j \in \mc{I}_{m}(i) } \widehat{W}^{(t, t)}_{h_m , ij}}}_{:=B_{m, it}}  +  \underbrace{ \frac{\sum_{j\in \mc{I}_{m}(i)}  u_{jt}\widehat{W}^{(t, t)}_{h_m , ij} }{\sum_{j \in \mc{I}_{m}(i) } \widehat{W}^{(t, t)}_{h_m , ij}}}_{:=V_{m, it}}, 
        \end{aligned}
        \end{equation}
        and 
        \begin{equation}\label{eq:decomposition_p_it}
        \begin{aligned}
            \widehat{\pi}_{it} -  \pi_{it}  
            = & \underbrace{ \frac{\sum_{j\in \mc{I}_{\pi}(i)}  (\pi_{jt} - \pi_{it})\widehat{W}^{(t-1, t)}_{h_\pi , ij} }{\sum_{j \in \mc{I}_{\pi}(i) } \widehat{W}^{(t-1, t)}_{h_\pi , ij}}}_{:=B_{\pi, it}}  +  \underbrace{ \frac{\sum_{j\in \mc{I}_{\pi}(i)}  e_{jt}\widehat{W}^{(t-1, t)}_{h_\pi , ij} }{\sum_{j \in \mc{I}_{\pi}(i) } \widehat{W}^{(t-1, t)}_{h_\pi , ij}}}_{:=V_{\pi, it}}. 
        \end{aligned}
        \end{equation}
    
        The second-order series expansion of $Q_4$ reads\footnote{
            It is straightforward to verify that higher-order expansions are asymptotically negligible under conditions in Theorem~\ref{thm:estimation_ATT_extension}. 
        }: 
        \begin{align*}
            Q_4 =  \underbrace{\frac{1}{n_t} \sum_{i=1}^{n}  \frac{\widehat{\pi}_{it} - \pi_{it}}{1 - \pi_{it}}(\widehat{m}_{i, t\mid t} - m_{i, t\mid t} )}_{:=S_1} + \underbrace{\frac{1}{n_t} \sum_{i =1}^{n}  \frac{(\widehat{\pi}_{it} - \pi_{it})^2 }{(1 - \pi_{it})^2}(\widehat{m}_{i, t\mid t} - m_{i, t\mid t} )}_{:=S_2}  + o_P\left(S_1 \right) + o_P\left( S_2\right)
        \end{align*}
        To establish the error bound for $S_1$, note that, 
        \begin{align*}
            S_1 = \frac{1}{n_t}\left( \sum_{i=1}^{n} \frac{B_{m, it} B_{\pi, it}}{1 - \pi_{it}}  +   \sum_{i=1}^{n} \frac{B_{m, it}V_{\pi, it}}{1 - \pi_{it}}  +   \sum_{i=1}^{n} \frac{V_{m, it}B_{\pi, it}}{1 - \pi_{it}}  +   \sum_{i=1}^{n} \frac{V_{m, it}V_{\pi, it}}{1 - \pi_{it}} \right)
        \end{align*}
        For the first term, using the same argument in obtaining~\eqref{eq:proposition_entry_simultaneous_B}, we have 
        \begin{align}\label{eq:thm_ATT_S_1_1}
            \frac{1}{n_t} \sum_{i=1}^{n} \frac{B_{m, it} B_{\pi, it}}{1 - \pi_{it}} = O_P\left(h_m h_{\pi}. \right) 
        \end{align}
        For the second term (here, it suffices to focus on $i\in \mc{I}_1$ where $1 \in \mc{I}_1$ to simplify the analysis), since (i) $\mb{E}\left(e_{jt} \mid \{X_{it}\}_{i=1}^{n}, \mc{F}_{t-1}, \{B_{m, it}\}_{i\in \mc{I}_1}, \{\widehat{W}^{(t-1, t)}_{h_\pi , ij} \}_{i\in \mc{I}_1, j\in \mc{I}_{\pi}(i)}\right) = 0$ by sample-splitting, Assumption~\ref{assumption:selection_extension}, and~\ref{assumption:surrogacy_extension}, (ii) conditional on $(\{X_{it}\}_{i=1}^{n}, \mc{F}_{t-1}, \{B_{m, it}\}_{i\in \mc{I}_1}, \{\widehat{W}^{(t-1, t)}_{h_\pi , ij} \}_{i\in \mc{I}_1, j\in \mc{I}_{\pi}(i)})$, $e_{jt}$ is independent across $j$ because of  Assumption~\ref{assumption:selection_extension}\ref{item:estimation_extension_panel_weak_dependent},  and (iii) $e_{jt}$ is conditionally sub-Gaussian given $\left(\{X_{it}\}_{i=1}^{n}, \mc{F}_{t-1}, \{B_{m, it}\}_{i\in \mc{I}_1}, \{\widehat{W}^{(t-1, t)}_{h_\pi , ij} \}_{i\in \mc{I}_1, j\in \mc{I}_{\pi}(i)}\right)$ by Markov and sub-Gaussian assumptions (Assumption~\ref{assumption:surrogacy_extension} and \ref{assumption:estimation_extension}\ref{item:estimation_extension_finite}), we then have  
        \begin{align*}
            & \mr{Var}\left( \frac{1}{n_t}  \sum_{i\in \mc{I}_1} \frac{B_{m, it}V_{\pi, it}}{1 - \pi_{it}}\mid \{X_{it}\}_{i=1}^{n}, \mc{F}_{t-1}, \{B_{m, it}\}_{i\in \mc{I}_1}, \{\widehat{W}^{(t-1, t)}_{h_\pi , ij} \}_{i\in \mc{I}_1, j\in \mc{I}_{\pi}(i)}\right)    \\
            \leq &   \frac{1}{n_t^2}  \sum_{j\in \mc{I}_{\pi} (1)}  \left( \sum_{i\in \mc{I}_1}\frac{B_{m, it}   \widehat{W}^{(t-1, t)}_{h_\pi , ij} }{(1 - \pi_{it})\sum_{\ell \in \mc{I}_{\pi}(i) } \widehat{W}^{(t-1, t)}_{h_\pi , i\ell} } \frac{ }{}\right)^2  \\
            \leqtext{(i)} &  \frac{\max_{i=1, \ldots, n} |B_{m, it}|^2 }{\left((1 - \overline{p}) \underline{C}(t-1, t) nh_{\pi}^{d_{\alpha} + d_{X}}\right)^2}  \frac{1}{n_t^2}  \sum_{j\in \mc{I}_{\pi} (1)}  \left( \sum_{i\in \mc{I}_1}  \widehat{W}^{(t-1, t)}_{h_\pi , ij}   \right)^2.  
        \end{align*}
        Here, (i) follows from Lemma~\ref{lemma:numerator}. We then employ the same argument in obtaining~\eqref{eq:proposition_entry_simultaneous_B} to show that $\max_{i=1, \ldots, n}B_{m, it} = O_P(h_m)$, and together with Lemma~\ref{lemma:numerator}, we have 
        \begin{align}\label{eq:thm_ATT_S_1_2}
            \mr{Var}\left( \frac{1}{n_t}  \sum_{i=1}^{n} \frac{B_{m, it}V_{\pi, it}}{1 - \pi_{it}}  \right) = O_P\left(\frac{h_m^2}{n}\right)=o_P(1/n) \Rightarrow \frac{1}{n_t}  \sum_{i=1}^{n} \frac{B_{m, it}V_{\pi, it}}{1 - \pi_{it}} = o_P(1/\sqrt{n}). 
        \end{align}

        The third term can be bounded similarly:
        \begin{align}\label{eq:thm_ATT_S_1_3}
            \frac{1}{n_t}\sum_{i=1}^{n} \frac{V_{m, it}B_{\pi, it}}{1 - \pi_{it}} = O_P\left(\frac{h_\pi}{\sqrt{n}}\right) = o_P(1/\sqrt{n}). 
        \end{align}
        The argument is identical to that used for the second term and is therefore omitted.
        
        For the last term, since (i) $\mb{E}\left(e_{jt} \mid \{X_{it}\}_{i=1}^{n}, \mc{F}_{t-1}, \{V_{m, it}\}_{i\in \mc{I}_1}, \{\widehat{W}^{(t-1, t)}_{h_\pi , ij} \}_{i\in \mc{I}_1, j\in \mc{I}_{\pi}(i)}\right) = 0$ by sample-splitting, Assumption~\ref{assumption:selection_extension}, and~\ref{assumption:surrogacy_extension}, (ii) $e_{jt}$ is independent across $j$ conditional on $(\{X_{it}\}_{i=1}^{n}, \mc{F}_{t-1}, \{V_{m, it}\}_{i\in \mc{I}_1}, \{\widehat{W}^{(t-1, t)}_{h_\pi , ij} \}_{i\in \mc{I}_1, j\in \mc{I}_{\pi}(i)})$ because of  Assumption~\ref{assumption:selection_extension}\ref{item:estimation_extension_panel_weak_dependent},  and (iii) $e_{jt}$ is conditionally sub-Gaussian given $\left(\{X_{it}\}_{i=1}^{n}, \mc{F}_{t-1}, \{V_{m, it}\}_{i\in \mc{I}_1}, \{\widehat{W}^{(t-1, t)}_{h_\pi , ij} \}_{i\in \mc{I}_1, j\in \mc{I}_{\pi}(i)}\right)$,  by Markov and sub-Gaussian assumptions (Assumption~\ref{assumption:surrogacy_extension} and \ref{assumption:estimation_extension}\ref{item:estimation_extension_finite}), we then have  
        \begin{align*}
            & \mr{Var}\left(\frac{1}{n_t}\sum_{i \in \mc{I}_1} \frac{V_{m, it}V_{\pi, it}}{1 - \pi_{it}} \mid \{X_{it}\}_{i=1}^{n}, \mc{F}_{t-1}, \{V_{m, it}\}_{i\in \mc{I}_1}, \{\widehat{W}^{(t-1, t)}_{h_\pi , ij} \}_{i\in \mc{I}_1, j\in \mc{I}_{\pi}(i)}  \right) \\
            \leq &   \frac{1}{n_t^2}  \sum_{j\in \mc{I}_{\pi} (1)}  \left( \sum_{i\in \mc{I}_1}\frac{V_{m, it}   \widehat{W}^{(t-1, t)}_{h_\pi , ij} }{(1 - \pi_{it})\sum_{\ell \in \mc{I}_{\pi}(i) } \widehat{W}^{(t-1, t)}_{h_\pi , i\ell} } \frac{ }{}\right)^2  \\
            \leqtext{(i)} &  \frac{\max_{i=1, \ldots, n} |V_{m, it}|^2 }{\left((1 - \overline{p}) \underline{C}(t-1, t) nh_{\pi}^{d_{\alpha} + d_{X}}\right)^2 }  \frac{1}{n_t^2}  \sum_{j\in \mc{I}_{\pi} (1)}  \left( \sum_{i\in \mc{I}_1}  \widehat{W}^{(t-1, t)}_{h_\pi , ij}   \right)^2.   
        \end{align*}
        Here, (i) follows from Lemma~\ref{lemma:numerator}. We have already shown that (see inequality~\eqref{eq:proposition_entry_simultaneous_V}) to show that $\max_{i=1, \ldots, n} \left| V_{m, it} \right| = O_P\left(\left(nh_m^{d_\alpha+ d_{X}}\right)^{-1/2}\sqrt{\log (n)} \right)$. 
        Combining it with Lemma~\ref{lemma:numerator} yields
        \begin{align}\label{eq:thm_ATT_S_1_4}
            \mr{Var}\left( \frac{1}{n_t}  \sum_{i=1}^{n} \frac{V_{m, it}V_{p, it}}{1 - \pi_{it}}  \right) = O_P\left(\frac{\sqrt{\log (n)} }{\sqrt{n} nh_m^{d_\alpha+ d_{X}} }\right) \Rightarrow \frac{1}{n_t}  \sum_{i=1}^{n} \frac{V_{m, it}V_{p, it}}{1 - \pi_{it}} = o_P(1\sqrt{n}). 
        \end{align}

        Combining~\eqref{eq:thm_ATT_S_1_1}, \eqref{eq:thm_ATT_S_1_2}, \eqref{eq:thm_ATT_S_1_3}, and \eqref{eq:thm_ATT_S_1_4} gives 
        \begin{align}\label{eq:thm_ATT_S_1}
            S_1 = O_P\left(h_m h_\pi\right) + o_P\left(1\sqrt{n}\right). 
        \end{align}

        To establish the error bound for $S_2$, note that by Proposition~\ref{prop:estimation_entry_extension}, we have 
        \begin{align*}
            \max_{i: G_i >t-1} |\widehat{\pi}_{it} - \pi_{it} |^2 = O_P\left(h_\pi^2 + \left(nh_\pi^{d_\alpha+ d_{X}}\right)^{-1}\log (n) \right). 
        \end{align*}
        It follows that, by the same argument in deriving~\eqref{eq:thm_ATT_S_1_2},  
        \begin{align*}
            S_2 = & \frac{1}{n_t} \sum_{i =1}^{n}  \frac{(\widehat{\pi}_{it} - \pi_{it})^2 }{(1 - \pi_{it})^2}B_{m, it} +  \frac{1}{n_t} \sum_{i =1}^{n}  \frac{(\widehat{\pi}_{it} - \pi_{it})^2 }{(1 - \pi_{it})^2}V_{m, it} \\
            = & O_P\left(h_m h_\pi^2 + h_m \left(nh_\pi^{d_\alpha+ d_{X}}\right)^{-1}\log (n) \right) + o_P\left(1/\sqrt{n}\right). 
        \end{align*}
        Therefore, we have
        \begin{align*}
            Q_4 = & O_P\left(h_mh_\pi + h_m h_\pi^2 + h_m \left(nh_\pi^{d_\alpha+ d_{X}}\right)^{-1}\log (n) \right) + o_P(1/\sqrt{n}) \\
            = &  O_P\left(h_mh_\pi +  h_m \left(nh_\pi^{d_\alpha+ d_{X}}\right)^{-1}\log (n)  \right) + o_P(1/\sqrt{n})
        \end{align*}
        Lastly, since $h_m \left(nh_\pi^{d_\alpha+ d_{X}}\right)^{-1}\log (n) = o_P(1)$, we obtain a looser by more interpretable bound 
        \begin{align}\label{thm_ATT_contemporaneous_Q_4}
            Q_4 = O_P\left(h_mh_\pi +  h_m \left(nh_\pi^{d_\alpha+ d_{X}}\right)^{-1/2} \sqrt{\log (n)}  \right) + o_P(1/\sqrt{n}). 
        \end{align}
        Combining it with~\eqref{eq:thm_ATT_contemporaneous_Q_2} and~\eqref{eq:thm_ATT_contemporaneous_Q_3} completes the proof for contemporaneous ATT.

        \paragraph{Dynamic ATT} We focus on $\mr{ATT}(t-1,t)$ for brevity. The extension to treatment effects over longer horizons follows from exactly the same argument and is omitted to avoid unnecessary repetition. 

        \subparagraph{Decomposition of $\widehat{m}_{i, t\mid t - 1}$ and $\widehat{\pi}_{i t - 1}$} $\widehat{m}_{i, t\mid t-1}$ admits the following decomposition 
        \begin{equation}\label{eq:decomposition_m_it-1}
        \begin{aligned}
            \widehat{m}_{i, t\mid t-1} - m_{i, t\mid t-1} 
            = B^{(1)}_{m, it-1} + B^{(2)}_{m, it-1} + V^{(1)}_{m, it-1} + V^{(2)}_{m, it-1},   
        \end{aligned}
        \end{equation}
        where 
        \begin{align*}
            B^{(1)}_{m, it-1} := & \sum_{j \in \mc{I}_{m}(i)} \widehat{W}_{h_m,  ij}^{(t-1, t-1)}\frac{\sum_{j' \in \mc{I}_m(i)}  \widehat{W}_{m,  jj'}^{(t, t)} (m_{j', t\mid t} - m_{j, t\mid t}) }{\sum_{\ell' \in \mc{I}_m(i)}  \widehat{W}_{m,  j\ell'}^{(t, t)}  }   \big / \sum_{\ell \in \mc{I}_{m}(i)} \widehat{W}_{h_m,  i\ell }^{(t-1, t-1)},    \\
            B^{(2)}_{m, it-1} := & \sum_{j \in \mc{I}_{m}(i)} \widehat{W}_{h_m,  ij}^{(t-1, t-1)}\left(m_{j, t\mid t - 1} - m_{i, t\mid t-1}\right) \big / \sum_{\ell \in \mc{I}_{m}(i)} \widehat{W}_{h_m,  i\ell }^{(t-1, t-1)},   \\
            V^{(1)}_{m, it-1} := & \sum_{j \in \mc{I}_{m}(i)} \widehat{W}_{h_m,  ij}^{(t-1, t-1)}\left(m_{j, t\mid t} - m_{j, t\mid t-1}\right)  \big / \sum_{\ell \in \mc{I}_{m}(i)} \widehat{W}_{h_m,  i\ell }^{(t-1, t-1)},   \\ 
            V^{(2)}_{m, it-1} := &   \sum_{j \in \mc{I}_{m}(i)} \widehat{W}_{h_m,  ij}^{(t-1, t-1)}\frac{\sum_{j' \in \mc{I}_m(i)} \widehat{W}_{m,  jj'}^{(t, t)}  u_{j't} }{\sum_{\ell' \in \mc{I}_m(i)}  \widehat{W}_{m,  j\ell'}^{(t, t)}  }  \big / \sum_{\ell \in \mc{I}_{m}(i)} \widehat{W}_{h_m,  i\ell }^{(t-1, t-1)}. 
        \end{align*}
        Similarly, $\widehat{\pi}_{i t - 1}$ admits the following decomposition: 
        \begin{equation}\label{eq:decomposition_p_it-1}
        \begin{aligned}
            \widehat{\pi}_{it-1} - \pi_{i t-1} 
            = & \underbrace{\frac{\sum_{k \in \mc{I}_\pi(i) }  \widehat{W}_{h_\pi,  ik}^{(t-2, t-1)} (\pi_{kt-1}- \pi_{it-1}) }{\sum_{q \in \mc{I}_\pi(i)  }  \widehat{W}_{h_\pi,  iq}^{(t-2, t-1)} }}_{:=B_{\pi, it-1}} + \underbrace{\frac{\sum_{k \in \mc{I}_\pi(i) }  \widehat{W}_{h_\pi,  ik}^{(t-2, t-1)} e_{kt-1} }{\sum_{q \in \mc{I}_\pi(i)  }  \widehat{W}_{h_\pi,  iq}^{(t-2, t-1)} }}_{:=V_{\pi, it-1}} .  
        \end{aligned}
        \end{equation}

        \subparagraph{Decomposition of $\widehat{\mr{ATT}(t-1, t)}$}  Let $n_{t-1} : =\sum_{i=1}^{n}\bs{1}(G_i = t-1)$, the decomposition of $\widehat{\mr{ATT}(t-1, t)}$ reads: 
        \begin{align*}
            & \widehat{\mr{ATT}(t-1, t)} - \widehat{\mr{ATT}(t-1, t)}^{\mr{oracle}} \\
            = & - \underbrace{\frac{1}{n_{t-1}} \sum_{i=1}^{n}  \frac{\bs{1}(G_i >t-2)}{1 - \pi_{i t-1}} \left( \widehat{m}_{i, t\mid t-1} - m_{i, t\mid t-1} \right)  e_{it-1}}_{:=Q_1}  - \underbrace{ \frac{1}{n_{t-1}} \sum_{i=1}^{n} \frac{\bs{1}(G_i >t-1) \pi_{i t-1}}{(1 - \pi_{i t-1})(1 - \pi_{i t})} \left( \widehat{m}_{i, t\mid t} - m_{i, t\mid t} \right)  e_{it}}_{:=Q_2} \\
            & - \underbrace{ \frac{1}{n_{t-1}} \sum_{i=1}^{n}  \bs{1}(G_i > t)\left(\frac{\widehat{\pi}_{i t-1}}{1 - \widehat{\pi}_{i t-1}}\frac{1}{1 - \widehat{\pi}_{i t}} - \frac{\pi_{i t-1}}{1 - \pi_{i t-1}}\frac{1}{1 - \pi_{i t}}\right)u_{it} }_{:=Q_3} \\
            & - \underbrace{\frac{1}{n_{t-1}} \sum_{i=1}^{n} \ \frac{\bs{1}(G_i >t-1) (\widehat{\pi}_{i t-1} - \pi_{i t-1})}{(1 - \pi_{i t-1})(1 - \widehat{\pi}_{i t-1})}(m_{i, t\mid t} - m_{i, t\mid t-1})}_{:=Q_4} \\
            & - \underbrace{ \frac{1}{n_{t-1}} \sum_{i=1}^{n} \frac{\bs{1}(G_i >t-1)}{1 - \pi_{it-1} } \frac{\widehat{\pi}_{it-1} - \pi_{it-1}}{1 - \widehat{\pi}_{it-1} }\left(\widehat{m}_{i, t\mid t} - m_{i, t\mid t}\right)}_{:=Q_5} \\
            & + \underbrace{ \frac{1}{n_{t-1}} \sum_{i=1}^{n} \frac{\bs{1}(G_i >t-1)}{1 - \pi_{it-1} } \frac{\widehat{\pi}_{it-1} - \pi_{it-1}}{1 - \widehat{\pi}_{it-1} }  \left(\widehat{m}_{i, t\mid t-1} - m_{i, t\mid t-1}\right)}_{:=Q_6} \\
            & + \underbrace{ \frac{1}{n_{t-1}} \sum_{i=1}^{n} \bs{1}(G_i >t) \left(\frac{\widehat{\pi}_{i t-1}}{1 - \widehat{\pi}_{i t-1}}\frac{1}{1 - \widehat{\pi}_{i t}} - \frac{\pi_{i t-1}}{1 - \pi_{i t-1}}\frac{1}{1 - \pi_{i t}}\right) \left(\widehat{m}_{i, t\mid t} - m_{i, t\mid t}\right)}_{:=Q_7}. 
        \end{align*}

        Applying the same argument used to derive~\eqref{eq:thm_ATT_contemporaneous_Q_2} and \eqref{eq:thm_ATT_contemporaneous_Q_3}, we conclude that 
        \begin{align}\label{eq:thm_ATT_dynamic_1}
            Q_1, Q_2, Q_3 = o_P\left(1\sqrt{n}\right). 
        \end{align}
        For $Q_4$, by selection mechanism in Assumption~\ref{assumption:selection_extension} and mean-independence condition in Assumption~\ref{assumption:surrogacy_extension}, we have $\mb{E}\left(m_{i, t\mid t} - m_{i, t\mid t-1}\mid \{X_{it-1}\}_{i=1}^{n}, \mc{F}_{t-2}, \left\{\frac{\bs{1}(G_i >t-1) (\widehat{\pi}_{i t-1} - \pi_{i t-1})}{(1 - \pi_{i t-1})(1 - \widehat{\pi}_{i t-1})}\right\}_{G_i >t-1} \right) = 0$. Therefore, Applying the same argument used to derive~\eqref{eq:thm_ATT_contemporaneous_Q_2} and \eqref{eq:thm_ATT_contemporaneous_Q_3}, we conclude that 
        \begin{align}\label{eq:thm_ATT_dynamic_2}
            Q_4 = o_P\left(1\sqrt{n}\right). 
        \end{align}

        To derive the bound for $Q_5, Q_6, Q_7$, applying the same argument used to derive~\eqref{eq:thm_ATT_contemporaneous_Q_2} and \eqref{eq:thm_ATT_contemporaneous_Q_3}, we conclude that 
        \begin{align}\label{eq:thm_ATT_dynamic_3}
            Q_5, Q_6, Q_7 = O_P\left(h_mh_\pi +  h_m \left(nh_\pi^{d_\alpha+ d_{X}}\right)^{-1/2} \sqrt{\log (n)}  \right) + o_P(1/\sqrt{n}). 
        \end{align}
        Combining~\eqref{eq:thm_ATT_dynamic_1}-\eqref{eq:thm_ATT_dynamic_3} completes the proof for dynamic ATT.     
    \end{proofthm}

    \begin{lemma}\label{lemma:numerator}
        Under the conditions in Theorem~\ref{thm:estimation_ATT_extension},  for any $t \geq t' \geq T_0 $, $h = h_m, h_p$, and for any $s \in \{1, 2, 3\}$, there exist constants (depending on $t$ and $t'$) $0 < \underline{C}(t, t' + 1) <\overline{C}(t, t' + 1)  < \infty$  such that the following inequalities 
        \begin{equation}\label{eq:lemma_numerator}
        \begin{aligned}
            \underline{C}(t, t' + 1)  nh^{d_{\alpha} + d_{X}} \leq  \min_{i= 1, \ldots,n } \sum_{j \in \mc{I}_s } \widehat{W}^{(t, t' + 1)}_{h, ij}   \leq  \max_{i= 1, \ldots,n } \sum_{j \in \mc{I}_s }  \widehat{W}^{(t, t' + 1)}_{h, ij}  \leq \bar{C}(t, t' + 1) nh^{d_{\alpha} + d_{X}}  
        \end{aligned}
        \end{equation}
        hold with probability approaching to $1$. 
    \end{lemma}
    \begin{prooflmm}{lemma:numerator}
        Define $q_{j, t\mid t' + 1} : = \mb{P}_T\left(G_j >t \mid H_{jt'}, X_{jt'+1}, \alpha_j, G_j >t'\right)$.  We write 
        \begin{equation}\label{eq:lemma_numerator_decomposition}
        \begin{aligned}
            \sum_{j \in \mc{I}_s }\widehat{W}^{(t, t' + 1)}_{h, ij} = & \underbrace{ \sum_{j \in \mc{I}_{s} } q_{j, t\mid t' + 1} \bs{1}(G_j > t') K^{(t' + 1)}_{h, ij}  }_{: = A_{1i}}\\
            & +\underbrace{ \sum_{j \in \mc{I}_s } q_{j, t\mid t' + 1} \bs{1}(G_j > t') (\widehat{K}^{(t' + 1)}_{h, ij} - K^{(t' + 1)}_{h, ij})  }_{: = A_{2i}} \\
            & + \underbrace{ \sum_{j \in \mc{I}_s } \left(  \bs{1}(G_j > t) - q_{j, t\mid t' + 1} \right) \bs{1}(G_j > t') \widehat{K}^{(t' + 1)}_{h, ij}  }_{A_{3i}}
        \end{aligned}
        \end{equation}
        \paragraph{Bound for $A_{1i}$} 
        Fix $(\Gamma_{T}, \alpha_i, X_{it' + 1})$. Then, by the compact condition in Assumption~\ref{assumption:estimation_extension}\ref{item:estimation_extension_compact} and the kernel function is bounded and compact supported (see Assumption~\ref{assumption:estimation_extension}\ref{item:estimation_extension_kernel}) it is straightforward to verify that
        \begin{align*}
            \max_{i=1, \ldots, n,  j \in \mc{I}_{s}  } \left| q_{j, t\mid t' + 1} \bs{1}(G_j > t') K^{(t' + 1)}_{h, ij}  \right| \leq \overline{K}, 
        \end{align*}
        and there exists a constant $ v >0$ such that
        \begin{align*}
            \max_{i=1, \ldots, n, j \in \mc{I}_{s}   } \mb{E}\left(q^2_{j, t\mid t' + 1} \bs{1}(G_j > t') \left(K^{(t' + 1)}_{h, ij}\right)^2 \mid \Gamma_{T}, \alpha_i,  X_{it' + 1} \right) \leq v h^{d_{\alpha} + d_{X}}.   
        \end{align*}
        Since individuals are independent conditional on $\Gamma_{T}$ by Assumption~\ref{assumption:estimation_extension}\ref{item:estimation_extension_panel_weak_dependent}, for each $i$, we apply Bernstein's inequality for independent random variables (Lemma~\ref{lemma:bernstein}) to show that there exists a constant $M>0$ (independent of $i$) such that, for any $\epsilon>0$, 
        \begin{align*}
            \mb{P}\left( \left| A_{1i} - \mb{E}\left(A_{1i} \mid \Gamma_{T}, \alpha_i,  X_{it' + 1} \right)\right| \geq \epsilon \mid \Gamma_{T}, \alpha_i,  X_{it' + 1} \right) \leq 2 \exp\left(\frac{ -\epsilon^2 }{M^2 n h^{d_{\alpha} + d_{X}} + \frac{1}{3} M\epsilon }\right). 
        \end{align*}
        Since the inequality above does not depend on realization of $(\alpha_i,  X_{it' + 1})$, we integrate over $(\alpha_i,  X_{it' + 1})$ and obtain 
        \begin{align*}
            \mb{P}\left( \left| A_{1i} - \mb{E}\left(A_{1i} \mid \Gamma_{T}, \alpha_i,  X_{it' + 1} \right)\right| \geq \epsilon \mid \Gamma_{T} \right) \leq 2 \exp\left(\frac{ -\epsilon^2 }{M^2 n h^{d_{\alpha} + d_{X}} + \frac{1}{3} M\epsilon }\right). 
        \end{align*}
        Therefore, we obtain 
        \begin{align*}
            \mb{P}\left(\max_{i =1, \ldots, n} \left| A_{1i} - \mb{E}\left(A_{1i} \mid \Gamma_{T}, \alpha_i,  X_{it' + 1} \right) \right| \geq \epsilon \mid \Gamma_{T} \right) = & \mb{P}\left(\bigcup_{i=1}^{n} \left\{\left| A_{1i} - \mb{E}\left(A_{1i} \mid \Gamma_{T}, \alpha_i,  X_{it' + 1} \right) \right| \geq \epsilon \right\}  \mid \Gamma_{T} \right) \\
            \leq & \sum_{i =1, \ldots, n} \mb{P}\left( \left| A_{1i} - \mb{E}\left(A_{1i} \mid \Gamma_{T}, \alpha_i,  X_{it' + 1} \right)\right| \geq \epsilon \mid \Gamma_{T} \right) \\
            \eqtext{(i)} & \sum_{i =1}^{n} 2 \exp\left(\frac{ -\epsilon^2 }{M^2 n h^{d_{\alpha} + d_{X}} + \frac{1}{3} M\epsilon }\right) \\
            = & 2n \exp\left(\frac{ -\epsilon^2 }{M^2 n h^{d_{\alpha} + d_{X}} + \frac{1}{3} M\epsilon }\right),  
        \end{align*}
        where (i) follows from Bernstein's inequality. It follows that there exists a sufficiently large constant $D_1 >0$ such that when $\epsilon = D_1 \sqrt{nh^{d_{\alpha} + d_{X}} \log (n)}$, 
        \begin{equation}\label{eq:lemma_numerator_1}
        \begin{aligned}
            \lim_{n\rightarrow \infty}\mb{P}\left(\max_{i =1, \ldots, n} \left| A_{1i} - \mb{E}\left(A_{1i} \mid \Gamma_{T}, \alpha_i,  X_{it' + 1} \right) \right| \geq D_1 \sqrt{nh^{d_{\alpha} + d_{X}} \log (n)}  \mid \Gamma_{T}  \right)  = 0
        \end{aligned}
        \end{equation}

        \vspace{0.5cm}

        For the bounds on $\min_{i =1, \ldots, n}\mb{E}\left(A_{1i} \mid \Gamma_{T}, \alpha_i,  X_{it' + 1} \right)$ and $\max_{i =1, \ldots, n}\mb{E}\left(A_{1i} \mid \Gamma_{T}, \alpha_i,  X_{it' + 1} \right)$. Since $|\mc{I}_{s}| = n/3$, we have 
        \begin{align}\label{eq:lemma_numerator_2}
            \mb{E}\left(A_{1i} \mid \Gamma_{T}, \alpha_i,  X_{it' + 1} \right) = \frac{n}{3} \mb{E}\left( q_{j, t\mid t' + 1} \bs{1}(G_j > t') K^{(t' + 1)}_{h, ij}   \mid \Gamma_{T}, \alpha_i,  X_{it' + 1} \right). 
        \end{align}
        Note that 
        \begin{equation}\label{eq:lemma_numerator_3}
        \begin{aligned}
            & \min_{i=1, \ldots, n } \mb{E}\left( q_{j, t\mid t' + 1}  \bs{1}(G_j > t') K^{(t' + 1)}_{h, ij}   \mid \Gamma_{T}, \alpha_i,  X_{it' + 1} \right) \\
            \geqtext{(i)} & (1 - \overline{p})^{t - t'} \min_{i=1, \ldots, n}  \mb{E}\left( \bs{1}(G_j > t') K^{(t' + 1)}_{h, ij}  \mid  \Gamma_{T}, \alpha_i,  X_{it' + 1}\right) \\
            = & (1 - \overline{p})^{t - t'} \min_{i=1, \ldots, n} \mb{E}
            \left( K^{(t' + 1)}_{h, ij} \mid \Gamma_{T}, \alpha_i, X_{it' + 1}, G_j>t'\right)\mb{P}\left(G_j>t'  \mid \Gamma_{T}, \alpha_i, X_{it' + 1}\right) \\
            \geqtext{(ii)} & (1 - \overline{p})^{t - T_0} \min_{i=1, \ldots, n} \mb{E}\left( K^{(t'+ 1)}_{h, ij}  \mid \Gamma_{T}, \alpha_i,  X_{it' + 1}, G_j>t'\right), 
        \end{aligned}   
        \end{equation}
        where inequalities (i) and (ii) follow from the common overlap condition (Assumption~\ref{assumption:estimation_extension}\ref{item:estimation_extension_overlap}). In addition, we have 
        \begin{equation}\label{eq:lemma_numerator_4}
        \begin{aligned}
            & \min_{i=1, \ldots, n} \mb{E}\left( K^{(t' + 1)}_{h, ij}  \mid  \Gamma_{T}, \alpha_i,  X_{it' + 1}, D_j > t' \right)\\
            = & \min_{i=1, \ldots, n} \int K\left(\frac{X_{jt' + 1}- X_{it' + 1} }{h}\right)K\left(\frac{d(\alpha_i, \alpha_j)}{h}\right)\mr{d}\mb{P}( (\alpha_j, X_{jt' + 1})\mid \Gamma_{T}, G_j > t') \\
            \eqtext{(i)} & \min_{i=1, \ldots, n} \int_{ \substack{\|X_{jt' + 1}- X_{it' + 1}\|_{\infty}, \\ d(\alpha_j - \alpha_i)\leq h }    } K\left(\frac{X_{jt' + 1}- X_{it' + 1}}{h}\right)K\left(\frac{d(\alpha_i, \alpha_j)}{h}\right) \mr{d} \mb{P}( (\alpha_j, X_{it' + 1})\mid \Gamma_{T}, G_j > t') \\
            \geqtext{(ii)} & \min_{i=1, \ldots, n} \int_{ \substack{ \|X_{jt' + 1} - X_{it' + 1} \|_{\infty} , \\ d(\alpha_j - \alpha_i)\leq \frac{K(0)}{2L_K} h }  } K\left(\frac{X_{jt' + 1}- X_{it' + 1}}{h}\right)K\left(\frac{d(\alpha_i, \alpha_j)}{h}\right) \mr{d} \mb{P}( (\alpha_j, X_{jt' + 1})\mid \Gamma_{T}, G_j > t')  \\
            \geqtext{(iii)} &  \underbrace{\left(\min_{ |u| \leq \frac{K(0)}{2L_K} } K(u) \right)^{p+1}}_{:=\tilde{K}} \min_{i=1, \ldots, n} \mb{P}\left(\|X_{jt' + 1}- X_{it' + 1}\|_{\infty}, d(\alpha_j - \alpha_i)\leq \frac{K(0)}{2L_K} h  \mid  \Gamma_{T}, \alpha_i,  X_{it' + 1}, G_j > t'\right) \\
            \geqtext{(iv)} & 
            \tilde{K} \min_{i=1, \ldots, n} \mb{P}\left(\|X_{jt' + 1} - X_{it' + 1} \|_{\infty} \leq 
            \frac{K(0)}{2L_K}h, \|\alpha_j - \alpha_i\|\leq \frac{\underline{\eta} K(0)}{2  L_K} h  \mid  \Gamma_{T}, \alpha_i,  X_{it' + 1}, G_j > t'\right) \\ 
            \geqtext{(v)} & \underbrace{\underline{c}_1 \underline{c}_2 \tilde{K}   \left(\frac{\min\{1, \underline{\eta}\}K(0)}{2 L_K}\right)^{d_{\alpha} + d_{X}}}_{:=C_1} h^{d_{\alpha} + d_{X}}. 
        \end{aligned}
        \end{equation}
        Here, (i) holds because $K(\cdot)$ is supported on $[-1, 1]$ by Assumption~\ref{assumption:estimation_extension}\ref{item:estimation_extension_kernel},  (ii) and (iii) follow from $K(0) >0$ and the Lipschitz continuity of $K(\cdot)$ (Assumption~\ref{assumption:estimation_extension}\ref{item:estimation_extension_kernel}), which together ensure that $K(\cdot)$ is strictly positive on $[-K(0)/2L_K, K(0)/2L_K]$. Inequality (iv) follows from the informativeness condition (see Assumption~\ref{assumption:informativeness_estimation}).  Finally, (v) follows from Assumption~\ref{assumption:estimation_extension}\ref{item:estimation_extension_compact}. Then, by~\eqref{eq:lemma_numerator_2}, \eqref{eq:lemma_numerator_3}, and \eqref{eq:lemma_numerator_4},  we have 
        \begin{equation}\label{eq:lemma_numerator_5}
        \begin{aligned}
            \min_{i=1, \ldots, n}\mb{E}\left(A_{1i} \mid \Gamma_{T}, \alpha_i,  X_{it'+1} \right) \geq \underbrace{\frac{C_1}{3} (1 - \overline{p})^{t - T_0}}_{:=C_1(t, t' + 1)} nh^{d_{\alpha} + d_{X}} = C_1(t, t' + 1)nh^{d_{\alpha} + d_{X}}.  
        \end{aligned}
        \end{equation}

        To establish the upper bound $\max_{i\in \mc{I}_{s_1}}\mb{E}\left(A_{1i} \mid \Gamma_{T}, \alpha_i,  X_{it'+1} \right) $, note that 
        \begin{equation}\label{eq:lemma_numerator_6}
        \begin{aligned}
            & \max_{i=1, \ldots, n} \mb{E}\left(  q_{j, t\mid t' + 1} \bs{1}(G_j > t') K^{(t' + 1)}_{h, ij}   \mid \Gamma_{T}, \alpha_i,  X_{it'+1} \right) \\
            \leqtext{(i)} & (1-\underline{p})^{t - t'} \max_{i=1, \ldots, n} \mb{E}\left( \bs{1}(G_j > t') K^{(t' + 1)}_{h, ij}  \mid  \Gamma_{T}, \alpha_i,  X_{it'+1}\right) \\
            = & (1-\underline{p})^{t - t'} \max_{i=1, \ldots, n} \mb{E}
            \left( K^{(t'+ 1)}_{h, ij} \mid \Gamma_{T}, \alpha_i,  X_{it'+1}, G_j>t'\right)\mb{P}\left(G_j>t' \mid   \Gamma_{T}, \alpha_i,  X_{it'+1}\right) \\
            \leqtext{(ii)} & (1-\underline{p})^{t - T_0} \max_{i=1, \ldots, n} \mb{E}\left( K^{(t' + 1)}_{h, ij}  \mid \Gamma_{T}, \alpha_i,  X_{it'+1}, G_j>t'\right), 
        \end{aligned}
        \end{equation}
        where inequalities (i) and (ii) follow from the common overlap condition (Assumption~\ref{assumption:estimation_extension}\ref{item:estimation_extension_overlap}). In addition, we have 
        \begin{equation}\label{eq:lemma_numerator_7}
        \begin{aligned}
            & \max_{i=1, \ldots, n} \mb{E}\left( K^{(t' + 1)}_{h, ij}  \mid  \Gamma_{T}, \alpha_i,  X_{it'+1}, D_j > t' \right)\\
            = & \max_{i=1, \ldots, n} \int K\left(\frac{X_{jt'+1} - X_{it'+1} }{h}\right)K\left(\frac{d(\alpha_i, \alpha_j)}{h}\right)\mr{d}\mb{P}( (\alpha_j, X_{jt'+1})\mid \Gamma_{T}, G_j > t') \\
            \eqtext{(i)} & \max_{i=1, \ldots, n} \int_{ \substack{\|X_{jt'+1} - X_{it'+1} \|_{\infty}\leq h, \\  d(\alpha_j - \alpha_i) \leq h  }  } K\left(\frac{X_{jt'+1}- X_{it'+1}}{h}\right)K\left(\frac{d(\alpha_i, \alpha_j)}{h}\right) \mr{d} \mb{P}( (\alpha_j, X_{jt'+1})\mid \Gamma_{T}, G_j > t') \\
            \leqtext{(ii)} & \overline{K}^{p + 1} \max_{i=1, \ldots, n} \int_{ \|X_{jt'+1}- X_{it'+1}\|_{\infty}, d(\alpha_j - \alpha_i) \leq  h  } \mr{d} \mb{P}( (\alpha_j, X_{jt'+1})\mid \Gamma_{T}, G_j > t')  \\
            \leqtext{(iii)} & \overline{K}^{p + 1}  \max_{i=1, \ldots, n} \mb{P}\left(\|X_{jt'+1}- X_{it'+1}\|_{\infty} \leq 
             h, \|\alpha_j - \alpha_i\|\leq \overline{\eta} h  \mid  \Gamma_{T}, \alpha_i,  X_{it'+1}, G_j > t'\right) \\
            \leqtext{(iv)} & \underbrace{\overline{c}_1 \overline{c}_2\overline{K}^{p + 1}    \left( 1+ \overline{\eta} \right)^{d_{\alpha} + d_{X}}}_{:=C_2 } h^{d_{\alpha} + d_{X}}. 
        \end{aligned}
        \end{equation}
        Here, (i) holds because $K(\cdot)$ is supported on $[-1, 1]$ by Assumption~\ref{assumption:estimation_extension}\ref{item:estimation_extension_kernel}, (ii) follows from the boundedness kernel function  (Assumption~\ref{assumption:estimation_extension}\ref{item:estimation_extension_kernel}), and (iii) follows from the information condition (see Assumption~\ref{assumption:informativeness_estimation}). Finally, inequality (iv) holds because of Assumption~\ref{assumption:estimation_extension}\ref{item:estimation_extension_compact}.
        Then, by~\eqref{eq:lemma_numerator_2}, \eqref{eq:lemma_numerator_6}, and \eqref{eq:lemma_numerator_7},  we have 
        \begin{equation}\label{eq:lemma_numerator_8}
        \begin{aligned}
            \max_{i=1, \ldots, n}\mb{E}\left(A_{1i} \mid \Gamma_{T}, \alpha_i,  Y_{it'}^{(p)} \right) \leq \underbrace{\frac{C_2}{3} (1 - \underline{p})^{t - T_0}}_{: = C_2(t, t' + 1)} nh^{d_{\alpha} + d_{X}} = C_2(t, t' + 1) nh^{d_{\alpha} + d_{X}}.  
        \end{aligned}
        \end{equation}

        Therefore, by \eqref{eq:lemma_numerator_1}, \eqref{eq:lemma_numerator_5}, and \eqref{eq:lemma_numerator_8}, we have that
        \begin{align*}
            \lim_{n\rightarrow \infty}\mb{P}\bigg(  C_1(t, t' + 1)nh^{d_{\alpha} + d_{X}} -D_1 \sqrt{nh^{d_{\alpha} + d_{X}} \log (n)} \leq \min_{i=1, \ldots, n} A_{1i} \leq \max_{i=1, \ldots, n} A_{1i} & \\
            \leq  C_2(t, t' + 1) nh^{d_{\alpha} + d_{X}} + D_1 \sqrt{nh^{d_{\alpha} + d_{X}} \log (n)} \mid \Gamma_{T} & \bigg) = 0
        \end{align*}
        Since $nh^{d_{\alpha} + d_{X}} \log(n)\rightarrow \infty$ as required in Theorem~\ref{thm:estimation_ATT_extension}, we have 
        \begin{equation}\label{eq:lemma_numerator_A_i1}
        \begin{aligned}
            \lim_{n\rightarrow \infty}\mb{P}\bigg(  \frac{C_1(t, t' + 1)}{2}nh^{d_{\alpha} + d_{X}} \leq \min_{ii=1, \ldots, n} A_{1i} \leq \max_{i=1, \ldots, n} A_{1i} \leq  2 C_2(t, t' + 1)  nh^{d_{\alpha} + d_{X}}  \bigg) = 0
        \end{aligned}
        \end{equation}

        \paragraph{Bound for $A_{2i}$} Note that $\max_{i=1, \ldots, n} | A_{2i}|$ is bounded by 
        \begin{align*}
            & \max_{i=1, \ldots, n} \sum_{j \in \mc{I}_{s}  }  \bs{1}(G_j > t') K\left(\frac{X_{jt'+ 1} -  X_{it'+ 1}}{h}\right)\left| K\left(\frac{\widehat{d}_{ij}}{h}\right) - K\left(\frac{d(\alpha_i, \alpha_j)}{h}\right) \right|  \\
            \leqtext{(i)}  &  \max_{i=1, \ldots, n} \sum_{j \in \mc{I}_{s}  }  \bs{1}(G_j > t') \bs{1}\left( \widehat{d}_{ij} \leq h \vee d(\alpha_i, \alpha_j) \leq h \right)K\left(\frac{ X_{jt'+ 1} -   X_{it'+ 1}}{h}\right) L_K \frac{| \widehat{d}_{ij} - d(\alpha_i, \alpha_j) |}{h} , 
        \end{align*}
        where (i) follows from that the kernel function is Lipschitz continuous and has bounded support  (Assumption~\ref{assumption:estimation_extension}\ref{item:estimation_extension_kernel}). 
        Then, by informativeness condition (Assumption~\ref{assumption:informativeness_estimation}), we have that, wpa1, 
        \begin{equation}\label{eq:lemma_numerator_A_9}
        \begin{aligned}
            \max_{i=1, \ldots, n} | A_{2i}| \leq & \frac{\delta_{n, T_0}}{h} \max_{i=1, \ldots, n} \sum_{j \in \mc{I}_{s}  }   \bs{1}(G_j > t')  \bs{1}\left( \widehat{d}_{ij} \leq h \vee d(\alpha_i, \alpha_j) \leq h \right) K\left(\frac{ X_{jt'+ 1} -   X_{it'+ 1}}{h}\right) \\
            \leqtext{(i)} & \frac{\delta_{n, T_0}}{h} \max_{i=1, \ldots, n} \underbrace{\sum_{j \in \mc{I}_{s}  }   \bs{1}(G_j > t')  \bs{1}\left( \|\alpha_i - \alpha_j\|\leq 2 \overline{\eta} h \right) K\left(\frac{ X_{jt'+ 1} -   X_{it'+ 1}}{h}\right)}_{: = B_{2i}}, 
        \end{aligned}
        \end{equation}
        where (i) follows from informativeness condition (Assumption~\ref{assumption:informativeness_estimation}) and Lemma~\ref{lemma:kernel_hat_d}. We employ the same argument used to prove \eqref{eq:lemma_numerator_1} to  show that there exists a sufficiently large constant $D_2 >0$ such that
        \begin{equation}\label{eq:lemma_numerator_A_10}
        \begin{aligned}
            \lim_{n\rightarrow \infty }\mb{P}\left(\max_{i=1, \ldots, n} \left|B_{2i} - \mb{E}\left(B_{2i}\mid \Gamma_{T}, \alpha_i,   X_{it'+ 1} \right)\right| \geq  D_2 \sqrt{nh^{d_{\alpha} + d_{X}} \log (n)} \mid \Gamma_{T} \right) = 0. 
        \end{aligned}
        \end{equation}
        Also, the same argument used to prove \eqref{eq:lemma_numerator_8}, we have
        \begin{equation}\label{eq:lemma_numerator_A_11}
        \begin{aligned}
            \max_{i=1, \ldots, n} \mb{E}\left( B_{2i} \mid \Gamma_{T}, \alpha_i,   X_{it'+ 1}  \right) \leq \underbrace{\frac{1}{3} (1 - \overline{p})^{t' - T_0} \overline{c}_1 \overline{c}_2\overline{K}^{p + 1}    \left(1 +  2 \overline{\eta} \right)^{d_{\alpha} + d_{X}}}_{:=C_3(t, t' + 1)} n h^{d_{\alpha} + d_{X}} . 
        \end{aligned}
        \end{equation}
        Therefore, combining \eqref{eq:lemma_numerator_A_9}, \eqref{eq:lemma_numerator_A_10}, \eqref{eq:lemma_numerator_A_11}, and $ nh^{d_{\alpha} + d_{X}}/\log (n) \rightarrow \infty $ as required in Theorem~\ref{thm:estimation_ATT_extension}, we have 
        \begin{equation}\label{eq:lemma_numerator_A_i2}
        \begin{aligned}
            \lim_{n, T_0 \rightarrow \infty}\mb{P}\left(\max_{i=1, \ldots, n}  | A_{2i}| \leq 2C_3(t, t' + 1) nh^{d_{\alpha} + d_{X}}\frac{\delta_{n, T_0}}{h}\right) = 0
        \end{aligned}
        \end{equation}

        \paragraph{Bound for $A_{3i}$} Conditional on $\{X_{it'+1}\}_{i=1}^{n}$ and $\mc{F}_{t'}$, it is easy to check that $\widehat{K}^{(t' + 1)}_{h, ij}$ is deterministic. In addition, by Assumption~\ref{assumption:selection_extension}, we obtain 
        \begin{align*}
            \mb{E}\left(\left(  \bs{1}(G_j>t) -  q_{j, t\mid t' + 1} \right) \bs{1}(G_j > t') \widehat{K}^{(t' + 1)}_{h, ij} \mid \{X_{it'+1}\}_{i=1}^{n}, \mc{F}_{t'}\right) = 0. 
        \end{align*} 
        By the compact condition (see Assumption~\ref{assumption:estimation_extension}\ref{item:estimation_extension_compact}) and the kernel function is bounded and compact supported (see Assumption~\ref{assumption:estimation_extension}\ref{item:estimation_extension_kernel}) it is straightforward to verify that
        \begin{align*}
            \max_{i=1, \ldots, n, j \in \mc{I}_{s}  } \left| \left(  \bs{1}(G_j>t) -  q_{j, t\mid t' + 1} \right) \bs{1}(G_j > t') \widehat{K}^{(t'+ 1)}_{h, ij} \right| \leq \overline{K}, 
        \end{align*}
        and there exists a constant $ v >0$ such that
        \begin{align*}
            \max_{ \substack{i=1, \ldots, n, \\ j \in \mc{I}_{s} }   } \mb{E}\left(\left(  \bs{1}(G_j>t) -  q_{j, t\mid t' + 1} \right)^2 \bs{1}(G_j > t') \left(\widehat{K}^{(t' + 1)}_{h, ij} \right)^2 \mid \{X_{it'+1}\}_{i=1}^{n}, \mc{F}_{t'} \right) \leq v h^{d_{\alpha} + d_{X}}.   
        \end{align*}
        Since individuals are independent conditional on $\{X_{it'+1}\}_{i=1}^{n}$ and $\mc{F}_{t'}$ by Assumption~\ref{assumption:estimation_extension}\ref{item:estimation_extension_panel_weak_dependent}, for each $i$, we apply Bernstein's inequality for independent random variables (Lemma~\ref{lemma:bernstein}) to show that there exists a constant $M>0$ (independent of $i$) such that, for any $\epsilon>0$, 
        \begin{align*}
            \mb{P}\left( \left| A_{3i} \right| \geq \epsilon \mid \{X_{it'+1}\}_{i=1}^{n}, \mc{F}_{t'} \right) \leq 2 \exp\left(\frac{ -\epsilon^2 }{M^2 n h^{d_{\alpha} + d_{X}} + \frac{1}{3} M\epsilon }\right). 
        \end{align*}
        Since the inequality above does not depend on realization of $\{\{\alpha_i\}_{i, 1\ldots, n}, \{Y_{i\tau}, X_{i\tau}\}_{i=1, \ldots, n, \tau\leq t' }\}$ and $\{X_{it'+1}\}_{i=1}^{n}$, we integrate over $\{\{\alpha_i\}_{i, 1\ldots, n}, \{Y_{i\tau}, X_{i\tau}\}_{i=1, \ldots, n, \tau\leq t' }\}$ and $\{X_{it'+1}\}_{i=1}^{n}$ to obtain 
        \begin{align*}
            \mb{P}\left( \left| A_{3i}\right| \geq \epsilon \mid \Gamma_{T}\right) \leq 2 \exp\left(\frac{ -\epsilon^2 }{M^2 n h^{d_{\alpha} + d_{X}} + \frac{1}{3} M\epsilon }\right). 
        \end{align*}
        Therefore, we obtain 
        \begin{align*}
            \mb{P}\left(\max_{i=1, \ldots, n} \left| A_{3i}  \right| \geq \epsilon \mid \Gamma_{T}\right) = & \mb{P}\left(\bigcup_{i=1, \ldots, n} \left\{\left| A_{3i}  \right| \geq \epsilon \right\}  \mid \Gamma_{T} \right) \\
            \leq & \sum_{i =1}^{n} \mb{P}\left( \left| A_{3i} \right| \geq \epsilon \mid \Gamma_{T} \right) \\
            \eqtext{(i)} & \sum_{i =1}^{n} 2 \exp\left(\frac{ -\epsilon^2 }{M^2 n h^{d_{\alpha} + d_{X}} + \frac{1}{3} M\epsilon }\right) \\
            = & 2n \exp\left(\frac{ -\epsilon^2 }{M^2 n h^{d_{\alpha} + d_{X}} + \frac{1}{3} M\epsilon }\right),  
        \end{align*}
        where (i) follows from Bernstein's inequality. It follows that there exists a sufficiently large constant $D_3 >0$ such that when $\epsilon = D_3 \sqrt{nh^{d_{\alpha} + d_{X}} \log (n)}$, 
        \begin{equation}\label{eq:lemma_numerator_A_i3}
        \begin{aligned}
            \lim_{n\rightarrow \infty}\mb{P}\left(\max_{i=1, \ldots, n} \left| A_{3i} \right| \geq D_3 \sqrt{nh^{d_{\alpha} + d_{X}} \log (n)}  \mid \Gamma_{T}  \right)  = 0
        \end{aligned}
        \end{equation}   

        Therefore, combining \eqref{eq:lemma_numerator_A_i1}, \eqref{eq:lemma_numerator_A_i2}, and \eqref{eq:lemma_numerator_A_i3}, 
        \begin{align*}
            \lim_{n, T_0\rightarrow\infty}\mb{P}\bigg(  & \frac{C_1(t, t' + 1)}{2}nh^{d_{\alpha} + d_{X}} - 2C_3(t, t' + 1) nh^{d_{\alpha} + d_{X}}\frac{\delta_{n, T_0}}{h} -D_3 \sqrt{nh^{d_{\alpha} + d_{X}} \log (n)} \\
            & \leq \min_{i=1, \ldots, n} \sum_{j \in \mc{I}_{s} } \widehat{W}^{(t, t' + 1)}_{h, ij} 
            \leq \max_{i=1, \ldots, n} \sum_{j \in \mc{I}_{s}   } \widehat{W}^{(t, t' + 1)}_{h, ij} \leq \\  
            &  2 C_2(t, t' + 1) nh^{d_{\alpha} + d_{X}} + 2C_3(t, t' + 1) nh^{d_{\alpha} + d_{X}}\frac{\delta_{n, T_0}}{h} + D_3 \sqrt{nh^{d_{\alpha} + d_{X}} \log (n)} \mid \Gamma_{T}  \bigg) = 0
        \end{align*}
        Since $nh^{d_{\alpha} + d_{X}} / \log (n) \rightarrow \infty$ and $h/\delta_{n, T_0}\rightarrow\infty$, we have 
        \begin{align*}
            \lim_{n, T_0\rightarrow\infty}\mb{P}\bigg(  & \frac{C_1(t, t' + 1)}{4}nh^{d_{\alpha} + d_{X}}  \leq \min_{i=1, \ldots, n} \sum_{j \in \mc{I}_{s} } \widehat{W}^{(t, t' + 1)}_{h, ij}  \\
            & \leq \max_{i=1, \ldots, n} \sum_{j \in \mc{I}_{s}  } \widehat{W}^{(t, t' + 1)}_{h, ij}   
            \leq  4 C_2(t, t' + 1) nh^{d_{\alpha} + d_{X}}  \mid \Gamma_{T} \bigg) = 0
        \end{align*}
        Let $\underline{C}(t, t' + 1): =  \frac{1}{4} C_1(t, t' + 1)$ and $\overline{C}(t, t' + 1): =  4 C_2(t, t' + 1)$. We complete the proof. 
    \end{prooflmm}

    \section{Proof for Pseudo distance and Informativeness}\label{appendix:proof_informativeness}

\subsection{Proof for Lemma~\ref{lemma:sufficient_estimation_consistency_d}}\label{appendix_sub:proof_sufficient_consistency_d}

    \begin{prooflmm}{lemma:sufficient_estimation_consistency_d}
        Let $\rho_g = \sup_{\alpha, \Gamma}|g(\alpha, \Gamma)|$. It is straightforward to see that $\rho_g<\infty$ by Assumpion~\ref{assumption:sufficient_estimation_consistency_d}\ref{item:sufficient_estimation_consistency_d_finite}. For notationally simplicity, for any $\alpha_1, \alpha_2 \in \mc{A}$  and any $\Gamma \in \mr{supp}(\Gamma_{T})$,  define $\Delta(\alpha_1, \alpha_2, \Gamma) := (g(\alpha_1, \Gamma) - g(\alpha_2, \Gamma))$ and 
        \begin{align*}
            \varphi(\alpha_1, \alpha_2, \alpha_3, \alpha_4) : =  \int \Delta(\alpha_{1}, \alpha_{2}, \Gamma)  \Delta(\alpha_{3}, \alpha_{4}, \Gamma) \mr{d}\mb{P}(\Gamma) . 
        \end{align*}
        Recall that 
        \begin{align*}
            \widehat{d}_{ij} = \max_{k_1, k_2 \in \{1, \ldots, n\}\setminus \{i, j\} }\frac{1}{T_0} \sum_{t=0}^{T_0-1} (Y_{k_1 t} - Y_{k_2 t} ) (Y_{it} - Y_{jt}), 
        \end{align*}
        and 
        \begin{align*}
            d(\alpha_i, \alpha_{j}) = d(\alpha_i,\alpha_j) : = \sup_{\alpha_1, \alpha_2 \in \mc{A}} \left| \int \Delta(\alpha_1, \alpha_2, \Gamma)\Delta(\alpha_i, \alpha_j, \Gamma) \mr{d}\mb{P}(\Gamma) \right|
        \end{align*}
        In addition, for notational simplicity, define  
        \begin{align*}
            \bar{d}_{ij}  := \max_{k_1, k_2 \in \{1, \ldots, n\}\setminus \{i, j\} }\frac{1}{T_0} \sum_{t=0}^{T_0-1} \Delta(\alpha_{k_1}, \alpha_{k_2}, \Gamma_t) \Delta(\alpha_i, \alpha_j, \Gamma_t). 
        \end{align*}
        Then 
        \begin{align*}
            \max_{i, j \in \{1, \ldots, n\} } | \widehat{d}_{ij} - d(\alpha_i, \alpha_{j})|  \leq \max_{i, j \in \{1, \ldots, n\} } | \widehat{d}_{ij} - \bar{d}_{ij}|  + \max_{i, j \in \{1, \ldots, n\} } | \bar{d}_{ij} - d(\alpha_i, \alpha_{j})|. 
        \end{align*}

        \paragraph{Bound for $\max_{i, j \in \{1, \ldots, n\} } | \widehat{d}_{ij} - \bar{d}_{ij}|$} Note that 
        \begin{align*}
            \max_{i, j \in \{1, \ldots, n\} } \left | \widehat{d}_{ij} - \bar{d}_{ij} \right|  = & \underbrace{\max_{i, j \in \{1, \ldots, n\} } \max_{k_1, k_2 \in \{1, \ldots, n\}\setminus \{i, j\} } \left| \frac{1}{T_0} \sum_{t=0}^{T_0-1} \Delta(\alpha_{k_1}, \alpha_{k_2}, \Gamma_t)  (\tilde{\epsilon}_{it} - \tilde{\epsilon}_{jt})\right| }_{A_1}  \\
            & + \underbrace{\max_{i, j \in \{1, \ldots, n\} } \max_{k_1, k_2 \in \{1, \ldots, n\}\setminus \{i, j\} } \left| \frac{1}{T_0} \sum_{t=0}^{T_0-1} \Delta(\alpha_i, \alpha_j, \Gamma_t) (\tilde{\epsilon}_{k_1t} - \tilde{\epsilon}_{k_1t}) \right| }_{A_2} \\
            & +  \underbrace{ \max_{i, j \in \{1, \ldots, n\} } \max_{k_1, k_2 \in \{1, \ldots, n\}\setminus \{i, j\} }  \left|\frac{1}{T_0} \sum_{t=0}^{T_0-1} (\tilde{\epsilon}_{k_1t} - \tilde{\epsilon}_{k_1t})(\tilde{\epsilon}_{it} - \tilde{\epsilon}_{jt}) \right| }_{A_3}
        \end{align*}
        It is straightforward to verify that for any constant $\delta >0$,  
        \begin{align*}
            & \mb{P}\left( A_1  \geq 2\rho_g\delta  \mid  \{\alpha_i\}_{i=1}^{n}, \Gamma_{T} \right) \\
            = &\mb{P}\left(  \bigcup_{i, j \in \{1, \ldots, n\} } \bigcup_{k_1, k_2 \in \{1, \ldots, n\}\setminus \{i, j\} } \left\{ \left| \frac{1}{T_0} \sum_{t=0}^{T_0-1} \Delta(\alpha_{k_1}, \alpha_{k_2}, \Gamma_t) (\tilde{\epsilon}_{it} - \tilde{\epsilon}_{jt})\right| \geq   2\rho_g\delta  \right\}   \mid  \{\alpha_i\}_{i=1}^{n}, \Gamma_{T} \right) \\
            \leq & \sum_{i, j \in \{1, \ldots, n\} } \sum_{k_1, k_2 \in \{1, \ldots, n\}\setminus \{i, j\} }  \mb{P}\left(    \left| \frac{1}{T_0} \sum_{t=0}^{T_0-1} \Delta(\alpha_{k_1}, \alpha_{k_2}, \Gamma_t)  (\tilde{\epsilon}_{it} - \tilde{\epsilon}_{jt})\right| \geq 2\rho_g\delta   \mid  \{\alpha_i\}_{i=1}^{n}, \Gamma_{T} \right) \\
            \leqtext{(i)} & n^4 \max_{\substack{ i, j \in \{1, \ldots, n\}, \\ k_1, k_2 \in \{1, \ldots, n\}\setminus \{i, j\}}}\mb{P}\left(    \left| \frac{1}{T_0} \sum_{t=0}^{T_0-1}  (\tilde{\epsilon}_{it} - \tilde{\epsilon}_{jt})\right| \geq \delta    \mid  \{\alpha_i\}_{i=1}^{n}, \Gamma_{T} \right), 
        \end{align*}
        where (i) follows from $\left| \Delta(\alpha_{k_1}, \alpha_{k_2}, \Gamma_t)\right|  \leq 2\rho_g$. 
        
        Then, since $\mb{E}\left(\Delta(\alpha_{k_1}, \alpha_{k_2}, \Gamma_t) (\tilde{\epsilon}_{it} - \tilde{\epsilon}_{jt}) \mid \{\alpha_i\}_{i=1}^{n}, \Gamma_{T}  \right) = 0$,  by Assumption~\ref{assumption:estimation_consistency_d}\ref{item:sufficient_estimation_consistency_d_weak_dependent} and~\ref{item:sufficient_estimation_consistency_d_finite}, we employ Lemma~\ref{lemma:concentration_weak_dependent_reference} to show that there exist positive constants $C_1, C_2, C_3, C_4 $ that do not depend on $(n, T_0)$, such that
        \begin{align*}
            & \max_{\substack{ i, j \in \{1, \ldots, n\}, \\ k_1, k_2 \in \{1, \ldots, n\}\setminus \{i, j\}}} \mb{P}\left(    \left| \frac{1}{T_0} \sum_{t=0}^{T_0-1}  (\tilde{\epsilon}_{it} - \tilde{\epsilon}_{jt})\right| \geq \delta    \mid  \{\alpha_i\}_{i=1}^{n}, \Gamma_{T} \right) \\
            \leq &   T_0 \exp\left(-\frac{T_0^{2/3} \delta^{2/3}}{C_1}\right) + \exp\left(- \frac{T_0 \delta^2}{C_2 }\right) + \exp\left( - \frac{T_0  \delta^2}{C_3 }\exp\left(\frac{T_0^{2/9} \delta^{2/9}}{C_4 \left(\log T_0 \right)^{2/3}}\right) \right). 
        \end{align*} 
        Therefore, there exists a sufficiently large constant $M_1 >0$ that is independent of $n, T_0$ such that when $\delta = M_1 \sqrt{\frac{\log (n T_0) }{T_0}}$, we have 
        \begin{align*}
            \mb{P}\left( A_1  \geq M_1 \sqrt{\frac{\log (n T_0) }{T_0}} \mid  \{\alpha_i\}_{i=1}^{n}, \Gamma_{T} \right) \leq 1/\sqrt{T_0}. 
        \end{align*}
        Since the inequality above does not depend on the realization $(\{\alpha_i\}_{i=1}^{n}, \Gamma_{T})$, we integrate over $(\{\alpha_i\}_{i=1}^{n}, \Gamma_{T})$ to obtain 
        \begin{align*}
            \mb{P}\left( A_1  \geq M_1 \sqrt{\frac{\log (n T_0) }{T_0}}  \right) \leq 1/\sqrt{T_0}. 
        \end{align*}
        By the similar argument, we can also show that there exist constants $M_2, 
        M_3 >0$ such that 
        \begin{gather*}
            \mb{P}\left( A_2  \geq M_2 \sqrt{\frac{\log (n T_0) }{T_0}}  \right) \leq 1/\sqrt{T_0} , \quad \text{and }
            \mb{P}\left( A_3  \geq M_3 \sqrt{\frac{\log (n T_0) }{T_0}}  \right) \leq 1/\sqrt{T_0}. 
        \end{gather*}
        Therefore, we obtain 
        \begin{align*}
            \mb{P}\left(  \max_{i, j \in \{1, \ldots, n\} } | \widehat{d}_{ij} - \bar{d}_{ij}|   \geq  (M_1 + M_2 +M_3) \sqrt{\frac{\log (n T_0) }{T_0}} \right) \leq 3 / \sqrt{T_0}. 
        \end{align*} 
        
        \paragraph{Bound for $\max_{i, j \in \{1, \ldots, n\} } |  \bar{d}_{ij} - d(\alpha_i, \alpha_j) |$}  Note that 
        \begin{align*}
            & |  \bar{d}_{ij} - d(\alpha_i, \alpha_j) | \\
            \leq & \max_{k_1, k_2 \in \{1, \ldots, n\}\setminus \{i, j\} } \left| \frac{1}{T_0} \sum_{t=0}^{T_0-1} \Delta(\alpha_{k_1}, \alpha_{k_2}, \Gamma_t)  \Delta(\alpha_{i}, \alpha_{j}, \Gamma_t)  - \varphi(\alpha_{k_1}, \alpha_{k_2}, \alpha_i, \alpha_j)\right|   \\
            & + \sup_{\alpha_1, \alpha_2 \in \mc{A}} \left|  \varphi(\alpha_{1}, \alpha_{2}, \alpha_i, \alpha_j)  \right| - \max_{k_1, k_2 \in \{1, \ldots, n\}\setminus \{i, j\}} \left| \varphi(\alpha_{k_1}, \alpha_{k_2}, \alpha_i, \alpha_j) \right|. 
        \end{align*}
        It follows that $\max_{i, j \in \{1, \ldots, n\} } |  \bar{d}_{ij} - d(\alpha_i, \alpha_j) |$ is bounded by 
        \begin{align*}
            & \underbrace{\max_{i, j \in \{1, \ldots, n\} }\max_{k_1, k_2 \in \{1, \ldots, n\}\setminus \{i, j\} } \left| \frac{1}{T_0} \sum_{t=0}^{T_0-1} \Delta(\alpha_{k_1}, \alpha_{k_2}, \Gamma_t)  \Delta(\alpha_{i}, \alpha_{j}, \Gamma_t)  - \varphi(\alpha_{k_1}, \alpha_{k_2}, \alpha_i, \alpha_j)\right|  }_{A_4}  \\
            & + \underbrace{\max_{i, j \in \{1, \ldots, n\} } \left(\sup_{\alpha_1, \alpha_2 \in \mc{A}} \varphi\left(\alpha_1, \alpha_2, \alpha_i, \alpha_j\right) - \max_{k_1, k_2 \in \{1, \ldots, n\}\setminus \{i, j\}} \varphi\left(\alpha_{k_1}, \alpha_{k_2}, \alpha_i, \alpha_j\right)\right)}_{A_5}. 
        \end{align*}
        By Assumption~\ref{assumption:sufficient_estimation_consistency_d}\ref{item:sufficient_estimation_consistency_d_gamma}, there exists a constant $M_4>0$ such that $A_4\leq M_4 \sqrt{\frac{\log(T_0)}{T_0}}$ wpa1. To establish an error bound for $A_5$, let $(\alpha_1^*, \alpha_2^*)$ be the maximizer of $d(\alpha_i, \alpha_j)$ such that 
        \begin{align*}
            d(\alpha_i, \alpha_j) = \varphi\left(\alpha_1^*, \alpha_2^*, \alpha_i, \alpha_j\right). 
        \end{align*}
        Then, we have 
        \begin{align*}
            A_5 \leq &  \max_{i, j \in \{1, \ldots, n\} }\left( \varphi\left(\alpha_1^*, \alpha_2^*, \alpha_i, \alpha_j\right) - \max_{k_1, k_2 \in \{1, \ldots, n\}\setminus \{i, j\}} \varphi\left(\alpha_{k_1}, \alpha_{k_2}, \alpha_i, \alpha_j\right) \right) \\
            = & \max_{i, j \in \{1, \ldots, n\} }\min_{k_1, k_2 \in \{1, \ldots, n\}\setminus \{i, j\}}  \left( \varphi\left(\alpha_1^*, \alpha_2^*, \alpha_i, \alpha_j\right) - \varphi\left(\alpha_{k_1}, \alpha_{k_2}, \alpha_i, \alpha_j\right) \right) \\
            \leq & \max_{i, j \in \{1, \ldots, n\} }\min_{k_1, k_2 \in \{1, \ldots, n\}\setminus \{i, j\}} 2 \rho_g L_g  \left( \|\alpha_{k_1} - \alpha_1^*\|_{\infty} + \|\alpha_{k_2} - \alpha_2^*\|_{\infty}\right) \\ 
            \leq & \max_{i, j \in \{1, \ldots, n\} }\min_{k_1, k_2 \in \{1, \ldots, n\}\setminus \{i, j\}} 2 \rho_g L_g d_{\alpha}^{1/2}\left( \| \alpha_{k_1} - \alpha^*_{1}\|  + \| \alpha_{k_2} - \alpha^*_{2}\| \right) \\
            \leq & 4 \rho_g L_g d_{\alpha}^{1/2} \max_{i, j \in \{1, \ldots, n\} } \sup_{\alpha \in \mc{A}} \min_{k  \in \{1, \ldots, n\}\setminus \{i, j\}}  \| \alpha_{k} - \alpha\|  
        \end{align*}
        By Assumption~\ref{assumption:sufficient_estimation_consistency_d}\ref{item:sufficient_estimation_consistency_d_compact}, there exists a $\delta$-covering of the support of $\alpha$ consisting of $\mc{J} \leq 4^{d_{\alpha}} d_{\alpha}^{d_{\alpha}/2} \delta^{-d_{\alpha}}$ open balls with radius $\delta$, denoted by $\{\bar{\alpha}_{\ell}\}_{\ell = 1}^{\mc{J}}$. We obtain 
        \begin{align*}
            \sup_{\alpha \in \mc{A}} \min_{k  \in \{1, \ldots, n\}\setminus \{i, j\}}  \| \alpha_{k} - \alpha\|  \leq \max_{\ell = 1, \ldots, \mc{J}} \min_{k  \in \{1, \ldots, n\}\setminus \{i, j\}}  \| \alpha_{k} - \bar{\alpha}_{\ell}\| + \delta. 
        \end{align*}
        It follows that, for any $\delta >0 $, 
        \begin{align*}
            & \mb{P}\left(\max_{i, j \in \{1, \ldots, n\} } \sup_{\alpha \in \mc{A}} \min_{k  \in \{1, \ldots, n\}\setminus \{i, j\}}  \| \alpha_{k} - \alpha\| \geq 2\delta \right)  \\
            = & \mb{P}\left(\bigcup_{i, j \in \{1, \ldots, n\} } \left\{\sup_{\alpha \in \mc{A}} \min_{k  \in \{1, \ldots, n\}\setminus \{i, j\}}  \| \alpha_{k} - \alpha\| \geq 2\delta \right\} \right)\\
            \leq & \sum_{i, j \in \{1, \ldots, n\} } \mb{P}\left(  \sup_{\alpha \in \mc{A}} \min_{k  \in \{1, \ldots, n\}\setminus \{i, j\}}  \| \alpha_{k} - \alpha\| \geq 2\delta \right) \\
            \leq & \sum_{i, j \in \{1, \ldots, n\} } \mb{P}\left(  \max_{\ell = 1, \ldots, \mc{J}} \min_{k  \in \{1, \ldots, n\}\setminus \{i, j\}}  \| \alpha_{k} - \bar{\alpha}_{\ell}\| \geq \delta \right) \\
            = & \sum_{i, j \in \{1, \ldots, n\} } \mb{P}\left(  \bigcup_{\ell = 1, \ldots, \mc{J}} \left\{\min_{k  \in \{1, \ldots, n\}\setminus \{i, j\}}  \| \alpha_{k} - \bar{\alpha}_{\ell}\| \geq \delta\right\} \right) \\
            \leq & \sum_{i, j \in \{1, \ldots, n\} } \sum_{\ell = 1, \ldots, \mc{J}} \mb{P}\left(   \min_{k  \in \{1, \ldots, n\}\setminus \{i, j\}}  \| \alpha_{k} - \bar{\alpha}_{\ell}\| \geq \delta  \right) \\
            \leq & \sum_{i, j \in \{1, \ldots, n\} } \sum_{\ell = 1, \ldots, \mc{J}} \left(\prod_{k  \in \{1, \ldots, n\}\setminus \{i, j\}} \mb{P}\left(   \| \alpha_{k} - \bar{\alpha}_{\ell}\| \geq \delta  \right)  \right) \\
            \leqtext{(i)} & n^2 \mc{J} \left(1 - \underline{c}_1 \delta^{d_{\alpha}}\right)^{n-2} \\
            \leq &  4^{d_{\alpha}} d_{\alpha}^{d_{\alpha}/2} n^2 \delta^{-d_{\alpha}} \left(1 - \underline{c} \delta^{d_{\alpha}}\right)^{n-2}
        \end{align*}
        Therefore, there exists a sufficiently large constant $M_5 >0$ such that, when $\delta = M_5 \frac{\log (n)}{n^{1/d_{\alpha}}}$, 
        \begin{align*}
            \mb{P}\left(A_5 \geq  M_5 \frac{\log (n)}{n^{1/d_{\alpha}}} \right) \leq 1/\sqrt{n}. 
        \end{align*}
        Therefore, let $\lambda_1 = M_5$ and $\lambda_2 = M_1 + M_2 + M_3 + M_4$, we have that the following inequality wpa1: 
        \begin{align*}
            \max_{i, j \in \{1, \ldots, n\} } | \widehat{d}_{ij} - d(\alpha_i, \alpha_{j})| \leq \gamma_1 \frac{\log (n)}{n^{1/d_{\alpha}}} + \gamma_2 \sqrt{\frac{\log(nT_0)}{T_0}}.  
        \end{align*}
        This completes the proof. 
    \end{prooflmm}

\subsection{Proof for informativeness condition}\label{appendix_sub:proof_sufficient_informativeness}

    \begin{prooflmm}{lemma:sufficient_informativeness_linear_IFE}
    By backward iteration,
        \begin{align*}
            g(\alpha, \Gamma_t)
            = \alpha'v_t,
            \qquad
            v_t := \sum_{\tau=0}^{\infty}\rho^\tau\gamma_{t-\tau}.
        \end{align*}
        The infinite series converges in mean square because $|\rho|<1$ and
        $s_{\max}(\mb{E}(\gamma_t\gamma_t'))<\infty$.
        Moreover, $\{v_t\}_{t\in\mb{Z}}$ is stationary and ergodic.
        Let $\Omega := \mb{E}(v_tv_t')$.

        We first establish upper and lower bounds for $\mb{E}(v_tv_t')$.
        For any $q\in\mb{R}^{d_\alpha}$, the Cauchy--Schwarz inequality
        and stationarity imply that
        \begin{align*}
            q'\mb{E}(v_tv_t') q
            &= \mb{E}\left(
                \left(\sum_{\tau=0}^{\infty}
                \rho^\tau q'\gamma_{t-\tau}\right)^2
            \right) \leq
            \left(\sum_{\tau=0}^{\infty}|\rho|^\tau\right)
            \sum_{\tau=0}^{\infty}|\rho|^\tau
            \mb{E}\left((q'\gamma_{t-\tau})^2\right) \\
            &= (1-|\rho|)^{-2}
            q'\mb{E}(\gamma_t\gamma_t')q.
        \end{align*}
        To obtain the lower bound, note that
        $\gamma_t=v_t-\rho v_{t-1}$. Hence, again by the
        Cauchy--Schwarz inequality and stationarity,
        \begin{align*}
            q'\mb{E}(\gamma_t\gamma_t')q
            &= \mb{E}\left((q'v_t-\rho q'v_{t-1})^2\right) \\
            &= (1+\rho^2)q'\mb{E}(v_tv_t') q
            -2\rho\mb{E}\left((q'v_t)(q'v_{t-1})\right) \\
            &\leq (1+\rho^2)q'\mb{E}(v_tv_t') q
            +2|\rho|
            \sqrt{
                \mb{E}\left((q'v_t)^2\right)
                \mb{E}\left((q'v_{t-1})^2\right)
            } \\
            &= (1+|\rho|)^2q'\mb{E}(v_tv_t') q.
        \end{align*}
        Since these inequalities hold for every $q\in\mb{R}^{d_\alpha}$,
        \begin{align*}
            (1+|\rho|)^{-2}\mb{E}(\gamma_t\gamma_t')
            \preceq \mb{E}(v_tv_t')
            \preceq (1-|\rho|)^{-2}\mb{E}(\gamma_t\gamma_t').
        \end{align*}
        It follows that $\mb{E}(v_tv_t')$ is positive definite and has a finite largest eigenvalue. Then, I substitute the factor representation into the definition of the population pseudo-distance: 
        \begin{align*}
            d(\alpha_i,\alpha_j)
            = \sup_{\alpha_1,\alpha_2\in\mc{A}}
            \left|
                \mb{E}\left(
                    (\alpha_1-\alpha_2)'v_tv_t'
                    (\alpha_i-\alpha_j)
                \right)
            \right| = \sup_{\alpha_1,\alpha_2\in\mc{A}}
            \left|
                (\alpha_1-\alpha_2)'\mb{E}(v_tv_t')(\alpha_i-\alpha_j)
            \right|.
        \end{align*}
        This quantity is finite because $\mc{A}$ is compact and
        $s_{\max}(\mb{E}(v_tv_t'))<\infty$.

        Fix any $\alpha_0\in\mc{A}$. Since $\mb{E}(v_tv_t')$ is positive definite,
        Lemma~\ref{lemma:basic_Informativeness_1} implies that there exists
        a constant $c>0$ such that, for all $\alpha_i,\alpha_j\in\mc{A}$,
        \begin{align*}
            d(\alpha_i,\alpha_j) \geq \sup_{\alpha\in\mc{A}} \left| (\alpha-\alpha_0)'\mb{E}(v_tv_t')(\alpha_i-\alpha_j) \right| \geq c\|\alpha_i-\alpha_j\|.
        \end{align*}
        Thus, letting $\overline{\eta}=c^{-1}$ yields $\|\alpha_i-\alpha_j\| \leq \overline{\eta} d(\alpha_i,\alpha_j)$, which completes the proof.
    \end{prooflmm}

    \begin{prooflmm}{lemma:sufficient_informativeness_nonlinear_IFE}
        Recall 
        \begin{align*}
            Y_{it}(\infty) = \bs{1}\left(\beta Y_{it-1}(\infty) + \gamma_t'\alpha_i -u_{it}\geq 0\right), 
        \end{align*}
        and let $F(\cdot)$ denote the cumulative distribution function of $u_{it}$. 
        
        It is straightforward to verify that, since $s_{\min}(\mb{E}(\gamma_t\gamma_t'))$ is positive definite, 
        \begin{align}
            \int (g(\alpha_1, \Gamma) -g(\alpha_1, \Gamma) )^2\mr{d}\mb{P}(\Gamma) >0, \quad \forall \alpha_1, \alpha_2 \in \mc{A}, \text{ and } \alpha_1 \neq \alpha_2. 
        \end{align}
        Since $\mc{A}$ is compact and $g$ is continuous,  
        for any $\delta_1 >0$ and any $\alpha_1, \alpha_2\in \mc{A}$ such that, there exists a constant $\eta(\delta_1) >0$ such that 
        \begin{align*}
            \inf_{ \alpha_1, \alpha_2 \in \mc{A}, \|\alpha_1 - \alpha_2\| \geq  \delta_1} \int (g(\alpha_1, \Gamma) -g(\alpha_1, \Gamma) )^2\mr{d}\mb{P}(\Gamma) \geq \eta(\delta_1). 
        \end{align*} 
        It follows that 
        \begin{equation}\label{eq:sufficient_informativeness_nonlinear_IFE_large_delta}
         \begin{aligned}
            \inf_{ \alpha_1, \alpha_2 \in \mc{A}, \|\alpha_1 - \alpha_2\| \geq  \delta_1} d(\alpha_1, \alpha_2) \geq \inf_{ \alpha_1, \alpha_2 \in \mc{A}, \|\alpha_1 - \alpha_2\| \geq  \delta_1} \int (g(\alpha_1, \Gamma) -g(\alpha_1, \Gamma) )^2\mr{d}\mb{P}(\Gamma) \geq \eta(\delta_1). 
        \end{aligned}   
        \end{equation}
        
        \bigskip 
        
        For notational simplicity, define $g_{t}(\alpha_i): = g(\alpha_i, \Gamma_t) = \mb{E}\left(Y_{it}(\infty) \mid \alpha_i, \Gamma_t\right)$. 
        In addition, let $\rho_t(\alpha_i) := F\left(\beta  + \gamma_t'\alpha_i \right) - F\left( \gamma_t'\alpha_i \right)$ denote the stochastic discount factor. Then,   
        \begin{align*}
            g_t(\alpha_i) = & (1 - g_{t-1}(\alpha_i))F\left(\gamma_t'\alpha_i \right)  + g_{t-1}(\alpha_i) F\left(\beta  + \gamma_t'\alpha_i \right) = F\left(\gamma_t'\alpha_i \right) + \rho_t( \alpha_i ) g_{t-1}(\alpha_i). 
        \end{align*}
        Taking first derivative with respect to $\alpha_i$ on both sides yields
        \begin{align*}
            \nabla_{\alpha} g_{t}(\alpha_i) = F^{(1)}(\gamma'_t\alpha_i )\gamma_t + g_{t-1}(\alpha_i) \rho^{(1)}_{t} ( \alpha_i) \gamma_t  + \rho_t(\alpha_i )\nabla_{\alpha} g_{t-1}(\alpha_i) .   
        \end{align*}
        
        \paragraph{Step 2: Positive definiteness of $\mb{E}\left( \nabla_{\alpha}  g_{t}(\alpha_i)  \nabla_{\alpha} g_{t}(\alpha_i) ' \right)$} The critical step for the proof is to show that $\mb{E}\left( \nabla_{\alpha}  g_{t}(\alpha_i)  \nabla_{\alpha} g_{t}(\alpha_i) ' \right)$  is positive definite. 
        The expectation is taken over the distribution of path $\Gamma_t$. 
        First, note that for any $u\in \mb{R}^{d_{\alpha}}$ with $\|u\| = 1$, we have 
        \begin{align*}
            u'\mb{E}\left(\nabla_{\alpha} g_t(\alpha_i) \nabla_{\alpha} g_t(\alpha_i)'\right)u 
            = \mb{E} \left((A + B)^2\right),
        \end{align*}
        where 
        \begin{align*}
            A: = \left(F^{(1)}(\gamma'_t\alpha_i ) +  g_{t-1}(\alpha_i)\rho^{(1)}_{t}(\alpha_i)\right) \gamma'_t u, \quad B := \rho_t (\alpha_i )\nabla_{\alpha} g_{t-1}(\alpha_i)' u. 
        \end{align*}
        Then for any $\mu  \in (0, 1)$, since $ 2 |AB| \leq \mu A^2 + 1/\mu B^2$, \footnote{
            \begin{align*}
                \left(\sqrt{\mu}A - \frac{1}{\sqrt{\mu}}B\right)^2 = \mu A^2 + \frac{1}{\mu}B^2 - 2AB \geq 0 \quad \Rightarrow  2AB\leq \mu A^2 + \frac{1}{\mu}B^2
            \end{align*}
            \begin{align*}
                \left(\sqrt{\mu}A + \frac{1}{\sqrt{\mu}}B\right)^2 = \mu A^2 + \frac{1}{\mu}B^2 + 2AB \geq 0 \quad \Rightarrow  2AB \geq -\mu A^2 - \frac{1}{\mu}B^2
            \end{align*}
            Thus, $2 |AB| \leq \mu A^2 + 1/\mu B^2$. 
        }  
        it follows that 
        \begin{align}\label{eq:sufficient_informativeness_nonlinear_IFE_1}
            u'\mb{E}\left(\nabla_{\alpha} g_{t}(\alpha_i) g_{t}(\alpha_i) '\right) u \geq (1-\epsilon)\mb{E}A^2 - \left(\frac{1}{\epsilon} - 1 \right)\mb{E} B^2. 
        \end{align}
        Also, since $\{\gamma_t\}_{t\in \mb{Z}}$ is i.i.d. across time, we have 
        \begin{equation}\label{eq:sufficient_informativeness_nonlinear_IFE_2}
        \begin{aligned}
            \mb{E} \left(B^2\right) \eqtext{(i)} & \mb{E}(\rho^2_t (\alpha_i))u'\mb{E}\left(\nabla_{\alpha} g_{t-1}(\alpha_i) \nabla_{\alpha} g_{t-1}(\alpha_i)' \right) u \\
            \eqtext{(ii)} & \mb{E}(\rho^2_t (\alpha_i))u'\mb{E}\left(\nabla_{\alpha} g_{t}(\alpha_i) \nabla_{\alpha} g_{t}(\alpha_i)' \right) u . 
        \end{aligned}
        \end{equation}
        Here, (i) follows from the fact that $\gamma_t$ is independent of $\Gamma_{t-1}$, and (ii) follows because $\Gamma_t$ and $\Gamma_{t-1}$ are identically distributed. Combining~\eqref{eq:sufficient_informativeness_nonlinear_IFE_1} and~\eqref{eq:sufficient_informativeness_nonlinear_IFE_2} gives   
        \begin{equation}\label{eq:sufficient_informativeness_nonlinear_IFE_3}
        \begin{aligned}
            \left(1 + \frac{\epsilon}{1-\epsilon} \mb{E}(\rho_t^2(\alpha_i)) \right) u'\mb{E}\left(\nabla_{\alpha} g_{t}(\alpha_i) \nabla_{\alpha} g_{t}(\alpha_i)' \right) u \geq & (1 -\epsilon) \mb{E}A^2 \\
            \Rightarrow u'\mb{E}\left(\nabla_{\alpha} g_{t}(\alpha_i) \nabla_{\alpha} g_{t}(\alpha_i)' \right) u \geq & \frac{1 -\epsilon}{1 + \frac{\epsilon}{1-\epsilon} \mb{E}(\rho^2_t(\alpha_i)) } \mb{E}(A^2) \\
            \geq & C_1 u'\mb{E}(\gamma_t \gamma_t') u
        \end{aligned}
        \end{equation}
        Since $A$ can be rewritten as
        \begin{align*}
            A = & \left(F^{(1)}(\gamma'_t\alpha_i ) +  g_{t-1}(\alpha_i) \left(F^{(1)}\left(\beta  + \gamma_t'\alpha_i \right) - F^{(1)}\left( \gamma_t'\alpha_i \right)\right)\right) \gamma'_t u \\
            = &  \left((1 - g_{t-1}(\alpha_i) )F^{(1)}(\gamma'_t\alpha_i ) +  g_{t-1}(\alpha_i) F^{(1)}\left(\beta  + \gamma_t'\alpha_i \right) \right) \gamma'_t u, 
        \end{align*}
        where the first part is a weighted average of 
        Since (i) $\alpha_i, \gamma_t$ are uniformly bounded over $\alpha_i$ and $\gamma_t$, and (ii) $F^{(1)}$ is a positive continuous function on its support, then $F^{(1)}(\gamma'_t\alpha_i )$ and $F^{(1)}\left(\beta  + \gamma_t'\alpha_i \right)$ are bounded above and uniformly bounded away from $0$ uniformly over $\alpha_i$ and $\gamma_t$. It immediately follows that, by definition, $|\rho_t(\alpha_i)| < 1$ uniformly over $\alpha_i$ and $\gamma_t$, and we let $C_1 >0$ denote such upper bound. Last, by the representation~\eqref{eq:nonlinear_model_recursive_transformation}, we obtain that $g_{t}(\alpha_i)$  is 
        bounded above and uniformly bounded away from $0$ uniformly over $i$ and $t$. There, there exists a constant $C_2 >0$, that is independent of $n, T_0$, such that $\alpha_i$ and $\gamma_t$, 
        \begin{align*}
             \left((1 - g_{t-1}(\alpha_i) )F^{(1)}(\gamma'_t\alpha_i ) +  g_{t-1}(\alpha_i) F^{(1)}\left(\beta  + \gamma_t'\alpha_i \right) \right) \geq C_2. 
        \end{align*}
        Letting $\epsilon = 1/2$. Then, combining the lower bound with~\eqref{eq:sufficient_informativeness_nonlinear_IFE_3} yields that,  for any $\alpha_i$ and $\gamma_t$, 
        \begin{equation}\label{eq:sufficient_informativeness_nonlinear_IFE_4}
        \begin{aligned}
            u'\mb{E}\left(\nabla_{\alpha} g_{t}(\alpha_i) \nabla_{\alpha} g_{t}(\alpha_i)' \right) u \geq & \frac{1 -\epsilon}{1 + \frac{\epsilon}{1-\epsilon} C_1^2 } C_2^2 u' \mb{E}\left(\gamma_t\gamma_t'\right) u \\
            \geq & \underbrace{\frac{C_2^2}{2 + 2C_1^2}  s_{\min}\left(\mb{E}\left(\gamma_t\gamma_t'\right)\right)}_{C_3}.  
        \end{aligned}
        \end{equation}
        
        \paragraph{Step 2: Taylor expansion} Define $\rho_{g'} := \sup_{\alpha \in \mc{A}, \Gamma \in \mr{supp}(\Gamma_t)} \|\nabla_{\alpha} g(\alpha, \Gamma)\|$, and define 
        \begin{align*}
            \rho_{g''} := \sup_{\alpha \in \mc{A}, \Gamma \in \mr{supp}(\Gamma_t)} s_{\max}\left(\nabla^2_{\alpha\alpha'} g(\alpha, \Gamma)\right) 
        \end{align*}
        as the upper bound of $g$'s largest eigenvalue of the second-order derivative matrix. It is straightforward to verify that $\rho_{g'}, \rho_{g''} <\infty$, because the compactness of the support of $(\alpha_i, \gamma_t)$, the  smoothness of $F$, and~\eqref{eq:nonlinear_model_recursive_transformation}.   
        For any $\alpha_i, \alpha_j \in \mc{A}$ and for any $\delta-2 >0$,  consider the second order Taylor expansions of $g(\alpha, \Gamma)$ and $g(\alpha_j, \Gamma)$ around $\alpha_i$:  
        \begin{align*}
            d(\alpha_i, \alpha_j) \geq & \sup_{\alpha \in \mc{A}}   \left| \int (g(\alpha, \Gamma) - g(\alpha_i, \Gamma))(g(\alpha_i, \Gamma) - g(\alpha_j, \Gamma)) \mr{d}\mb{P}(\Gamma)   \right| \\
            \geq & \sup_{\alpha \in \mc{A}} \bigg\{ \left| (\alpha - \alpha_i)'\int \nabla_{\alpha} g(\alpha_i, \Gamma) \nabla_{\alpha} g(\alpha_i, \Gamma)' \mr{d}\mb{P}(\Gamma)(\alpha_i - \alpha_j)\right|  \\
            & - \frac{1}{2} \rho_{g'}\rho_{g''} \left( \|\alpha - \alpha_i\|^2 \|\alpha_i - \alpha_j\| + \|\alpha - \alpha_i\| \|\alpha_i - \alpha_j\|^2\right) - \frac{1}{4} \rho^2_{g''} \|\alpha - \alpha_i\|^2 \|\alpha_i - \alpha_j\|^2 \bigg\}  \\
            \geq & \sup_{\alpha \in \mc{A} \cap \mc{B}(\alpha_i, \delta_2) } \bigg\{ \left| (\alpha - \alpha_i)''\mb{E}\left(\nabla_{\alpha} g_{t}(\alpha_i) \nabla_{\alpha} g_{t}(\alpha_i)' \right) (\alpha_i - \alpha_j)\right|  \\
            & - \frac{1}{2} \rho_{g'}\rho_{g''} \left( \|\alpha - \alpha_i\|^2 \|\alpha_i - \alpha_j\| + \|\alpha - \alpha_i\| \|\alpha_i - \alpha_j\|^2\right) - \frac{1}{4} \rho^2_{g''} \|\alpha - \alpha_i\|^2 \|\alpha_i - \alpha_j\|^2 \bigg\}  \\
            \geqtext{(i)} & C_3 \min\left\{\frac{1}{2}, \frac{\delta_2}{2 D} \right\}\overline{\rho} \|\alpha_1 - \alpha_2 \|  - \frac{1}{2}  \rho_{g'}\rho_{g''}   \|\alpha_i - \alpha_j\|^2 \delta_2 \\
            & - \left(\frac{1}{2}  \rho_{g'}\rho_{g''} \|\alpha_i - \alpha_j\| + \frac{1}{4} \rho^2_{g''} \|\alpha_i - \alpha_j\|^2 \right)\delta_2^2. 
        \end{align*}
        Here, (i) follows from~\eqref{eq:sufficient_informativeness_nonlinear_IFE_4} and  Lemma~\ref{lemma:basic_Informativeness_2}, 
        where $D:= \sup_{\alpha_1, \alpha_2 \in \mc{A}} \|\alpha_1 - \alpha_2\| <\infty$, and $\overline{\rho} : = \sup_{\alpha \in \mc{A}}\left\{\rho \geq 0:\ B(\alpha,\rho)\subset \mc{A} \right\}$. One can choose a sufficiently small $\delta_2^* >0$ such that 
        \begin{align*}
            C_3 \min\left\{\frac{1}{2}, \frac{\delta^*_2}{2 D} \right\}\overline{\rho} \|\alpha_1 - \alpha_2 \| \geq \left( \rho_{g'}\rho_{g''} + \frac{1}{2} \rho_{g''}^2 D \right)  \|\alpha_i - \alpha_j\| \delta_2^{*2}, 
        \end{align*}
        and consequently, 
        \begin{align*}
            d(\alpha_i, \alpha_j) \geq \frac{1}{2} C_3 \min\left\{\frac{1}{2}, \frac{\delta_2^{*}}{2 D} \right\}\overline{\rho} \|\alpha_1 - \alpha_2 \| - \frac{1}{2}  \rho_{g'}\rho_{g''}   \|\alpha_i - \alpha_j\|^2 \delta_2^{*}. 
        \end{align*}
        Also, we choose a sufficiently small $\delta_1^* >0$ such that $\delta_1^{*} \leq \frac{C_3 \min\left\{\frac{1}{2}, \frac{\delta_2^{*}}{2 D} \right\}\overline{\rho} }{2 \rho_{g'}\rho_{g''}\delta_2^*}$. It is straightforward to verify that 
        \begin{equation}\label{eq:sufficient_informativeness_nonlinear_IFE_small_delta}
        \begin{aligned}
            d(\alpha_i, \alpha_j) \geq \frac{1}{4} C_3 \min\left\{\frac{1}{2}, \frac{\delta_2^{*}}{2 D} \right\}\overline{\rho} \|\alpha_1 - \alpha_2 \|, \quad \forall \alpha_1, \alpha_2 \in \mc{A}, \text{ and } \|\alpha_1 - \alpha_2\| \leq \delta_1^*. 
        \end{aligned}
        \end{equation}
        Combining~\eqref{eq:sufficient_informativeness_nonlinear_IFE_large_delta} and~\eqref{eq:sufficient_informativeness_nonlinear_IFE_small_delta}, we have that, for any $\alpha_1, \alpha_2 \in \mc{A}$, 
        \begin{align*}
            d(\alpha_i, \alpha_j) \geq \min\left\{ \frac{\eta(\delta_1^*)}{D}, \frac{1}{4} C_3 \min\left\{\frac{1}{2}, \frac{\delta_2^{*}}{2 D} \right\}\overline{\rho}  \right\}\|\alpha_1 - \alpha_2 \|. 
        \end{align*}
        This completes the proof. 
    \end{prooflmm}

    \begin{lemma}\label{lemma:basic_Informativeness_1}
        Suppose $\mc{A} \subset \mb{R}^{d_{\alpha}}$ is nonredundant, compact,  and convex. Given a symmetric positive definite matrix $\Omega \in \mb{R}^{d_{\alpha} \times d_{\alpha}}$. Then, there exists a constant $c > 0 $ such that for any $\alpha_1, \alpha_2, \alpha_3 \in \mc{A}$, 
        \begin{align*}
            \sup_{\alpha \in \mc{A}} \left| (\alpha - \alpha_3)' \Omega (\alpha_2 - \alpha_1)\right| \geq c\|\alpha_1 - \alpha_2 \| 
        \end{align*}  
    \end{lemma}
    \begin{prooflmm}{lemma:basic_Informativeness_1}
        By the nonredundancy of $\mc{A}$, there exist $\{\tilde{\alpha}_0, \tilde{\alpha}_1, \ldots, \tilde{\alpha}_d\}\subset \mc{A}$ such that $\{\tilde{\alpha}_1 - \tilde{\alpha}_0, \ldots, \tilde{\alpha}_d - \tilde{\alpha}_0\}$ are linear independent. We collect $\{\tilde{\alpha}_1 - \tilde{\alpha}_0, \ldots, \tilde{\alpha}_d - \tilde{\alpha}_0\}$ into matrix 
        \begin{align*}
            A = 
            \begin{pmatrix}
                \tilde{\alpha}'_1 - \tilde{\alpha}_0' \\
                \vdots \\
                \tilde{\alpha}'_d - \tilde{\alpha}_0'
            \end{pmatrix}, 
        \end{align*}
        and it is straightforward to verify that $A$ is invertible. In addition, 
        \begin{align*}
            & \sup_{\alpha \in \mc{A}} \left| (\alpha - \alpha_3)'   \Omega (\alpha_2 - \alpha_1)\right|\\
            = & \max\{\sup_{\alpha \in \mc{A}}\alpha'\Omega(\alpha_2 - \alpha_1) - \alpha_3'\Omega(\alpha_2 - \alpha_1), \alpha_3'\Omega(\alpha_2 - \alpha_1) - \inf_{\alpha \in \mc{A}}\alpha'\Omega(\alpha_2 - \alpha_1)\} \\
           \geqtext{(i)} & \frac{1}{2} \left(\sup_{\alpha \in \mc{A}}\alpha' \Omega (\alpha_2 - \alpha_1) - \inf_{\alpha \in \mc{A}}\alpha' \Omega (\alpha_2 - \alpha_1)\right) \\
           \geq & \frac{1}{2} \left(\max_{\alpha \in \{\tilde{\alpha}_0, \tilde{\alpha}_1, \ldots, \tilde{\alpha}_d\}}\alpha' \Omega (\alpha_2 - \alpha_1) - \min_{\alpha \in \{\tilde{\alpha}_0, \tilde{\alpha}_1, \ldots, \tilde{\alpha}_d\}}\alpha' \Omega (\alpha_2 - \alpha_1)\right) \\
           = & \frac{1}{2} \left(\max_{\alpha \in \{\tilde{\alpha}_0, \tilde{\alpha}_1, \ldots, \tilde{\alpha}_d\}}(\alpha - \tilde{\alpha}_0 )'\Omega (\alpha_2 - \alpha_1) - \min_{\alpha \in \{\tilde{\alpha}_0, \tilde{\alpha}_1, \ldots, \tilde{\alpha}_d\}}(\alpha - \tilde{\alpha}_0 )'\Omega (\alpha_2 - \alpha_1)\right) \\
           = & \frac{1}{2} \max\{0, \tilde{\alpha}_1'\Omega (\alpha_2 - \alpha_1), \ldots, \tilde{\alpha}_d'\Omega(\alpha_2 - \alpha_1) \} - \frac{1}{2} \min\{0, \tilde{\alpha}_1'\Omega(\alpha_2 - \alpha_1), \ldots, \tilde{\alpha}_d'\Omega(\alpha_2 - \alpha_1) \} \\
           \geq & \frac{1}{2} \|A\Omega (\alpha_2 - \alpha_1)\|_{\infty}. 
        \end{align*}
        Here, (i) holds because $\max\{a, b\} \geq \frac{a + b}{2}$. 
        In addition, $\frac{1}{2}\|A\Omega (\alpha_2 - \alpha_1)\|_{\infty} \geq \frac{1}{2 \sqrt{d}}\|A \Omega(\alpha_2 - \alpha_1)\| \geq \frac{\sigma_{\min}(A\Omega)}{2\sqrt{d}}\|(\alpha_2 - \alpha_1)\|$. Let $c = \frac{\sigma_{\min}(A\Omega)}{2 \sqrt{d}}$. Since $\Omega$ is positive definite and $A$ is invertible, $c >0$. This completes the proof. 
    \end{prooflmm}

    \begin{lemma}\label{lemma:basic_Informativeness_2}
        Suppose $\mc{A} \subset \mb{R}^{d_{\alpha}}$ is nonredundant, compact,  and convex. Given a symmetric positive definite matrix $\Omega \in \mb{R}^{d_{\alpha} \times d_{\alpha}}$ such that the smallest eigenvalue of $\Omega$ satisfies $s_{\min}(\Omega) \geq \underline{s} >0$. Then,  for any given $\delta>0$ and any $\alpha_1, \alpha_2\in \mc{A}$, there exists a constant $c(\delta) > 0 $ such that 
        \begin{align*}
            \sup_{\alpha \in \mc{A} \cap \mc{B}(\alpha_1, \delta)} \left| (\alpha - \alpha_1)' \Omega (\alpha_1 - \alpha_2)\right| \geq \min\left\{\frac{1}{2}, \frac{\delta}{2 D} \right\}\overline{\rho} \underline{s} \|\alpha_1 - \alpha_2 \|, 
        \end{align*} 
        where $D:= \sup_{\alpha_1, \alpha_2 \in \mc{A}} \|\alpha_1 - \alpha_2\| <\infty$, and $\overline{\rho} : = \sup_{\alpha \in \mc{A}}\left\{\rho \geq 0:\ B(\alpha,\rho)\subset \mc{A} \right\}$. 
    \end{lemma}
    \begin{prooflmm}{lemma:basic_Informativeness_2}
        By definition, $\overline{\rho}$ is the largest radius of a Euclidean ball that can be placed entirely inside $\mc{A}$. Since $\mc{A}$ is compact, convex, and $\mr{affine}(\mr{supp}) = \mb{R}^{d_{\alpha}}$, such balls exist  and $ 0 <\overline{\rho} < \infty$. Pick one of these balls and suppose its center is $\alpha_0$, then $\mc{B}(\alpha_0, \overline{\rho}) \subset \mc{A}$. Define  $\ell := \min\{1/2, \delta/2D\}$ and $\tilde{\alpha}_{1} = (1-\ell)\alpha_1 + \ell \alpha_0$. In addition, define a new ball as
        \begin{align*}
            \mc{B}(\tilde{\alpha}_1, \ell  \overline{\rho} ):=\left\{  (1-\ell)\alpha_1 + \ell \alpha \mid  \alpha_0 \in \mc{B}(\alpha_0, \overline{\rho}) \right\}. 
        \end{align*}
        It is straightforward to verify that $\mc{B}(\tilde{\alpha}_1, \ell  \overline{\rho} ) \subset \mc{B}(\alpha_1, \delta)$, and by the convexity of $\mc{A}$, $\mc{B}(\tilde{\alpha}_1, \ell  \overline{\rho} ) \subset \mc{A}$ as well. Then, 
        \begin{align*}
            & \sup_{\alpha \in \mc{A} \cap \mc{B}(\alpha_1, \delta)} \left| (\alpha - \alpha_1)'\Omega (\alpha_1 - \alpha_2)\right| \\
            \geq &  \sup_{ \alpha \in \mc{B}(\tilde{\alpha}_1, \ell  \overline{\rho} ) } \left| (\alpha - \alpha_1)' \Omega(\alpha_1 - \alpha_2)\right| \\ 
            \geq &  \max\bigg \{\left| \ell(\alpha_1 - \alpha_0)'\Omega (\alpha_2 - \alpha_1) + \frac{\ell \overline{\rho} }{\|\alpha_1 - \alpha_2\| }(\alpha - \alpha_1)'\Omega (\alpha_1 - \alpha_2) \right|, \\
            & \qquad \qquad \left| \ell (\alpha_1 - \alpha_0)'(\alpha_2 - \alpha_1) - \frac{\ell \overline{\rho} }{\|\alpha_1 - \alpha_2\| }(\alpha - \alpha_1)'\Omega (\alpha_1 - \alpha_2)  \right| \bigg \} \\
            \geq &  \ell \overline{\rho} \underline{s} \|\alpha_2 - \alpha_1\| \\
            \geqtext{(i)} & \min\left\{\frac{1}{2}, \frac{\delta}{2 D} \right\}\overline{\rho} \underline{s} \|\alpha_1 - \alpha_2 \|. 
        \end{align*}
        Here, (i) follows from the fact that $\max\{|a + b|, |a - b|\} \geq |b|$.  
        This completes the proof. 
    \end{prooflmm}

    \clearpage

    \section{Supplementary Proofs}\label{appendix:supplementary}

    \begin{lemma}[Based on Theorem~1 of \citet{merlevede2009bernstein}]\label{lemma:concentration_weak_dependent_reference}
            When (i) $(Y_t)_{t=1}^{T}$ is mean-zero and sub-Gaussian, i.e., there exist constant $\sigma >0$ such that $\max_{t=1, \ldots, T}\mb{E}(\exp(\lambda Y_t)) \leq \exp(\frac{\lambda^2\sigma^2}{2})$,  and (ii) $(Y_t)_{t=1}^{T}$ is $\alpha$-mixing, and for a constant $c >0$, $\sup_{\tau\geq 1}\alpha(\tau) \leq \exp(-c\tau)$. Then, for any $\epsilon >0$, there exists constants $C_1, C_2, C_3, C_4 >0$ such that  
            \begin{equation}\label{eq:concentration_weak_dependent_reference}
            \begin{aligned}
                \mb{P}\left(\left|\frac{1}{T}\sum_{t=1}^{T} Y_t\right| \geq \epsilon \right) \leq &  T \exp\left(-\frac{T^{2/3} \epsilon^{2/3}}{C_1}\right) + \exp\left(- \frac{T \epsilon^2}{C_2 }\right) \\
                & + \exp\left( - \frac{T \epsilon^2}{C_3 }\exp\left(\frac{T^{2/9} \epsilon^{2/9}}{C_4 \left(\log T \right)^{2/3}}\right) \right). 
            \end{aligned}
            \end{equation}
        \end{lemma}

    \begin{lemma}[Hoeffding inequality]\label{lemma:hoeffding}
        Suppose that the $\{X_i\}_{i=1}^{n}$ are conditionally independent across $i$ given $\{Z_i\}_{i=1}^{n}$. Assume that $\mb{E}(X_i \mid \{Z_i\}_{i=1}^{n} ) = 0$ for each $i$, and that, for some constant $\sigma>0$,
        \begin{align*}
            \mb{E}\left(\exp(\lambda X_i)\mid \{Z_i\}_{i=1}^{n} \right)
            \leq \exp\left(\frac{\sigma^2\lambda^2}{2}\right),  
        \end{align*}
        for all $\lambda\in\mb{R}$ and all $i$. 
        Then, for every $\epsilon>0$,
        \begin{align*}
            \mb{P}\left(
            \left|\sum_{i=1}^{n} Z_i X_i\right|\geq \epsilon \mid \{Z_i\}_{i=1}^{n}\right) \leq 2 \exp\left( -\frac{\epsilon^2}{2\sigma^2\sum_{i=1}^{n}Z_i^2} \right). 
        \end{align*}
    \end{lemma}
    \begin{lemma}[Bernstein inequality]\label{lemma:bernstein}
        Suppose that, conditional on random variables $(Z_i)_{i=1}^{n}$, $\{X_i\}_{i=1}^n$ are independent and satisfy $\mb{E}(X_i \mid (Z_i)_{i=1}^{n})=0$ for each $i=1,\ldots,n$. Define 
        \begin{align*}
            B_n : = \max_{i = 1, \ldots, n }|Z_i X_i|, \quad  V_n :=\frac{1}{n} \sum_{i= 1}^{n}\mb{E}\left(Z_i^2X_i^2 \mid  (Z_i)_{i=1}^{n} \right) 
        \end{align*}
        Then, for any $\epsilon>0$,
        \begin{equation}\label{eq:bernstein_1}
        \begin{aligned}
            \mb{P}\left(\left| \sum_{i=1}^{n} Z_iX_i \right|\geq \epsilon \mid (Z_i)_{i=1}^{n} \right)
            \leq
            2\exp\left(-\frac{\epsilon^2} {2\left(nV_n +  \frac{B_n}{3} \epsilon \right)} \right).
        \end{aligned}
        \end{equation}
        In addition, when there exist constants $ b, v \in (0, \infty)$ such that $B_n \leq b $ and $V_n \leq v$ with probability approaching to $1$, then 
        \begin{equation}\label{eq:bernstein_2}
        \begin{aligned}
            \mb{P}\left(\left| \sum_{i=1}^{n} Z_iX_i \right|\geq \epsilon \right)
            \leq
            2\exp\left(-\frac{\epsilon^2} {2\left(nv + \frac{b}{3} \epsilon \right)} \right) + r_n, 
        \end{aligned}
        \end{equation}
        with $r_n\rightarrow 0$ as $n\rightarrow \infty$. 
    \end{lemma}
    \begin{prooflmm}{lemma:bernstein}
        Conditional on $(Z_i)_{i=1}^{n}$, the variables $(Z_i X_i )_{i=1}^{n}$ are independent, have conditional mean zero, and satisfy the stated boundedness and conditional second moment restrictions. Therefore, the Bernstein inequality (\citet[Proposition~2.14]{wainwright2019high}) implies the conditional bound~\eqref{eq:bernstein_1}. 
        Define $\mc{A} := \left\{B_n \leq b , V_n \leq v \right\}$. Taking expectations with respect to $(Z_i)_{i=1}^{n}$ (on $\mc{A}$ and $\mc{A}^c$ separately) yields the stated unconditional bound~\eqref{eq:bernstein_2}. 
    \end{prooflmm}

\end{document}